\documentclass[aps,prd,reprint,nofootinbib,groupedaddress,showpacs,eqsecnum,floatfix]{revtex4-1}

\usepackage{amsmath}
\usepackage{amssymb}
\usepackage{graphicx}
\usepackage{bm}
\usepackage{braket}
\usepackage{enumitem}
\usepackage{tensor}
\usepackage{color}
\usepackage{combelow}
\usepackage{slashed}
\usepackage{color}
\usepackage{epstopdf}

\def\Et{\widetilde{E}}
\def\Ew{\overline{E}}
\def\rhoo{\overline{\rho}}
\def\hatt{{\hat{t}}}
\def\hati{{\hat{i}}}
\def\hatz{{\hat{z}}}
\def\hrho{{\hat{\rho}}}
\def\hvarphi{{\hat{\varphi}}}
\def\halpha{{\hat{\alpha}}}
\def\hbeta{{\hat{\beta}}}

\def\hsigma{{\hat{\sigma}}}
\def\htau{{\hat{\tau}}}
\def\acomm#1#2{\left\{#1,#2\right\}}
\def\comm#1#2{\left[#1,#2\right]}
\def\abs#1{\left|#1\right|}
\def\wj{{\overline{\jmath}}}

\def\tr{{\text{tr}}}
\def\psibar{\overline{\psi}}
\def\rhoo{\overline{\rho}}
\def\casx{\mathtt{x}}
\def\casI{\mathcal{I}}
\def\casIo{\overline{\mathcal{I}}}
\def\casU{\mathtt{U}}
\def\casW{\mathtt{W}}
\def\vx{{\bm{x}}}
\def\current{{\mathsf{J}}}

\begin{document}

\title{Rotating fermions inside a cylindrical boundary}

\author{Victor E. \surname{Ambru\cb{s}}}
\email{Victor.Ambrus@gmail.com}
\affiliation{Department of Physics, West University of Timi\cb{s}oara,
Bd.~Vasile P\^arvan 4, Timi\cb{s}oara, 300223, Romania}
\author{Elizabeth \surname{Winstanley}}
\email{E.Winstanley@sheffield.ac.uk}
\affiliation{Consortium for Fundamental Physics, School of Mathematics and Statistics, The University of Sheffield,
Hicks Building, Hounsfield Road, Sheffield, S3 7RH, United Kingdom}

\date{\today}

\begin{abstract}
We study a quantum fermion field inside a cylinder in Minkowski space-time.
On the surface of the cylinder, the fermion field satisfies either spectral or MIT bag boundary conditions.
We define rigidly-rotating quantum states in both cases, assuming that the radius of the cylinder is sufficiently small that the speed-of-light surface is excluded from the space-time.
With this assumption, we calculate rigidly-rotating thermal expectation values of the fermion condensate, neutrino charge current and stress-energy tensor relative to the bounded vacuum state.
These rigidly-rotating thermal expectation values are finite everywhere inside and on the surface of the cylinder and
their detailed properties depend on the choice of boundary conditions.
We also compute the Casimir divergence of the expectation values of these quantities in the bounded vacuum state relative to the unbounded Minkowski vacuum.
We find that the rate of divergence of the Casimir expectation values depends on the conditions imposed on the boundary.
\end{abstract}


\maketitle

\section{Introduction}\label{sec:intro}

The definition of quantum states is of central importance in quantum field theory (QFT) on both flat and curved space-times.
Of the possible quantum states on a given space-time, defining a (not necessarily unique) vacuum state is essential, as states containing particles can be built up from a vacuum state using particle creation operators.
Even in flat space-time, the definition of a vacuum state is nontrivial when the space-time contains boundaries or one is interested in the definition of
particles as seen by a noninertial observer.

To define a vacuum state in the canonical quantization approach to QFT, one starts with an expansion of the quantum field in terms of a basis of orthonormal field modes.
These modes are split into ``positive'' and ``negative'' frequency modes.
For a quantum scalar field, this split is not completely arbitrary; it must be the case that positive frequency modes have positive Klein-Gordon norm and negative frequency modes have negative Klein-Gordon norm.
For a quantum fermion field, all modes have positive Dirac norm and the split between ``positive'' and ``negative'' frequency modes is much less constrained.

This difference between quantum scalar and fermion fields was explored in Ref.~\cite{art:rotunb} for rigidly-rotating fields on unbounded Minkowski space-time.
For a quantum scalar field, the norm of a field mode is proportional to the Minkowski energy $E$ of that mode. As a consequence, positive frequency modes must have positive Minkowski energy and the only possible vacuum state is the (nonrotating) Minkowski vacuum \cite{art:letaw}.
For a quantum fermion field, two possible vacua have been considered in the literature: the nonrotating (Vilenkin) vacuum \cite{art:vilenkin} and the rotating
(Iyer) vacuum \cite{art:iyer}.
To construct the nonrotating vacuum, positive frequency fermion modes have positive Minkowski energy $E$ as in the scalar field case. For the rotating vacuum, positive frequency fermion modes have positive corotating energy $\Et $ (the energy of the mode as seen by an observer rigidly rotating about the $z$-axis in Minkowski space-time with angular speed $\Omega $).  In general $E\neq \Et$ for a particular field mode.  On unbounded Minkowski space-time, there exist fermion field modes with $E\Et <0$, which means that the nonrotating and rotating vacua are not equivalent \cite{art:rotunb}.

Rigidly-rotating thermal states on unbounded Minkowski space-time can be defined from the above vacuum states.
The rigidly-rotating nature of these states means that the thermal factor in the thermal Green's functions and corresponding expectation values involves the
corotating energy $\Et $.
For a quantum scalar field, rigidly-rotating thermal states are divergent everywhere in the unbounded space-time \cite{art:vilenkin, art:duffy_ottewill}.
The density of states factor in the thermal expectation values (t.e.v.s) for a bosonic field is $\left[ e^{\beta {\Et}}-1 \right]^{-1}$, where $\beta $ is the inverse temperature.
This thermal factor diverges when the corotating energy $\Et $ vanishes, even though such modes are nonzero in general \cite{art:duffy_ottewill}.
Modes with vanishing corotating energy therefore make an infinite contribution to rigidly-rotating t.e.v.s, leading to divergences.

One way to resolve this difficulty is to enclose the system in an infinitely long cylinder of radius $R$, with the axis of the cylinder along the $z$-axis and $\Omega R<c$, where $c$ is the speed of light.
For this range of values of $R$, the boundary of the cylinder is inside the speed-of-light surface (SOL) (the surface on which an observer rigidly-rotating about the $z$-axis with angular speed $\Omega $ must travel at the speed of light).
With the SOL removed from the space-time, it can be shown that $E\Et >0$ for all scalar field modes, so that the modes which lead to divergences in t.e.v.s on unbounded Minkowski space-time are absent \cite{art:nicolaevici01}.
The resulting rotating t.e.v.s for a quantum scalar field on the space-time inside the cylinder are regular everywhere inside and on the boundary of the cylinder \cite{art:duffy_ottewill}.

Rigidly-rotating thermal states for a quantum fermion field on unbounded Minkowski space-time were studied in Ref.~\cite{art:rotunb} and exhibit different behaviour from those for a quantum scalar field.
Rigidly-rotating t.e.v.s are regular inside the SOL and diverge as the SOL is approached.
If the nonrotating (Vilenkin) vacuum is used, then t.e.v.s contain spurious temperature-independent terms \cite{art:rotunb, art:vilenkin78} which are unphysical since t.e.v.s with respect to the vacuum state should vanish in the limit of zero temperature.
These temperature-independent terms vanish if the rotating (Iyer) vacuum is used instead \cite{art:rotunb}.

In this paper, we study the fermionic analogues of the rotating thermal states inside a cylinder, studied for the scalar case in Ref.~\cite{art:duffy_ottewill}.
We construct rigidly-rotating quantum states for Dirac fermions enclosed inside an infinitely long cylinder in Minkowski space-time.
The axis of the cylinder is along the axis of rotation, the $z$-axis.
On the boundary of the cylinder, the fermions satisfy either spectral boundary conditions \cite{art:spectral} or one of two versions of the MIT bag
boundary conditions, the standard
\cite{art:MIT} and chiral \cite{art:lutken84} MIT bag models.
In each case, we find that the rotating and nonrotating vacua coincide when the boundary of the cylinder lies within the SOL.
We compute rigidly-rotating t.e.v.s of the fermion condensate, neutrino charge current and stress-energy tensor for each set of boundary conditions,
comparing the results with those in Ref.~\cite{art:rotunb} for the unbounded space-time~\footnote{A free Dirac field in thermal equilibrium within a rotating cylinder is also considered in Ref.~\cite{art:becattini}.}.
We also study Casimir expectation values, namely the expectation values for the bounded vacuum state relative to the (nonrotating) vacuum state on unbounded Minkowski space-time.
Our Casimir expectation values for a fermion field are compared with those in \cite{art:duffy_ottewill} and
\cite{art:bezerra08} for a quantum scalar field and for fermions obeying MIT bag boundary conditions, respectively.

The outline of this paper is as follows. In Sec.~\ref{sec:unb}, we review the construction of mode solutions of the Dirac equation
in unbounded Minkowski space-time, the second quantization procedure and the definition of the rotating and nonrotating vacuum states.
For the remainder of the paper we consider the bounded space-time.
For the spectral and MIT bag boundary conditions, in Sec.~\ref{sec:bcs}, we study mode solutions of the Dirac equation satisfying the boundary conditions, their energy spectra and the construction of the vacuum state.
Rigidly-rotating thermal expectation values are computed in Sec.~\ref{sec:tevs}
and the Casimir effect is analysed in Sec.~\ref{sec:cas}. Finally, Sec.~\ref{sec:conc} contains some further discussion.

\section{Unbounded space-time}\label{sec:unb}

In this section, we review the construction of mode solutions and vacuum states in a rigidly rotating, unbounded, Minkowski space-time~\cite{art:rotunb}.
The Dirac equation is introduced in Sec.~\ref{sec:diraceq}, while the construction of its solutions is presented in
Sec.~\ref{sec:modes}. The section closes with a discussion of the choice of vacuum state on the unbounded space-time in Sec.~\ref{sec:quantization}.

\subsection{Dirac equation in rotating Minkowski space-time}\label{sec:diraceq}
The world line of an observer rotating with a constant angular velocity $\Omega$ about the $z$-axis can
be parametrized in cylindrical coordinates as $x^\mu = (t, \rho, \Omega t, z)$ for fixed $\rho $ and $z$. The coordinate frame with respect to
which the observer is at rest can be obtained from the usual Minkowski coordinates $x^\mu_{\rm M}$ by setting
$\varphi = \varphi_{\rm M} - \Omega t$.  The Minkowski metric then takes the form:
\begin{equation}\label{eq:ds}
 ds^2 = -(1 - \rho^2\Omega^2) dt^2 + 2\rho^2 \Omega \, dt\,d\varphi + d\rho^2 +
 \rho^2 d\varphi^2 + dz^2.
\end{equation}
Throughout this paper we use units in which $c=\hbar =k_{B}=1$.
The Killing vector $\partial_t$, defining the corotating Hamiltonian
$H = i\partial_t$, becomes null on the SOL,
which is defined as the surface where $\rho = \Omega^{-1}$.

To construct the Dirac equation, we introduce the following tetrad in the Cartesian
gauge \cite{art:cota_external_symm}:
\begin{align}
 e_{\hatt} =& \partial_t - \Omega \partial_\varphi, &\qquad
 e_{\hati} =& \partial_i,\nonumber\\
 \omega^{\hatt} =& dt, &\qquad
 \omega^{\hati} =& dx^{i} + (\Omega \times \bm{x})^i dt,
 \label{eq:tetrad}
\end{align}
with respect to which the Dirac equation for fermions of mass $\mu$ reads:
\begin{equation}\label{eq:drot}
 \left(i \gamma^\halpha D_\halpha - \mu\right) \psi(x) = 0.
\end{equation}
The gamma matrices are in the Dirac representation \cite{book:itzykson_zuber}:
\begin{equation}
 \gamma^{\hatt} =
 \begin{pmatrix}
  1 & 0 \\ 0 & -1
 \end{pmatrix}, \qquad
 \gamma^{\hati} =
 \begin{pmatrix}
  0 & \sigma_i \\ -\sigma_i & 0
 \end{pmatrix},
 \label{eq:gamma_mink}
\end{equation}
where the Pauli matrices $\sigma_i$ are given by:
\begin{equation}
 \sigma_1 =
 \begin{pmatrix}
  0 & 1 \\ 1 & 0
 \end{pmatrix}, \quad
 \sigma_2 =
 \begin{pmatrix}
  0 & -i \\ i & 0
 \end{pmatrix}, \quad
 \sigma_3 =
 \begin{pmatrix}
  1 & 0 \\ 0 & -1
 \end{pmatrix}.
 \label{eq:Pauli}
\end{equation}
The gamma matrices obey the following canonical anti-commutation rules:
\begin{equation}\label{eq:gamma_acomm}
 \acomm{\gamma^{\halpha}}{\gamma^{\hbeta}} = -2\eta^{\halpha\hbeta},
\end{equation}
where $\eta^{\halpha\hbeta}$ is the inverse of the Minkowski metric
$\eta_{\halpha\hbeta} = \text{diag}(-1,1,1,1)$.
We use the convention
that hatted indices denote
tensor components with respect to the orthonormal tetrad introduced in
Eq.~\eqref{eq:tetrad} and
are raised and lowered using the Minkowski metric $\eta _{\halpha \hbeta}$.

The covariant derivatives $D_\halpha$ in Eq.~\eqref{eq:drot} are given by:
\begin{equation}\label{eq:covd}
 i D_{\hat{t}} = H + \Omega {\mathcal {M}}_z, \qquad -i D_{\hat{j}} = P_j.
\end{equation}
In the above, $H = i\partial_t$ is the corotating Hamiltonian, $P_j = -i \partial_j$ are the momentum
operators and
\begin{equation}
\label{eq:angmomz}
 {\mathcal{M}}_z = -i\partial_\varphi + \frac{1}{2}
 \begin{pmatrix}
  \sigma_3 & 0\\
  0 & \sigma_3
 \end{pmatrix}
\end{equation}
is the $z$-component of the angular momentum operator.
\subsection{Mode solutions}\label{sec:modes}
The rotating system under consideration is just Minkowski space-time written in terms
of corotating coordinates. Therefore mode solutions of Eq.~\eqref{eq:drot} can be
obtained from any complete set of mode solutions found on Minkowski space by applying a suitable
coordinate transformation. Mode solutions of the Dirac equation on Minkowski space
with respect to cylindrical coordinates have been reported in
Refs.~\cite{art:rotunb, art:vilenkin, art:iyer, art:bezerra08, art:bezerra_saharian, art:balweerd,
art:audretsch96, art:skarzhinsky97, art:bakke, art:manning15}.

In this paper, we follow Ref.~\cite{art:rotunb} and construct
the solutions of the Dirac equation \eqref{eq:drot} as simultaneous eigenvectors of the
complete set of commuting operators $\{H, P_z, {\mathcal {M}}_z, W_0\}$, where the helicity operator
$W_0 = \bm{P} \cdot \bm{{\mathcal {M}}} / p$ is the time component of the Pauli-Lubanski vector,
with $\bm{P}$ the momentum operator and
${\bm {\mathcal {M}}}$ the angular momentum operator.
The helicity operator $W_{0}$ has the following form:
\begin{equation}
\label{eq:helicity}
 W_0 = \begin{pmatrix}
  h & 0 \\
  0 & h
 \end{pmatrix}, \qquad
 h = \frac{\bm{\sigma}\cdot \bm{P}}{2p},
\end{equation}
where $p$ is the magnitude of the momentum.

To solve the eigenvalue equations corresponding to the above operators, the
eigenspinors $U_j$ can be put in the form:
\begin{equation}
 U_j(t, \rho, \varphi, z) = \frac{1}{2\pi}
 e^{-i \Et_j t + i k_j z} u_j(\rho, \varphi),\label{eq:Udef}
\end{equation}
where
\begin{equation}
 j = (\Et_j, k_j, m_j, \lambda_j) \label{eq:jdef}
\end{equation}
collects the eigenvalues of the set of operators
$(H, P_z, {\mathcal {M}}_z, W_0)$.
In this paper, sometimes we will explicitly keep the index $j$ (\ref{eq:jdef}) on various quantities; however, at other times, we shall suppress the index $j$ to keep expressions manageable.
Further, in some expressions it will be necessary to explicitly show individual eigenvalues in $j$ (\ref{eq:jdef}).  When this
is the case, we will use the notation $U_{Ekm}^{\lambda }$ for spinors.

In (\ref{eq:Udef}) the corotating energy $\Et_j$ is linked to the Minkowski energy $E_j$
through
\begin{equation}
 \Et_j = E_j - \Omega(m_j + \tfrac{1}{2}),
 \label{eq:energy}
\end{equation}
where $E_j = \pm \sqrt{p_j^2 + \mu^2}$ can be written in terms of the modulus $p_j$ of the momentum of the mode.
The four-spinors $u_j$ introduced in Eq.~\eqref{eq:Udef} are eigenvectors
of $W_0$ and ${\mathcal {M}}_z$, corresponding to the eigenvalues $\lambda_j = \pm \frac{1}{2}$ and
$m_j + \tfrac{1}{2}$, respectively, where $m_j = 0, \pm 1, \pm 2, \dots$.

Due to the diagonal form of $W_0$ and ${\mathcal {M}}_z$, the four-spinors $u_j$ can be written as:
\begin{equation}
 u_j(\rho, \varphi) =
 \begin{pmatrix}
  \mathcal{C}^{\text{up}}_j \phi_{j}(\rho, \varphi)\\
  \mathcal{C}^{\text{down}}_{j} \phi_j(\rho, \varphi)
 \end{pmatrix},\label{eq:udef}
\end{equation}
where $\mathcal{C}^{\text{up}}_{j}$ and $\mathcal{C}^{\text{down}}_{j}$ are constants.
The angular momentum equation,
\begin{equation}
 \begin{pmatrix}
  -i \partial_\varphi + \frac{1}{2} & 0\\
  0 & -i \partial_\varphi - \frac{1}{2}
 \end{pmatrix}
 \phi_{j}(\rho, \varphi) = (m_j + \tfrac{1}{2}) \phi_{j}(\rho, \varphi),
\end{equation}
can be solved by setting:
\begin{equation}
 \phi_j(\rho, \varphi) = e^{i(m_j + \frac{1}{2})\varphi}
 \begin{pmatrix}
  e^{-\frac{i}{2}\varphi} \phi^{-}_{j}(\rho)\\
  e^{\frac{i}{2}\varphi} \phi^{+}_{j}(\rho)
 \end{pmatrix},
 \label{eq:phijdef}
\end{equation}
where $\phi^{\pm}_{j}$ are scalar functions of $\rho$.
The two-spinors $\phi_{j}$ also obey the helicity eigenvalue equation:
\begin{equation}\label{eq:hel}
 \frac{1}{2p_j}
 \begin{pmatrix}
  k_j & P_- \\
  P_+ & -k_j
 \end{pmatrix}
 \phi_j(\rho, \varphi) =
 \lambda_j \phi_j(\rho, \varphi),
\end{equation}
where $P_{\pm} = P_x \pm i P_y$ are differential operators given by:
\begin{equation}\label{eq:P_shifters}
 P_{\pm} = -i e^{\pm i\varphi} (\partial_\rho \pm i \rho^{-1} \partial_{\varphi}).
\end{equation}
The helicity eigenvalue equation \eqref{eq:hel} can be used to show that the scalar functions
$\phi^{\pm}_{j}$ satisfy Bessel-type equations:
\begin{subequations}\label{eq:bess}
\begin{align}
 [{\mathfrak {z}}_j^2 \partial^2_{{\mathfrak {z}}_j} + {\mathfrak {z}}_j \partial_{{\mathfrak {z}}_j} + {\mathfrak {z}}_j^2 - (m_j + 1)^2] \phi^{+}_{j} =& 0,\label{eq:bessp}\\
 [{\mathfrak {z}}_j^2 \partial^2_{{\mathfrak {z}}_j} + {\mathfrak {z}}_j \partial_{{\mathfrak {z}}_j} + {\mathfrak {z}}_j^2 - m_j^2] \phi^{-}_{j} =& 0,\label{eq:bessm}
\end{align}
\end{subequations}
where ${\mathfrak {z}}_j = q_j \rho$ is written in terms of the transverse momentum
\begin{equation}
q_j = \sqrt{p_j^2 - k_j^2}.
\end{equation}
The solutions of Eqs.~(\ref{eq:bess}) which are regular at the origin have the form:
\begin{align}
 \phi^{+}_{j}(\rho) =& \mathcal{N}^{+}_j J_{m + 1}(q\rho),\nonumber\\
 \phi^{-}_{j}(\rho) =& \mathcal{N}^{-}_j J_m(q\rho),
 \label{eq:phipmdef}
\end{align}
where $m$ is understood to refer to $m_j$ and $q$ to $q_{j}$.
The constants $\mathcal{N}^\pm_j$ can be determined as follows.

The operators $P_{\pm}$ \eqref{eq:P_shifters}
act  like shift operators for the angular momentum quantum number $m$, i.e.:
\begin{equation}
 P_{\pm} e^{i m \varphi} J_m(q \rho) = \pm i q e^{i (m \pm 1) \varphi} J_{m \pm 1} (q \rho).
\end{equation}
Hence, the helicity equation \eqref{eq:hel} implies that
\begin{equation}
 \mathcal{N}^+_j = \frac{iq_j}{k_j + 2\lambda_j p_j} \mathcal{N}^-_j,
\end{equation}
enabling $\phi_j$ (\ref{eq:phijdef}) to be written as:
\begin{equation}\label{eq:phi}
 \phi_j(\rho, \varphi) = \frac{1}{\sqrt{2}}
 \begin{pmatrix}
 \mathsf{p}_{\lambda }e^{im\varphi} J_m(q\rho)\\
  2i\lambda \mathsf{p}_{-\lambda} e^{i(m+1)\varphi} J_{m+1}(q\rho)
 \end{pmatrix},
\end{equation}
where
\begin{equation}\label{eq:pdef}
 \mathsf{p}_\pm \equiv \mathsf{p}_{\pm 1/2} = \sqrt{1 \pm \frac{k}{p}}.
\end{equation}
For brevity, the index $j$ was dropped from the right-hand side of Eq.~\eqref{eq:phi}.
The overall $1/\sqrt{2}$ factor in Eq.~\eqref{eq:phi} comes from the generalized orthogonality
relation \cite{art:balweerd}:
\begin{equation}\label{eq:phi_ortho}
 \sum_{m = -\infty}^\infty \phi^{\lambda\dagger}_{Ekm}(\rho, \varphi)
 \phi^{\lambda'}_{Ekm}(\rho, \varphi) = \delta_{\lambda\lambda'},
\end{equation}
where ${}^{\dagger }$ denotes the Hermitian conjugate of the two-spinor.

Returning to the four-spinors (\ref{eq:udef}), the Dirac equation (\ref{eq:drot}) can be used to constrain the constants $\mathcal{C}^{\text{up}}_{j}$
and $\mathcal{C}^{\text{down}}_{j}$:
\begin{equation}
 \begin{pmatrix}
  E - \mu & -2p\lambda\\
 2p\lambda & -E - \mu
 \end{pmatrix}
 \begin{pmatrix}
  \mathcal{C}^{\text{up}}_{j}\\
  \mathcal{C}^{\text{down}}_{j}
 \end{pmatrix} = 0.
\end{equation}
Imposing the generalized completeness relation \cite{art:balweerd}
\begin{equation}
 \sum_{m = -\infty}^\infty u^{\lambda\,\dagger}_{Ekm}(x) u^{\lambda'}_{Ekm}(x) = \delta_{\lambda\lambda'},
 \label{eq:u_norm}
\end{equation}
gives the following expression for the spinor $u_{j}$ introduced into the mode $U_j$ in Eq.~\eqref{eq:Udef}:
\begin{equation}
 u_j(\rho, \varphi) = \frac{1}{\sqrt{2}}
 \begin{pmatrix}
  \mathsf{E}_+ \phi_j\\
  \frac{2\lambda E}{\abs{E}} \mathsf{E}_- \phi_j
 \end{pmatrix}, \label{eq:u}
\end{equation}
where
\begin{equation}\label{eq:Edef}
 \mathsf{E}_\pm = \sqrt{1 \pm \frac{\mu}{E}}.
\end{equation}
The normalization of $u_j$ means that the mode $U_j$ (\ref{eq:Udef}) has unit norm with respect to the
Dirac inner product, which for the metric (\ref{eq:ds}) takes the form \cite{art:rotunb}:
\begin{equation}
\label{eq:scprod}
 \braket{\psi, \chi} = \int_{-\infty}^\infty dz \int_{0}^{2\pi} d\varphi \int_0^R d\rho\, \rho \,
 \psi^\dagger(x) \chi(x).
\end{equation}

Anti-particle modes $V_{j}$ are obtained from the particle modes (\ref{eq:Udef}) through charge conjugation, i.e.:
\begin{equation}
 V_{j}(x) = i \gamma^{\hat{2}} U_{j}^*(x)
\end{equation}
and have the following expression:
\begin{subequations}\label{eq:Vsol}
\begin{equation}
 V_{j}(t, \rho, \varphi, z) = \frac{1}{2\pi}
 e^{i\Et_{j} t - ik_{j} z} v_{j}(\rho, \varphi),\label{eq:V}
\end{equation}
where $v_j(\rho, \varphi) \equiv v_{Ekm}^\lambda(\rho, \varphi)$ is given by:
\begin{equation}
 v_{Ekm}^\lambda(\rho, \varphi) = \frac{(-1)^m}{\sqrt{2}} \frac{iE}{\abs{E}}
 \begin{pmatrix}
  \mathsf{E}_- \phi^\lambda_{E,-k,-m-1}\\
  -\frac{2\lambda E}{|E|} \mathsf{E}_+ \phi^\lambda_{E,-k,-m-1}
 \end{pmatrix}.\label{eq:v}
\end{equation}
\end{subequations}
The $V_{j}$ modes can be written in terms of the $U_{j}$ modes, as follows:
\begin{equation}\label{eq:VU_conn}
 V_j = (-1)^{m_j} \frac{iE_j}{\abs{E_j}} U_{\wj},
\end{equation}
where
\begin{equation}
 \wj = (-E_j, -k_j, -m_j-1, \lambda_j).\label{eq:wjdef}
\end{equation}

\subsection{Second quantization}\label{sec:quantization}
As discussed in Refs.~\cite{art:rotunb,art:iyer}, the vacuum state
for the Dirac field on a rigidly-rotating space-time is not uniquely defined.
This nonuniqueness arises from the freedom to choose how fermion field modes are split into ``particle'' and ``anti-particle'' modes.
This freedom is constrained for a quantum scalar field by the requirement that particle modes must have positive Klein-Gordon norm (and anti-particle modes
must have negative Klein-Gordon norm) in order for the particle creation and annihilation operators to
obey canonical commutation relations.
For a quantum fermion field, all field modes have positive norm and so this split is unconstrained, leading to freedom in how particle creation and annihilation operators are defined, and, correspondingly, freedom in the definition of the vacuum state \cite{art:rotunb}.

Two possible choices for the vacuum state on unbounded rotating Minkowski space-time are the
(nonrotating) Minkowski vacuum, considered  by Vilenkin \cite{art:vilenkin}, and
the rotating vacuum, introduced by Iyer \cite{art:iyer}.
For the nonrotating Minkowski vacuum, particle modes have positive Minkowski energy $E>0$; for the rotating vacuum particle modes have positive corotating energy $\Et >0$, with these two energies linked by (\ref{eq:energy}).

Rigidly-rotating thermal
expectation values (t.e.v.s) constructed with respect to the nonrotating Minkowski vacuum state
contain spurious temperature-independent terms, due to the inclusion of modes satisfying
$\Et < 0$ in the set of particle modes \cite{art:rotunb}.
The temperature-independent terms disappear when the rotating vacuum is considered, where
modes with $\Et > 0$ (including modes with negative $E$) are interpreted as particle modes \cite{art:rotunb}.
Rigidly-rotating t.e.v.s of the fermion condensate, neutrino charge current and stress-energy tensor are computed for both the Iyer and Vilenkin quantizations in Ref.~\cite{art:rotunb}.
It is found that, using the Iyer quantization, these t.e.v.s are regular everywhere inside the SOL, but diverge as the SOL is approached.

The difference between the Iyer and Vilenkin quantization methods rests in the interpretation of the
modes for which $E\Et < 0$, namely whether such modes are considered to be particle or anti-particle modes.
For a quantum scalar field, enclosing the system inside a boundary of radius not greater than that of the SOL
eliminates energies satisfying $E \Et < 0$ from the particle spectrum \cite{art:nicolaevici01}.
Vilenkin \cite{art:vilenkin} argues that the same holds for fermions.
In Sec.~\ref{sec:bcs}, we show that this is indeed the case for  spectral and MIT bag boundary conditions (defined in Secs.~\ref{sec:spec} and \ref{sec:MIT} respectively), for a cylindrical boundary placed inside or on the SOL.
In this case the nonrotating (Vilenkin \cite{art:vilenkin}) and rotating (Iyer \cite{art:iyer}) vacua are therefore equivalent.

Assuming that there are no modes with $E\Et <0$ in the particle spectrum,
second quantization can be performed as in unbounded nonrotating Minkowski space, by expanding the field
operator $\psi(x)$ as:
\begin{equation}\label{eq:psi}
 \psi(x) = \sum_j \theta(E_j)
 \left[U_j(x) {\mathfrak {b}}_{j} + V_{j}(x) {\mathfrak {d}}^\dagger_{j}\right],
\end{equation}
where the step function $\theta(E_j)$ ensures that the Minkowski energy $E_j$ is positive and
\begin{equation}
 \sum_j \equiv \sum_{\lambda_j = \pm \frac{1}{2}} \sum_{m_{j} = -\infty}^\infty \int_{\abs{E_j} > \mu} dE_j
 \int_{-p_j}^{p_j} dk_j,\label{eq:sumj_def}
\end{equation}
where $p_j$ is the modulus of the momentum of a particle of Minkowski energy $E_j$.
The negative $E_j$ values, excluded by the step function $\theta(E_j)$ in Eq.~\eqref{eq:psi},
are included in the domain of integration in Eq.~\eqref{eq:sumj_def} for later convenience.
The one-particle operators ${\mathfrak {b}}_{j}$ and ${\mathfrak {d}}^\dagger_{j}$ in Eq.~\eqref{eq:psi} obey canonical anti-commutation
relations:
\begin{equation}
 \acomm{{\mathfrak {b}}_j}{{\mathfrak {b}}^\dagger_{j'}} = \delta(j,j'), \qquad
 \acomm{{\mathfrak {d}}_j}{{\mathfrak {d}}^\dagger_{j'}} = \delta(j,j'),
\end{equation}
where
\begin{equation}
 \delta(j, j') = \frac{\delta(E_j - E_{j'})}{\abs{E_j}} \delta(k_j - k_{j'})
 \delta_{m_j, m_{j'}} \delta_{\lambda _j, \lambda_{j'}}.\label{eq:deltajjp_def}
\end{equation}
The vacuum state $| 0 \rangle $ is defined as that state which is annihilated by the annihilation operators ${\mathfrak {b}}_{j}$ and ${\mathfrak {d}}_{j}$:
\begin{equation}
{\mathfrak {b}}_{j}| 0 \rangle = 0 = {\mathfrak {d}}_{j} | 0 \rangle .
\end{equation}
In the next section, we shall investigate the properties of rigidly-rotating t.e.v.s for thermal states constructed from this vacuum state, for a fermion field satisfying either spectral or MIT bag boundary conditions.

%%%%%%%%%%%%%%%%%%%%%%%%%%%%%%%%%%%%%%%%%%%%%%%%%%%%%%%%%%%%%%%%%%%%%%%%%%%%%%
\section{Boundary conditions}\label{sec:bcs}
%%%%%%%%%%%%%%%%%%%%%%%%%%%%%%%%%%%%%%%%%%%%%%%%%%%%%%%%%%%%%%%%%%%%%%%%%%%%%%

Our focus in this paper is a quantum fermion field on rotating Minkowski space-time, inside a cylinder
centered on the $z$-axis (the axis of rotation) and having radius $R$.
We exclude the space-time exterior to the cylinder from our considerations.
For $R\Omega <1$ (where $\Omega $ is the angular speed about the $z$-axis), the cylinder lies completely inside the SOL, which is therefore removed from our space-time.
For $R\Omega =1$, the boundary of the cylinder is the SOL.
For $R\Omega >1$, the SOL lies within the cylinder - we do not consider this possibility.

We consider two models for the implementation of boundary conditions for a quantum fermion field on the surface of the cylinder:
the spectral \cite{art:spectral} and MIT bag \cite{art:MIT} models.
In Sec.~\ref{sec:sadj}, the self-adjointness of the Hamiltonian is used to derive a constraint on the behaviour of the fermion field on the boundary. Secs.~\ref{sec:spec} and \ref{sec:MIT}
introduce the spectral and MIT bag models, respectively.
For each model, the energy
spectrum and corresponding vacuum states are discussed, confirming that if the SOL is
not inside the boundary, the rotating and Minkowski vacua coincide.

%%%%%%%%%%%%%%%%%%%%%%%%%%%%%%%%%%%%%%%%%%%%%%%%%%%%%%%%%%%%%%%%%%%%%%%%%%%%%%
\subsection{Self-adjointness of the Hamiltonian}\label{sec:sadj}
%%%%%%%%%%%%%%%%%%%%%%%%%%%%%%%%%%%%%%%%%%%%%%%%%%%%%%%%%%%%%%%%%%%%%%%%%%%%%%
The Hamiltonian is, by definition, a self-adjoint operator with respect to the Dirac inner product:
\begin{equation}\label{eq:sadjH}
 \braket{\psi, H \chi} = \braket{H \psi, \chi},
\end{equation}
for any combination of solutions $(\psi, \chi)$ of the Dirac equation (\ref{eq:drot}).
On a general background, the Dirac inner product is given by:
\begin{equation}\label{eq:scprod_gen}
 \braket{\psi, \chi} = \int_V d^3x \sqrt{-g}\, \psibar \gamma^t(x) \chi,
\end{equation}
where $\psibar = \psi^\dagger \gamma^{\hatt}$ and
$\gamma^\mu = e^\mu_\halpha \gamma^\halpha$ are the covariant versions of the
gamma matrices introduced in Eq.~\eqref{eq:gamma_mink}, satisfying
\begin{equation}
 \acomm{\gamma^\mu}{\gamma^\nu} = -2g^{\mu\nu}.
 \label{eq:gamma_cov}
\end{equation}
For $H = i\partial_t$, Eq.~\eqref{eq:sadjH} is equivalent to:
\begin{equation}
 \partial_t \braket{\psi, \chi} = 0.\label{eq:sadj_dt}
\end{equation}
This time derivative can be obtained from the Dirac equation \eqref{eq:drot}, which reads for
a general space-time as follows:
\begin{equation}
 i \gamma^\lambda \partial_\lambda \psi + i \gamma^\lambda \Gamma_\lambda \psi = \mu \psi,
 \label{eq:diraceq}
\end{equation}
where $\Gamma_\lambda$ is the spin connection \cite{book:birrell},
defined to preserve the general covariance of the gamma matrices:
\begin{equation}
 \comm{D_\mu}{\gamma^\nu} = \partial_\mu \gamma^\nu + \Gamma\indices{^\nu_\lambda_\mu} \gamma^\lambda +
 \comm{\Gamma_\mu}{\gamma^\nu} = 0.
\end{equation}
Taking into account the following properties:
\begin{align}
 \gamma^t \partial_t \chi =& -\gamma^i\partial_i \chi - \gamma^\lambda \Gamma_\lambda \chi  - i\mu \chi,\nonumber\\
 \overline{\partial_t \psi} \gamma^t  =& -\overline{\partial_i \psi} \gamma^i +
 \psibar \Gamma_\lambda \gamma^\lambda + i\mu \psibar,\nonumber\\
 \partial_t (\sqrt{-g} \gamma^t) =& - \partial_i(\gamma^i \sqrt{-g}) -
 \sqrt{-g} \comm{\Gamma_\lambda}{\gamma^\lambda},
\end{align}
an integration by parts in Eq.~\eqref{eq:sadj_dt} shows that
\begin{equation}\label{eq:sadjHsurf}
 \partial_t \braket{\psi, \chi} = - \int_{\partial V} d \Sigma_i \sqrt{-g}\, \psibar \gamma^i \chi,
\end{equation}
where $\partial V$ is the 2-boundary of the integration 3-surface $V$. In our case, the integration domain
is the volume contained inside an infinite cylinder of radius $R$ and its boundary is the
enclosing cylinder.
Thus, the Hamiltonian is self-adjoint only if:
\begin{equation}\label{eq:sadjHcyl}
 R \int_{-\infty}^\infty dz \int_0^{2\pi} d\varphi \,
 \left.\psibar \gamma^{\hat{\rho}} \chi\right\rfloor_{\rho = R} = 0.
\end{equation}
Eq.~\eqref{eq:sadjHcyl} provides  necessary and sufficient
conditions for a set of boundary conditions to yield a consistent quantization.
In the following two sections, two types of boundary conditions satisfying (\ref{eq:sadjHcyl}) are presented.
\subsection{Spectral boundary conditions}\label{sec:spec}
To implement spectral boundary conditions, the integral over $\varphi$ in Eq.~\eqref{eq:sadjHcyl} is
performed by considering the Fourier transform of the solutions $\psi$ of the Dirac equation with
respect to the polar angle $\varphi$:
\begin{multline}
 \psi(x) = \sum_{m = -\infty}^\infty e^{i\varphi(m + \tfrac{1}{2})} \\\times
 \begin{pmatrix}
  e^{-\frac{i}{2}\varphi} \psi_{m+\tfrac{1}{2}}^1 & e^{\frac{i}{2}\varphi} \psi_{m+\tfrac{1}{2}}^2 &
  e^{-\frac{i}{2}\varphi} \psi_{m+\tfrac{1}{2}}^3 & e^{\frac{i}{2}\varphi} \psi_{m+\tfrac{1}{2}}^4
 \end{pmatrix}^T.
\end{multline}
The inner product of any two solutions $\psi$ and $\chi$ is time-invariant if:
\begin{subequations}\label{eq:spec_integrals}
\begin{multline}
 R \int_{-\infty}^\infty dz \sum_{m = -\infty}^\infty \left( \psi_{m+\tfrac{1}{2}}^{4\,*} \chi_{m+\tfrac{1}{2}}^1 +
 \psi_{m+\tfrac{1}{2}}^{3\,*} \chi_{m+\tfrac{1}{2}}^2 \right. \\ \left. + \psi_{m+\tfrac{1}{2}}^{2\,*} \chi_{m+\tfrac{1}{2}}^3 +
 \psi_{m+\tfrac{1}{2}}^{1\,*} \chi_{m+\tfrac{1}{2}}^4 \right) = 0.
\end{multline}
The inner product of the  charge conjugate $\psi_c = i\gamma^{\hat {2}} \psi^*$ of $\psi$
and an arbitrary solution $\chi$ must also be time-invariant.  This is the case if:
\begin{multline}
 R \int_{-\infty}^\infty dz \sum_{m = -\infty}^\infty \left(\psi_{-m-\tfrac{1}{2}}^{1} \chi_{m+\tfrac{1}{2}}^1 -
 \psi_{-m-\tfrac{1}{2}}^{2} \chi_{m+\tfrac{1}{2}}^2 \right. \\ \left. -\psi_{-m-\tfrac{1}{2}}^{3} \chi_{m+\tfrac{1}{2}}^3 +
 \psi_{-m-\tfrac{1}{2}}^{4} \chi_{m+\tfrac{1}{2}}^4\right) = 0.
\end{multline}
\end{subequations}
To satisfy both equations (\ref{eq:spec_integrals}), the solution employed in the spectral model is to set equal to zero either the top
and third, or the second and fourth components of $\psi$, depending on their spectral index $m$, as follows \cite{art:balweerd}:
\begin{align}
 \psi_{m+\tfrac{1}{2}}^1\rfloor_{\rho = R} = \psi_{m+\tfrac{1}{2}}^3\rfloor_{\rho = R} =& 0, \qquad  {\mbox {for }} m + \tfrac{1}{2} > 0,\nonumber\\
 \psi_{m+\tfrac{1}{2}}^2\rfloor_{\rho = R} = \psi_{m+\tfrac{1}{2}}^4\rfloor_{\rho = R} =& 0, \qquad {\mbox {for }} m + \tfrac{1}{2} < 0.
 \label{eq:spec}
\end{align}
We note that it is also possible to satisfy Eqs.~\eqref{eq:spec_integrals} by letting the
second and fourth components of $\psi$ vanish for positive $m + \tfrac{1}{2}$, with the
first and third components vanishing when $m + \tfrac{1}{2} < 0$. For brevity, we only
consider the first implementation in this paper.  We would expect the second implementation to give physically similar results for expectation values.
\subsubsection{Discretization of the transverse momentum}\label{sec:spec_disc}
Applying the prescription \eqref{eq:spec} to the mode solutions \eqref{eq:Udef} requires that the
transverse momentum $q$ must be discretized according to:
\begin{equation}
 q_{m,\ell} R =
 \begin{cases}
  \xi_{m,\ell} & m + \tfrac{1}{2} > 0,\\
  \xi_{-m-1,\ell} & m + \tfrac{1}{2} < 0,
 \end{cases}
 \label{eq:specq}
\end{equation}
where $\xi_{m,\ell}$ is the $\ell$th nonzero root of the Bessel function $J_m$.
Hence, the mode solutions of the Dirac equation which satisfy spectral boundary conditions
can be written as:
\begin{equation}
 U^{\text{sp}}_{j}(x) = \mathcal{C}^{\text{sp}}_{j} U_j(x),\label{eq:Uspec}
\end{equation}
with $j$ defined by analogy to Eq.~\eqref{eq:jdef}, now including the new index $\ell $:
\begin{equation}
 j = (E_{j}, k_j, m_j, \lambda_j, \ell_j),\label{eq:jdef_sp}
\end{equation}
where $E_j = \pm \sqrt{q_j^2 + k_j^2 + \mu^2}$ is the Minkowski energy.
The constants $\mathcal{C}^{\text{sp}}_{j}$ in Eq.~\eqref{eq:Uspec} are calculated
in Sec.~\ref{sec:spec_norm} to ensure that the modes have unit norm.

\subsubsection{Energy spectrum} \label{sec:spec_energy}
As discussed in Sec.~\ref{sec:quantization}, if modes with $E\Et <0$ are not present in the particle spectrum, then
the  rotating and nonrotating Minkowski vacua are equivalent.
To show that this is the case for  the spectral boundary conditions, we start with the following
inequality for the first zero of the Bessel function $J_{m}$ \cite{book:watson}:
\begin{equation}
 \xi_{m,1} > m + \frac{1}{2}.
\end{equation}
Hence, for $E > 0$, we have
\begin{equation}
E R \ge qR > m + \frac{1}{2}
\end{equation}
 and therefore, using (\ref{eq:energy}),
\begin{equation}
 \Et R > (1 - \Omega R) (m + \tfrac{1}{2}),\label{eq:Epos_spec}
\end{equation}
showing that $E \Et > 0$ for all values of $\mu$, $k$, $m$ and $\ell$, as long as the boundary is inside
or on the SOL ($\Omega R \le 1$).
Thus, the rotating and nonrotating Minkowski vacua are equivalent.
This will enable us, in Sec.~\ref{sec:spec_quantization}, to perform second quantization for a fermion field satisfying spectral boundary conditions along the lines discussed in Sec.~\ref{sec:quantization}.

\subsubsection{Normalization}\label{sec:spec_norm}
Before we can proceed with second quantization, the modes~\eqref{eq:Uspec} must be normalized
with respect to the Dirac inner product \eqref{eq:scprod_gen}, which in the case under
consideration takes the form (\ref{eq:scprod}).
For the case of two particle modes (\ref{eq:Uspec}), Eq.~(\ref{eq:scprod}) reads:
\begin{widetext}
\begin{multline}
 \braket{U^{\text{sp}}_j, U^{\text{sp}}_{j'}} =
 \frac{1}{4} (\mathcal{C}^{\text{sp}}_{j})^* \mathcal{C}^{\text{sp}}_{j'}
 \delta(k-k') \delta_{mm'}
 e^{i \Delta \Et t}
 \left(
 \mathsf{E}_+ \mathsf{E'}_+ + 4\lambda\lambda' \frac{EE'}{\abs{EE'}} \mathsf{E}_- \mathsf{E'}_-\right)\\
 \times
 \left[\mathsf{p}_\lambda \mathsf{p}'_{\lambda'} \int_{0}^R J_m(q\rho) J_m (q'\rho) \rho\, d\rho +
 4\lambda\lambda' \mathsf{p}_{-\lambda} \mathsf{p}'_{-\lambda'} \int_{0}^R J_{m+1}(q\rho) J_{m+1} (q'\rho) \rho\, d\rho\right],
 \label{eq:spec_norm_int}
\end{multline}
\end{widetext}
where the labels $m$ and $\ell$ are implicit on $q$ and any quantities derived from it (e.g.~$E$).
The labels $j$ and $j'$ have also been dropped.
Furthermore, the quantities $\mathsf{p}$ and $\mathsf{E}$ are defined in (\ref{eq:pdef}) and (\ref{eq:Edef}) respectively, and $\Delta \Et = \Et_{j}-\Et_{j'}$

The modes (\ref{eq:Uspec}) are normalized if the constants $\mathcal{C}^{\text{sp}}_j$ are chosen such that
the right-hand side of Eq.~(\ref{eq:spec_norm_int}) equals
\begin{equation}
 \delta(j,j') \equiv \delta(k-k') \delta_{mm'} \delta_{\ell\ell'} \delta_{\lambda\lambda'} \theta(EE'),
 \label{eq:deltajjp_def_sp}
\end{equation}
where the step function $\theta(EE')$ ensures that the Minkowski energies $E_j$ and $E'=E_{j'}$ have the same
relative sign.
Since the boundary conditions (\ref{eq:spec}) preserve the self-adjointness of the Hamiltonian,
the time-independence of the inner product requires that modes of differing energies
(i.e.~$\Delta \Et=\Et - \Et' \neq 0$) are orthogonal.
For the evaluation of the integrals of the Bessel functions in Eq.~\eqref{eq:spec_norm_int}
when $q = q'$, it is convenient to use the following results \cite{book:gradshteyn}:
\begin{widetext}
\begin{align}
 {\mathfrak {I}}^+_{m + \frac{1}{2}} =& \int_0^R d\rho\, \rho\, \frac{1}{2}[J_m^2(q\rho) + J_{m+1}^2(q\rho)]
 = \frac{R^2}{2} \left[J_{m+1}^2(qR) - \frac{2m+1}{qR} J_m(qR) J_{m+1}(qR)  + J_{m}^2(qR) \right],\nonumber\\
 {\mathfrak {I}}^-_{m + \frac{1}{2}} =& \int_0^R d\rho\, \rho\, \frac{1}{2}[J_m^2(q\rho) - J_{m+1}^2(q\rho)]
 = \frac{R}{2q} J_m (qR) J_{m+1}(qR).\label{eq:Jint}
\end{align}
\end{widetext}
The spectral boundary conditions (\ref{eq:specq}) ensure that the product $J_m(qR) J_{m+1}(qR)$ vanishes for all $m$.
For positive $m + \tfrac{1}{2}$, the normalization constants $\mathcal {C}_{j}^{\text{sp}}$ (\ref{eq:Uspec}) take the following values:
\begin{equation}\label{eq:spec_norm}
 \mathcal{C}^{\lambda,\text{sp}}_{Ekm\ell} = \mathcal{C}^{\lambda,\text{sp}}_{E,k,-m-1,\ell} =
 \frac{\sqrt{2}}{R \abs{J_{m+1}(\xi_{m,\ell})}}.
\end{equation}

Using Eq.~\eqref{eq:VU_conn}, it can be seen that anti-particle and particle modes
obeying spectral boundary conditions are linked through:
\begin{equation}\label{eq:spec_VU_conn}
 V^{\text{sp}}_{j}(x) =(-1)^m \frac{iE_j}{\abs{E_j}} U^{\text{sp}}_{\wj}(x),
\end{equation}
where
\begin{equation}
 \wj = (-E_j,-k_j,-m_j-1,\lambda_j, \ell_{j}).\label{eq:wjdef_sp}
\end{equation}
Since the modes $U^{\text{sp}}_{\wj}$ are normalized (the above calculation is valid for $E_{j}<0$, as well as
for $E_{j}>0$), so too are the anti-particle
modes (\ref{eq:spec_VU_conn}).
\subsubsection{Second quantization}\label{sec:spec_quantization}
As shown in Sec.~\ref{sec:spec_energy}, the condition $E \Et > 0$ is satisfied by all modes
obeying spectral boundary conditions if the
boundary is placed on or inside the SOL.
We do not consider the case
when $R\Omega >1$ and the boundary is outside the SOL.
Thus, the rotating and Minkowski vacua are
identical and second quantization can be performed as outlined in Sec.~\ref{sec:quantization}.
First we expand the quantum fermion field in terms of the normalized modes (\ref{eq:Uspec}, \ref{eq:spec_VU_conn}):
\begin{equation}
 \psi_{\text{sp}} = \sum_j \theta(E_j)
 \left[U^{\text{sp}}_j {\mathfrak {b}}^{\text{sp}}_{j} + V^{\text{sp}}_j {\mathfrak {d}}^{\text{sp}\, \dagger}_{j}\right],\label{eq:psi_sp}
\end{equation}
where $j$ is defined in Eq.~\eqref{eq:jdef_sp} for the spectral case and
\begin{equation}\label{eq:sumj_sp}
 \sum_j \equiv \sum_{\lambda_j = \pm 1/2} \sum_{m_j = -\infty}^\infty
 \sum_{\ell_j = 1}^\infty \int_{-\infty}^\infty dk_j
 \sum_{E_j = \pm \abs{E_j}}.
\end{equation}
The vacuum for the spectral case, $| 0 ^{\text {sp}}\rangle$, is then defined as that state annihilated by the operators ${\mathfrak {b}}^{\text{sp}}_{j}$
and ${\mathfrak {d}}^{\text{sp}}_{j}$:
\begin{equation}
{\mathfrak {b}}^{\text {sp}}_{j}| 0^{\text {sp}} \rangle = 0 = {\mathfrak {d}}^{\text{sp}}_{j} | 0^{\text{sp}} \rangle  .
\end{equation}
In Sec.~\ref{sec:spec_tevs},  we will calculate expectation values for thermal states constructed from $| 0 ^{\text {sp}}\rangle$.

\subsection{MIT bag boundary conditions}\label{sec:MIT}
First introduced in Ref.~\cite{art:MIT}, the MIT boundary conditions are defined
in a purely local manner, by ensuring that the integrand in Eq.~\eqref{eq:sadjHsurf}
vanishes at any point $x_{\text{b}}$ on the boundary $\partial V$.
This is achieved by setting
\begin{equation}\label{eq:MIT}
 i \slashed{n} \psi(x_{\text{b}}) = \varsigma\, \psi(x_{\text{b}}),
\end{equation}
where $n_\mu$ represents the normal to the boundary and $\slashed{n}=\gamma ^{\mu }n_{\mu }$. The coefficient $\varsigma$ can take
the general form \cite{art:lutken84}:
\begin{equation}
 \varsigma = \exp(-i \gamma_5 \Theta) = \cos\Theta - i\gamma_5 \sin\Theta,
\end{equation}
where $\Theta$ is referred to as the chiral angle. In this paper, only the cases
$\Theta = 0$ (MIT) \cite{art:MIT} and $\Theta = \pi$ (chiral) \cite{art:lutken84} are considered, in which case the parameter
$\varsigma$ takes the following values:
\begin{equation}
 \varsigma = \begin{cases}
  1 & \text{(MIT)},\\
  -1 & \text{(chiral)}.
 \end{cases}
 \label{eq:varsigma}
\end{equation}
\subsubsection{Discretization of the transverse momentum}\label{sec:MIT_disc}
In the present case, $n = -d\rho$, and thus the boundary conditions \eqref{eq:MIT} are:
\begin{equation}\label{eq:MITcyl}
 i \gamma^{\hat{\rho}} \psi(x_{\text{b}}) = -\varsigma \psi(x_{\text{b}}).
\end{equation}
It can be checked that if $\psi(x)$ obeys the above boundary conditions, so does its charge conjugate
$i \gamma^{\hat{2}} \psi^*(x)$.

Mode solutions that satisfy MIT boundary conditions can be constructed starting
from the complete set of modes described in Sec.~\ref{sec:modes}. The desired solutions of the Dirac equation
can be simultaneous eigenvectors of the corotating Hamiltonian $H$, $z$-component of momentum $P_z$ and $z$-component of angular momentum
${\mathcal {M}}_z$ (\ref{eq:angmomz}), since these operators commute with $i\gamma^{\hat{\rho}}$.
However, the helicity
operator $W_0$  (\ref{eq:helicity}) does not commute with $i\gamma^{\hat{\rho}}$. Hence, $\psi(x)$ must be a linear combination of
solutions corresponding to $\lambda = \pm \frac{1}{2}$:
\begin{equation}\label{eq:MIT_U}
 U^{\text{MIT}}_{Ekm\ell}(x) = b^+_{Ekm\ell} U^+_{Ek m}(x) + b^-_{Ekm\ell} U^-_{E k m}(x),
\end{equation}
where $b^\pm_{Ekm\ell}$ are constants, $E$ is the Minkowski energy and
the index $\ell$ has been introduced anticipating the quantization of the transverse momentum $q$.
For a given value of $m$, the allowed values of the transverse momentum are labeled by $\ell$ in
increasing order, such that $q_{m,\ell} < q_{m,\ell+1}$.
To avoid cumbersome notation, the indices $m,\ell$ are omitted from the corresponding
momentum $p_{m,\ell}$ or Minkowski energy $E_{m,\ell}$ where there is no risk of confusion.

Thus, Eq.~\eqref{eq:MITcyl} becomes:
\begin{multline}
 \varsigma \mathsf{E}_+ (b^+_{Ekm\ell} \phi^+_{Ekm\ell} + b^-_{Ekm\ell} \phi^-_{Ekm\ell})\\
  = -\frac{iE}{\abs{E}} \mathsf{E}_- (b^+_{Ekm\ell} \sigma^\rho \phi^+_{Ekm\ell} -
 b^-_{Ekm\ell} \sigma^\rho \phi^-_{Ekm\ell}),\label{eq:MIT_aux}
\end{multline}
where $\mathsf{E}_\pm$ is defined in Eq.~\eqref{eq:Edef} and $\phi ^{\pm }$ are given in Eq.~(\ref{eq:phi}).
Eq.~\eqref{eq:MIT_aux} can be written as a set of linear equations in $b^\pm$:
\begin{widetext}
\begin{equation}\label{eq:bpmsys}
 \begin{pmatrix}
  \varsigma \mathsf{E}_+ \mathsf{p}_+ J_m - \frac{E}{\abs{E}} \mathsf{E}_- \mathsf{p}_- J_{m+1} &
  \varsigma \mathsf{E}_+ \mathsf{p}_- J_m - \frac{E}{\abs{E}} \mathsf{E}_- \mathsf{p}_+ J_{m+1} \\
  \varsigma \mathsf{E}_- \mathsf{p}_+ J_m + \frac{E}{\abs{E}} \mathsf{E}_+ \mathsf{p}_- J_{m+1} &
  -\varsigma \mathsf{E}_- \mathsf{p}_- J_m - \frac{E}{\abs{E}} \mathsf{E}_+ \mathsf{p}_+ J_{m+1}
 \end{pmatrix}
 \begin{pmatrix}
  b^+_{Ekm\ell}\\ b^-_{Ekm\ell}
 \end{pmatrix} = 0 ,
\end{equation}
\end{widetext}
where the argument of the Bessel functions is $q_{m,\ell}R$ and $\mathsf{p}_{\pm }$ are defined in (\ref{eq:pdef}).
The system \eqref{eq:bpmsys} has nontrivial solutions if:
\begin{equation}\label{eq:MITq}
 \mathsf{j}^2_{m\ell} + \frac{2\varsigma \mu}{q_{m,\ell}} \mathsf{j}_{m\ell} - 1 = 0,
\end{equation}
where
\begin{equation}
 \mathsf{j}_{m\ell} = \frac{J_{m}(q_{m,\ell} R)}{J_{m+1}(q_{m,\ell}R)}.\label{eq:MIT_x_def}
\end{equation}
Eq.~\eqref{eq:MITq} can be solved numerically to yield an infinite number of roots.
Eq.~\eqref{eq:MITq} is invariant under $E \rightarrow -E$, hence, $q_{m,\ell}$ does not
depend on the sign of $E$.
Moreover, the relation $J_{-m}(z) = (-1)^m J_{m}(z)$ (valid for all integer values of $m$) ensures that
\begin{equation}\label{eq:MIT_q_minv}
 q_{-m-1,\ell} = q_{m,\ell}.
\end{equation}
Eq.~\eqref{eq:bpmsys} fixes $\mathsf{b} \equiv \mathsf{b}_{Ekm\ell} = b^+_{Ekm\ell} / b^-_{Ekm\ell}$ to be
\begin{equation}\label{eq:MIT_beta}
 \mathsf{b} = -\frac{\frac{\varsigma E}{\abs{E}} \mathsf{E}_+ \mathsf{p}_- \mathsf{j}_{m\ell} - \mathsf{E}_- \mathsf{p}_+}
 {\frac{\varsigma E}{\abs{E}} \mathsf{E}_+ \mathsf{p}_+ \mathsf{j}_{m\ell} - \mathsf{E}_- \mathsf{p}_-}
 = \frac{\frac{\varsigma E}{\abs{E}} \mathsf{E}_- \mathsf{p}_- \mathsf{j}_{m\ell} + \mathsf{E}_+ \mathsf{p}_+}
 {\frac{\varsigma E}{\abs{E}} \mathsf{E}_- \mathsf{p}_+ \mathsf{j}_{m\ell} + \mathsf{E}_+ \mathsf{p}_-},
\end{equation}
where we have used the definitions (\ref{eq:pdef}, \ref{eq:Edef}, \ref{eq:MIT_x_def}).
The result (\ref{eq:MIT_beta}) is invariant under $(E,k,m) \rightarrow (-E,-k,-m-1)$.

For consistency of notation, we now write the modes (\ref{eq:MIT_U}) in a form analogous to that for the modes (\ref{eq:Uspec}) satisfying the spectral boundary conditions:
\begin{equation}
U^{\text {MIT}}_{j} = {\mathcal {C}}^{\text {MIT}}_{j} \left[ {\mathsf{b}} U_{Ekm}^{+} + U_{Ekm}^{-} \right] ,
\label{eq:UMIT}
\end{equation}
where $\mathsf{b}$ is given by (\ref{eq:MIT_beta}),
\begin{equation}
{\mathcal {C}}^{\text {MIT}}_{j} = b_{Ekm\ell }^{-}
\end{equation}
and the index $j$ on the modes is
\begin{equation}
 j = (E_{j}, k_j, m_j, \ell_j) .
 \label{eq:jdef_MIT}
\end{equation}
Note that the index $j$ (\ref{eq:jdef_MIT}) does not contain any explicit dependence on the helicity $\lambda $. This is because the MIT modes (\ref{eq:UMIT}) are a linear combination of positive and negative helicity spinors.
The normalization constants ${\mathcal {C}}_{j}^{\text {MIT}}$ will be found in Sec.~\ref{sec:MIT_norm} below.
\subsubsection{Energy spectrum}\label{sec:MIT_energy}
We now examine whether modes with $E\Et <0$ are excluded from the particle spectrum when we use MIT bag boundary conditions.

We begin with massless particles, $\mu =0$.
In this case, the equation satisfied by $\mathsf{j}_{m\ell}$ (\ref{eq:MITq}) does not depend on $\varsigma $.
Therefore, the energy spectrum is also independent of $\varsigma $.
The solutions of Eq.~\eqref{eq:MITq} when $\mu =0$ are simply $\mathsf{j}_{m\ell} = \pm 1$,
i.e.~the points where the graph of $J_m$ intersects either $J_{m+1}$ or $-J_{m+1}$.
According to Theorem 3.1 of Ref.~\cite{art:beneventano}, the values of $q_{m,\ell}R$ such that
$\mathsf{j}_{m\ell} = \pm 1$ (or, equivalently, $J_m(q_{m\ell} R) = \pm J_{m+1}(q_{m\ell}R)$ using (\ref{eq:MIT_x_def})) satisfy:
\begin{equation}
 \xi_{m,\ell}' < q_{m,2\ell-1} R < \xi_{m,\ell} < q_{m,2\ell} R < \xi'_{m,\ell+1},
\end{equation}
where $\xi_{m,\ell}$ and $\xi'_{m,\ell}$ are the $\ell $th zeroes of $J_m(z)$ and $J'_m(z)$, respectively.
The roots $q_{m,\ell} R$ are also staggered such that
\begin{equation}
 J_m(q_{m,\ell}R) = (-1)^{\ell + 1} J_{m+1}(q_{m,\ell} R).
\end{equation}
Using the following property \cite{book:watson}: % page 486
\begin{equation}
 \xi_{m, 1}' > \sqrt{m(m + 2)},
\end{equation}
the following lower bound can be established for the energy of the modes obeying MIT bag boundary conditions:
\begin{equation}
 \abs{E_{m,\ell}} R \ge q_{m,\ell} R > m + \tfrac{1}{2}.\label{eq:MIT_mu0_lowerb}
\end{equation}
The argument of Sec.~\ref{sec:spec_energy} then shows that, if the boundary is inside or on the SOL,
\begin{equation}
 E\Et > 0,
\end{equation}
where $\Et $ is given by (\ref{eq:energy}).

\begin{figure}
\begin{tabular}{c}
\includegraphics[width=0.95\columnwidth]{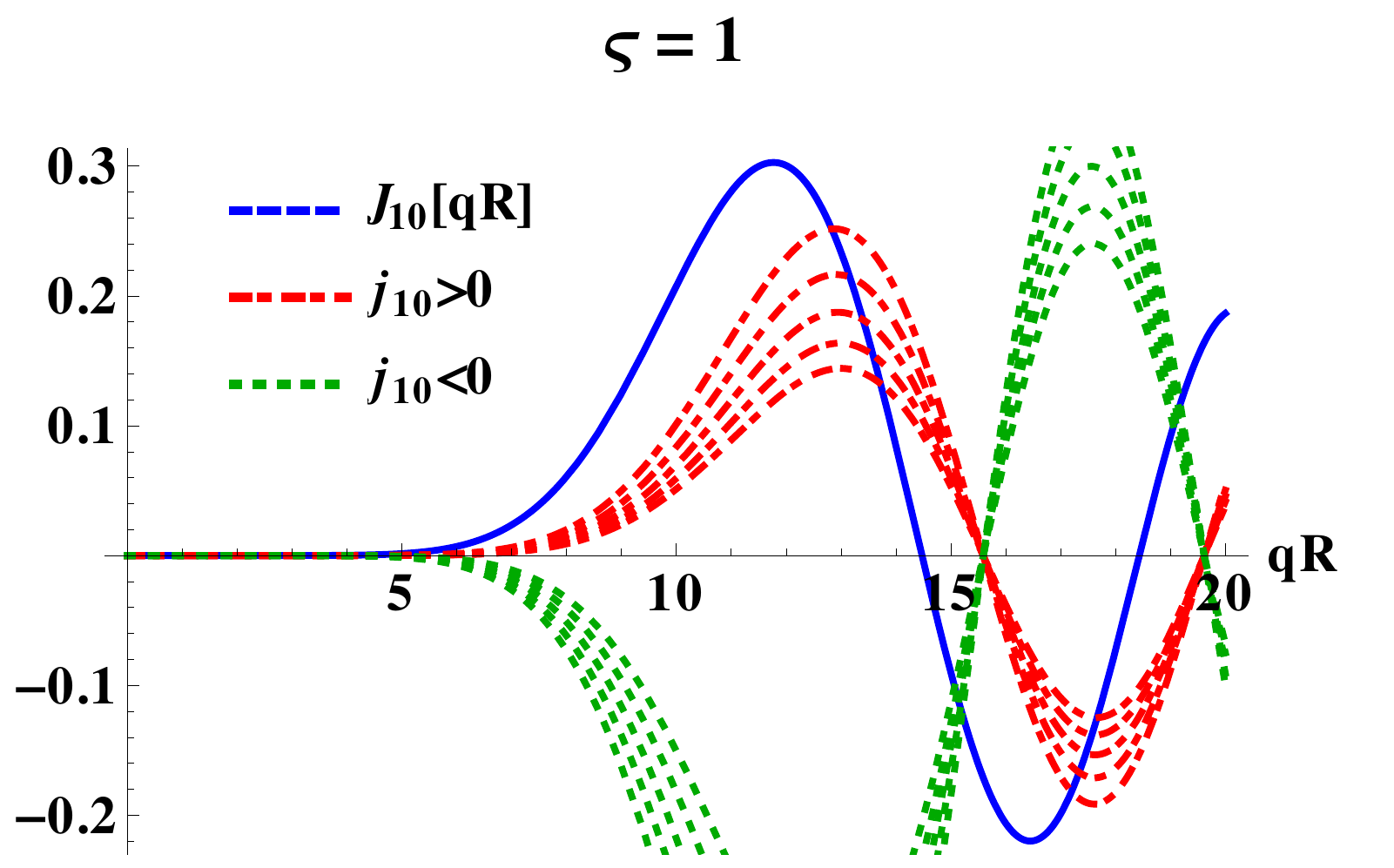}\\
(a)\\
\includegraphics[width=0.95\columnwidth]{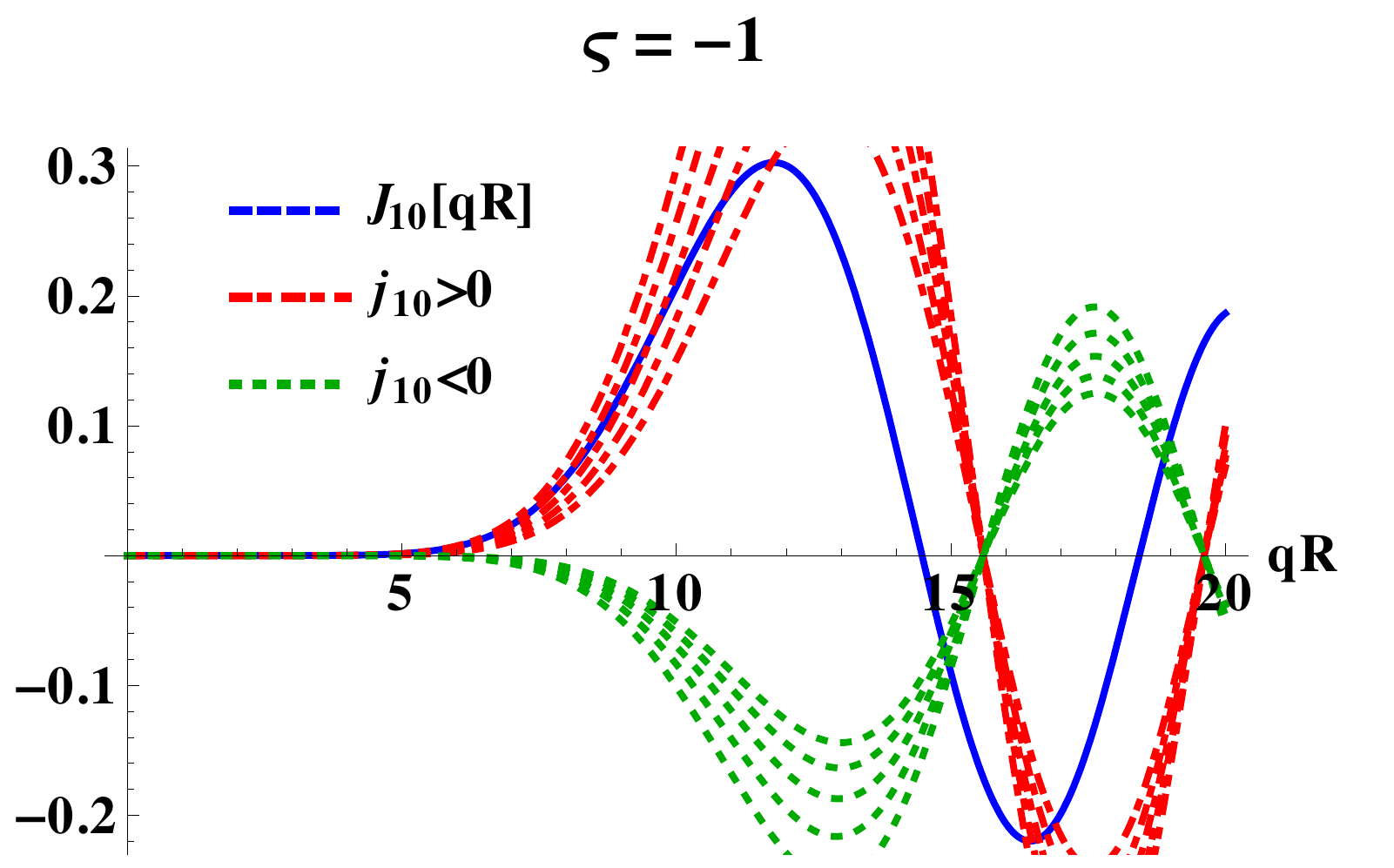}\\
(b)
\end{tabular}
\caption{Graphs for finding the first value of the transverse momentum $q_{m,1}$ allowed by the MIT bag boundary
conditions for $m = 10$. The roots of Eq.~(\ref{eq:MITq}) are located at the intersection between the solid line (representing
$J_m(qR)$) and the dashed lines (representing $J_{m}(qR)$ multiplied by the right-hand side of
Eq.~\eqref{eq:MITx}). The dashed lines correspond to masses $\mu R = 0, 2, 4, 6, 8$ and $10$, while
$\varsigma = 1$ in (a) and $\varsigma = -1$ in (b). The two sets of dashed lines correspond to
the sign of $\mathsf{j}_{m\ell}$, i.e.~the dash-dot lines (red curves, positive for small $qR$) correspond to $\mathsf{j}_{m\ell} > 0$
while the dashed lines (green curves, negative for small $qR$) represent the case $\mathsf{j}_{m\ell} < 0$.}
\label{fig:roots_MIT}
\end{figure}

When the mass $\mu $ is nonzero,  solving (\ref{eq:MITq}) enables us to write $\mathsf{j}_{m\ell }$ in terms of the (as yet unknown)
transverse momenta $q_{m,\ell }$ as follows:
\begin{equation}\label{eq:MITx}
 \mathsf{j}_{m\ell} = -\frac{\varsigma \mu}{q_{m,\ell}} \pm \sqrt{1 + \frac{\mu^2}{q^2_{m,\ell}}}.
\end{equation}
When $\varsigma = 1$ (the original MIT case), it can be seen that
\begin{equation}
\label{eq:qm1inequality}
 0 < -\frac{\varsigma \mu}{q_{m,\ell}} + \sqrt{1 + \frac{\mu^2}{q^2_{m,\ell}}} < 1.
\end{equation}
Now consider the lowest value of the transverse momentum, $q_{m,1}$.
We have $J_m(q_{m,1} R)=\mathsf{j}_{m1} J_{m+1}(q_{m,1}R)$,
in other words, when $q=q_{m,1}$ the graphs of the functions $J_{m}(qR)$ and $\mathsf{j}_{m1} J_{m+1}(qR)$ intersect.
The inequality (\ref{eq:qm1inequality}) tells that
$\mathsf{j}_{m1}<1$, so $J_m(q_{m,1} R) < J_{m+1}(q_{m,1}R)$.
Therefore the value of $q_{m,1}R$ is in the interval
where $J_m$ decreases towards its first zero,
after the first zero of $J_m'$. Fig.~\ref{fig:roots_MIT}(a) illustrates this behaviour.
Hence, in this case,
we can use the same argument as that given above in the massless case to show that
 the lowest allowed positive energy obeys $\Et R > (1 - \Omega R) (m + \frac{1}{2})$.
Therefore when $\varsigma = 1$ we again have $E \Et > 0$ for all $R \le \Omega^{-1}$.

In the chiral case ($\varsigma = -1$), from (\ref{eq:MITx}), $\mathsf{j}_{m\ell}$ increases as the mass increases and $q_{m,1} R$ approaches the origin, as
illustrated in Fig.~\ref{fig:roots_MIT}(b).
Rearranging Eq.~\eqref{eq:MITx} as:
\begin{equation}\label{eq:MITx-lin}
 q \frac{J_m(qR)}{J_{m+1}(qR)} = \mu + E(\mu),
\end{equation}
where $E(\mu) = \sqrt{\mu^2 + q^2}$ is the smallest positive Minkowski energy for a particle of
mass $\mu$ and transverse momentum $q$ (i.e.~corresponding to $k = 0$),
it can be seen that $qR = 0$ is a solution of (\ref{eq:MITq}) when $\mu R = m + 1$, by using:
\begin{equation}
 \lim_{z \rightarrow 0} z \frac{J_m(z)}{J_{m+1}(z)} = 2(m+1).
\end{equation}
If the mass $\mu $ increases further, the first root no longer corresponds to $\mathsf{j}_{m\ell} > 0$ (i.e.~the root
satisfying $q_{m,\ell} R < \xi_{m,1}$ disappears).
In this case, with $\mu R>1+m$, we have $E R > m + \frac{1}{2}$ just from the mass contribution to $E(\mu )$. Knowing that, by virtue of
Eq.~\eqref{eq:MIT_mu0_lowerb}, the same condition is satisfied  when $\mu = 0$, it remains
to investigate the behaviour of the
smallest allowed energy $E_{m,1}(\mu)$ for $q_{m,1}$ between $\mu  R= 0$ and $\mu R= m +1$.

To this end, let us consider the derivative of $E_{m,1}(\mu )$ with respect to $\mu $:
\begin{equation}\label{eq:MIT_Emuprime}
 E'_{m,1}(\mu) = \frac{1}{E_{m,1}(\mu)}[\mu + q_{m,1}(\mu) q'_{m,1}(\mu)],
\end{equation}
where the prime denotes differentiation with respect to the argument $\mu$. Since $q_{m,1}(\mu)$ decreases
as the mass increases, $q'_{m,1}(\mu) < 0$ for this range of $\mu R$ and $E'_{m,1}(\mu = 0) < 0$. The energy reaches a minimum when
\begin{equation}
 q_{m,1}(\mu_0) q'_{m,1}(\mu_0) = -\mu_0.
\end{equation}
A second expression for $E_{m,1}'(\mu)$ can be obtained by taking the derivative of
Eq.~\eqref{eq:MITx-lin} with respect to $\mu$:
\begin{equation}
 q' \frac{J_m(qR)}{J_{m+1}(qR)} \left[1 +
 R\frac{J_m'(qR)}{J_{m}(qR)} - R \frac{J_{m+1}'(qR)}{J_{m+1}(qR)}\right] = 1 + E'.
 \label{eq:MITx-lin-deriv}
\end{equation}
Using Eq.~\eqref{eq:MIT_Emuprime} to eliminate $q'_{m,1}$ in favour of
$E_{m,1}$, together with the following properties of the Bessel functions:
\begin{align}
 J_m'(z) =& -J_{m+1}(z) + \frac{m}{z} J_m(z),\nonumber\\
 J_{m+1}'(z) =& J_m(z) - \frac{m + 1}{z} J_{m+1}(z),\label{eq:BesselJprime_rec}
\end{align}
Eq.~\eqref{eq:MITx-lin-deriv} can be solved to yield:
\begin{equation}\label{eq:MIT_Eprime}
 E'(\mu) = \frac{\mu(2m+1) - 2\mu E R + E}{E(2m+1) - 2E^2 R + \mu}.
\end{equation}
Since $E'(\mu) < 0$ at $\mu = 0$, either $E_{m,1}$ reaches its minimum when $\mu R = m+1$ (in which case $q_{m,1} = 0$
and  $E_{m,1} = \mu = R^{-1}(m + 1)$),
or there must be at least one value $\mu = \mu_0$ between $0$ and $R^{-1}(m + 1)$ where $E'(\mu_0) = 0$.
At such a point, Eq.~\eqref{eq:MIT_Eprime} predicts that the value of the energy would be:
\begin{equation}\label{eq:MIT_mu_emin}
 E(\mu_0)R = \frac{2\mu_0 R}{2\mu_0 R - 1} (m + \tfrac{1}{2}).
\end{equation}
Since $E$ was assumed to be positive, Eq.~\eqref{eq:MIT_mu_emin} implies that $E$ cannot be minimized
with respect to the mass for $\mu _{0}R \le \frac{1}{2}$.
If a stationary point occurs for any $\mu _{0}R > \frac{1}{2}$, the corresponding
value of the energy will be greater than $R^{-1} (m + \frac{1}{2})$.
Since the energy is above $R^{-1}(m + \frac{1}{2})$ at the endpoints $\mu = 0$ and $\mu = m+1$ (where the corresponding
value of $q$ would be zero) and since
at its stationary points we also have $E>R^{-1}(m+\frac {1}{2})$, we can conclude that the energy will always satisfy:
\begin{equation}
E _{m,\ell }R > m + \frac {1}{2},
\label{eq:MITenergybound}
\end{equation}
and therefore, using (\ref{eq:energy}),
\begin{equation}
 \Et_{m,\ell} R > (1 - \Omega R) (m + \tfrac{1}{2}).
\end{equation}

\begin{figure}
\begin{tabular}{c}
\includegraphics[width=0.95\columnwidth]{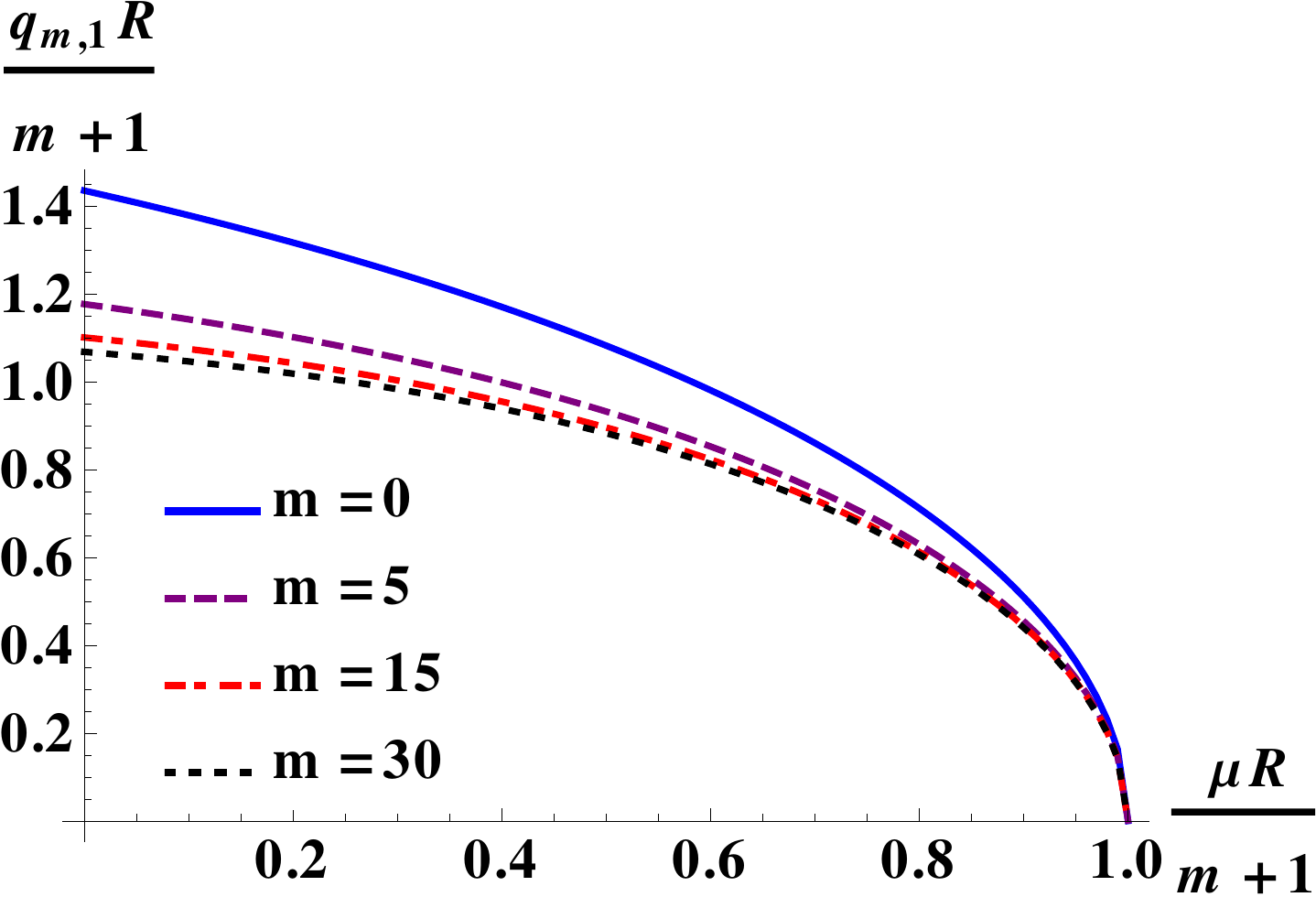}\\
(a)\\
\includegraphics[width=0.95\columnwidth]{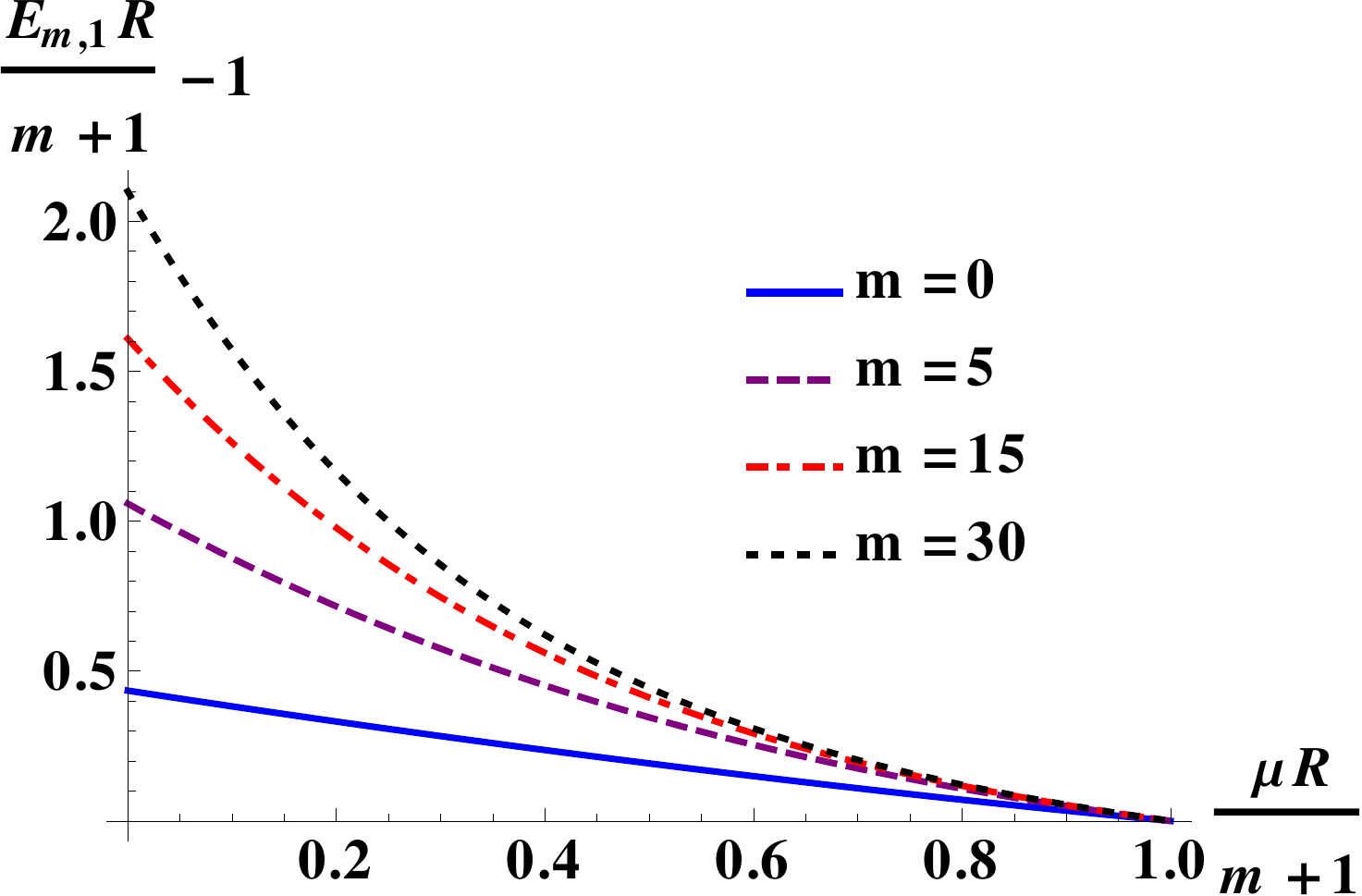}\\
(b)
\end{tabular}
\caption{The dependence of the smallest allowed transverse momentum (a) and energy (b)
in the MIT bag model corresponding to $\varsigma = -1$ for $\mu R$ between $0$ and $m + 1$ and
$m = 0, 5, 15, 30$. The $x$ axis represents the ratio $\mu R / (m + 1)$, normalizing the mass such
that for any value of $m$, the range of the $x$ axis is from $0$ to $1$. The transverse momentum
$q_{m,1}$ and energy $E_{m,1}$ are divided by $R^{-1}(m + 1)$. Plot (b) shows the difference
$\frac{E_{m,1} R}{m+1} -1$ in terms of $\mu R / (m + 1)$.
The energy $E_{m,1}$ is monotonically decreasing and has no stationary points for this range of values of $\mu R$.}
\label{fig:energy_MIT}
\end{figure}

Our numerical experiments confirm Eq.~\eqref{eq:MITenergybound}. Furthermore, the energy seems to decrease monotonically
towards its minimum value of $(m+1)/R$ as $\mu R$ increases from $0$ to $m + 1$, as shown in Fig.~\ref{fig:energy_MIT}.

Hence, the MIT bag boundary conditions with $\varsigma = \pm 1$ restrict the energy spectrum such that $E \Et > 0$ for all values of
$\mu$, $k$, $m$ and $\ell$, as long as the boundary of the cylinder is inside or on the SOL.
\subsubsection{Normalization}\label{sec:MIT_norm}
We now turn to the normalization of the MIT modes (\ref{eq:UMIT}). We require these modes to have
unit norm with respect to the Dirac inner product \eqref{eq:scprod}:
\begin{equation}\label{eq:MIT-norm}
 \braket{U^{\text{MIT}}_{j}, U^{\text{MIT}}_{j'}} = \delta(j, j'),
\end{equation}
where $\delta(j,j')$ is defined in analogy to Eq.~\eqref{eq:deltajjp_def_sp}:
\begin{equation}
\delta(j,j') = \delta(k_j - k_{j'}) \delta_{m_j, m_{j'}} \delta_{\ell_j, \ell_{j'}}
 \theta(E_j E_{j'}), \label{eq:deltajjp_def_MIT}
\end{equation}
where, as in the spectral case, $\theta(E_j E_{j'})$ ensures that $E_j$ and $E_{j'}$ have the same sign.
There is no helicity dependence in (\ref{eq:deltajjp_def_MIT}) because the MIT modes (\ref{eq:UMIT}) are linear combinations of positive and negative
helicity spinors.

The time invariance of the Dirac inner product \eqref{eq:sadjH},
guaranteed to hold in the MIT bag model by Eq.~\eqref{eq:MIT}, ensures that the result of the inner product
of modes with different corotating energies (i.e.~nonzero $\Delta \Et = \Et_j - \Et_{j'}$) vanishes.
Thus, the following result is obtained:
\begin{multline}\label{eq:MIT_norm_int}
 \braket{U^{\text{MIT}}_{j}, U^{\text{MIT}}_{j'}} =
 \frac{1}{4} \delta(k - k') \delta_{mm'} \delta_{\ell\ell'} \theta(EE')\\
\times \abs{{\mathcal {C}}_{Ekm\ell }^{\text {MIT}}}^2
 \left[ (\mathcal{S}^+_+ + \mathcal{S}^-_+) {\mathfrak {I}}^+_{m + \frac{1}{2}} +
 (\mathcal{S}^+_- + \mathcal{S}^-_-) {\mathfrak {I}}^-_{m + \frac{1}{2}}\right],
\end{multline}
where the integrals ${\mathfrak {I}}^\pm_{m + \frac{1}{2}}$ were introduced in Eq.~\eqref{eq:Jint} and their coefficients are given by:
\begin{subequations}\label{eq:Spm}
\begin{align}
 \mathcal{S}^+_\pm =& \mathsf{E}_+^2(\mathsf{b}_{Ekm\ell} \mathsf{p}_+ + \mathsf{p}_-)^2 \pm
 \mathsf{E}_-^2(\mathsf{b}_{Ekm\ell} \mathsf{p}_- + \mathsf{p}_+)^2,\label{eq:Sp}\\
 \mathcal{S}^-_\pm =& \mathsf{E}_-^2(\mathsf{b}_{Ekm\ell} \mathsf{p}_+ - \mathsf{p}_-)^2 \pm
 \mathsf{E}_+^2(\mathsf{b}_{Ekm\ell} \mathsf{p}_- - \mathsf{p}_+)^2,\label{eq:Sm}
\end{align}
\end{subequations}
where ${\mathsf{p}}_{\pm}$ are defined in (\ref{eq:pdef}), $\mathsf{E}_{\pm }$ are defined in (\ref{eq:Edef}) and $\mathsf{b}$ is given in (\ref{eq:MIT_beta}).
The combinations of $\mathcal{S}^+_\pm$ and $\mathcal{S}^-_\pm$ occuring in Eq.~\eqref{eq:MIT_norm_int}
can be evaluated using the following identities:
\begin{subequations}
\begin{align}
 \mathcal{S}^+_\pm =& \frac{4k^2}{E^2}
 \frac{1 \pm \mathsf{j}_{m\ell}^2}{(\frac{\varsigma E}{\abs{E}} \mathsf{E}_+ \mathsf{p}_+ \mathsf{j}_{m\ell} - \mathsf{E}_- \mathsf{p}_-)^2},\\
 \mathcal{S}^-_\pm =& \frac{4k^2}{E^2}
 \frac{1 \pm \mathsf{j}_{m\ell}^2}{(\frac{\varsigma E}{\abs{E}} \mathsf{E}_- \mathsf{p}_+ \mathsf{j}_{m\ell} + \mathsf{E}_+ \mathsf{p}_-)^2},
\end{align}
where $\mathsf{j}_{m\ell}$ is given by (\ref{eq:MIT_x_def}).
Then we have
\begin{equation}
 \mathcal{S}^+_\pm + \mathcal{S}^-_\pm = \frac{8 (1 \pm \mathsf{j}_{m\ell}^2)}{\mathsf{p}_+^2 \mathsf{j}_{m\ell}^2 + \mathsf{p}_-^2}.
\end{equation}
\end{subequations}
Hence, the modes \eqref{eq:UMIT}
are normalized according to Eq.~\eqref{eq:MIT-norm} if
\begin{multline}\label{eq:b-}
 {\mathcal {C}}_{j}^{\text{MIT}} = \frac{1}{R \abs{J_{m+1}(q_{m,\ell}R)}} \\
 \times \sqrt{\frac{\mathsf{p}_-^2 + \mathsf{p}_+^2 \mathsf{j}_{m\ell}^2}
 {(\mathsf{j}_{m\ell}^2 + 1)(\mathsf{j}_{m\ell}^2 + 1 - \frac{2m + 1}{q_{m,\ell}R} \mathsf{j}_{m\ell}) - (\mathsf{j}_{m\ell}^2 - 1)\frac{\mathsf{j}_{m\ell}}{q_{m,\ell}R}}}.
\end{multline}
In the massless limit, ${\mathcal {C}}_{j}^{\text{MIT}}$ (\ref{eq:b-}) simplifies to:
\begin{equation}\label{eq:MIT_b_mu0}
 {\mathcal {C}}_{j}^{\text {MIT}}\rfloor_{\mu = 0} = \frac{1}{R\sqrt{2} \abs{J_{m+1}(q_{m,\ell}R)}}
 \left[1 - \frac{\mathsf{j}_{m\ell}(m + \frac{1}{2})}{q_{m,\ell} R} \right]^{-\frac{1}{2}}.
\end{equation}

The normalization constant $\mathcal {C}_{j}^{\text {MIT}}$ (\ref{eq:b-}) is invariant under $(E,k,m) \rightarrow (-E,-k,-m-1)$.
The quantity $\mathsf{b}_{j}$, defined
in Eq.~\eqref{eq:MIT_beta}, is also invariant
under the same transformation.
Therefore, using the property \eqref{eq:VU_conn}, the relationship between particle and anti-particle spinors satisfying the MIT bag boundary conditions is:
\begin{equation}\label{eq:MIT_VU_conn}
 V^{\text{MIT}}_{j} = (-1)^{m_j} \frac{iE_j}{\abs{E_j}} U^{\text{MIT}}_{\wj},
\end{equation}
where
\begin{equation}
 \wj = (-E_{j},-k_j, -m_j - 1, \ell_j).\label{eq:wjdef_MIT}
\end{equation}
Since the particle modes $U^{\text{MIT}}_{\wj}$ are normalized, so too are the anti-particle modes (\ref{eq:MIT_VU_conn}).

\subsubsection{Second quantization}\label{sec:MIT_quantization}
For the remainder of this paper, we shall assume that $R\Omega \le 1$ and the boundary is located inside or on the SOL.
In this case, we have shown in Sec.~\ref{sec:MIT_energy} that $E\Et >0$ for all fermion field modes satisfying MIT bag boundary conditions.
As discussed in Sec.~\ref{sec:quantization}, this means that the rotating and nonrotating Minkowski vacua are the same and second quantization of the
field is straightforward.
The quantum fermion field is expanded in terms of the normalized modes (\ref{eq:UMIT}, \ref{eq:MIT_VU_conn}):
\begin{equation}
 \psi = \sum_j \theta(E_j) \left[U^{\text{MIT}}_j {\mathfrak {b}}^{\text{MIT}}_{j} +
 V^{\text{MIT}}_{j} {\mathfrak {d}}^{{\text{MIT}}\,\dagger}_{j}\right],\label{eq:psi_MIT_V}
\end{equation}
where $j$ is defined in Eq.~\eqref{eq:jdef_MIT} and the sum over $j$ is
defined as:
\begin{equation}\label{eq:sumj_MIT}
 \sum_j \equiv \sum_{m_j = -\infty}^\infty \sum_{\ell_j = 1}^\infty \int_{-\infty}^\infty dk_j
 \sum_{E_j = \pm \abs{E_j}}.
\end{equation}
There is no sum over helicity because the modes (\ref{eq:UMIT}) are linear combinations of positive and negative helicity spinors.

The vacuum for the MIT case, $| 0 ^{\text {MIT}}\rangle$, is then defined as that state
which is annihilated by the operators ${\mathfrak {b}}^{\text{MIT}}_{j}$
and ${\mathfrak {d}}^{\text{MIT}}_{j}$:
\begin{equation}
{\mathfrak {b}}^{\text {MIT}}_{j}| 0^{\text {MIT}} \rangle = 0 = {\mathfrak {d}}^{\text{MIT}}_{j} | 0^{\text{MIT}} \rangle  .
\end{equation}
In Sec.~\ref{sec:MIT_tevs},  we will calculate expectation values for thermal states constructed from $| 0 ^{\text {MIT}}\rangle$.

\subsection{Summary}\label{sec:bcs_summary}
In this section,
we have considered a quantum fermion field on rotating Minkowski space-time inside a cylinder of radius $R$ with the axis of the cylinder along the $z$-axis.
We have examined two boundary conditions for the fermion field on the surface of the cylinder: spectral \cite{art:spectral} and MIT bag \cite{art:MIT,art:lutken84}.
In each case we have studied the quantization condition for the transverse momentum, the resulting energy
spectrum and the corresponding normalized mode solutions.
An important conclusion pertaining to the energy spectrum, summarized in subsections~\ref{sec:spec_energy}
and \ref{sec:MIT_energy} for the spectral and MIT cases, respectively, was that modes with $E_j \Et_j < 0$ are
excluded from the energy spectrum if the boundary is placed inside the SOL, that is, $R\Omega \le 1$ where $\Omega $ is the angular speed about the $z$-axis.
In this case the rotating and nonrotating Minkowski vacua are identical and second quantization of the fermion field is straightforward.
\section{Thermal expectation values}\label{sec:tevs}
In this section, we calculate rigidly-rotating t.e.v.s of the fermion condensate $\psibar\psi$ (FC),
parity-violating neutrino charge current $\current^z_\nu$ (CC) and stress-energy tensor $T_{\mu\nu}$ (SET) for a quantum fermion field
inside a cylinder of radius $R$, where
$R\Omega \le 1$ and the boundary is inside or on the SOL.
We use the thermal Hadamard function and the point-splitting method, as outlined in Ref.~\cite{book:birrell}.
The spectral and MIT bag boundary conditions are considered separately.
We compare our results with those for rotating fermions on unbounded Minkowski space-time, as discussed in
Refs.~\cite{art:rotunb,art:vilenkin,art:vilenkin78}.

For completeness, the main steps for the construction of the thermal Hadamard function, presented in Ref.~\cite{book:birrell},
are summarized below. We start with the Pauli-Jordan (Schwinger) function,
\begin{equation}\label{eq:S_def}
 S(x,x') = \braket{0| \acomm{\psi(x)}{\psibar(x'}|0},
\end{equation}
Fourier transform of which can be written as:
\begin{equation}
 S(x,x') = \int_{-\infty}^\infty d\omega \,e^{-i\omega \Delta t} c(\omega; \bm{x}, \bm{x}'),
\end{equation}
where $\bm{x}$ is the spatial part of the space-time point $x$.
We note  that since $\acomm{\psi(x)}{\psibar(x')}$ is proportional to the identity operator,
the Schwinger function $S(x,x')$ (\ref{eq:S_def}) is state independent (i.e.~evaluates to the same number regardless of the
state $|0\rangle $ under consideration).
The Fourier coefficients $c(\omega; \bm{x}, \bm{x}')$ can be used to compute the thermal Hadamard function at inverse temperature $\beta $:
\begin{equation}\label{eq:S1beta_def}
 S^{(1)}_\beta(x,x') = \int_{-\infty}^\infty d\omega\, e^{-i\omega \Delta t} c(\omega; \bm{x}, \bm{x}')
 \tanh\frac{\beta \omega}{2}.
\end{equation}
The thermal Hadamard function $S^{(1)}_\beta(x,x')$ (\ref{eq:S1beta_def}) is
independent of the initial choice of vacuum $|0\rangle $ in (\ref{eq:S_def}).

Since we consider only the case where the boundary is inside the SOL, as discussed in Secs.~\ref{sec:quantization}, \ref{sec:spec_quantization} and \ref{sec:MIT_quantization}, the rotating and nonrotating Minkowski vacua inside the cylinder are identical for each set of boundary conditions.
However, the two vacua for the different boundary conditions, namely $| 0 ^{\text {sp}}\rangle$ (spectral) and $| 0 ^{\text {MIT}}\rangle$ (MIT)
are {\emph {not}} the same.
In this section, we compute rigidly-rotating t.e.v.s with respect to the $| 0 ^{\text {sp}}\rangle$ and $| 0 ^{\text {MIT}}\rangle$ vacuum states,
using the difference
$\Delta S^{(1)}_\beta(x,x')$ between the thermal Hadamard function $S^{(1)}_\beta(x,x')$ (\ref{eq:S1beta_def}) and
its vacuum counterpart, defined as:
\begin{equation}\label{eq:S1_def}
 S^{(1)}(x,x') = \braket{0^{*}| \comm{\psi(x)}{\psibar(x')}|0^{*}},
\end{equation}
where $|0^{*}\rangle $ is either $| 0 ^{\text {sp}}\rangle$ or $| 0 ^{\text {MIT}}\rangle$.
We first derive a general expression for the thermal Hadamard function (\ref{eq:S1beta_def}) in terms of fermion field modes, before considering separately the situations where the field satisfies spectral or MIT bag boundary conditions.
\subsection{Thermal Hadamard function}\label{sec:had}
Using the notation in Eq.~\eqref{eq:Udef}, the fermion field operator can be written as:
\begin{multline}
 \psi(x) = \frac{1}{2\pi} \sum_{j} \theta(E_j) \left[e^{-i\Et_j t + i k_j z} \mathcal{C}_j u_j(x) {\mathfrak {b}}_j\right.\\ +
 \left. e^{i\Et_j t - i k_j z} \mathcal{C}_j^* v_j(x) {\mathfrak {d}}^{\dagger}_j\right],
\end{multline}
where the sum over $j$, the normalization constants $\mathcal{C}_j$, the four-spinors $u_j$ and
their charge conjugates $v_j$ depend on the boundary conditions employed, and are described in detail in Sec.~\ref{sec:bcs}.
The corotating energy $\Et _{j}$ and the Minkowski energy $E_{j}$ are related by Eq.~(\ref{eq:energy}).
The Schwinger function \eqref{eq:S_def} takes the form:
\begin{equation}
\label{eq:Sdefgen}
 S(x,x') = \sum_j \theta(E_j) \left[U_j(x) \otimes \overline{U}_j(x') + V_j(x) \otimes \overline{V}_j(x')\right],
\end{equation}
where $\otimes $ denotes an outer product, the $U_{j}$ are particle modes and the $V_{j}$ are anti-particle modes.
The expression (\ref{eq:Sdefgen}) is valid irrespective of the state in which it is evaluated \cite{book:birrell}.
Thus, the Fourier coefficients of the Schwinger function take the form:
\begin{multline}
 c(\omega; \bm{x}, \bm{x}') = \sum_{j}  \frac{\abs{\mathcal{C}_j}^2 \theta(E_j)}{4\pi^2} \\\times
 \left[\delta(\omega - \Et_j) e^{ik_j \Delta z} u_j(\vx) \otimes \overline{u}_j(\vx')\right.\\
 \left. + \delta(\omega + \Et_j) e^{-ik_j\Delta z} v_j(\vx) \otimes \overline{v}_j(\vx')\right],
\end{multline}
where $\Delta z = z- z'$.
From these Fourier coefficients, the thermal Hadamard function \eqref{eq:S1beta_def} can be derived:
\begin{multline}
 S_\beta^{(1)}(x,x') = \sum_j \theta(E_j) \, \tanh\frac{\beta \Et_j}{2} \\\times\left[U_j(x)\otimes \overline{U}_j(x')
 - V_j(x) \otimes \overline{V}_j(x')\right].
\end{multline}
Subtracting the vacuum Hadamard function \eqref{eq:S1_def} from the above thermal Hadamard function gives:
\begin{multline}\label{eq:hadth_uv}
 \Delta S_\beta^{(1)}(x,x') = -\sum_j w_j \\\times \left[U_j(x)\otimes \overline{U}_j(x')
 - V_j(x) \otimes \overline{V}_j(x')\right],
\end{multline}
where the thermal factor $w_j$ takes the form:
\begin{equation}\label{eq:wj_def}
 w_{j} = \frac{2\theta(E_j)}{e^{\beta \Et_j} + 1}.
\end{equation}
In (\ref{eq:wj_def}), the step function $\theta(E_j)$ ensures that the sum over $j$ in Eq.~\eqref{eq:hadth_uv} runs
only over positive Minkowski energies (i.e.~$E_j > 0$).

In this section we calculate the  (rigidly-rotating) t.e.v.s for the FC $\braket{:\psibar \psi:}_\beta ^{*}$,
charge current $\braket{:\current^{\halpha}:}_\beta  ^{*}$ and SET $\braket{:T_{\halpha\hsigma}:}_\beta ^{*}$,
where all components are with respect to the tetrad (\ref{eq:tetrad}).
The notation $\braket{: {\mathcal {O}} :}_{\beta }^{*}$, for an operator ${\mathcal {O}}$, indicates that we are considering t.e.v.s relative to the vacuum state (either $| 0 ^{\text {sp}}\rangle$ or $| 0 ^{\text {MIT}}\rangle$ as applicable).
The superscript ${}^{*}$ will be either ${}^{\text {sp}}$ or ${}^{\text{MIT}}$ depending on which boundary conditions we are considering.
For the rest of this section, all expectation values will be for rotating thermal states, relative to the appropriate (bounded) vacuum state.
We will consider expectation values in the bounded vacuum state relative to the unbounded Minkowski vacuum state in Sec.~\ref{sec:cas}.

The t.e.v.s are calculated from the difference (\ref{eq:hadth_uv}) between the thermal Hadamard function and the vacuum Hadamard function, as follows:
\begin{subequations}
\begin{align}
\label{eq:ppsi}
 \braket{:\psibar \psi:}_\beta ^{*} = & -\frac{1}{2} \lim_{x'\rightarrow x}
 \text{tr}\left[\Delta S^{(1)}_\beta(x,x')\right],
 \\
\label{eq:J}
 \braket{:\current^{\halpha}:}_\beta ^{*}= &-\frac{1}{2} \lim_{x'\rightarrow x} \tr
 \left[\gamma^{\halpha} \Delta S^{(1)}_\beta(x,x')\right],
 \\
 \braket{:T_{\halpha\hsigma}:}_\beta ^{*}= &\frac{i}{4} \lim_{x'\rightarrow x} \tr \left[
 \gamma_{(\halpha} D_{\hsigma)} \Delta S^{(1)}_\beta(x,x')\right.
\nonumber \\
\label{eq:SET}
 & - \Delta S^{(1)}_\beta(x,x') \left.
 \overleftarrow{\overline{D}}_{(\hsigma} \gamma_{\halpha)} \right].
\end{align}

It will turn out, in Secs.~\ref{sec:spec_tev_CC} and \ref{sec:MIT_tev_CC}, that the expectation value (\ref{eq:J}) for the charge current vanishes identically for both spectral and MIT bag boundary conditions.
We will therefore also consider the charge current for fermions of negative chirality only.
It has been remarked by Vilenkin~\cite{art:vilenkin78} that the restriction of the particle spectrum to fermions of negative
chirality induces a nonvanishing charge current
anti-parallel to the rotation vector $\bm{\Omega}$.
Since these particles are traditionally called (in the massless case) neutrinos, we will use the term neutrino charge current (and abbreviate this to CC)
for this quantity.
The t.e.v.s of the CC $\current^\halpha_\nu$
of particles of negative chirality can be calculated using:
\begin{equation}\label{eq:Jnu}
 \braket{:\current^\halpha_\nu:}_\beta ^{*}= -\frac{1}{2}\lim_{x'\rightarrow x} \tr \left[
 \gamma^\halpha \frac{1 - \gamma^5}{2} \Delta S^{(1)}_\beta(x,x')\right] .
\end{equation}
\label{eq:tevs}\end{subequations}
Here $(1 - \gamma^5)/2$ projects onto the space of modes of negative chirality
with the help of the matrix $\gamma^5 = i \gamma^{\hat{0}} \gamma^{\hat{1}} \gamma^{\hat{2}}
\gamma^{\hat{3}}$, which in the Dirac representation has the form \cite{book:itzykson_zuber}:
\begin{equation}\label{eq:gamma5}
 \gamma^5 =
 \begin{pmatrix}
  0 & 1\\ 1 & 0
 \end{pmatrix}.
\end{equation}

We now turn to the computation of the t.e.v.s (\ref{eq:tevs}), considering the spectral and MIT bag boundary conditions separately. In each case, we first construct the thermal Hadamard function before computing the t.e.v.s and examining their properties.

\subsection{Spectral boundary conditions}\label{sec:spec_tevs}

Using the relation~\eqref{eq:spec_VU_conn} to write the anti-particle modes in terms of the particle modes,
the difference between
the thermal and vacuum Hadamard functions (\ref{eq:hadth_uv}) can be written as:
\begin{equation}\label{eq:Had_M}
 \Delta S^{(1)}_\beta(x,x') = -\sum_j \frac{\abs{\mathcal{C}_j^{\text{sp}}}^2}{4\pi^2}
 e^{-i\Et_j \Delta t + ik_j \Delta z} ( w_j - w_{\wj}) M_j^\lambda,
\end{equation}
where $\Et _{j}$ is the corotating energy, $\Delta t = t-t'$, $\Delta z = z-z'$ and the normalization constant
$\mathcal{C}_j^{\text{sp}} \equiv \mathcal{C}^{\lambda_j,\text{sp}}_{E_{j} k_j m_j \ell_j}$ is given in
Eq.~\eqref{eq:spec_norm}. The sum over $j$ can be found in (\ref{eq:sumj_sp}).
The thermal factors $w_j$ and $w_{\wj }$ are given by (\ref{eq:wj_def}) with the indices $j$ and $\wj$ in Eqs.~(\ref{eq:jdef_sp}) and (\ref{eq:wjdef_sp}) respectively.
The matrix $M_j^\lambda \equiv M_j^\lambda(x,x') =
u^{\lambda_j}_{E_{j} k_j m_j \ell_j}(x) \otimes \overline{u}^{\lambda_j}_{E_{j} k_j m_j \ell_j}(x')$ is
given explicitly by:
\begin{equation}\label{eq:spec_uu}
 M_j^\lambda = \frac{1}{2}
 \begin{pmatrix}
  \mathsf{E}_+^2 & -\frac{2\lambda E}{\abs{E}} \mathsf{E}_+ \mathsf{E}_-\\
  \frac{2\lambda E}{\abs{E}} \mathsf{E}_+ \mathsf{E}_- & -\mathsf{E}_-^2
 \end{pmatrix} \otimes \left[\phi_j (x)\otimes \phi_j^\dagger (x')\right],
\end{equation}
where $\mathsf{E}_{\pm }$ are given in (\ref{eq:Edef}) and the spinors $\phi _{j}$ in (\ref{eq:phi}).
In (\ref{eq:spec_uu}), the first occurrence of $\otimes$ has the meaning of a Kronecker product of two $2\times 2$ matrices,
i.e.:
\begin{equation}
 \begin{pmatrix}
  a_{11} & a_{12} \\
  a_{21} & a_{22}
 \end{pmatrix} \otimes B =
 \begin{pmatrix}
  a_{11} B & a_{12} B \\
  a_{21} B & a_{22} B
 \end{pmatrix}.
 \label{eq:Kronecker}
\end{equation}
In other words, the outer product $\phi_j (x)\otimes \phi_j^\dagger(x')$
is to be copied into each of the four matrix elements to the left of the Kronecker $\otimes$ sign, thus
producing a $4 \times 4$ matrix.

Introducing the notation:
\begin{equation}\label{eq:spec_M_bblocks}
 M_j \equiv \sum_{\lambda_j = \pm 1/2} M_j^\lambda = \frac{1}{2}
 \begin{pmatrix}
  M_j^{\text{up}} & -M_j^{\times}\\
  M_j^{\times} & -M_j^{\text{down}}
 \end{pmatrix},
\end{equation}
the following expressions can be found for the $2\times 2$ matrices introduced on the right-hand-side of (\ref{eq:spec_M_bblocks}), by using the explicit form \eqref{eq:phi} of the $\phi _{j}$ spinors:
\begin{align}
\label{eq:specMmatrices}
 M_j^{\text{up}} =& \mathsf{E}_+^2
 \begin{pmatrix}
  1 & 0\\
  0 & 1
 \end{pmatrix} \circ \mathcal{M}_j,\nonumber\\
 M_j^{\text{down}} =& \mathsf{E}_-^2
 \begin{pmatrix}
  1 & 0\\
  0 & 1
 \end{pmatrix} \circ \mathcal{M}_j,\nonumber\\
 M_j^{\times} =& \frac{1}{E}
 \begin{pmatrix}
  k & q \\
  q & -k
 \end{pmatrix} \circ \mathcal{M}_j .
\end{align}
In (\ref{eq:specMmatrices}), the Hadamard (Schur) product symbol $\circ$ has been used for the element-wise product of
two matrices of the same size, defined for two $2\times 2$ matrices $A$, $B$ as:
\begin{equation}
 A \circ B =
 \begin{pmatrix}
  a_{11} b_{11} & a_{12} b_{12} \\
  a_{21} b_{21} & a_{22} b_{22}
 \end{pmatrix}.
 \label{eq:Hadamard}
\end{equation}
The matrix $\mathcal{M}_j$ on the right of the Hadamard product symbol $\circ$ in (\ref{eq:specMmatrices}) is defined as:
\begin{equation}\label{eq:M_struct}
 \mathcal{M}_j = \begin{pmatrix}
  J_m J_m e^{im\Delta \varphi} &
  -i J_m J_{m+1} e^{i(m+1)\Delta \varphi - i \varphi} \\
  i J_{m+1} J_m e^{im\Delta \varphi + i \varphi} &
  J_{m+1} J_{m+1} e^{i(m+1)\Delta \varphi}
 \end{pmatrix},
\end{equation}
where $\Delta \varphi = \varphi - \varphi '$ and the arguments of the first and second Bessel functions in the products above are $q\rho$ and $q\rho'$,
respectively, e.g.~$J_m J_{m+1} \equiv J_m(q\rho) J_{m+1}(q\rho')$.

For the purpose of computing t.e.v.s, it is advantageous to write $M_j$ (\ref{eq:spec_M_bblocks}) as:
\begin{multline}\label{eq:M_symm}
 2 M_j = \frac{1}{2} {\mathbb {I}}_2 \otimes (M_j^{\text{up}} - M_j^{\text{down}})\\
 + \frac{1}{2} \sigma_3 \otimes (M_j^{\text{up}} + M_j^{\text{down}}) +
 \begin{pmatrix}
  0 & -1\\
  1 & 0
 \end{pmatrix}
 \otimes M_j^{\times} ,
\end{multline}
where ${\mathbb {I}}_{2}$ is the $2\times 2$ identity matrix and the Pauli matrix $\sigma _{3}$ can be found in Eq.~(\ref{eq:Pauli}).
Thus, the following form is obtained for $M_j$:
\begin{multline}\label{eq:spec_M}
 M_j = \left[\frac{\mu}{2E} {\mathbb {I}}_2 + \frac{1}{2}\sigma_3\right] \otimes
 \left[\begin{pmatrix}
  1 & 0 \\ 0 & 1
 \end{pmatrix} \circ \mathcal{M}_j\right]\\
 + \frac{1}{2E}
 \begin{pmatrix}
  0 & -1 \\ 1 & 0
 \end{pmatrix} \otimes
 \left[\begin{pmatrix}
  k & q \\ q & -k
 \end{pmatrix} \circ \mathcal{M}_j\right].
\end{multline}
Having computed above the explicit form of $M_j$ appearing in Eq.~\eqref{eq:Had_M}, t.e.v.s
can now be calculated, as described in the following sections.

\begin{figure*}
\begin{tabular}{cc}
 \includegraphics[width=0.95\columnwidth]{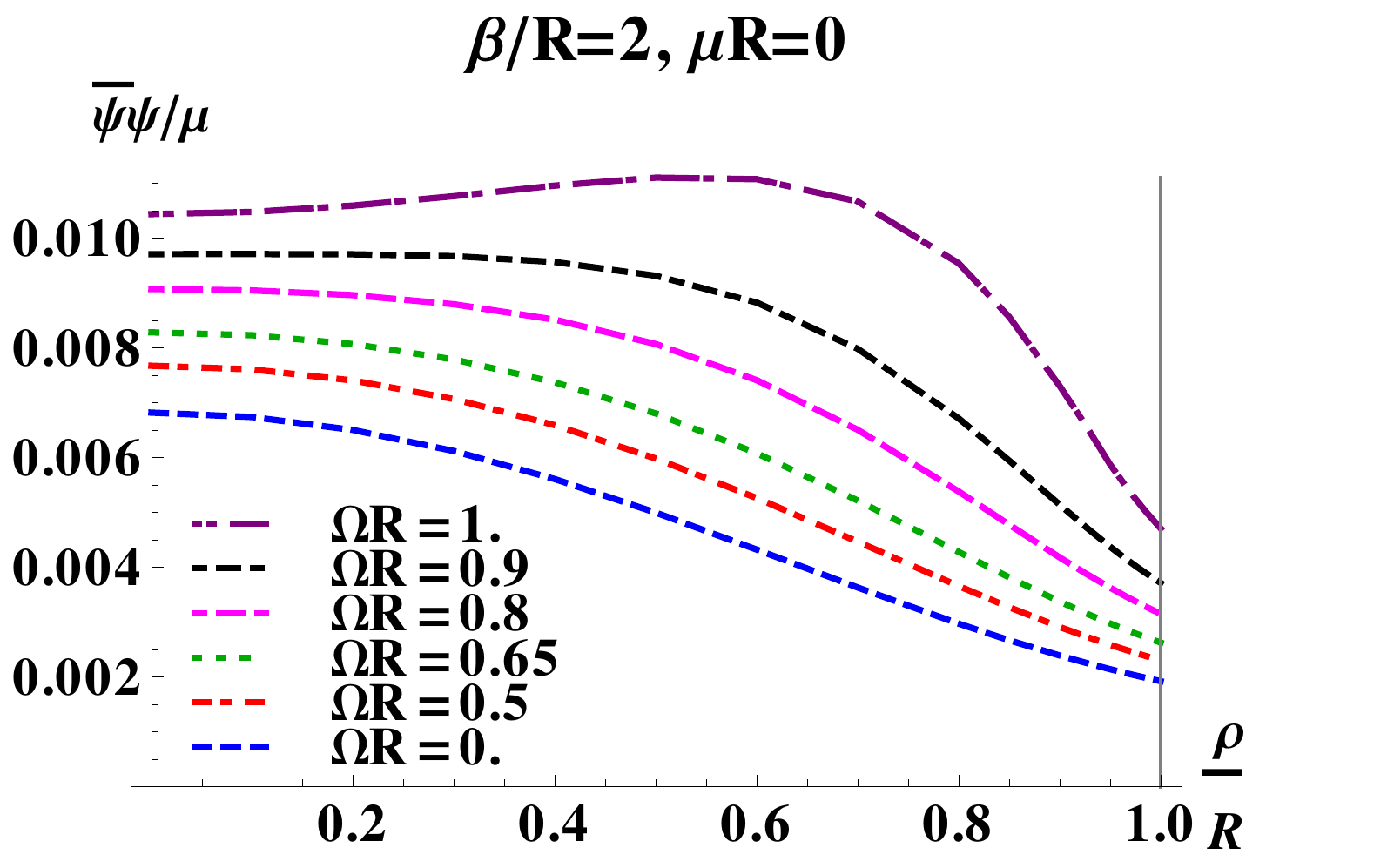} &
 \includegraphics[width=0.95\columnwidth]{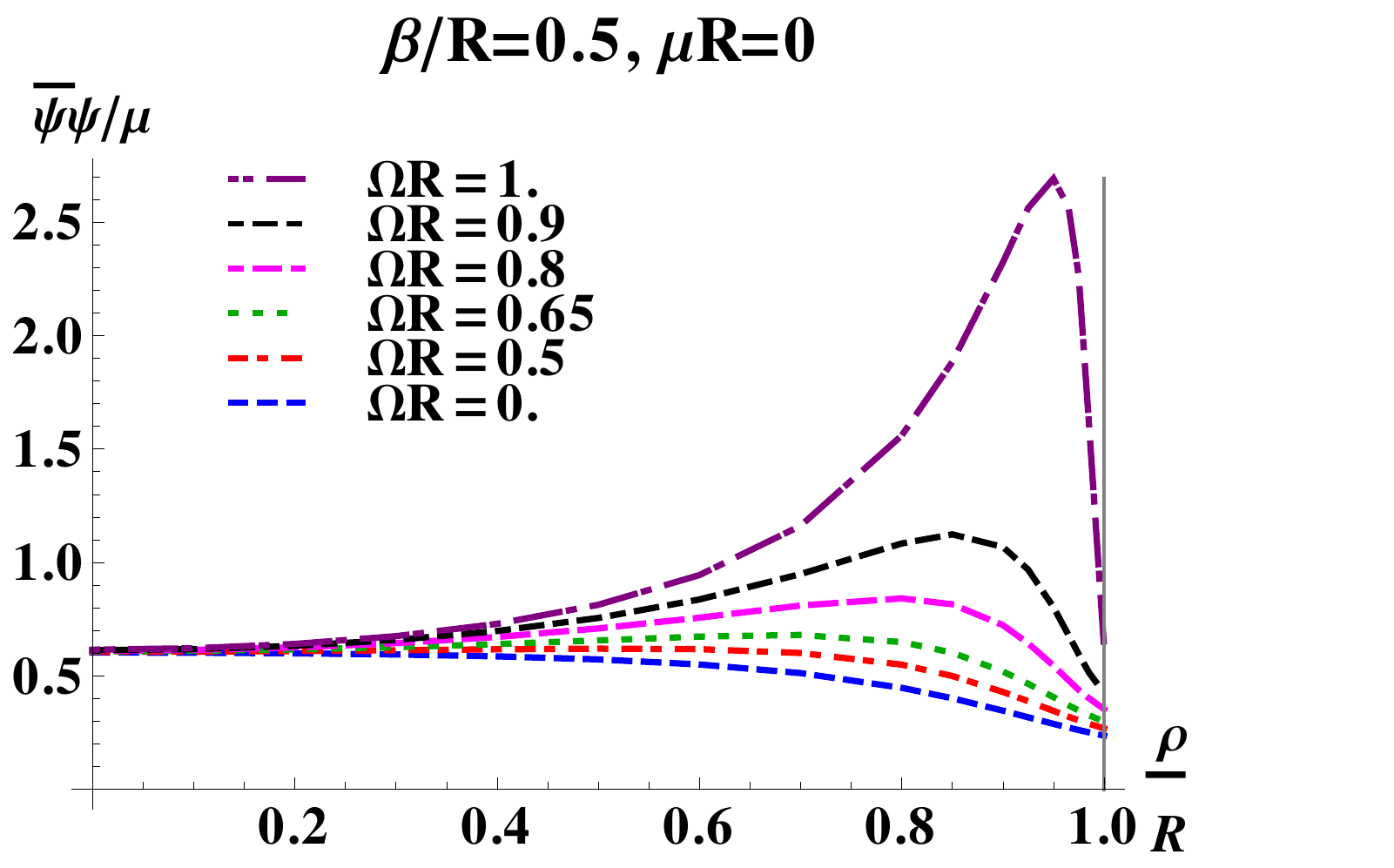} \\
 (a) & (b) \\
 \includegraphics[width=0.95\columnwidth]{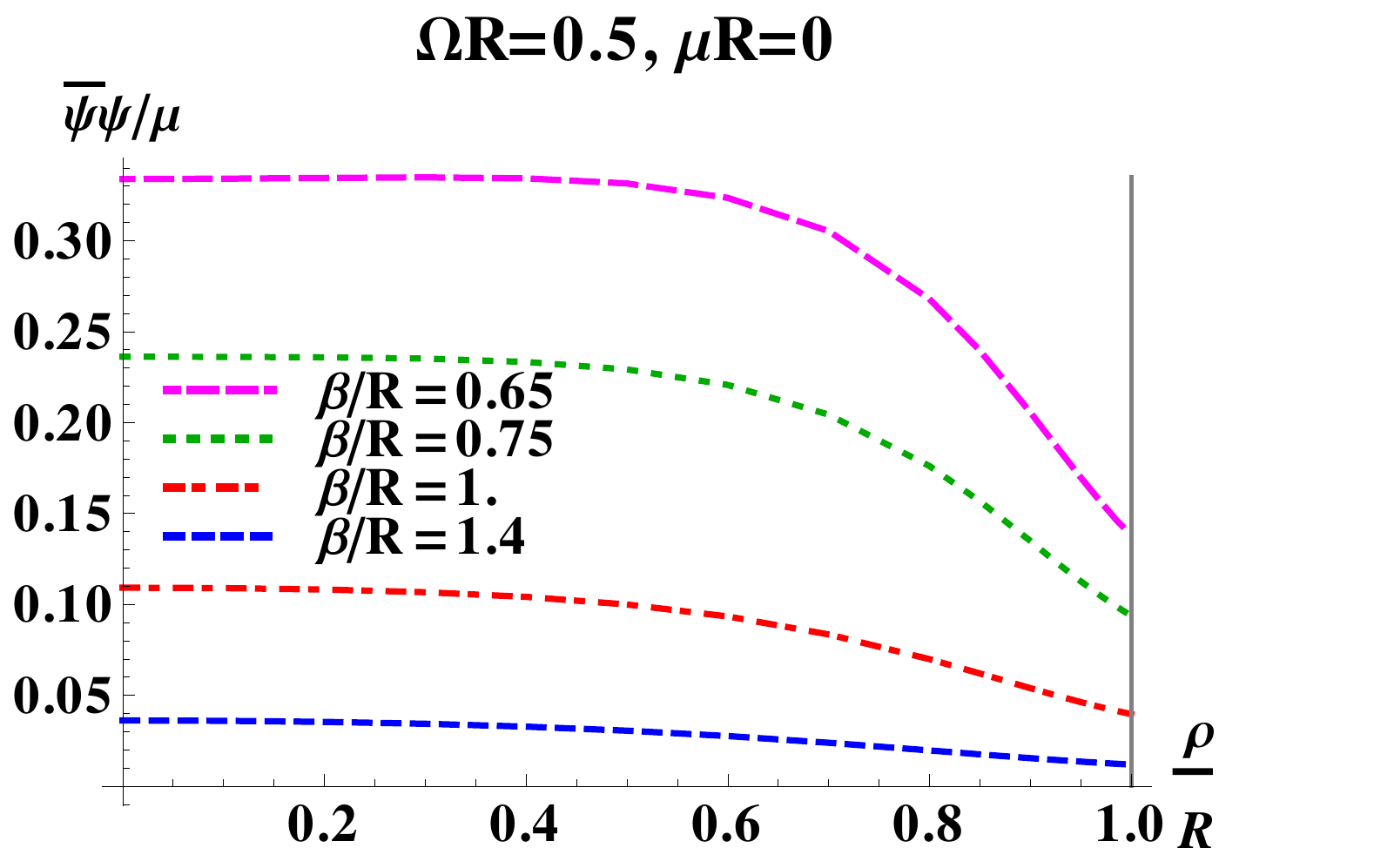} &
 \includegraphics[width=0.95\columnwidth]{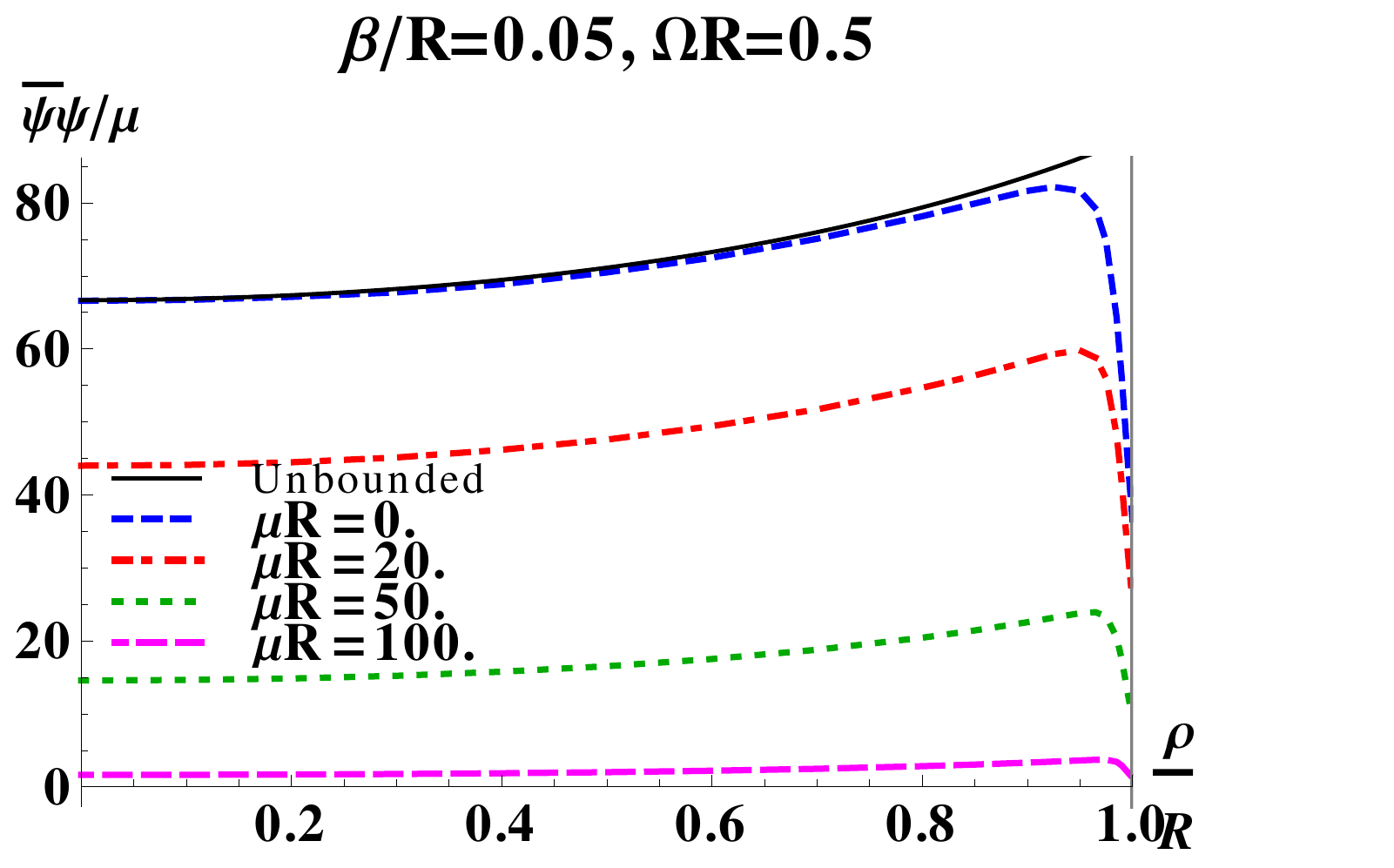} \\
 (c) & (d)
\end{tabular}
\caption{Thermal expectation values (t.e.v.s) of the fermion condensate (FC) $\braket{:\psibar\psi:}_\beta^{\text{sp}}$
(\ref{eq:spec_ppsi}) for spectral boundary conditions divided by the fermion mass $\mu $, as a function of the scaled radial coordinate $\rho /R$, so that the boundary of the cylinder
is at $\rho /R=1$.
(a) Massless fermions $\mu =0$, fixed inverse temperature $\beta =2R$ and various values of the angular speed $\Omega $.
(b) Massless fermions $\mu =0$, fixed inverse temperature $\beta =0.5R$ and various values of the angular speed $\Omega $.
(c) Massless fermions $\mu =0$, fixed angular speed $\Omega = 0.5/R$ and various values of the inverse temperature $\beta $.
(d) Fixed inverse temperature $\beta = 0.05R$, fixed angular speed $\Omega = 0.5/R$ and various values of the fermion mass $\mu $.
The solid curve in (d) shows the t.e.v.~of
the FC for massless fermions in the unbounded case, given in Eq.~\eqref{eq:rotunb_ppsi}, for comparison.}
\label{fig:spec_ppsiomu}
\end{figure*}

\subsubsection{Fermion condensate}\label{sec:spec_tev_FC}
The t.e.v.~of the FC $\braket{:\psibar \psi:}_\beta ^{\text {sp}}$ is computed from the difference between the thermal
and vacuum Hadamard functions (\ref{eq:Had_M}) using (\ref{eq:ppsi}).
Looking at Eq.~\eqref{eq:spec_M}, it is clear that only the first term (the one involving ${\mathbb{I}}_2$ on the left
of the direct product sign $\otimes$) contributes, giving:
\begin{equation}
 \braket{:\psibar \psi:}_\beta ^{\text {sp}} = \sum_j \frac{\abs{\mathcal{C}_j^{\text {sp}}}^2}{8\pi^2} (w_j - w_{\wj}) \frac{\mu}{E_j}
 J_m^+(q \rho),
\label{eq:spec_ppsi_temp}
\end{equation}
where the notation $J_m^+(z)$ is the same as in Ref.~\cite{art:rotunb}:
\begin{align}
 J_m^\pm(z) =& J_m^2(z) \pm J_{m+1}^2(z),\nonumber\\
 J_m^\times(z) =& 2 J_m(z) J_{m+1}(z).\label{eq:BesselJ*}
\end{align}
It is convenient to express the sum over $j$ as a sum over
positive energies:
\begin{equation}\label{eq:spec_ppsi}
 \braket{:\psibar \psi:}_\beta^{\text{sp}} = \sum_{m = 0}^\infty \sum_{\ell = 1}^\infty \int_0^\infty
 \frac{\mu\,dk}{E \pi^2 R^2} \frac{w(\Et) + w(\Ew)}{J^2_{m+1}(q R)} J_m^+(q\rho),
\end{equation}
where we have used (\ref{eq:spec_norm}) for the normalisation constants
$\mathcal{C}_j^{\text {sp}}$ and the thermal weight factor $w(x)$ is:
\begin{equation}\label{eq:spec_w}
 w(x) = \frac{2}{e^{\beta x} + 1},
\end{equation}
while its arguments $\Et$ and $\Ew$ are defined as:
\begin{equation}
 \Et = E - \Omega(m + \tfrac{1}{2}), \qquad
 \Ew = E + \Omega(m + \tfrac{1}{2}).
 \label{eq:Ew_def}
\end{equation}

Thus, in the spectral model, the t.e.v.~of the FC vanishes for massless fermions with $\mu =0$.
In Fig.~\ref{fig:spec_ppsiomu} we have therefore plotted $\mu^{-1} \braket{:\psibar\psi :}_\beta^{\text{sp}}$ to facilitate comparisons between the
t.e.v.s for different values of the mass $\mu $.
It can be seen from Fig.~\ref{fig:spec_ppsiomu} that the t.e.v.~of the FC is  positive everywhere, including on the boundary,
where its value is finite.  This is true for all $R$ provided
that the boundary of the cylinder is either inside or on the SOL.
In Figs.~\ref{fig:spec_ppsiomu}(a) and \ref{fig:spec_ppsiomu}(b), we have fixed the inverse temperature $\beta $ and the fermion mass $\mu =0$,
and show the t.e.v.s of
$\mu^{-1} \braket{:\psibar\psi :}_\beta^{\text{sp}}$
for various values of the angular speed $\Omega $.
The t.e.v.~of the FC increases for each fixed value of $\rho $ as $\Omega R$ increases. This is particularly marked in the higher-temperature plot Fig.~\ref{fig:spec_ppsiomu}(b).
When $\Omega R =1$ and the boundary is on the SOL, the FC increases rapidly as we move away from the axis of rotation, with a large peak just inside the boundary.
However, even in this case, the FC is finite on the boundary.
In Fig.~\ref{fig:spec_ppsiomu}(c) we have fixed the angular speed $\Omega $ and again consider massless fermions $\mu =0$, varying the inverse temperature $\beta $.
As expected, the t.e.v.s decrease as $\beta $ increases and the temperature decreases.
Finally, in Fig.~\ref{fig:spec_ppsiomu}(d) we fix the inverse temperature $\beta $ and angular speed $\Omega $ and vary the fermion mass $\mu $.
We see that $\mu^{-1} \braket{:\psibar\psi :}_\beta^{\text{sp}}$ decreases as $\mu $ increases.

For comparison,  in Fig.~\ref{fig:spec_ppsiomu}(d) we also plot
the t.e.v.~of the FC corresponding to the massless unbounded case \cite{art:rotunb}:
\begin{equation}
 \left.\frac{1}{\mu} \braket{:[\psibar\psi]:}_{\beta,I}^{\text{unb}}\right\rfloor_{\mu = 0} =
 -\frac{1}{6\beta^2 \varepsilon}, \label{eq:rotunb_ppsi}
\end{equation}
where $\varepsilon = 1 - \rho^2 \Omega^2$. The subscript $I$ indicates that the
above t.e.v.~is given with respect to the rotating (Iyer) vacuum \cite{art:rotunb,art:iyer}.
In the interior of the cylinder, we see that the rigidly-rotating t.e.v.~of the FC with spectral boundary conditions and a massless fermion is almost identical to that for a massless fermion on unbounded Minkowski space-time.
They differ significantly only near the boundary.  The t.e.v.~on unbounded Minkowski space-time continues to increase as the boundary is approached, while that for spectral boundary conditions decreases near the boundary.

\subsubsection{Neutrino charge current}\label{sec:spec_tev_CC}
Next, we consider the t.e.v.~of the charge current operator $\braket{:\current^{\halpha}:}_\beta ^{\text{sp}}$, defined in (\ref{eq:J}).
It is straightforward to see that the t.e.v.s of all the components of $\braket{:\current^{\halpha}:}_\beta ^{\text{sp}}$ vanish.
This is because the expression for $\braket{:\current^{\halpha}:}_\beta ^{\text{sp}}$ analogous to (\ref{eq:spec_ppsi_temp}) contains a summand
which is odd under either $m \rightarrow -m-1$ (for $\alpha \in \{t, \rho, \varphi\}$)
or $k \rightarrow -k$ (for $\alpha = z$). To illustrate this point, let us consider the time component:
\begin{equation}
 \braket{:\current^\hatt:}_\beta^{\text{sp}} = -\sum_j (w_j - w_\wj) \frac{\abs{\mathcal{C}_j^{\text {sp}}}^2}{8\pi^2} J_m^+(q\rho) ,
\label{eq:spec_current_temp}
\end{equation}
where the various quantities are defined in (\ref{eq:spec_norm}, \ref{eq:wj_def}, \ref{eq:BesselJ*}).
After restricting the energy to positive values, Eq.~(\ref{eq:spec_current_temp}) reduces to:
\begin{equation}
 \braket{:\current^\hatt:}_\beta^{\text{sp}} = \sum_{m = -\infty}^\infty \sum_{\ell = 1}^\infty \int_{-\infty}^\infty
 \frac{dk}{2\pi^2 R^2} \frac{w(\Et) - w(\Ew)}{J_{m+1}^2(qR)} J_m^+(q\rho),
\label{eq:spec_current_temp1}
\end{equation}
where the thermal weight factors and their arguments are given in (\ref{eq:spec_w}, \ref{eq:Ew_def}).
Since the summand in (\ref{eq:spec_current_temp1}) is odd with respect to $m \rightarrow -m-1$, we can conclude that
$\braket{:\current ^\hatt:}_\beta ^{\text{sp}} = 0$.
Similar arguments apply to the other components of $\braket{:\current^\halpha:}_\beta^{\text{sp}}$.

\begin{figure}[t]
\begin{tabular}{c}
 \includegraphics[width=0.86\columnwidth]{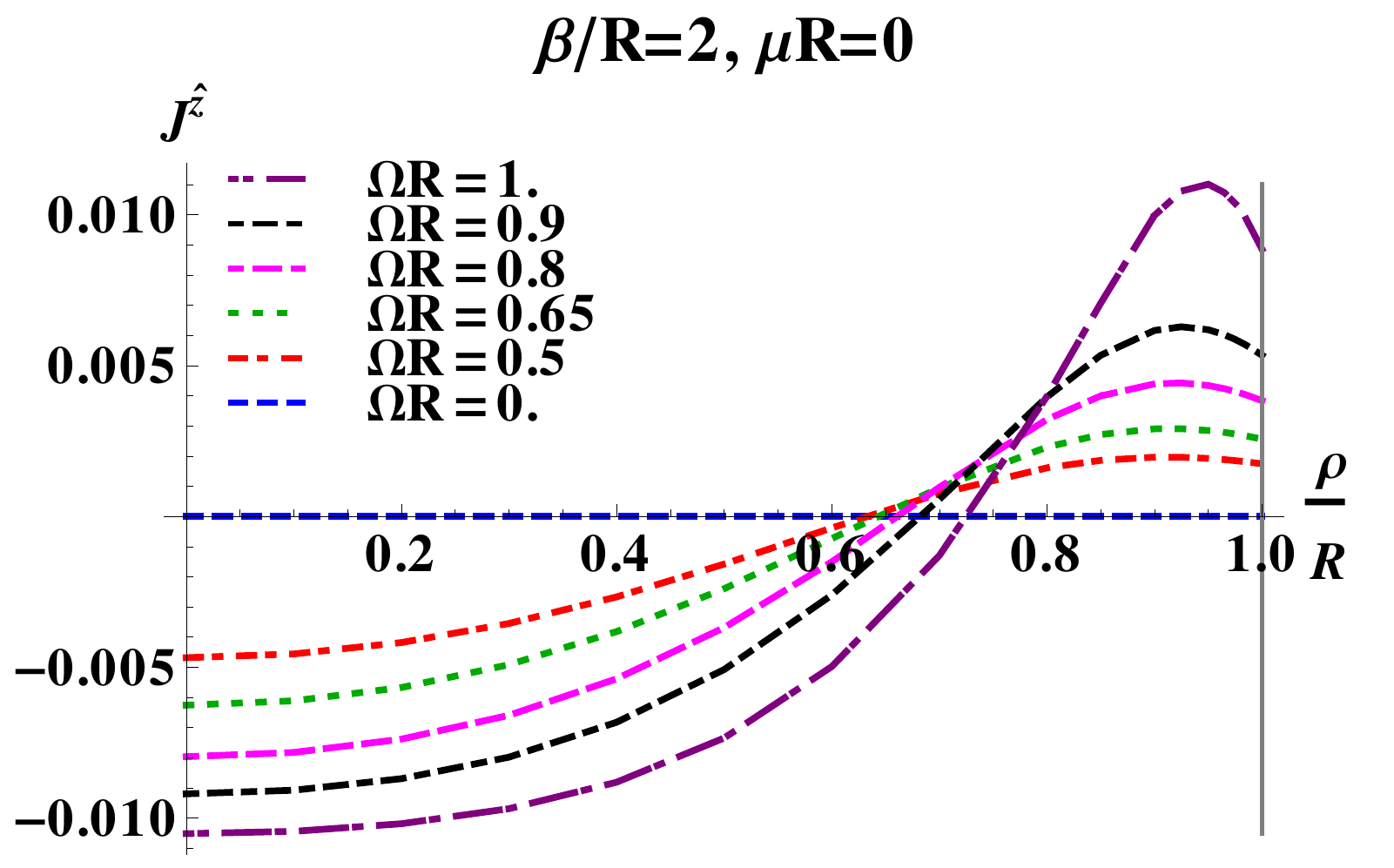}\\
 (a)\\
 \includegraphics[width=0.86\columnwidth]{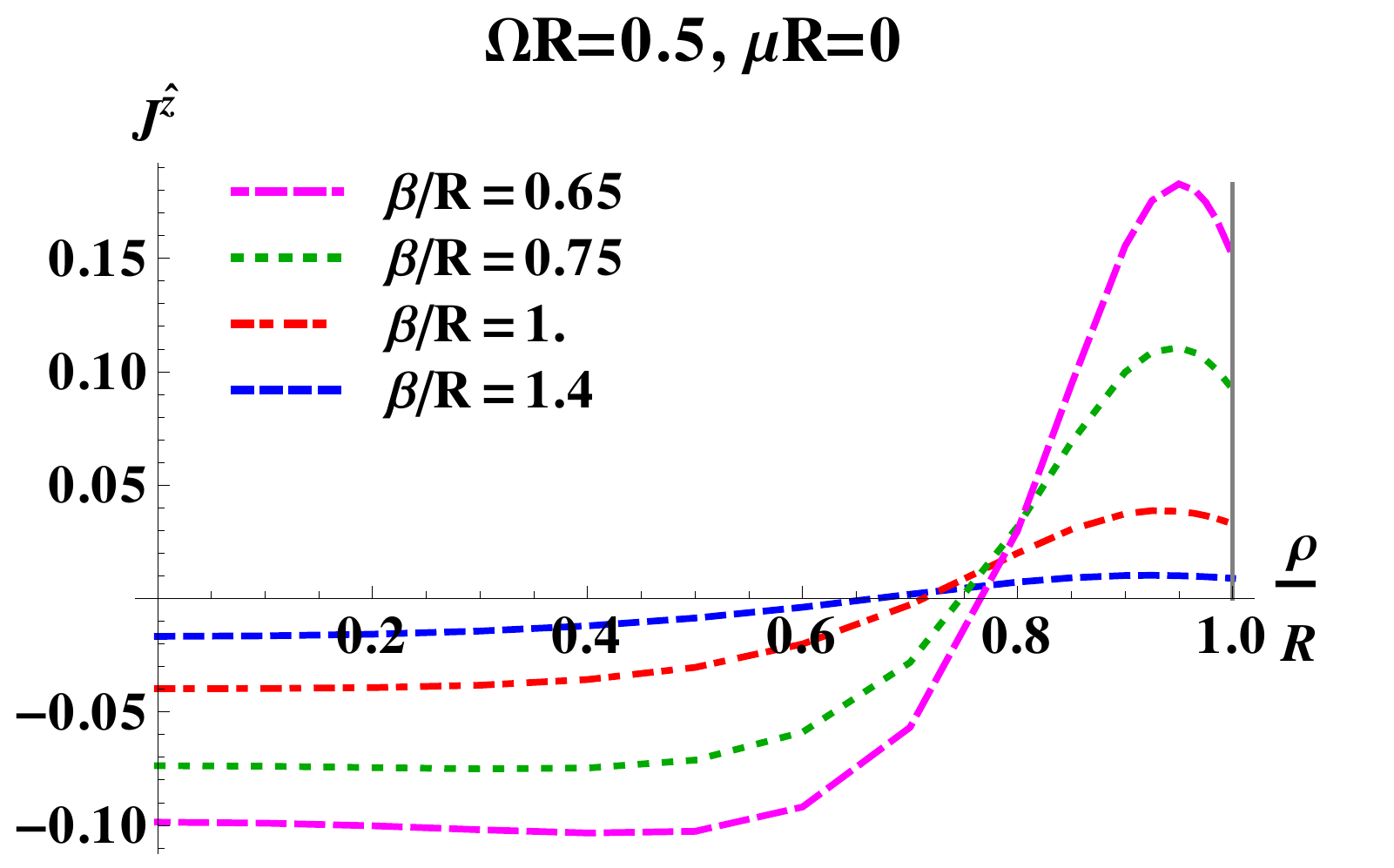}\\
 (b)\\
 \includegraphics[width=0.86\columnwidth]{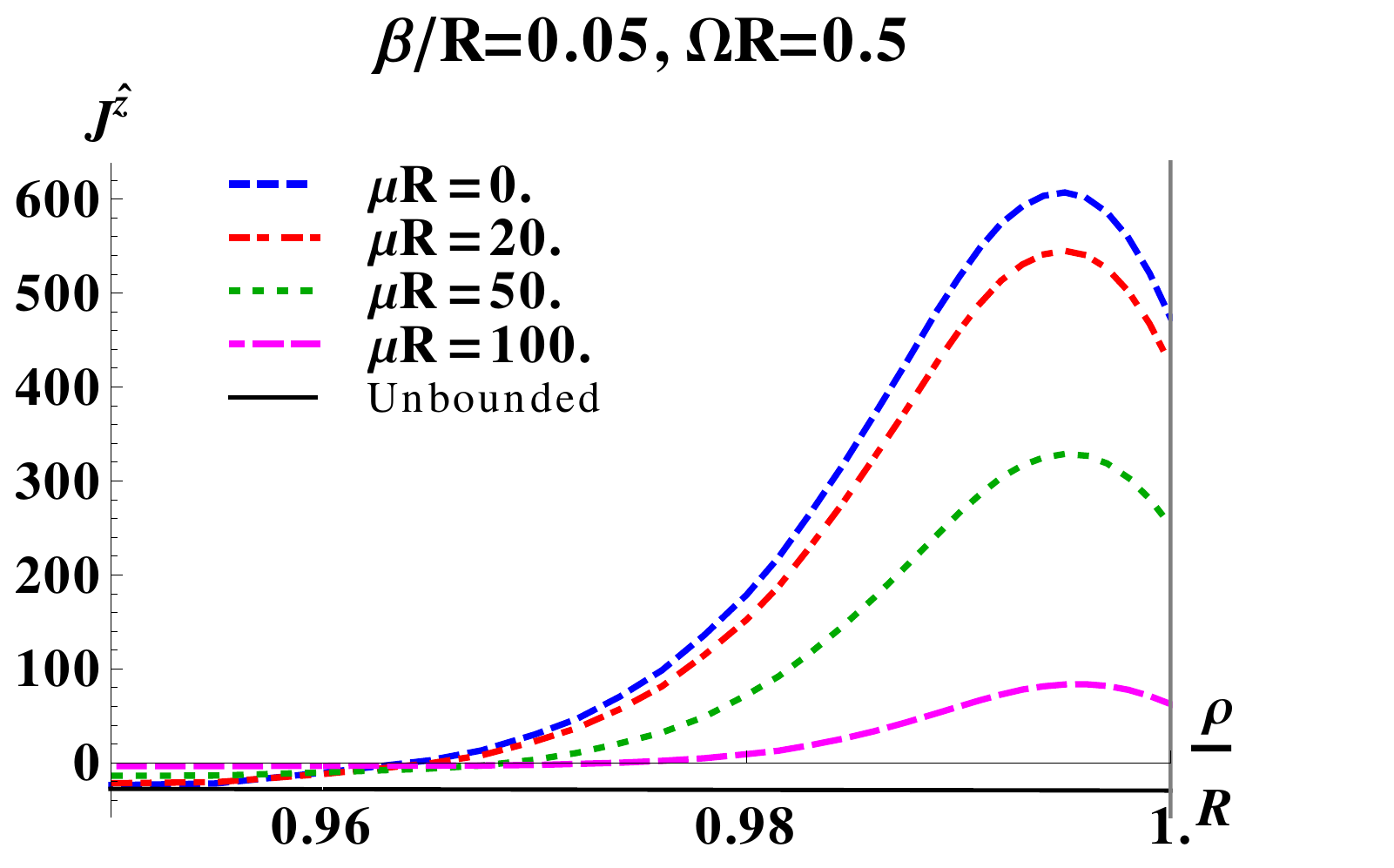}\\
 (c)
\end{tabular}
\caption{Thermal expectation values (t.e.v.s) of the neutrino charge current (CC) $\braket{:\current^{\halpha}_{\nu }:}_\beta^{\text{sp}}$
(\ref{eq:spec_Jnuz}) for spectral boundary conditions, as a function of the scaled radial coordinate $\rho /R$, so that the boundary of the cylinder
is at $\rho /R=1$.
(a) Massless fermions $\mu =0$, fixed inverse temperature $\beta =2R$ and various values of the angular speed $\Omega $.
(b) Massless fermions $\mu =0$, fixed angular speed $\Omega = 0.5/R$ and various values of the inverse temperature $\beta $.
(c)
Zoom of the region close to the boundary at
fixed inverse temperature $\beta = 0.05R$, fixed angular speed
$\Omega = 0.5/R$ and various values of the fermion mass $\mu $.
The solid curve in (c) shows the t.e.v.~of
the CC for massless fermions in the unbounded case, given in Eq.~\eqref{eq:rotunb_jz}, for comparison.}
\label{fig:spec_Jnuz}
\end{figure}

We therefore consider the neutrino charge current (CC), whose t.e.v.~is given by (\ref{eq:Jnu}).
While the $t$, $\rho$ and $\varphi$ components of the CC vanish, the $z$ component is nonzero (in accordance with \cite{art:vilenkin78}):
\begin{subequations}
\begin{equation}
 \braket{:\current^{\hat{z}}_\nu:}_\beta^{\text{sp}} = -\sum_{m = 0}^\infty \sum_{\ell = 1}^\infty \int_{0}^\infty
 \frac{dk}{2\pi^2 R^2} \frac{w(\Et) - w(\Ew)}{J_{m+1}^2(qR)} J_m^-(q\rho) ,
 \label{eq:spec_Jnuz}
\end{equation}
where $J_m^-(q\rho)$ is defined in (\ref{eq:BesselJ*}).

In Fig.~\ref{fig:spec_Jnuz} we plot the t.e.v.~(\ref{eq:spec_Jnuz}) for a range of values of the fermion mass $\mu $, inverse temperature $\beta $ and
angular speed $\Omega $.
For all values of the parameters we studied, it can be seen in Fig.~\ref{fig:spec_Jnuz} that the t.e.v.~of the CC changes sign from negative on the axis of rotation $\rho =0$ to positive  on the boundary $\rho =R$.
This can be explicitly checked by considering the value of $\braket{:\current^{\hat{z}}_\nu:}_\beta ^{\text {sp}}$
on the rotation axis $\rho = 0$,
\begin{multline}
 \left.\braket{:\current^{\hat{z}}_\nu:}_\beta^{\text{sp}}\right\rfloor_{\rho = 0} = -\sum_{\ell = 1}^\infty \int_{0}^\infty
 \frac{dk}{2 \pi^2 R^2 J_{1}^2(qR)} \\\times
 \left[w\left(E - \frac{\Omega}{2}\right) - w\left(E + \frac{\Omega}{2}\right)\right] < 0,
\end{multline}
and on the boundary $\rho = R$ (recall that $\Et $ and $\Ew$ are given in (\ref{eq:Ew_def})):
\begin{equation}
 \left.\braket{:\current^{\hat{z}}_\nu:}_\beta^{\text{sp}}\right\rfloor_{\rho = R} = \sum_{m = 0}^\infty \sum_{\ell = 1}^\infty \int_{0}^\infty
 \frac{dk}{2\pi^2 R^2} [w(\Et) - w(\Ew)] > 0.
 \label{eq:spec_Jnuz_boundary}
\end{equation}
\end{subequations}
As the angular speed $\Omega $ or temperature $\beta ^{-1}$ increases, the t.e.v.~$\braket{:\current^{\hat{z}}_\nu:}_\beta^{\text{sp}}$
decreases on the axis of rotation and increases on the boundary. It remains finite everywhere inside and on the boundary.
In Fig.~\ref{fig:spec_Jnuz} (a) we see that $\braket{:\current^{\hat{z}}_\nu:}_\beta^{\text{sp}}$ vanishes when the angular speed $\Omega =0$.
This is also the case on unbounded Minkowski space-time \cite{art:rotunb}.
As the fermion mass $\mu $ increases, $\braket{:\current^{\hat{z}}_\nu:}_\beta^{\text{sp}}$ also decreases close to the boundary.
 Fig.~\ref{fig:spec_Jnuz} (c) also shows the t.e.v.~of the CC for the massless unbounded case, which is given by \cite{art:rotunb}:
\begin{equation}
 \braket{:\current^{{\hat {z}}}:}_{\beta, I}^{\text{unb}} = -\frac{\Omega}{12\beta^2\varepsilon^2}.\label{eq:rotunb_jz}
\end{equation}
Close to the boundary, $\braket{:\current^{\hat{z}}_\nu:}_\beta^{\text{sp}}$ changes sign and increases to values which
are an order of magnitude higher than the absolute value of $\braket{:\current^{\hat{z}}:}_{\beta, I}^{\text{unb}}$, which is negative everywhere.
In Fig.~\ref{fig:spec_Jnuz} (c), note that $\braket{:\current^{\hat{z}}:}_{\beta, I}^{\text{unb}}$ is not constant, as it might appear.
It changes only by a small amount in the region shown, whereas $\braket{:\current^{\hat{z}}_\nu:}_\beta^{\text{sp}}$ changes very rapidly in this region.

\begin{figure*}
\begin{tabular}{cc}
 \includegraphics[width=0.95\columnwidth]{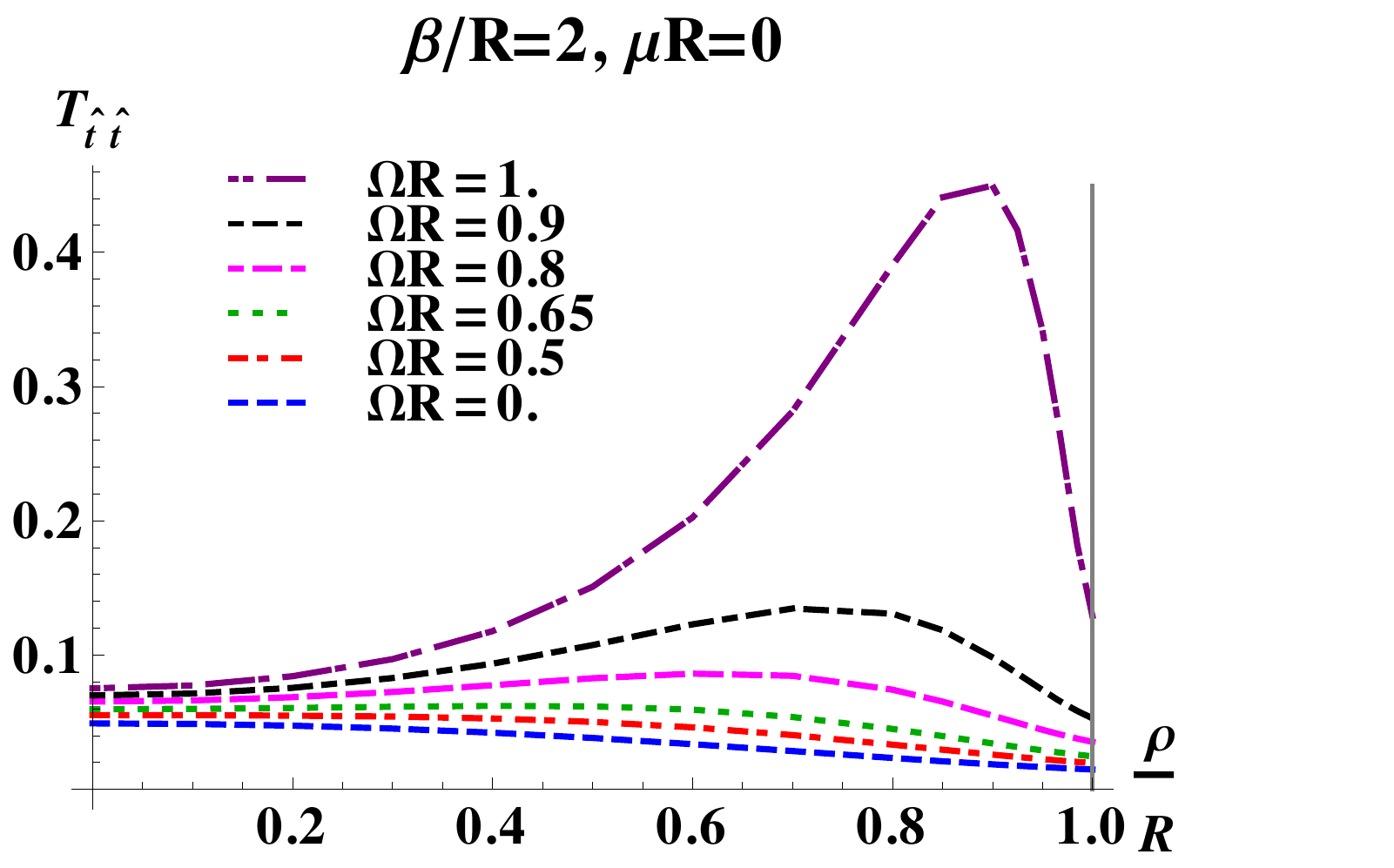} &
 \includegraphics[width=0.95\columnwidth]{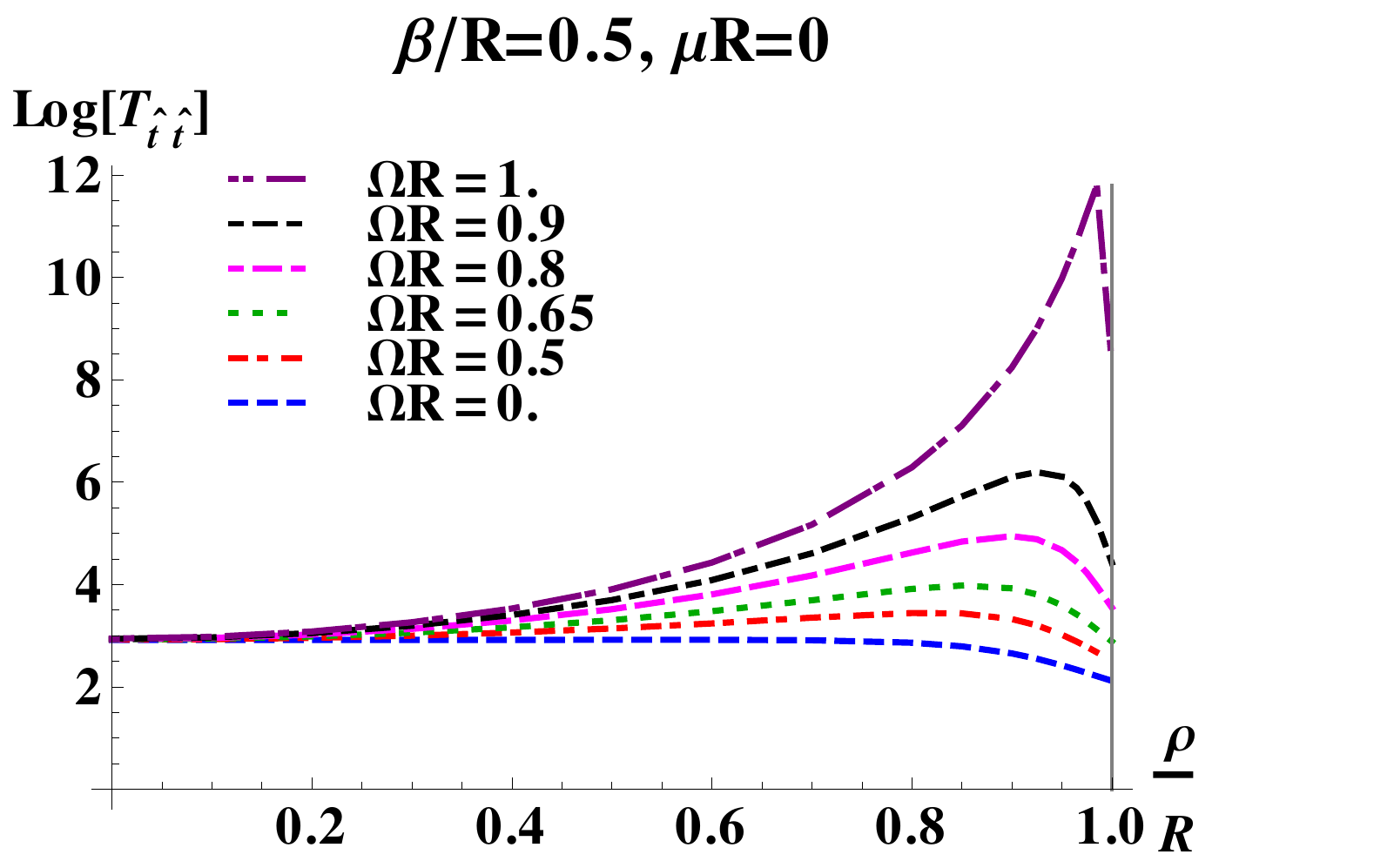}\\
 (a) & (b)\\
 \includegraphics[width=0.95\columnwidth]{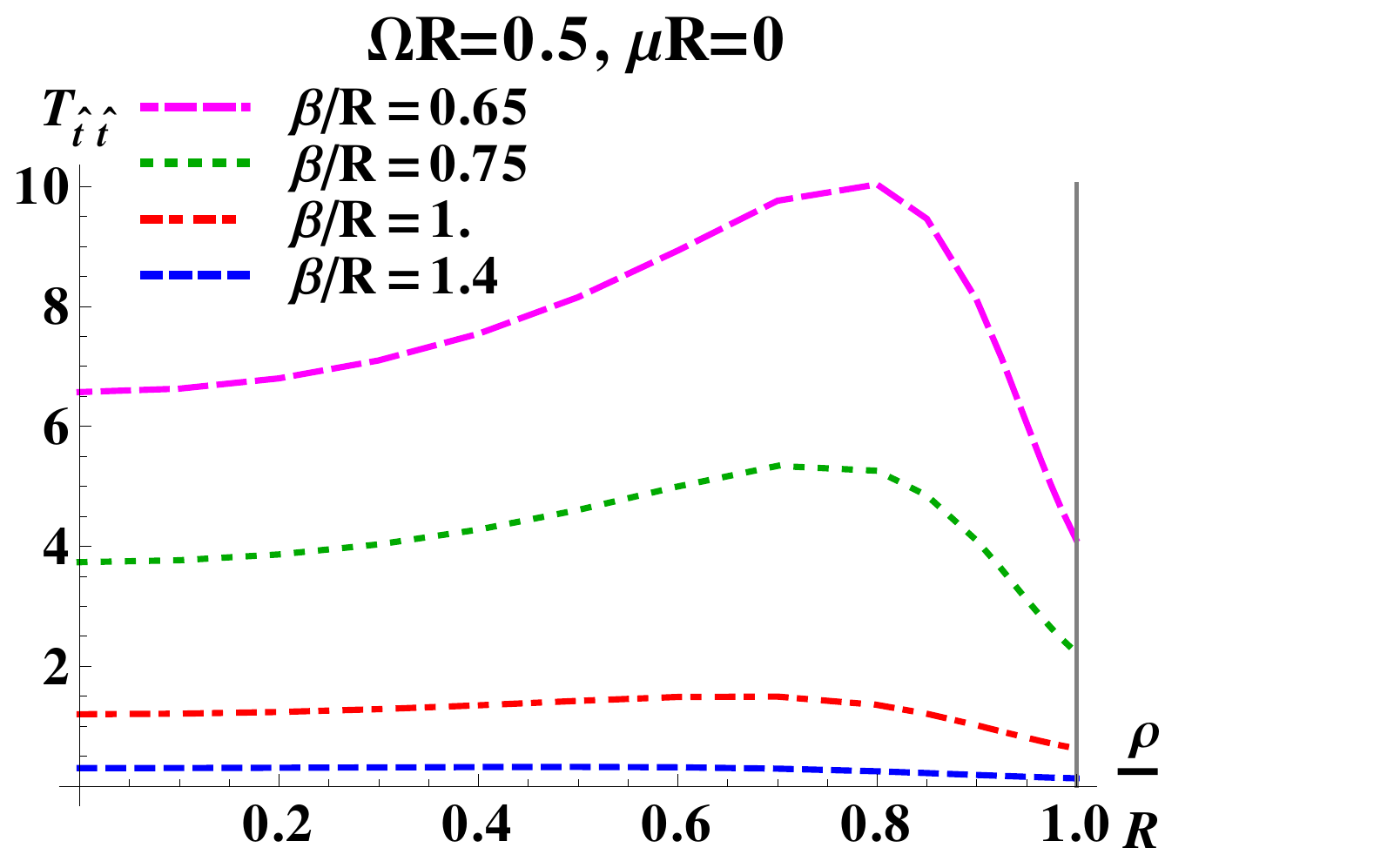} &
 \includegraphics[width=0.95\columnwidth]{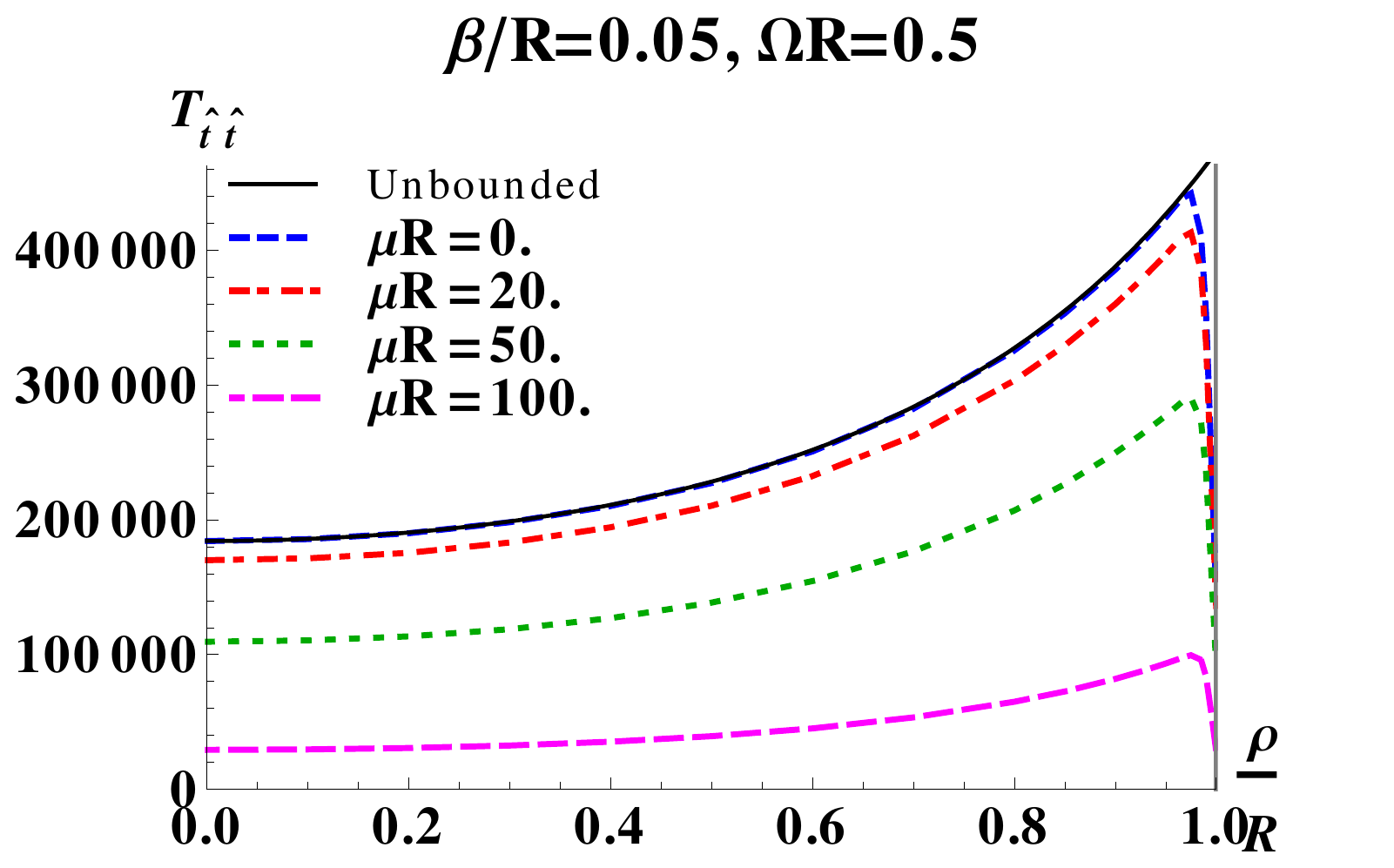} \\
 (c) & (d)
\end{tabular}
\caption{Thermal expectation values (t.e.v.s) of the SET component $\braket{:T_{\hatt\hatt}:}_\beta^{\text{sp}}$
(\ref{eq:spec_ttt}) for spectral boundary conditions, as a function of the scaled radial coordinate $\rho /R$, so that the boundary of the cylinder
is at $\rho /R=1$.
(a) Massless fermions $\mu =0$, fixed inverse temperature $\beta =2R$ and various values of the angular speed $\Omega $.
(b) Massless fermions $\mu =0$, fixed inverse temperature $\beta =0.5R$ and various values of the angular speed $\Omega $.
(c) Massless fermions $\mu =0$, fixed angular speed $\Omega = 0.5/R$ and various values of the inverse temperature $\beta $.
(d) Fixed inverse temperature $\beta = 0.05R$, fixed angular speed $\Omega = 0.5/R$ and various values of the fermion mass $\mu $.
The solid curve in (d) shows the t.e.v.~of
$\braket{:T_{\hat{t}\hat{t}}:}^{\text{unb}}_{\beta ,I}$
 for massless fermions in the unbounded case, given in Eq.~\eqref{eq:rotunb_ttt}, for comparison.}
\label{fig:spec_SET}
\end{figure*}

\subsubsection{Stress-energy tensor}\label{sec:spec_tev_SET}

The t.e.v.~of the SET $\braket{:T_{\halpha\hsigma}:}_\beta ^{\text{sp}}$ with respect to the tetrad (\ref{eq:tetrad}) can be calculated using the formula (\ref{eq:SET}),
with the difference between the thermal and vacuum Hadamard functions given by (\ref{eq:Had_M}).
By construction, the action of $iD_\hatt$ on $e^{-i\Et_j t} M_j$ (with the matrix $M_{j}$ given in (\ref{eq:spec_uu})) gives the energy $E_j$:
\begin{equation}
 i D_\hatt e^{-i\Et_j t} M_j = E_j e^{-i\Et_j t} M_j,
\end{equation}
while for the derivatives with respect to $\rho$ and $\varphi$, the inner structure \eqref{eq:M_struct}
of the ${\mathcal {M}}_j$ matrix must be taken into account.
Using the quantities defined in (\ref{eq:BesselJ*}, \ref{eq:spec_w}, \ref{eq:Ew_def}),
together with the relation
\begin{equation}
 J_{m+1}'(z) J_m(z) - J_m'(z) J_{m+1}(z) = J_m^+(z) - \frac{m+\tfrac{1}{2}}{z} J_m^\times(z) ,
\end{equation}
we find the following expressions for the components of the t.e.v.~$\braket{:T_{\halpha\hsigma}:}_\beta ^{\text{sp}}$ relative to the tetrad (\ref{eq:tetrad}):
\begin{widetext}
\begin{subequations}\label{eq:spec_SET}
\begin{align}
 \braket{:T_{\hatt\hatt}:}_\beta^{\text{sp}} =& \sum_{m =0}^\infty \sum_{\ell = 1}^\infty \int_0^\infty
 \frac{E \, dk}{\pi^2 R^2} \frac{w(\Et) + w(\Ew)}{J_{m+1}^2(qR)} J_m^+(q\rho),\label{eq:spec_ttt}\\
 \braket{:T_{\hrho\hrho}:}_\beta^{\text{sp}} =& \sum_{m =0}^\infty \sum_{\ell = 1}^\infty \int_0^\infty
 \frac{q^2dk}{E \pi^2 R^2} \frac{w(\Et) + w(\Ew)}{J_{m+1}^2(qR)}
 \left[J_m^+(q\rho) - \frac{m + \tfrac{1}{2}}{qR} J_m^\times(q\rho)\right],\label{eq:spec_trr}\\
 \braket{:T_{\hvarphi\hvarphi}:}_\beta^{\text{sp}} =& \sum_{m =0}^\infty \sum_{\ell = 1}^\infty \int_0^\infty
 \frac{q \,dk}{\rho E \pi^2 R^2} \frac{w(\Et) + w(\Ew)}{J_{m+1}^2(qR)}
 (m + \tfrac{1}{2}) J_m^\times(q\rho),\label{eq:spec_tpp}\\
 \braket{:T_{\hatz\hatz}:}_\beta^{\text{sp}} =& \sum_{m =0}^\infty \sum_{\ell = 1}^\infty \int_0^\infty
 \frac{k^2 dk}{E \pi^2 R^2} \frac{w(\Et) + w(\Ew)}{J_{m+1}^2(qR)} J_m^+(q\rho),\label{eq:spec_tzz}\\
 \braket{:T_{\hatt\hvarphi}:}_\beta^{\text{sp}} =& -\sum_{m =0}^\infty \sum_{\ell = 1}^\infty \int_0^\infty
 \frac{dk}{\rho \pi^2 R^2} \frac{w(\Et) - w(\Ew)}{J_{m+1}^2(qR)}
\left[(m + \tfrac{1}{2}) J_m^+(q\rho) - \tfrac{1}{2} J_m^-(q\rho) + q\rho J_m^\times(q\rho)\right] .
 \label{eq:spec_ttp}
\end{align}
\end{subequations}
\end{widetext}
Eqs.~\eqref{eq:spec_SET} can be used to check the identity:
\begin{equation}\label{eq:Ttrace}
 \braket{:T\indices{^\halpha_\halpha}:}_\beta = -\mu \braket{:\psibar\psi:}_\beta.
\end{equation}

In Fig.~\ref{fig:spec_SET} we have plotted the t.e.v.~$\braket{:T_{\hatt\hatt}:}_\beta ^{\text {sp}}$ (\ref{eq:spec_ttt}) for a range of values of the
inverse temperature $\beta $, angular speed $\Omega $ and fermion mass $\mu $. Other components of (\ref{eq:spec_SET}) are discussed in Sec.~\ref{sec:num}.
As was observed earlier for the FC and CC,
if the angular speed $\Omega $ or temperature $\beta ^{-1}$ increases with the other parameters fixed, then the t.e.v.~$\braket{:T_{\hatt\hatt}:}_\beta ^{\text {sp}}$ also increases.
It is finite everywhere inside and on the boundary, including in the case where $\Omega R=1$ and the boundary is on the SOL.
When $\Omega R=1$, in Figs.~\ref{fig:spec_SET} (a) and \ref{fig:spec_SET} (b), we see a large peak in $\braket{:T_{\hatt\hatt}:}_\beta ^{\text {sp}}$ close to the boundary.
Fig.~\ref{fig:spec_SET} (d) shows that $\braket{:T_{\hatt\hatt}:}_\beta ^{\text {sp}}$ decreases as the fermion mass increases with the other parameters fixed.
Also in Fig.~\ref{fig:spec_SET} (d), we have plotted for comparison the t.e.v.~of this component of the SET for the unbounded Minkowski space-time.  The components of the t.e.v.~of the SET in this case are \cite{art:rotunb}:
\begin{subequations}\label{eq:rotunb_SET}
\begin{align}
 \braket{:T_{\hatt\hatt}:}_{\beta, I}^{\text{unb}} =&
 \frac{7\pi^2}{60\beta^4\varepsilon^3}\left(\tfrac{4}{3} - \tfrac{1}{3} \varepsilon\right)
 + \frac{\Omega^2}{8\beta^2\varepsilon^4}\left(\tfrac{8}{3} - \tfrac{16}{9}\varepsilon +
 \tfrac{1}{9} \varepsilon^2 \right),
 \label{eq:rotunb_ttt}\\
 \braket{:T_{\hvarphi\hatt}:}_{\beta, I}^{\text{unb}} =& -\rho\Omega \left[\frac{7\pi^2}{45\beta^4 \varepsilon^3} +
 \frac{2\Omega^2}{9\beta^2\varepsilon^4}\left(\tfrac{3}{2} - \tfrac{1}{2}\varepsilon\right)\right],
 \label{eq:rotunb_tpt}\\
 \braket{:T_{\hrho\hrho}:}_{\beta, I}^{\text{unb}} =& \frac{7\pi^2}{180\beta^4 \varepsilon^2} +
 \frac{\Omega^2}{24\beta^2\varepsilon^3} \left(\tfrac{4}{3} - \tfrac{1}{3} \varepsilon\right),
 \label{eq:rotunb_trr}\\
 \braket{:T_{\hvarphi\hvarphi}:}_{\beta ,I}^{\text{unb}} =& \frac{7\pi^2}{180\beta^4\varepsilon^3} \left(4 - 3\varepsilon\right)
 + \frac{\Omega^2}{24\beta^2\varepsilon^4}\left(8 - 8\varepsilon + \varepsilon^2\right),\label{eq:rotunb_tpp}
\end{align}
and
\begin{equation}
 \braket{:T_{\hatz\hatz}:}_{\beta ,I}^{\text{unb}} = \braket{:T_{\hrho\hrho}:}_{\beta ,I}^{\text{unb}}.\label{eq:rotunb_tzz}
\end{equation}
\end{subequations}
Fig.~\ref{fig:spec_SET} (d) shows that, for massless fermions at least (in this high-temperature case), the t.e.v.~$\braket{:T_{\hatt\hatt}:}_\beta ^{\text {sp}}$ with spectral boundary conditions is very close to that on the unbounded space-time except in a region close to the boundary.

\subsection{MIT bag model}\label{sec:MIT_tevs}
The method employed for the spectral model in the previous section can be applied also for the MIT bag model.
An expression for the difference between the thermal and vacuum Hadamard functions, equivalent to Eq.~\eqref{eq:Had_M}, can be written for the MIT case: \begin{equation}
\label{eq:deltaS_MIT}
 \Delta S^{(1)}_\beta(x,x') = -\sum_j \frac{\abs{{\mathcal {C}}_j^{\text {MIT}}}^2}{4\pi^2} e^{-i\Et_j \Delta t + ik_j \Delta z} ( w_j - w_{\wj}) M_j,
\end{equation}
where $\Et _{j}$ is the corotating energy, $\Delta t=t-t'$, $\Delta z = z-z'$, and the normalization constant ${\mathcal {C}}_{j}^{\text {MIT}}$ is given in (\ref{eq:b-}).
The sum over $j$ in this case is defined in Eq.~\eqref{eq:sumj_MIT}.
The thermal factors $w_j$ and $w_{\wj }$ are given by (\ref{eq:wj_def}) with the indices $j$ and $\wj$ in Eqs.~(\ref{eq:jdef_MIT}) and (\ref{eq:wjdef_MIT}) respectively.
The matrix $M_j$ now takes the form:
\begin{equation}
 M_j = \mathsf{b}_j^2 u^+_j \otimes \overline{u}^{\,+}_j + \mathsf{b}_j(u^+_j \otimes \overline{u}^{\,-}_j +
 u^-_j \otimes \overline{u}^{\,+}_j) + u^-_j \otimes \overline{u}^{\,-}_j.
\label{eq:MIT_uu}
\end{equation}
The superscripts $\pm$ indicate the sign of the helicity and the quantity $\mathsf{b}$ is given in (\ref{eq:MIT_beta}).
As in Sec.~\ref{sec:had}, it is understood that the spinors on the left and
right of the direct product symbol $\otimes$ depend on $x$ and $x'$, respectively.

To find $M_{j}$, we start with the following results:
\begin{align}
 \frac{u^\pm_j \otimes \overline{u}^{\,\pm}_j}{\abs{{\mathcal {C}}_{j}^{\text {MIT}}}^2} =&
 \frac{1}{2}\begin{pmatrix}
  \mathsf{E}_+^2 & \mp \frac{E}{\abs{E}} \mathsf{E}_- \mathsf{E}_+\\
  \pm \frac{E}{\abs{E}} \mathsf{E}_- \mathsf{E}_+ & -\mathsf{E}_-^2
 \end{pmatrix} \otimes
 \left[\phi^\pm_j \otimes \phi^{\pm\,\dagger}_j\right],\nonumber\\
 \frac{u^\pm_j \otimes \overline{u}^{\,\mp}_j}{\abs{{\mathcal {C}}_{j}^{\text {MIT}}}^2} =&
 \frac{1}{2}\begin{pmatrix}
  \mathsf{E}_+^2 & \pm \frac{E}{\abs{E}} \mathsf{E}_- \mathsf{E}_+\\
  \pm \frac{E}{\abs{E}} \mathsf{E}_- \mathsf{E}_+ & \mathsf{E}_-^2
 \end{pmatrix} \otimes
 \left[\phi^\pm_j \otimes \phi^{\mp\,\dagger}_j\right],
\end{align}
where $\mathsf{E}_{\pm }$ are given in (\ref{eq:Edef}).
The spinors $\phi _{j}^{\pm }$ can be found in (\ref{eq:phi}), and the $\phi _{j}^{\pm }$ have the argument $x$, while their Hermitian conjugates have the argument $x'$.
Using Eq.~\eqref{eq:phi}, the direct products of the $\phi _{j}^{\pm }$ two-spinors can be written as:
\begin{align}
 \phi^\pm_j \otimes \phi^{\pm\,\dagger}_j =& \frac{1}{2}
 \begin{pmatrix}
  \mathsf{p}_\pm^2 & \pm {\mathsf{p}}_- \mathsf{p}_+ \\
  \pm \mathsf{p}_- \mathsf{p}_+ & \mathsf{p}_\mp^2
 \end{pmatrix} \circ \mathcal{M}_j,\nonumber\\
 \phi^\pm_j \otimes \phi^{\mp\,\dagger}_j =& \frac{1}{2}
 \begin{pmatrix}
  {\mathsf{p}}_+ {\mathsf{p}}_- & \mp {\mathsf{p}}_\pm^2\\
  \pm \mathsf{p}_\mp^2 & - \mathsf{p}_+ \mathsf{p}_-
 \end{pmatrix}\circ \mathcal{M}_j,
 \label{eq:pp}
\end{align}
where $\mathsf{p}_{\pm }$ can be found in  (\ref{eq:pdef}) and Eq.~\eqref{eq:M_struct} gives the matrix $\mathcal{M}_j$.
Next, $M_j$  (\ref{eq:MIT_uu}) can be written in a manner similar to Eq.~\eqref{eq:spec_M_bblocks}:
\begin{equation}
 M_j = \frac{1}{2}
 \begin{pmatrix}
  M_j^{\text{up}} &
  -M_{j,-}^{\times}\\
  M_{j,+}^{\times} &
  -M_j^{\text{down}}
 \end{pmatrix},
\end{equation}
where
\begin{widetext}
\begin{align}
 M_j^{\text{up}} =& \frac{\mathsf{E}_+^2}{2}
 \begin{pmatrix}
  (\mathsf{b} \mathsf{p}_+ + \mathsf{p}_-)^2 &
  (\mathsf{b} \mathsf{p}_+ + \mathsf{p}_-) (\mathsf{b} \mathsf{p}_- - \mathsf{p}_+) \\
  (\mathsf{b} \mathsf{p}_+ + \mathsf{p}_-) (\mathsf{b} \mathsf{p}_- - \mathsf{p}_+) &
  (\mathsf{b} \mathsf{p}_- - \mathsf{p}_+)^2
 \end{pmatrix} \circ \mathcal{M}_j,\nonumber\\
 M_j^{\text{down}} =& \frac{\mathsf{E}_-^2}{2}
 \begin{pmatrix}
  (\mathsf{b} \mathsf{p}_+ - \mathsf{p}_-)^2 &
  (\mathsf{b} \mathsf{p}_+ - \mathsf{p}_-) (\mathsf{b} \mathsf{p}_- + \mathsf{p}_+)\\
  (\mathsf{b} \mathsf{p}_+ - \mathsf{p}_-) (\mathsf{b} \mathsf{p}_- + \mathsf{p}_+) &
  (\mathsf{b} \mathsf{p}_- + \mathsf{p}_+)^2
 \end{pmatrix} \circ \mathcal{M}_j,\nonumber\\
 M_{j,\pm}^{\times} =& \frac{E}{2\abs{E}} \mathsf{E}_+ \mathsf{E}_-
 \begin{pmatrix}
  (\mathsf{b}^2 \mathsf{p}_+^2 - \mathsf{p}_-^2) &
  (\mathsf{b} \mathsf{p}_+ \mp \mathsf{p}_-)(\mathsf{b} \mathsf{p}_- \mp \mathsf{p}_+) \\
  (\mathsf{b} \mathsf{p}_+ \pm \mathsf{p}_-)(\mathsf{b} \mathsf{p}_- \pm \mathsf{p}_+)&
  (\mathsf{b}^2 \mathsf{p}_-^2 - \mathsf{p}_+^2)
 \end{pmatrix} \circ \mathcal{M}_j.\label{eq:MIT_Mbblocks}
\end{align}
In the above, the Hadamard product $\circ$ is taken with the matrix $\mathcal{M}_j$ defined in Eq.~\eqref{eq:M_struct}.
Using the result \eqref{eq:MIT_beta} for $\mathsf{b}$, the following identities can be established:
\begin{equation}
 \mathsf{b} = \frac{2\varsigma E}{p} \frac{\mathsf{j}}{\mathsf{p}_+^2 \mathsf{j}^2 + \mathsf{p}_-^2}, \qquad
 \mathsf{b}^2 + 1 = \frac{2(\mathsf{j}^2 + 1)}{\mathsf{p}_+^2 \mathsf{j}^2 + \mathsf{p}_-^2}, \qquad
 \mathsf{b}^2 - 1 = -\frac{2k}{p} \frac{\mathsf{j}^2 - 1}{\mathsf{p}_+^2 \mathsf{j}^2 + \mathsf{p}_-^2},
\end{equation}
where $\varsigma = \pm 1$ and $\mathsf{j}=\mathsf{j}_{m\ell}$ is in Eq.~(\ref{eq:MIT_x_def}).
Thus, the matrices introduced in \eqref{eq:MIT_Mbblocks}
can be put in the form:
\begin{align}
 M_j^{\text{up}} =& \frac{\mathsf{E}_+^2}{\mathsf{p}_+^2 \mathsf{j}^2 + \mathsf{p}_-^2}
 \begin{pmatrix}
  \mathsf{j}^2 + 1 - \frac{k^2}{p^2} (\mathsf{j}^2 - 1) + \frac{2\varsigma q E}{p^2} \mathsf{j} &
  -\frac{kq}{p^2}\left(\mathsf{j}^2 - 1 + \frac{2\varsigma E}{q} \mathsf{j}\right) \\
  -\frac{kq}{p^2}\left(\mathsf{j}^2 - 1 + \frac{2\varsigma E}{q} \mathsf{j}\right) &
  \mathsf{j}^2 + 1 + \frac{k^2}{p^2} (\mathsf{j}^2 - 1) - \frac{2\varsigma q E}{p^2} \mathsf{j}
 \end{pmatrix} \circ \mathcal{M}_j,\nonumber\\
 M_j^{\text{down}} =& \frac{\mathsf{E}_-^2}{\mathsf{p}_+^2 \mathsf{j}^2 + \mathsf{p}_-^2}
 \begin{pmatrix}
  \mathsf{j}^2 + 1 - \frac{k^2}{p^2} (\mathsf{j}^2 - 1) - \frac{2\varsigma q E}{p^2} \mathsf{j} &
  -\frac{kq}{p^2}\left(\mathsf{j}^2 - 1 - \frac{2\varsigma E}{q} \mathsf{j}\right) \\
  -\frac{kq}{p^2}\left(\mathsf{j}^2 - 1 - \frac{2\varsigma E}{q} \mathsf{j}\right) &
  \mathsf{j}^2 + 1 + \frac{k^2}{p^2} (\mathsf{j}^2 - 1) + \frac{2\varsigma q E}{p^2} \mathsf{j}
 \end{pmatrix} \circ \mathcal{M}_j,\nonumber\\
 M_{j,\pm}^{\times} =& \frac{1}{\mathsf{p}_+^2 \mathsf{j}^2 + \mathsf{p}_-^2}
 \begin{pmatrix}
  \frac{2k}{E} &
  \frac{q}{E}(\mathsf{j}^2 + 1) \mp 2\varsigma \mathsf{j} \\
  \frac{q}{E}(\mathsf{j}^2 + 1) \pm 2\varsigma \mathsf{j} &
  - \frac{2k}{E} \mathsf{j}^2
 \end{pmatrix} \circ \mathcal{M}_j.
 \label{eq:MIT_M}
\end{align}
We can alternatively write $M_j$ in terms of the Pauli matrices $\sigma _{1}$, $\sigma _{2}$, $\sigma _{3}$ (\ref{eq:Pauli}):
\begin{multline}\label{eq:MIT_M_symm}
 M_j = \frac{1}{\mathsf{p}_+^2 \mathsf{j}^2 + \mathsf{p}_-^2} \left\{\frac{1}{2E} {\mathbb {I}}_2 \otimes
 \left[\begin{pmatrix}
  \mu(\mathsf{j}^2 + 1) + \frac{2\varsigma}{q} (\mu^2 + q^2) \mathsf{j} &  E\\
  E & \mu(\mathsf{j}^2 + 1) - \frac{2\varsigma}{q} (\mu^2 + q^2) \mathsf{j}
 \end{pmatrix} \circ \mathcal{M}_j\right]\right.\\
 \left. + \sigma_3 \otimes
 \left[\begin{pmatrix}
  1 & 0\\
  0 & \mathsf{j}^2
 \end{pmatrix} \circ \mathcal{M}_j\right] - \frac{i}{2E} \sigma_2 \otimes
 \left[\begin{pmatrix}
  2k & q(\mathsf{j}^2 + 1)\\
  q(\mathsf{j}^2 + 1) & -2k
 \end{pmatrix} \circ \mathcal{M}_j\right] - \varsigma \mathsf{j} \sigma_1 \otimes
 \left[\begin{pmatrix}
  0 & 1\\
  -1 & 0
 \end{pmatrix} \circ \mathcal{M}_j\right]
 \right\} ,
\end{multline}
where ${\mathbb {I}}_{2}$ is the $2\times 2 $ identity matrix.
\subsubsection{Fermion condensate}\label{sec:MIT_tev_FC}
\begin{figure*}
\begin{tabular}{cc}
\includegraphics[width=0.475\columnwidth]{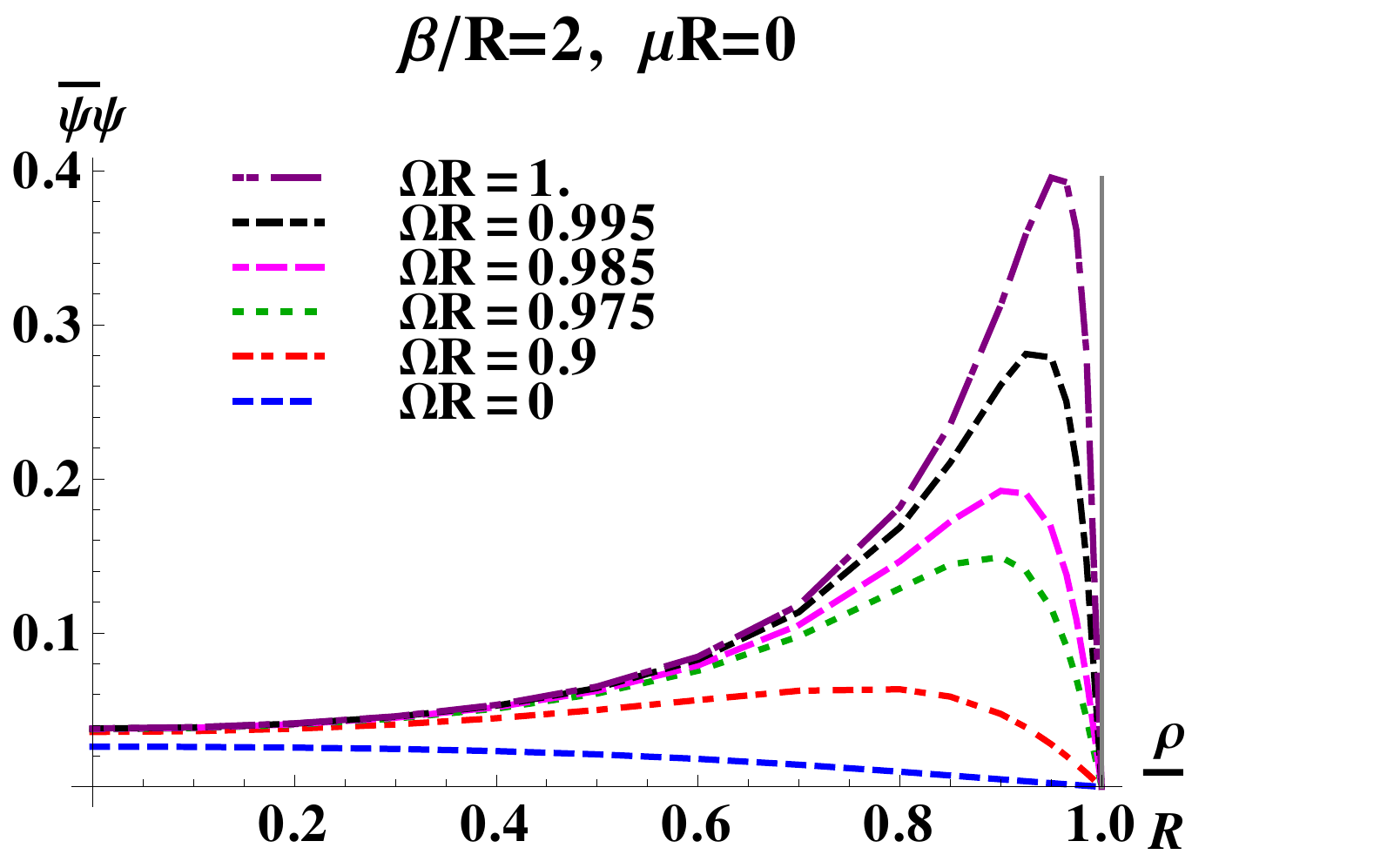} &
\includegraphics[width=0.475\columnwidth]{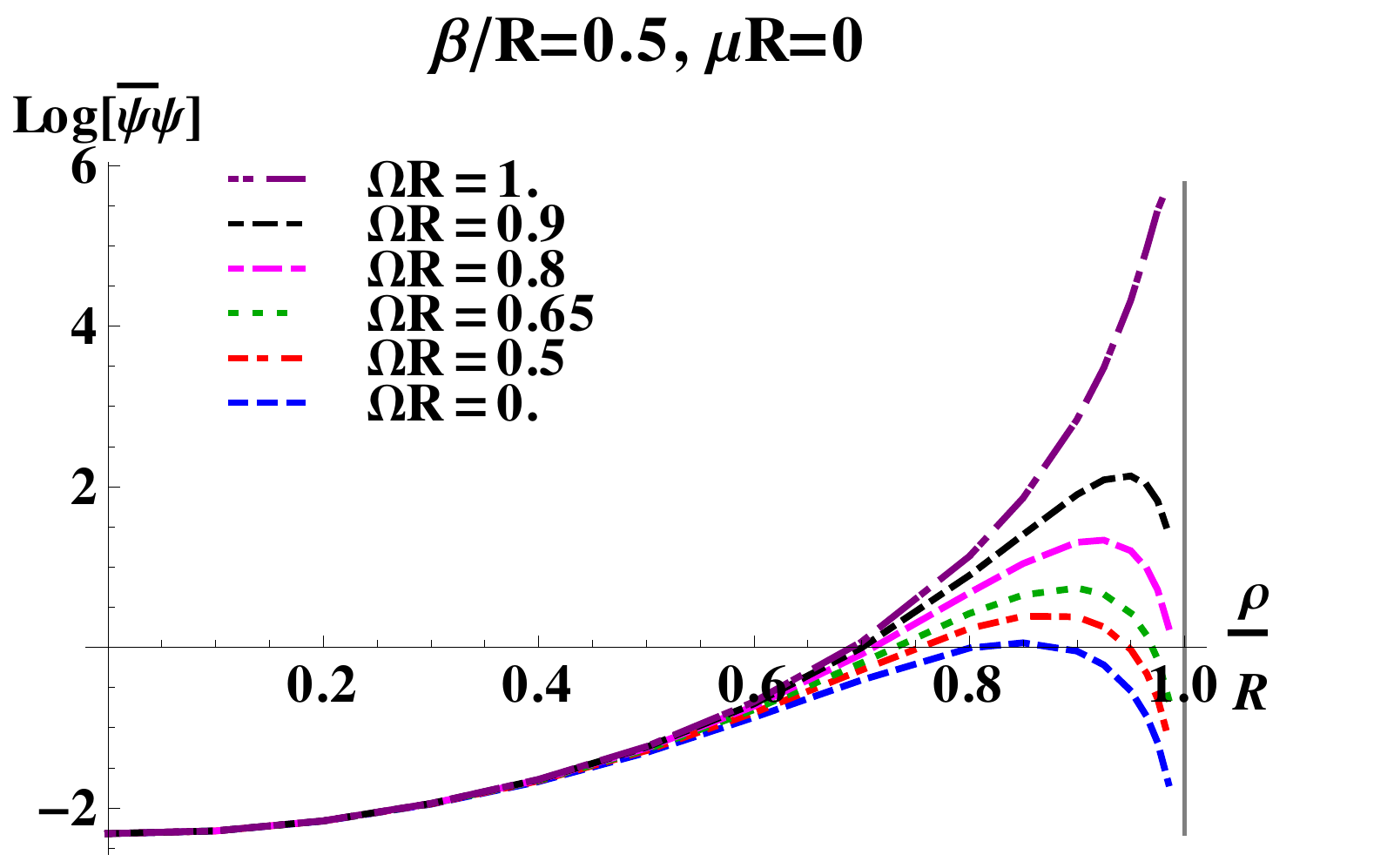}\\
(a) & (b) \\
\includegraphics[width=0.475\columnwidth]{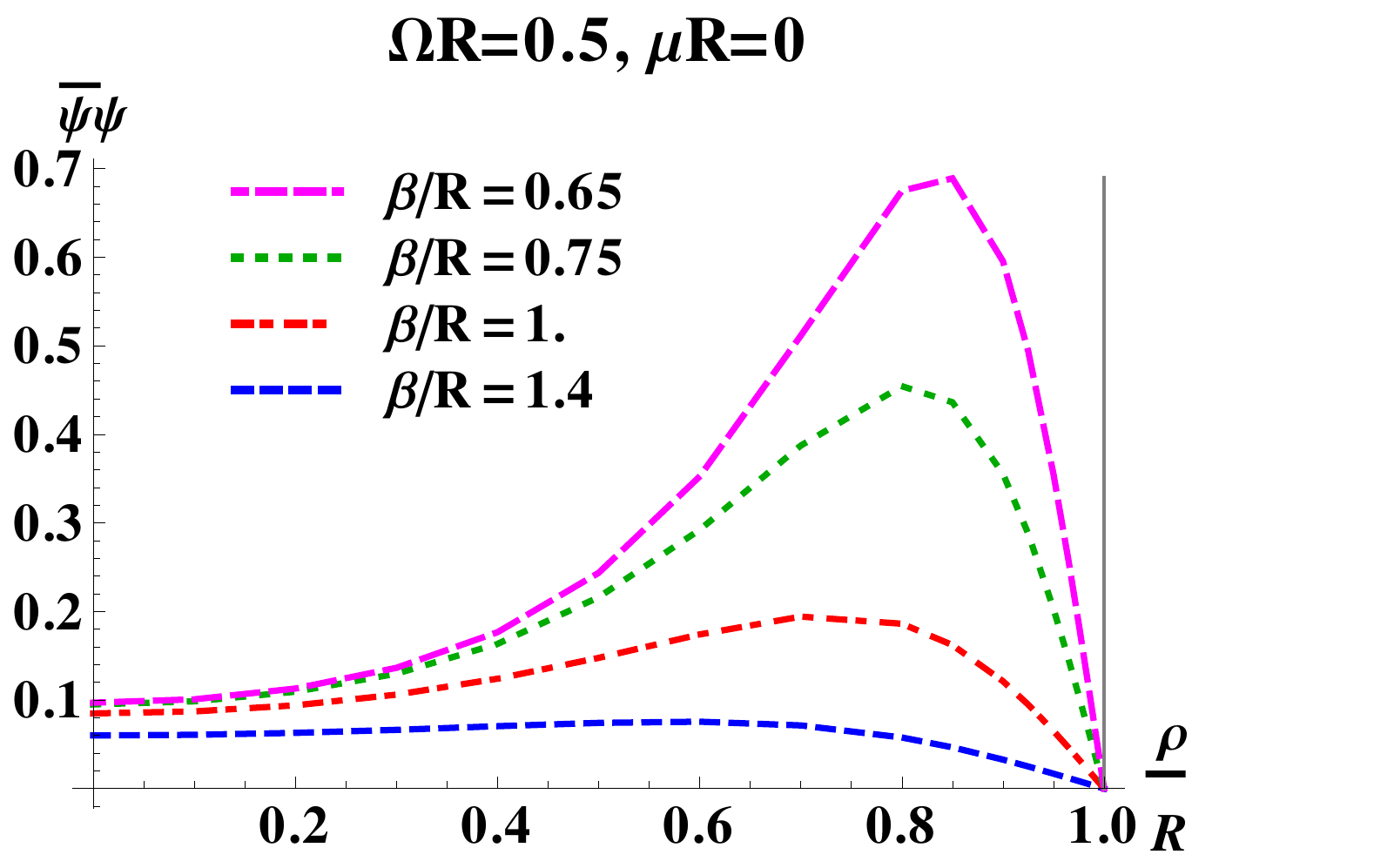} &
\includegraphics[width=0.475\columnwidth]{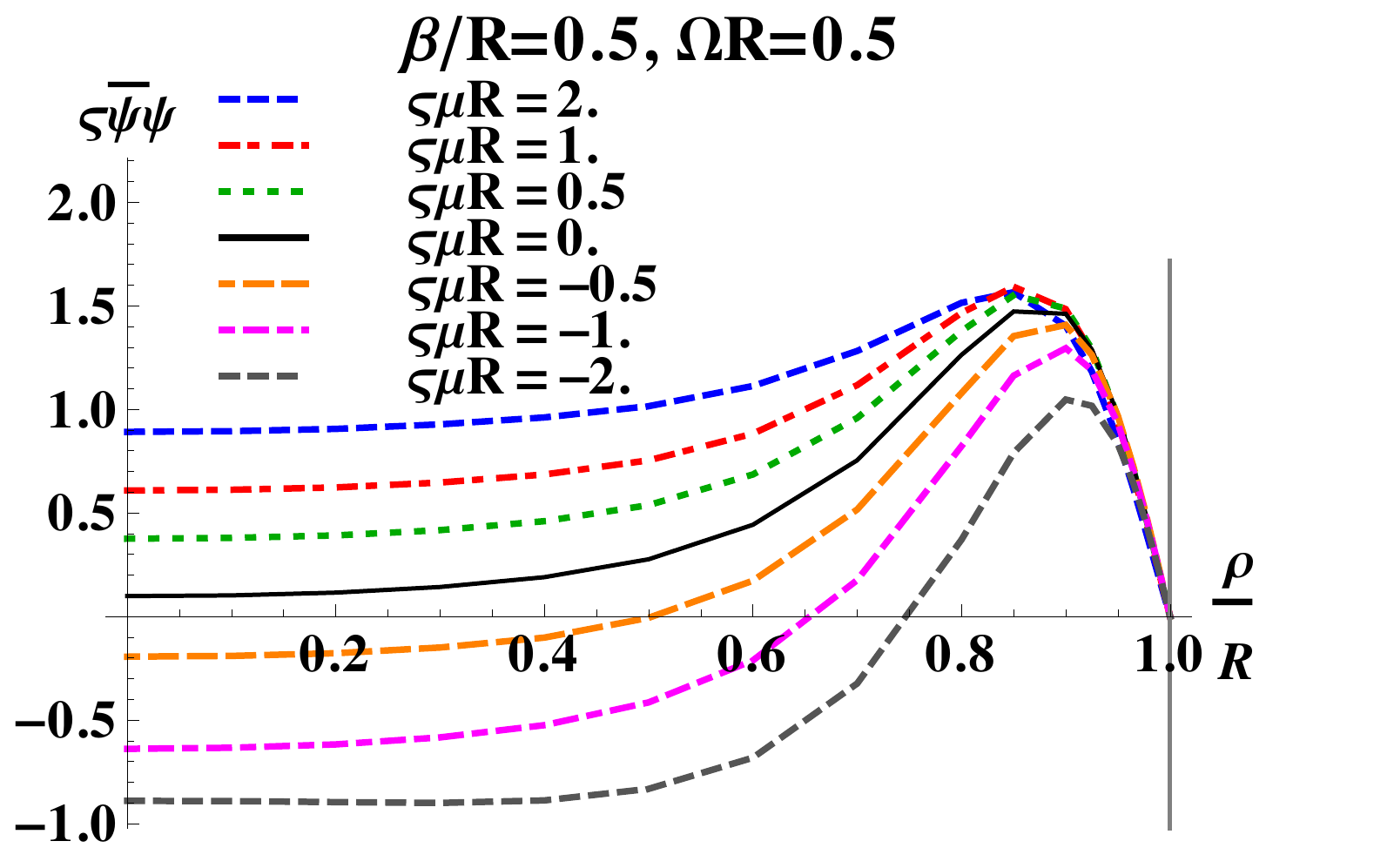} \\
(c) & (d)
\end{tabular}
\caption{Thermal expectation values (t.e.v.s) of the FC $\braket{:\psibar\psi:}_\beta^{\text{MIT}}$
(\ref{eq:MIT_ppsi}) for MIT bag boundary conditions, as a function of the scaled radial coordinate $\rho /R$, so that the boundary of the cylinder
is at $\rho /R=1$.
(a) Massless fermions $\mu =0$, fixed inverse temperature $\beta =2R$ and various values of the angular speed $\Omega $.
(b) Massless fermions $\mu =0$, fixed inverse temperature $\beta =0.5R$ and various values of the angular speed $\Omega $.
(c) Massless fermions $\mu =0$, fixed angular speed $\Omega = 0.5/R$ and various values of the inverse temperature $\beta $.
(d) Fixed inverse temperature $\beta = 0.5R$, fixed angular speed $\Omega = 0.5/R$ and various values of the fermion mass $\mu $.
Note that (b) has a logarithmic vertical scale, but a linear horizontal scale.
In (d) we have considered $\varsigma = \pm 1$ for each
value of the mass, while, in the massless case, the quantity
$\varsigma \braket{:\psibar\psi:}_\beta^{\text{MIT}}$
does not depend on $\varsigma$.
}
\label{fig:MIT_ppsi}
\end{figure*}

As in the case of spectral boundary conditions, the first t.e.v.~we consider for MIT bag boundary conditions is the FC $\braket{:\psibar\psi:}_\beta^{\text{MIT}}$, evaluated from (\ref{eq:deltaS_MIT}) via (\ref{eq:ppsi}).
Only the term containing ${\mathbb{I}}_2$ on the right-hand-side of Eq.~\eqref{eq:MIT_M_symm} contributes to the t.e.v.~of the FC \eqref{eq:ppsi}, giving:
\begin{equation}
 \braket{:\psibar\psi:}_\beta^{\text{MIT}} =
 \sum_{m = 0}^\infty \sum_{\ell = 1}^\infty \int_0^\infty \frac{dk}{2\mathcal{D}_{m\ell}^{\text{MIT}}}
 \left[w(\Et) + w(\Ew)\right] \left[\frac{\mu}{E}(\mathsf{j}^2 + 1)J_m^+(q\rho) +
 \frac{2\varsigma \mathsf{j}}{qE} (q^2 + \mu^2) J_m^-(q\rho)\right],
 \label{eq:MIT_ppsi}
\end{equation}
where $J^{\pm }$ are defined in (\ref{eq:BesselJ*}), the thermal factors can be found in (\ref{eq:spec_w}), their arguments can be found in (\ref{eq:Ew_def}) and the quantity $\mathsf{j}$ can be found in (\ref{eq:MIT_x_def}).
In Eq.~(\ref{eq:MIT_ppsi}), the term $\mathcal{D}_{m\ell}^{\text{MIT}}$ in the denominator is given by:
\begin{equation}
 \mathcal{D}_{m\ell}^{\text{MIT}} = \pi^2 R^2 J_{m+1}^2(qR)\left[(\mathsf{j}^2+1)\left(\mathsf{j}^2 + 1 -
 \frac{2m+1}{qR} \mathsf{j}\right) - \frac{\mathsf{j}}{qR} (\mathsf{j}^2 - 1)\right].
 \label{eq:MIT_D}
\end{equation}
\end{widetext}

In Fig.~\ref{fig:MIT_ppsi}, we plot the t.e.v.~of the FC $\braket{:\psibar\psi:}_\beta^{\text{MIT}}$ for various values of the parameters $\beta $, $\mu $ and $\Omega $ and both $\varsigma = \pm 1$.
For a massless fermion field, as discussed in Sec.~\ref{sec:MIT_energy}, the energy spectra
for $\varsigma = \pm 1$ are identical and therefore changing the sign of $\varsigma$ changes
only the sign of $\braket{:\psibar\psi:}_\beta^{\text{MIT}}$, without changing its magnitude.

The plots in Fig.~\ref{fig:MIT_ppsi} (a--c) are for $\mu =0$. They show many qualitative features similar to those in Fig.~\ref{fig:spec_ppsiomu} for spectral boundary conditions. In particular,  the t.e.v.s increase with increasing angular speed $\Omega $ for fixed inverse temperature $\beta $; there is a sharp peak near the boundary for $\Omega R=1$ but the t.e.v.s remain finite everywhere inside and on the boundary; and the t.e.v.s also increase as the temperature $\beta ^{-1}$ increases for fixed $\Omega $.

In Fig.~\ref{fig:MIT_ppsi} (d), we show the effect of varying the fermion mass $\mu $ and $\varsigma $ (\ref{eq:varsigma}).
Note that in Fig.~\ref{fig:MIT_ppsi} we have plotted $\varsigma \braket{: \psibar\psi:}_\beta^{\text{MIT}}$ rather than
$\braket{: \psibar\psi:}_\beta^{\text{MIT}}$.
It can be seen that increasing the fermion mass $\mu $ when $\varsigma = -1$ also increases
$\varsigma \braket{: \psibar\psi:}_\beta^{\text{MIT}}$ on the rotation axis.
When $\varsigma = 1$, the FC decreases on the rotation axis to negative values as $\mu$ is increased.

There are also some differences between the results in Fig.~\ref{fig:MIT_ppsi} for MIT bag boundary conditions and those in Fig.~\ref{fig:spec_ppsiomu} for
spectral boundary conditions.
In particular, the massless limit of the FC in the
MIT model is finite and nonzero, whereas, for spectral boundary conditions, the FC vanishes when the fermions are massless (see Sec.~\ref{sec:spec_tev_FC}). Furthermore, from (\ref{eq:MIT_ppsi}), the t.e.v.~of the FC vanishes on the boundary:
\begin{align}
 \left.\braket{:\psibar\psi:}_\beta^{\text{MIT}}\right\rfloor_{\rho = R} =&
 \sum_{m = 0}^\infty \sum_{\ell = 1}^\infty \int_0^\infty \frac{\mu\,dk}{2E \mathcal{D}_{m\ell}^{\text{MIT}}}
 \left[w(\Et) + w(\Ew)\right]
 \nonumber \\
 & \times \left[(\mathsf{j}^2 + 1)^2 -
 4\mathsf{j}^2 \left(1 + \frac{\mu^2}{q^2}\right)\right]\nonumber\\
 =& 0, \label{eq:MIT_ppsi_boundary}
\end{align}
where the last equality follows from using Eq.~\eqref{eq:MITq} to eliminate the $\mathsf{j}^4$ term.
Again, this feature is not present for spectral boundary conditions, when the t.e.v.~of the FC is finite (but in general nonzero) on the boundary.

\begin{figure*}
\begin{tabular}{cc}
 \includegraphics[width=0.95\columnwidth]{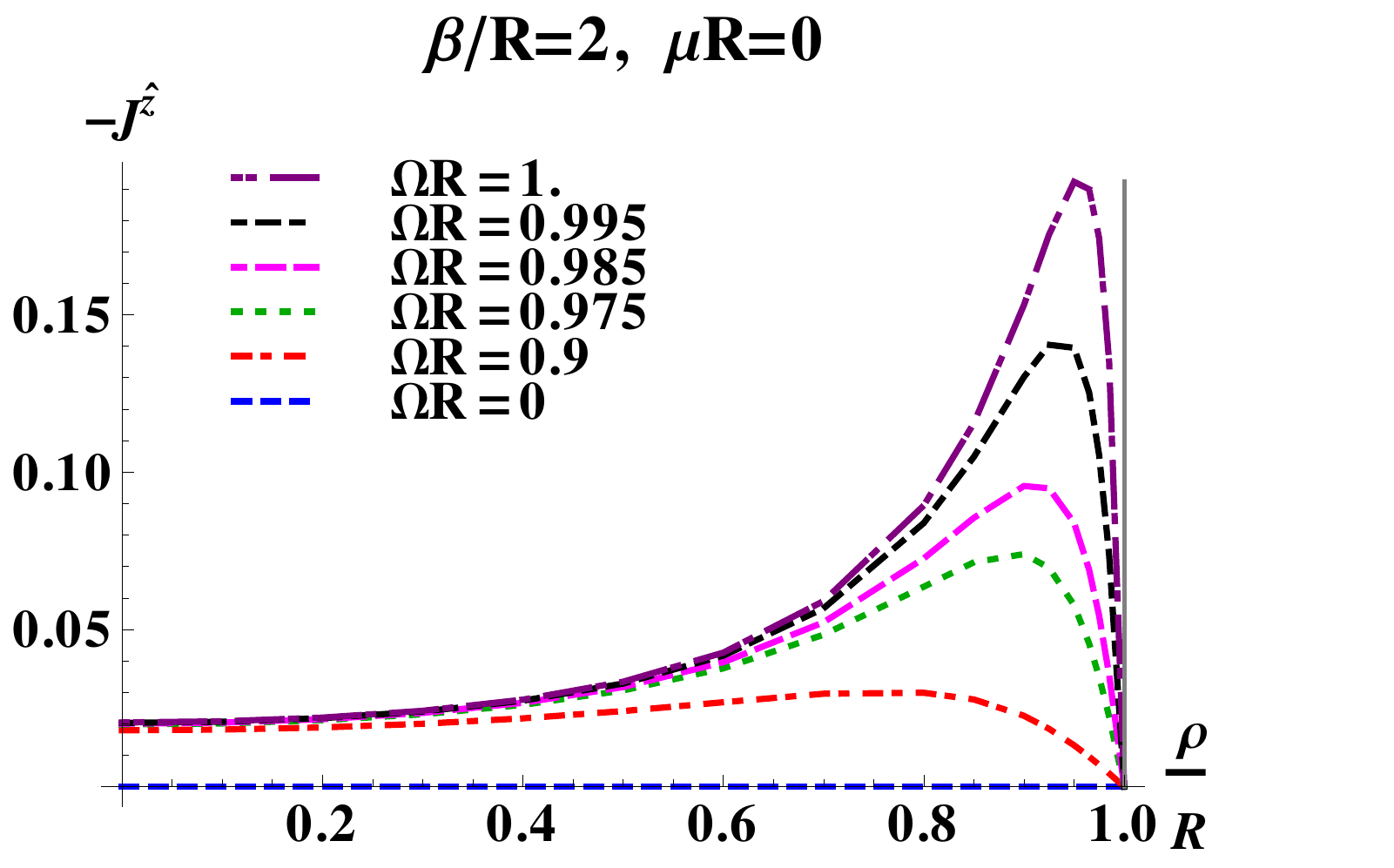} &
 \includegraphics[width=0.95\columnwidth]{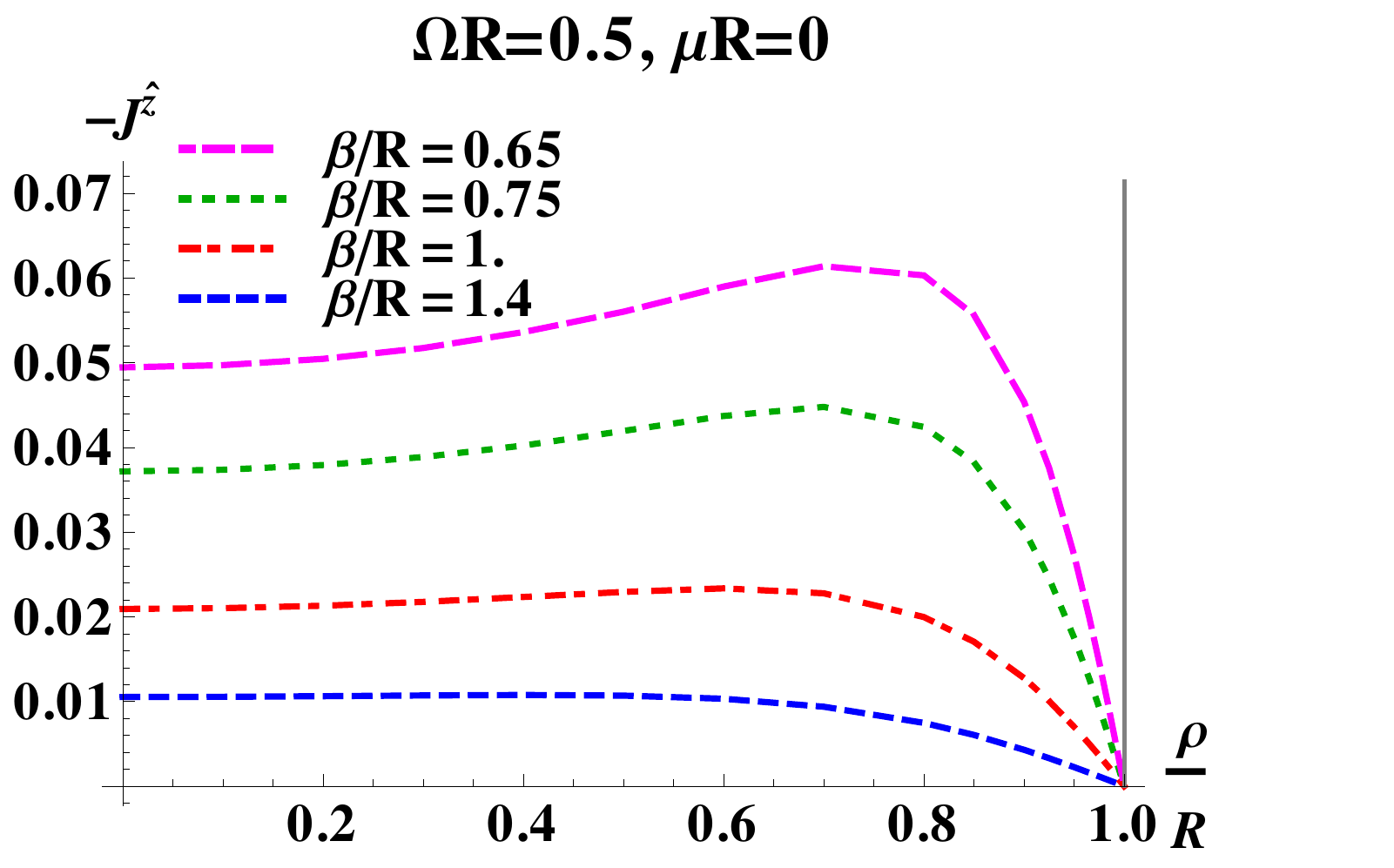}\\
 (a) & (b)\\
 \includegraphics[width=0.95\columnwidth]{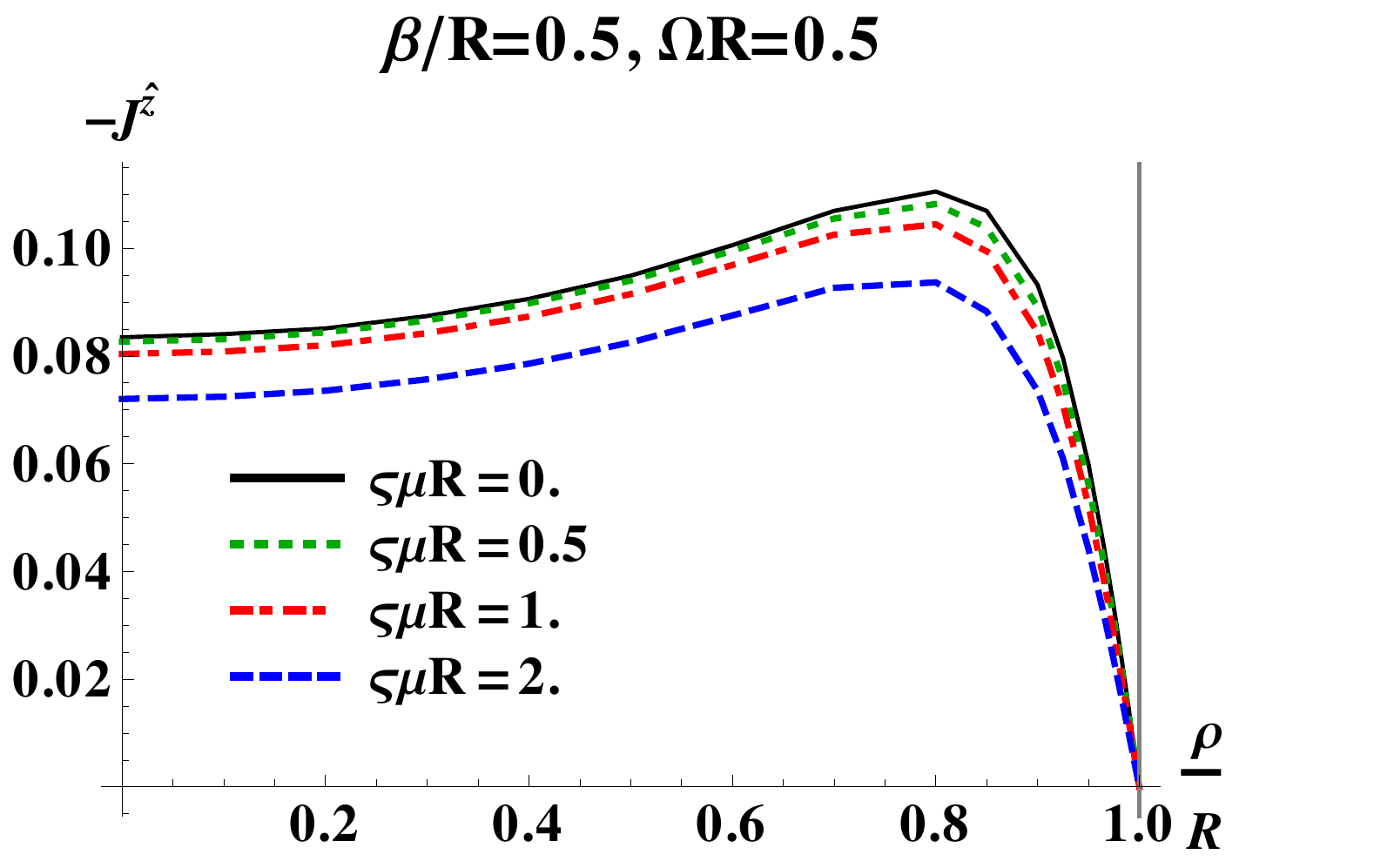} &
 \includegraphics[width=0.95\columnwidth]{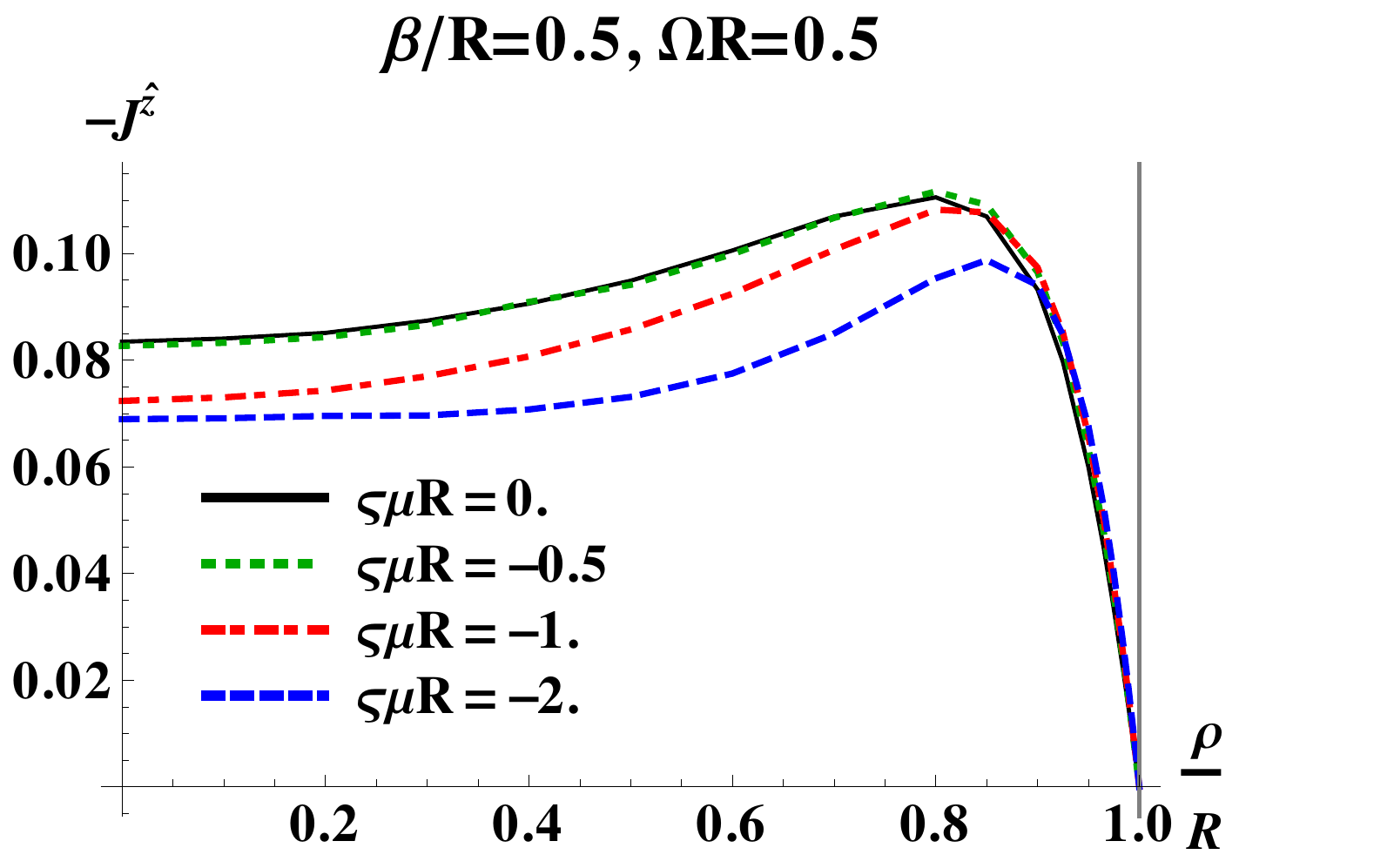}\\
 (c) & (d)
 \end{tabular}
\caption{Thermal expectation values (t.e.v.s) of the CC $\braket{:\current^{\hatz}_{\nu }:}_\beta^{\text{MIT}}$
(\ref{eq:MIT_Jnu}) for MIT bag boundary conditions, as a function of the scaled radial coordinate $\rho /R$, so that the boundary of the cylinder
is at $\rho /R=1$.
This expectation value is always negative, so we show $-\braket{:\current^{\hatz}_{\nu }:}_\beta^{\text{MIT}}$ in all plots.
(a) Massless fermions $\mu =0$, fixed inverse temperature $\beta =2R$ and various values of the angular speed $\Omega $.
(b) Massless fermions $\mu =0$, fixed angular speed $\Omega = 0.5/R$ and various values of the inverse temperature $\beta $.
(c--d) Fixed inverse temperature $\beta = 0.5R$, fixed angular speed $\Omega = 0.5/R$ and various values of the fermion mass $\mu $
for $\varsigma = 1$ (c) and $\varsigma = -1$ (d).
}
\label{fig:MIT_Jnuz}
\end{figure*}

\subsubsection{Neutrino charge current}\label{sec:MIT_tev_CC}
As in the spectral case, the t.e.v.s of all the components of the charge current \eqref{eq:J} vanish.
For the $t$, $\rho$ and $\varphi$ components, the summands are odd with respect to $m \rightarrow -m - 1$;
for the $z$ component the summand is odd under the transformation $k \rightarrow -k$.
The rules for checking the required transformation properties under $m \rightarrow -m- 1$ are,
using (\ref{eq:MITx}, \ref{eq:BesselJ*}, \ref{eq:spec_w}, \ref{eq:Ew_def}):
\begin{align}
 \mathsf{j} \rightarrow &-\frac{1}{\mathsf{j}}, \qquad
 m + \tfrac{1}{2} \rightarrow -m -\tfrac{1}{2}, \qquad
 J_m^\pm \rightarrow \pm J_m^\pm,
 \nonumber \\
 J_m^\times \rightarrow &-J_m^\times, \qquad
 w(\Et) \pm w(\Ew) \rightarrow \pm [w(\Et) \pm w(\Ew)].
\end{align}

The only nonvanishing component of the neutrino charge current (CC) \eqref{eq:Jnu} is, as in the spectral model case,
the $z$ component (see Eqs.~(\ref{eq:MIT_x_def}, \ref{eq:BesselJ*}, \ref{eq:MIT_D}) for the definitions of various quantities):
\begin{multline}
 \braket{:\current ^{\hatz}_\nu:}_\beta^{\text{MIT}} = -\sum_{m =0}^\infty \sum_{\ell = 1}^\infty \int_0^\infty
 \frac{dk}{4 \mathcal{D}_{m\ell}^{\text{MIT}}} \left[w(\Et) - w(\Ew)\right]
 \\
 \times
 \left[(\mathsf{j}^2 + 1) J_m^-(q\rho) - (\mathsf{j}^2 - 1) J_m^+(q\rho)\right].
 \label{eq:MIT_Jnu}
\end{multline}
In Fig.~\ref{fig:MIT_Jnuz} we illustrate the behaviour of $\braket{:\current^{\hatz}_\nu:}_\beta^{\text{MIT}}$.
We find that $\braket{:\current ^{\hatz}_\nu:}_\beta^{\text{MIT}}$ is negative everywhere, and hence in Fig.~\ref{fig:MIT_Jnuz} we plot $-\braket{:\current ^{\hatz}_\nu:}_\beta^{\text{MIT}}$, in contrast with the spectral case where $\braket{:\current^{\hatz}_\nu:}_\beta^{\text{sp}}$ is positive near the boundary.
We also see from Fig.~\ref{fig:MIT_Jnuz} that $\braket{:\current^{\hatz}_\nu:}_\beta^{\text{MIT}}$ vanishes on the boundary, and this can be verified analytically:
\begin{equation}
 \left.\braket{:\current^{\hatz}_\nu:}_\beta^{\text{MIT}}\right\rfloor_{\rho = R} = 0 .
\end{equation}
Again this is not the same behaviour as found in the case of spectral boundary conditions, when $\braket{:\current^{\hatz}_\nu:}_\beta^{\text{sp}}$ was found to be positive on the boundary.

For fixed inverse temperature $\beta $, we see in Fig.~\ref{fig:MIT_Jnuz} (a) that $-\braket{:\current^{\hatz}_\nu:}_\beta^{\text{MIT}}$ increases as the angular speed $\Omega $ increases.
As seen in previous figures, when the boundary is on the SOL, there is a large peak in $-\braket{:\current^{\hatz}_\nu:}_\beta^{\text{MIT}}$
close to the boundary, but the t.e.v.~remains finite everywhere inside and on the cylinder.
For fixed angular speed $\Omega $, Fig.~\ref{fig:MIT_Jnuz} (b) confirms our expectations that the absolute value of
the t.e.v.~$\braket{:\current^{\hatz}_\nu:}_\beta^{\text{MIT}}$
increases as the temperature $\beta ^{-1}$ increases.
Varying the mass of the fermion field with fixed inverse temperature $\beta $ and angular speed $\Omega $ does not alter the t.e.v.~of the CC very much, as can be seen in Figs.~\ref{fig:MIT_Jnuz} (c) and \ref{fig:MIT_Jnuz} (d).
When $\varsigma =1$, as the mass $\mu $ increases the magnitude of $\braket{:\current^{\hatz}_\nu:}_\beta^{\text{MIT}}$ decreases everywhere inside the boundary.
For $\varsigma = -1$, the magnitude of $\braket{:\current^{\hatz}_\nu:}_\beta^{\text{MIT}}$ decreases as $\mu $ increases apart from close to the boundary,
where the magnitude of $\braket{:\current^{\hatz}_\nu:}_\beta^{\text{MIT}}$ appears to be increasing.

\subsubsection{Stress-energy tensor}\label{sec:MIT_tev_SET}
\begin{figure*}
\begin{tabular}{cc}
\includegraphics[width=0.9\columnwidth]{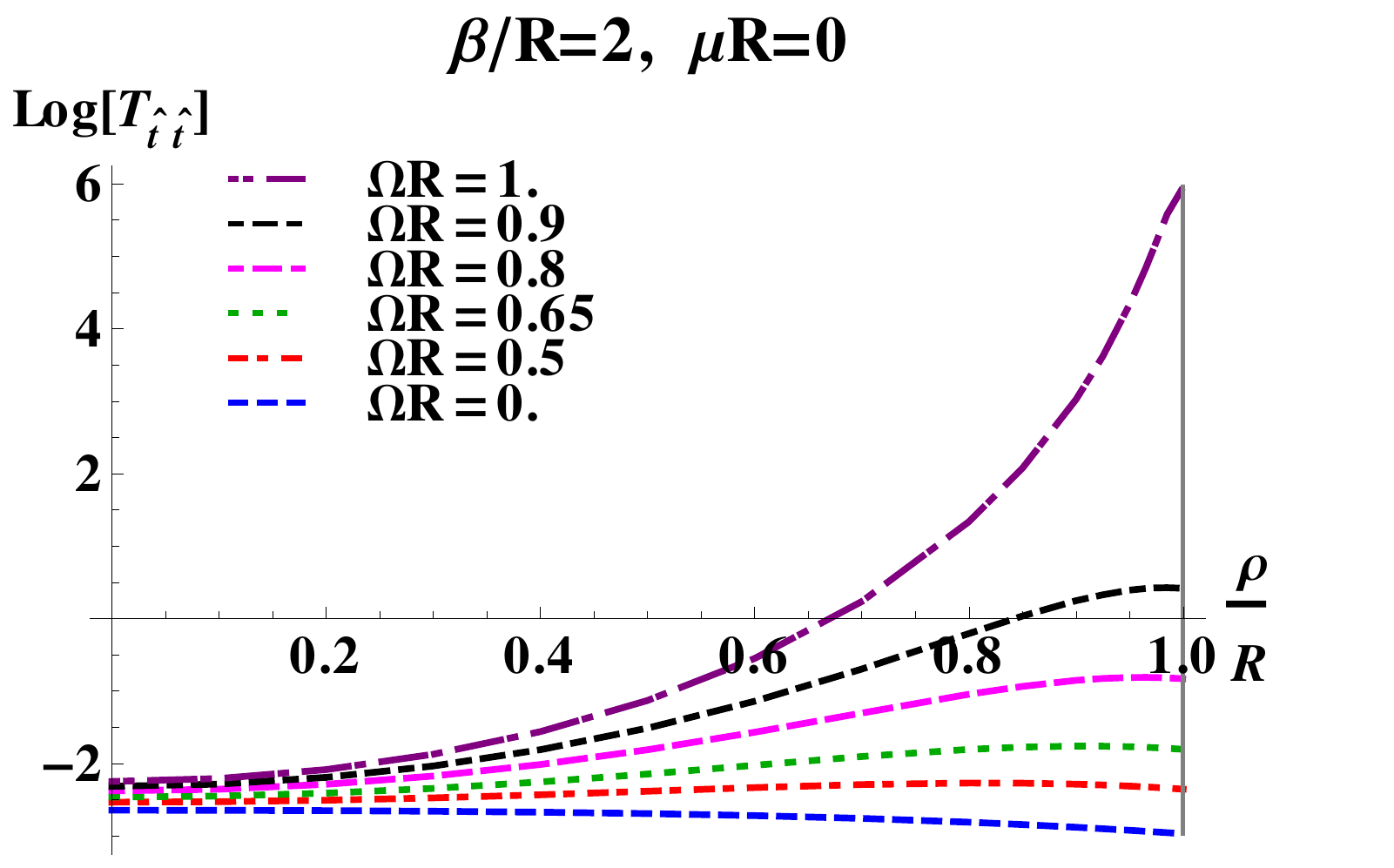} &
\includegraphics[width=0.9\columnwidth]{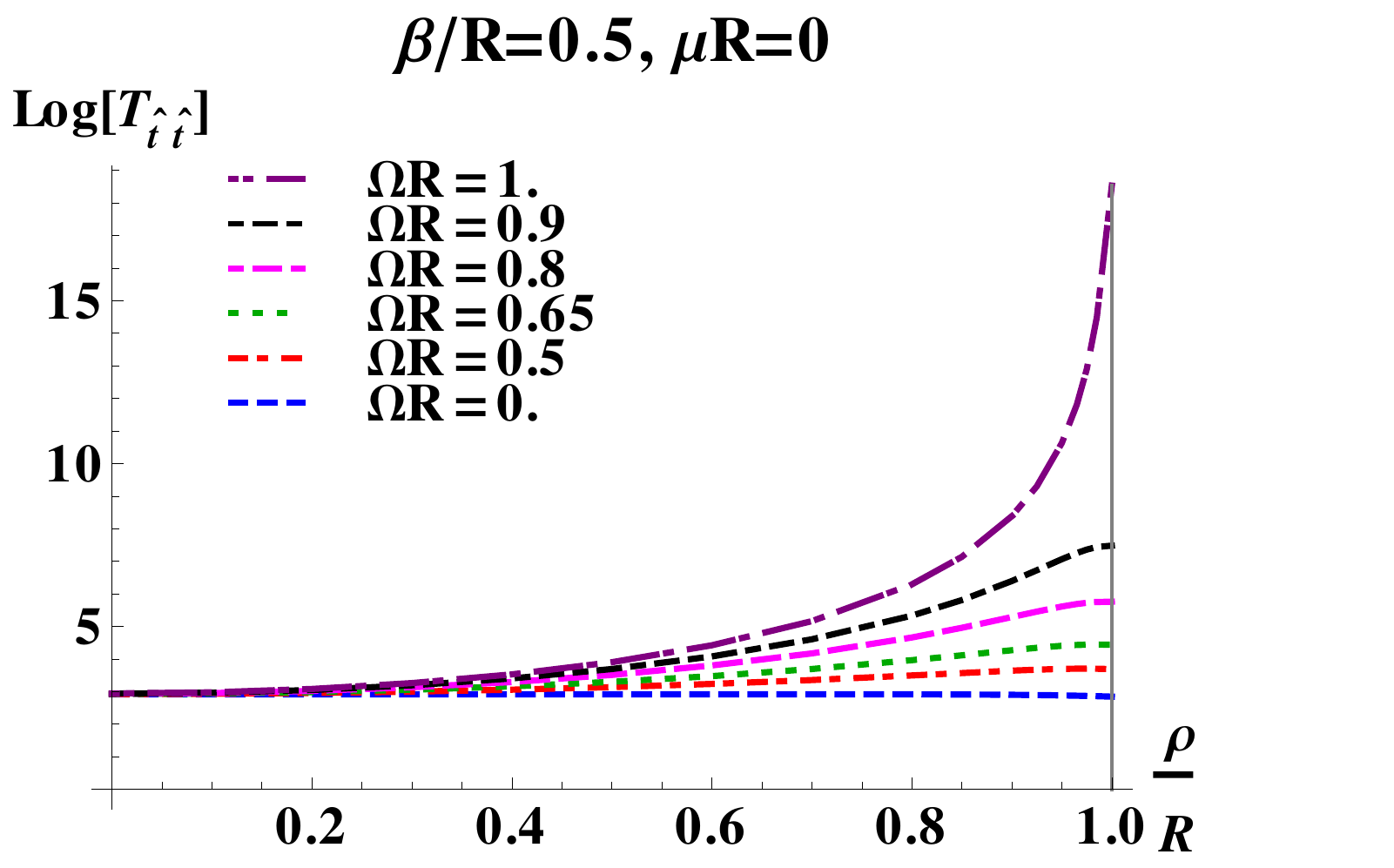}\\
(a) & (b) \\
\includegraphics[width=0.9\columnwidth]{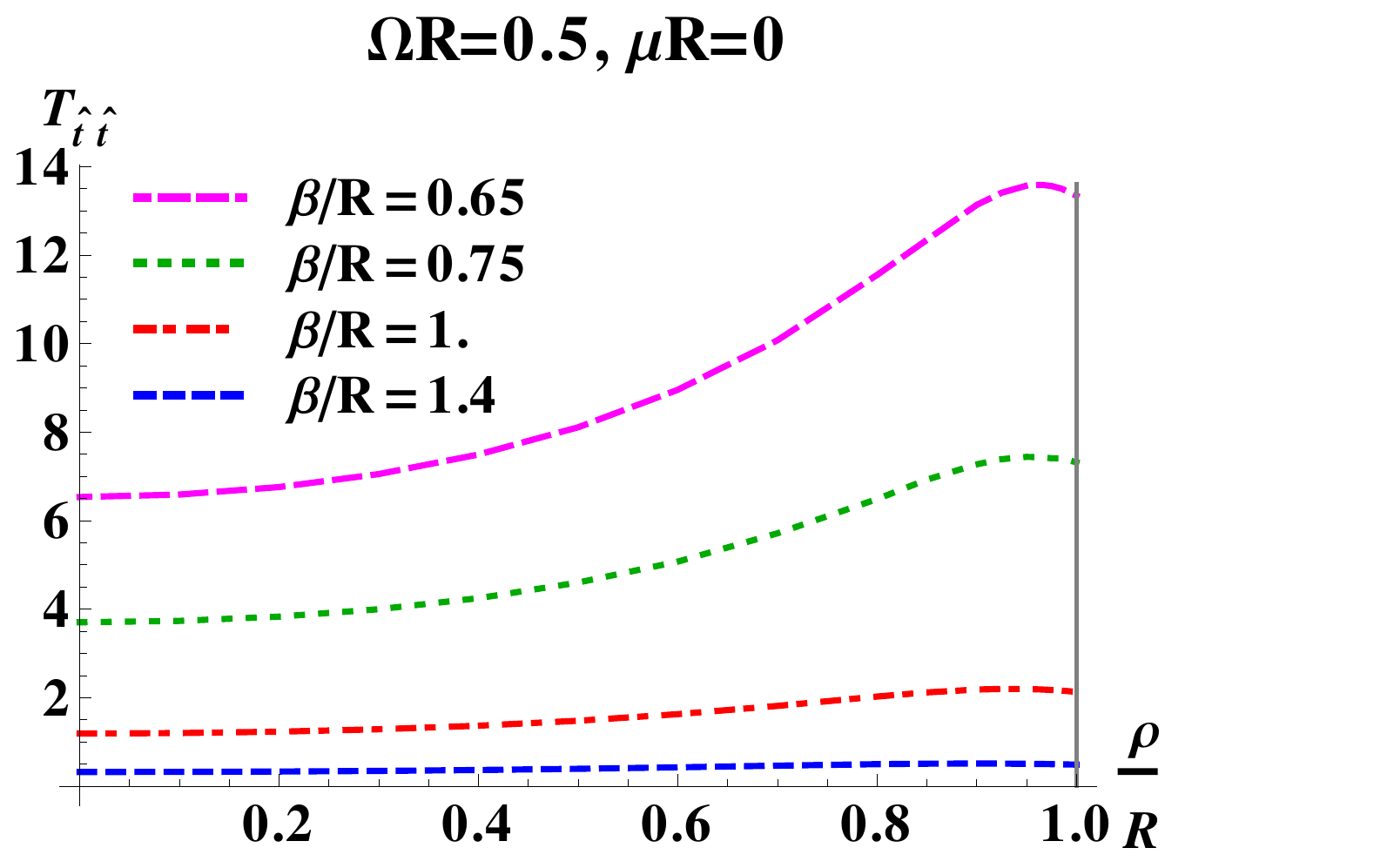} &
\includegraphics[width=0.9\columnwidth]{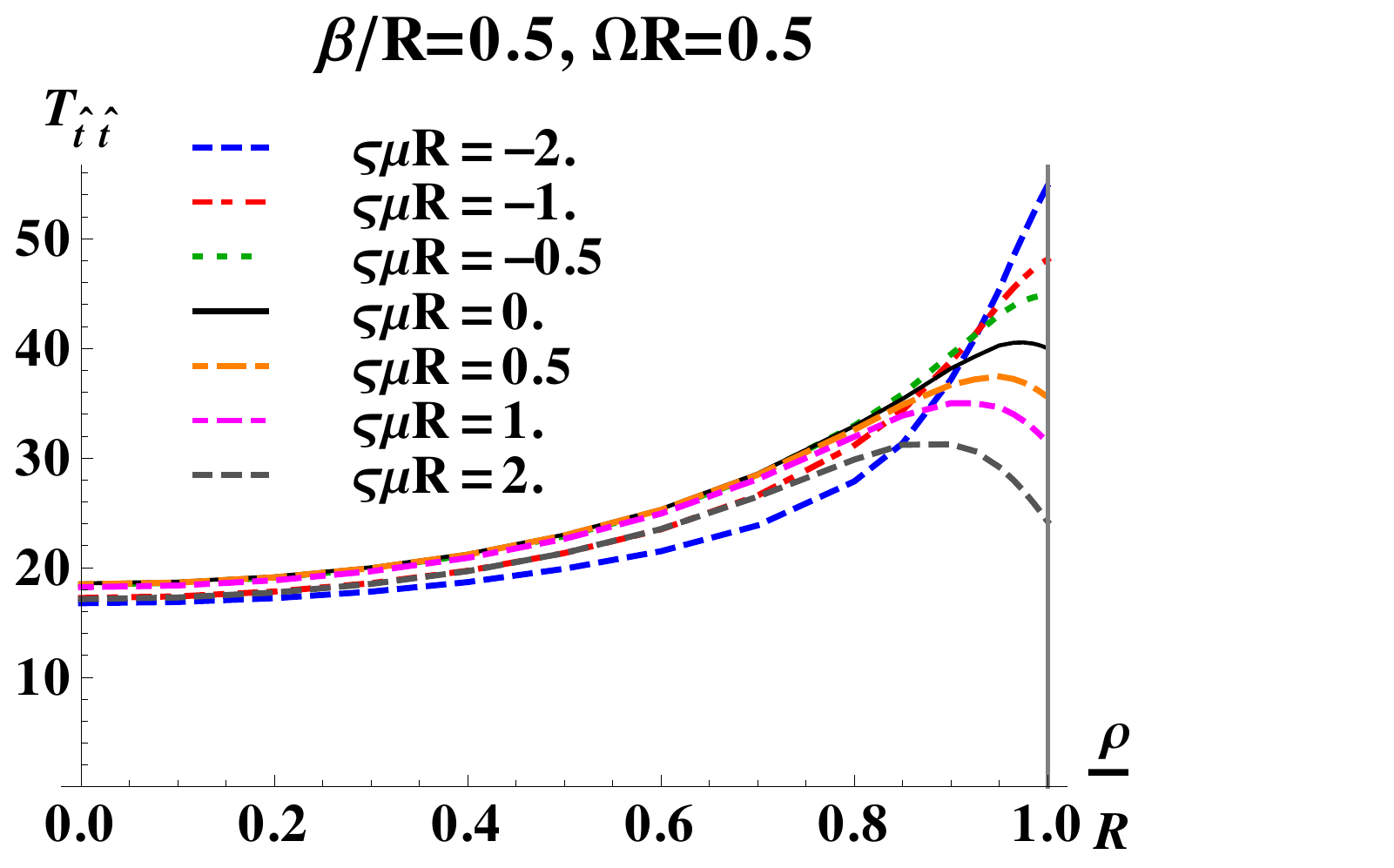} \\
(c) & (d)
\end{tabular}
\caption{Thermal expectation values (t.e.v.s) of the SET component $\braket{:T_{\hatt\hatt}:}_\beta^{\text{MIT}}$
(\ref{eq:MIT_ttt}) for MIT bag boundary conditions, as a function of the scaled radial coordinate $\rho /R$, so that the boundary of the cylinder
is at $\rho /R=1$.
(a) Massless fermions $\mu =0$, fixed inverse temperature $\beta =2R$ and various values of the angular speed $\Omega $.
(b) Massless fermions $\mu =0$, fixed inverse temperature $\beta =0.5R$ and various values of the angular speed $\Omega $.
(c) Massless fermions $\mu =0$, fixed angular speed $\Omega = 0.5/R$ and various values of the inverse temperature $\beta $.
(d) Fixed inverse temperature $\beta = 0.5R$, fixed angular speed $\Omega = 0.5/R$ and various values of the fermion mass $\mu $.
In (a) and (b) we use a logarithmic vertical scale.
In (d) we have considered both $\varsigma = \pm 1$ (for a massless field the t.e.v.s are independent of the value of $\varsigma $).}
\label{fig:MIT_ttt}
\end{figure*}

We now turn to the t.e.v.s of the SET for MIT bag boundary conditions.
The nonvanishing t.e.v.s of the components of the SET with respect to the tetrad (\ref{eq:tetrad}), calculated using (\ref{eq:SET}, \ref{eq:deltaS_MIT}), are:
\begin{widetext}
\begin{subequations}
\label{eq:MIT_SET}
\begin{align}
 \braket{:T_{\hatt\hatt}:}_\beta^{\text{MIT}} =& \sum_{m =0}^\infty \sum_{\ell = 1}^\infty \int_0^\infty
 \frac{E\,dk}{2\mathcal{D}_{m\ell}^{\text{MIT}}} \left[w(\Et) + w(\Ew)\right] \left[(\mathsf{j}^2 + 1)J_m^+(q\rho) -
 (\mathsf{j}^2 - 1) J_m^-(q\rho)\right],
 \label{eq:MIT_ttt} \\
 \braket{:T_{\hrho\hrho}:}_\beta^{\text{MIT}} =& \sum_{m =0}^\infty \sum_{\ell = 1}^\infty \int_0^\infty
 \frac{q^2\,dk}{2E\mathcal{D}_{m\ell}^{\text{MIT}}} \left[w(\Et) + w(\Ew)\right] \left\{(\mathsf{j}^2 + 1)\left[J_m^+(q\rho) -
 \frac{m + \frac{1}{2}}{q\rho} J_m^\times(q\rho)\right]\right\},
 \label{eq:MIT_trr} \\
 \braket{:T_{\hvarphi\hvarphi}:}_\beta^{\text{MIT}} =& \sum_{m =0}^\infty \sum_{\ell = 1}^\infty \int_0^\infty
 \frac{q^2\,dk}{2 E\mathcal{D}_{m\ell}^{\text{MIT}}} \left[w(\Et) + w(\Ew)\right]
 (\mathsf{j}^2 + 1) \frac{m + \tfrac{1}{2}}{q\rho} J_m^\times(q\rho),
 \label{eq:MIT_tpp}\\
 \braket{:T_{\hatz\hatz}:}_\beta^{\text{MIT}} =& \sum_{m =0}^\infty \sum_{\ell = 1}^\infty \int_0^\infty
 \frac{k^2\,dk}{2E \mathcal{D}_{m\ell}^{\text{MIT}}} \left[w(\Et) + w(\Ew)\right] \left[(\mathsf{j}^2 + 1)J_m^+(q\rho) -
 (\mathsf{j}^2 - 1) J_m^-(q\rho)\right],
 \label{eq:MIT_tzz}\\
% %
 \braket{:T_{\hatt\hvarphi}:}_\beta^{\text{MIT}} =& -\sum_{m =0}^\infty \sum_{\ell = 1}^\infty \int_0^\infty
 \frac{dk}{4\rho \mathcal{D}_{m\ell}^{\text{MIT}}} \left[w(\Et) - w(\Ew)\right] \nonumber\\
 & \times\left\{(\mathsf{j}^2 + 1)\left[(m+\tfrac{1}{2})J_m^+(q\rho) - \tfrac{1}{2}J_m^-(q\rho) + q\rho J_m^\times(q\rho)\right]
 + (\mathsf{j}^2 - 1)\left[\tfrac{1}{2} J_m^+(q\rho) - (m + \tfrac{1}{2}) J_m^-(q\rho)\right]\right\} ,
 \label{eq:MIT_ttp}
\end{align}
\end{subequations}
\end{widetext}
where we refer the reader to Eqs.~(\ref{eq:MIT_x_def}, \ref{eq:BesselJ*}, \ref{eq:spec_w}, \ref{eq:Ew_def}, \ref{eq:MIT_D}) for the definitions of the quantities appearing in (\ref{eq:MIT_SET}).
As in the spectral case, the relation~\eqref{eq:Ttrace} between the trace of the SET and the FC
can be directly verified.

Fig.~\ref{fig:MIT_ttt} illustrates how the
energy density $\braket{:T_{\hatt\hatt}:}_\beta^{\text{MIT}}$ changes with $\Omega$, $\beta$ and $\mu$.
Other components of (\ref{eq:MIT_SET}) are discussed in Sec.~\ref{sec:num}.
As expected, $\braket{:T_{\hatt\hatt}:}_\beta^{\text{MIT}}$ increases as either the temperature $\beta ^{-1}$ or angular speed $\Omega $ increases with
the other parameters fixed.
The energy density is finite and positive everywhere inside and on the boundary of the cylinder, including the case when $\Omega R=1$ and the boundary is on the SOL.
For some values of the parameters, $\braket{:T_{\hatt\hatt}:}_\beta^{\text{MIT}}$ increases monotonically as $\rho $ increases from $0$ (the axis of rotation)
to $R$ (the boundary); in other cases there is a peak in the energy density close to the boundary.
Figure~\ref{fig:MIT_ttt} (d) illustrates the effect of changing the mass on the profile of
$\braket{:T_{\hatt\hatt}:}_\beta^{\text{MIT}}$. When $\varsigma = 1$ (original MIT case),
$\braket{:T_{\hatt\hatt}:}_\beta^{\text{MIT}}$ behaves as expected, its value decreasing everywhere in the domain
as $\mu$ is increased. A notable feature of the chiral case (when $\varsigma = -1$) is that the value on the boundary
of $\braket{:T_{\hatt\hatt}:}_\beta^{\text{MIT}}$ increases as $\mu$ increases.

\subsection{Comparison between the spectral and MIT models}\label{sec:num}
\begin{figure*}
\begin{tabular}{cc}
 \includegraphics[width=0.475\linewidth]{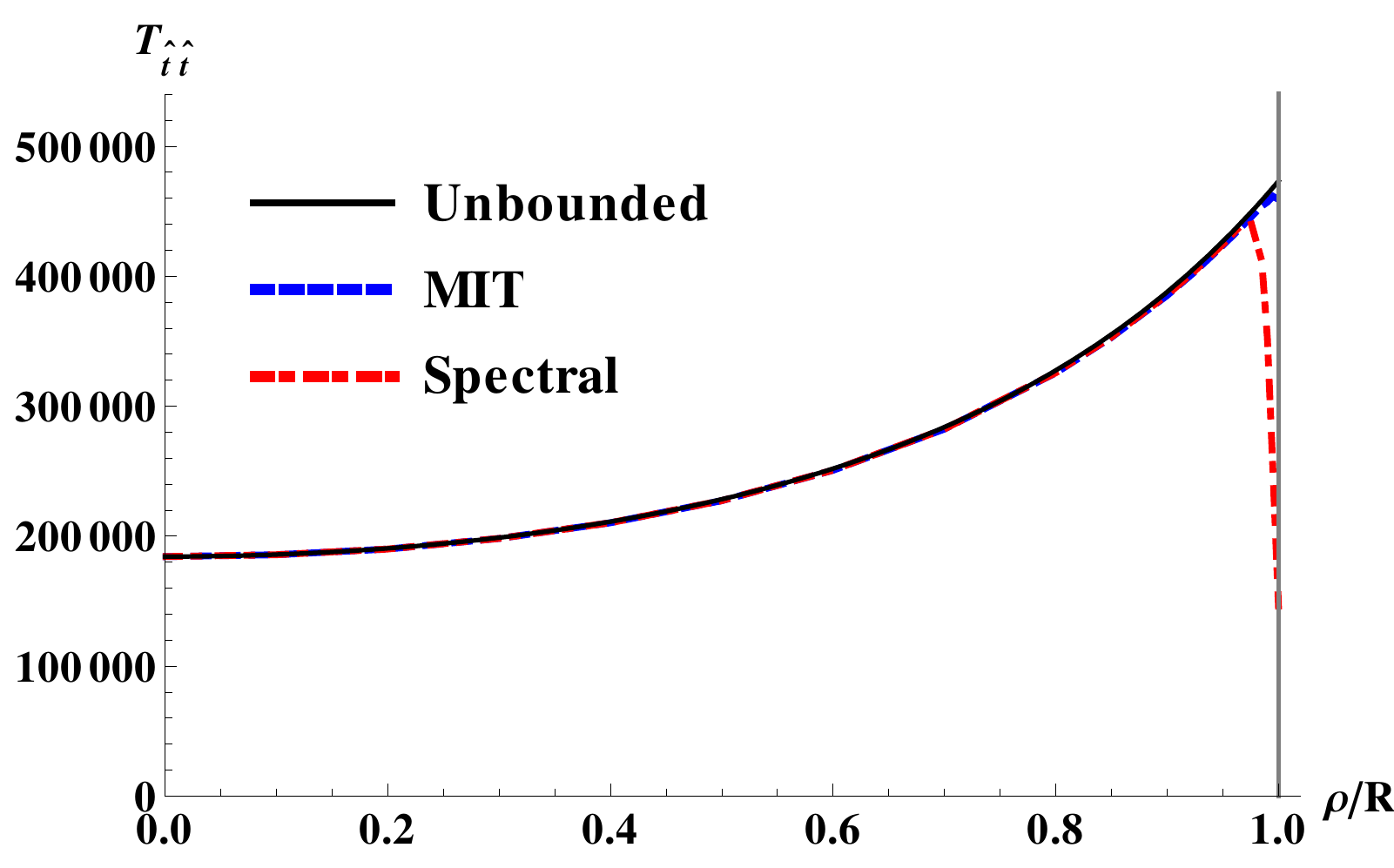} &
 \includegraphics[width=0.475\linewidth]{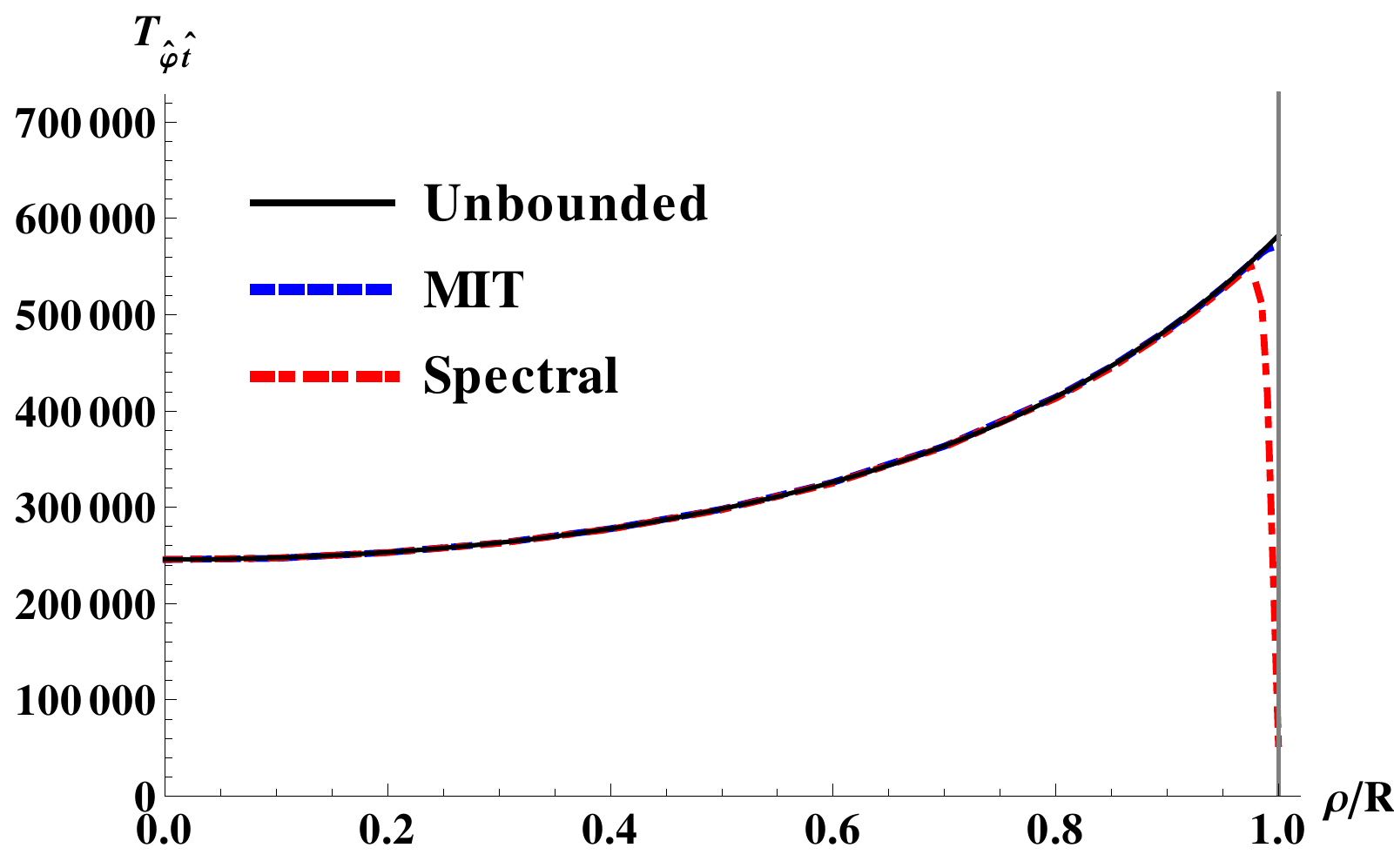} \\
 (a) & (b)\\
 \includegraphics[width=0.475\linewidth]{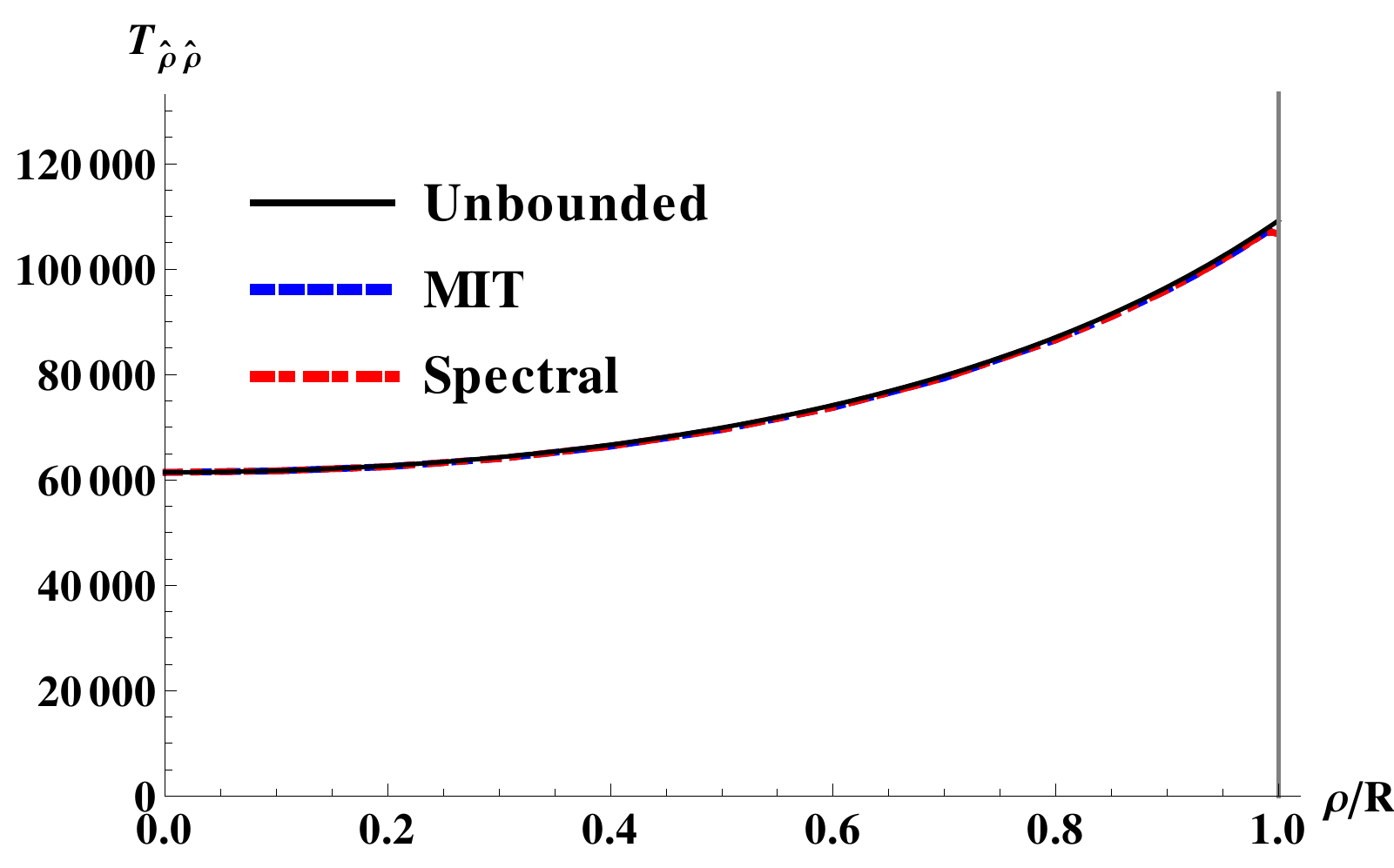} &
 \includegraphics[width=0.475\linewidth]{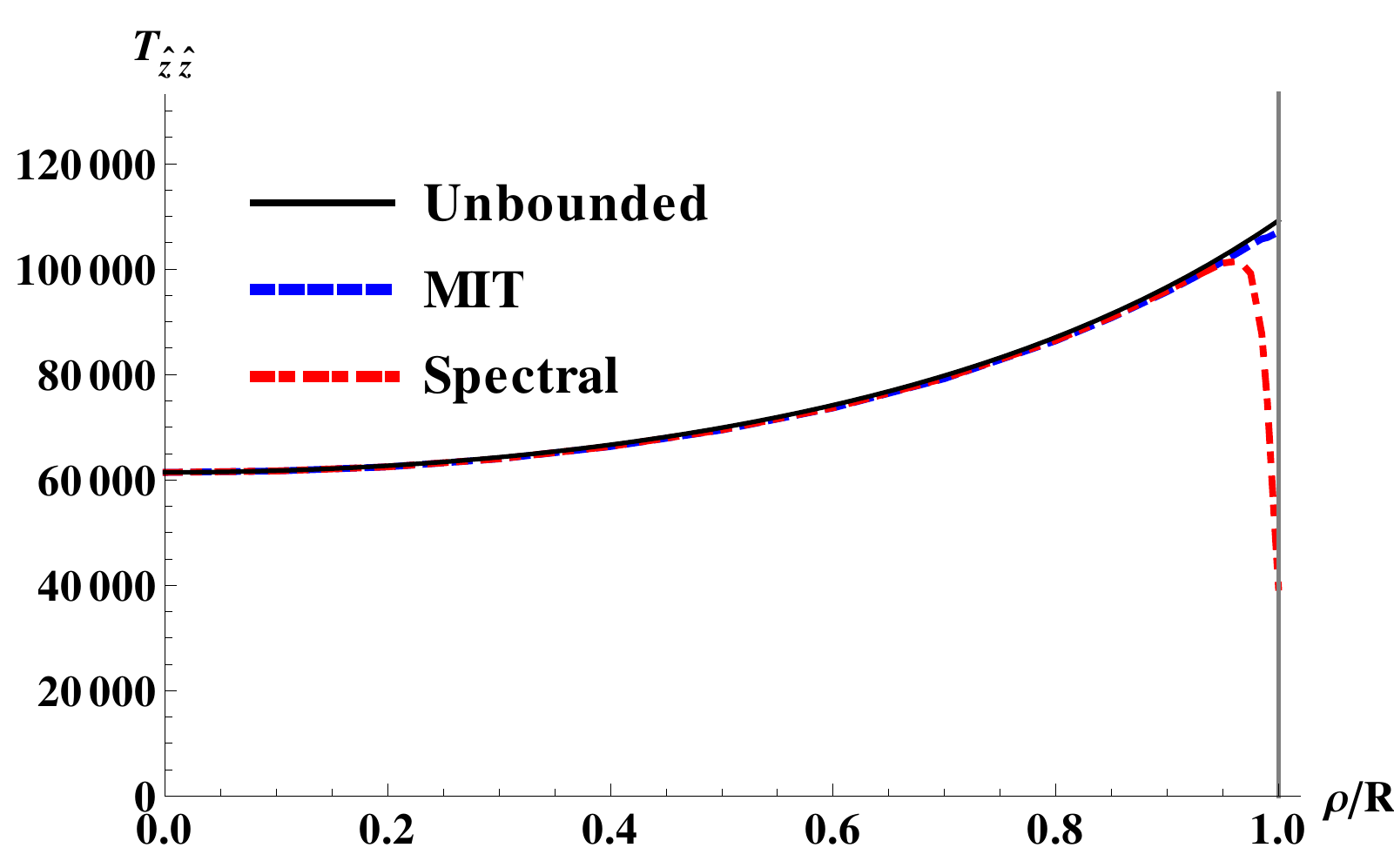} \\
 (c) & (d)\\
 \includegraphics[width=0.475\linewidth]{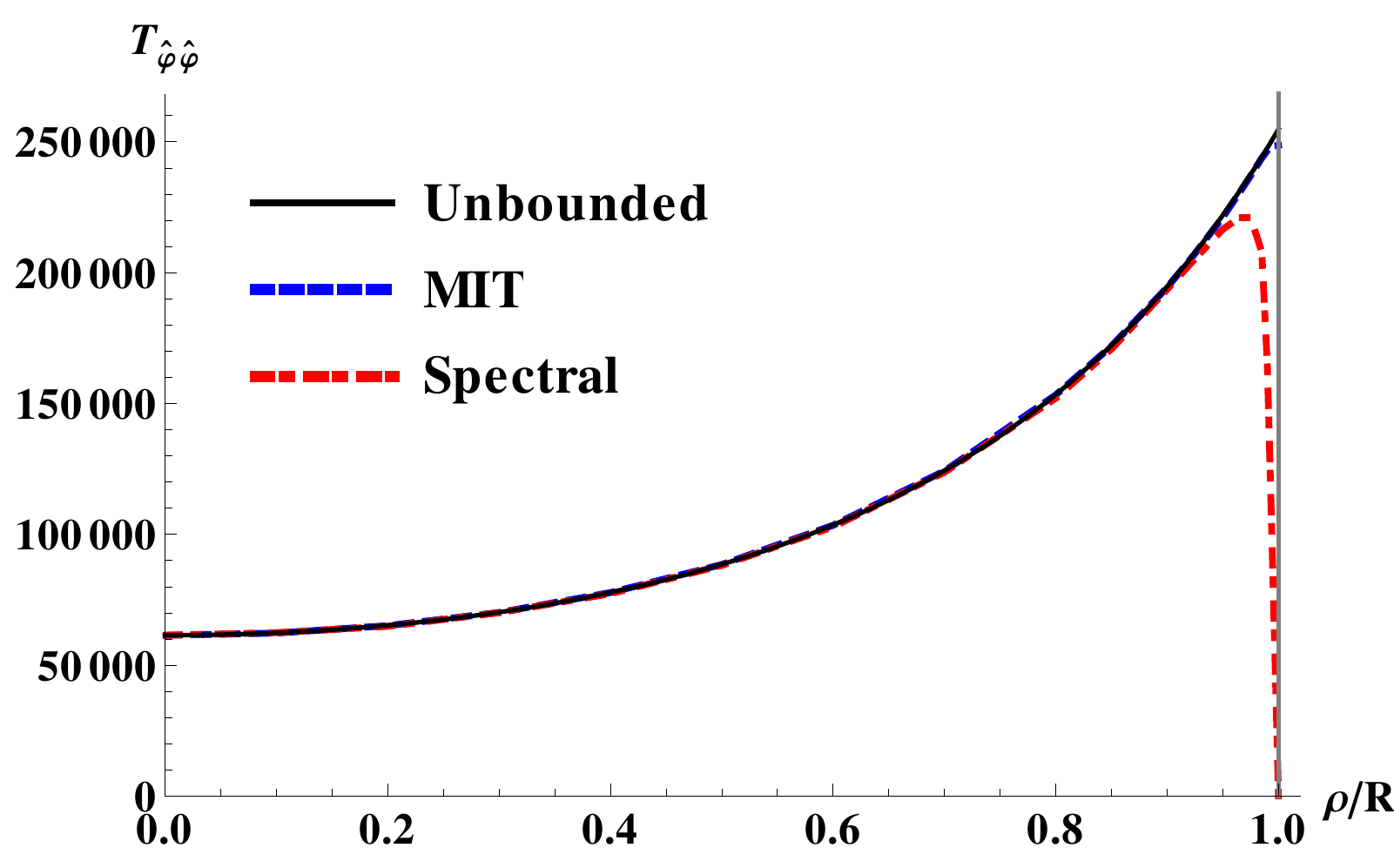} &
 \includegraphics[width=0.475\linewidth]{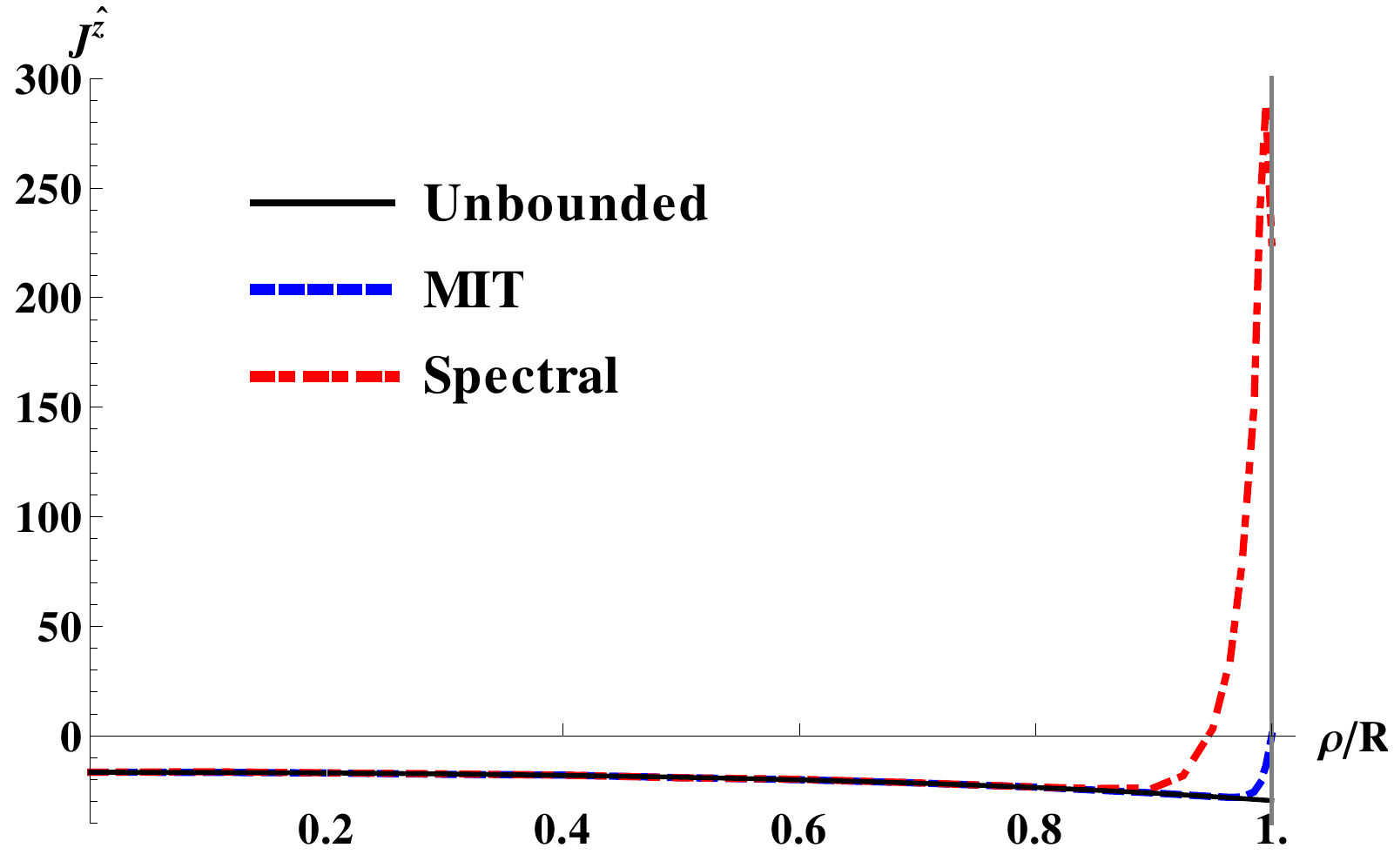} \\
 (e) & (f)
\end{tabular}
\caption{Thermal expectation values (t.e.v.s) of the SET $\braket{:T_{\halpha\hsigma}:}_\beta $ components and the CC $\braket{:\current^{\hatz}_\nu:}_\beta $ for MIT bag (blue, dashed lines) and spectral (red, dot-dashed lines) boundary conditions.
Our results are compared to those for unbounded Minkowski space-time (thin lines) \cite{art:rotunb}.
The plots show t.e.v.s as functions of the scaled radial coordinate $\rho /R$, so that the boundary of the cylinder
is at $\rho /R=1$.
The angular speed is taken to be $\Omega = 0.5R^{-1}$, the inverse temperature is $\beta = 0.05 R$ and the fermion field
is massless.
The profiles obtained in the three setups (spectral, MIT and unbounded)
agree very well, except in the vicinity of the boundary, where the results obtained with the spectral model
present visible deviations. The MIT model yields results for the SET which closely follow the unbounded case,
differing from the latter only slightly on the boundary.}
\label{fig:MITvsspec}
\end{figure*}

In this section we have computed rigidly-rotating t.e.v.s (thermal expectation values) of the fermion condensate (FC)
$\braket{:\psibar\psi:}_\beta $, and the nonzero components of the neutrino charge current (CC) $\braket{:\current^{\hatz}_\nu:}_\beta $
and stress-energy tensor (SET) $\braket{:T_{\halpha\hsigma}:}_\beta $ for a massive fermion
field satisfying either spectral (\ref{eq:spec})
or MIT bag (\ref{eq:MIT}) boundary conditions.
All components are computed with respect to the tetrad (\ref{eq:tetrad}).
We have considered only the case where the boundary is inside or on the speed-of-light surface (SOL).
All expectation values computed are finite everywhere inside and on the boundary. This is true even when the boundary is on the SOL.
The t.e.v.s with these two boundary conditions share many features. Typically their absolute values increase as either the temperature $\beta ^{-1}$ or angular speed $\Omega $ increases, with other parameters held fixed.
In the spectral case, increasing the fermion mass $\mu R$ appears to decrease the magnitude of the t.e.v.s
throughout the domain. A similar effect can be observed for the original MIT boundary conditions (when $\varsigma = 1$).
In the chiral case ($\varsigma = -1$), the values of the t.e.v.s appear to be decreasing close to the rotation
axis as $\mu$ increases, while close to the boundary, the t.e.v.s appear to increase with $\mu$.

In Fig.~\ref{fig:MITvsspec} we compare our results for the nonzero components of the SET $\braket{:T_{\halpha\hsigma}:}_\beta $ and CC $\braket{:\current^{\hatz}_\nu:}_\beta $
for the spectral and MIT bag boundary conditions with those for rotating states on unbounded Minkowski space-time \cite{art:rotunb}.
In Fig.~\ref{fig:MITvsspec}, the temperature is very high $\beta ^{-1}=20R^{-1}$, and the boundary is far inside the SOL ($\Omega R=0.5$). For these values of the parameters there is very little difference between the t.e.v.s in the unbounded, spectral and MIT bag cases.  The only noticeable variation between these three t.e.v.s is close to the boundary.
The MIT bag t.e.v.s are still very similar to those in the unbounded case but those for spectral boundary conditions show a marked difference from the unbounded case.

Combining our results in Fig.~\ref{fig:MITvsspec} with those earlier in this section,
we find the following qualitative differences between the spectral and MIT models:
\begin{itemize}
\item
The t.e.v.~of the fermion condensate vanishes everywhere for massless fermions in the spectral case,
while in the MIT case, it is finite and depends on the value of $\varsigma$.
For massless fermions, $\braket{:\psibar\psi:}_\beta^{\text{MIT}}$ has the sign of $\varsigma$ everywhere,
while
in the case of massive fermions, $\braket{:\psibar\psi:}_\beta^{\text{MIT}}$ can start with opposite sign on the
rotation axis.
\item
The t.e.v.~of the neutrino charge current is negative on the rotation axis and becomes positive on the
boundary in the spectral case, while in the MIT case, it stays negative throughout the space-time, except
on the boundary, where it vanishes.
\item
$\braket{:T_{\hvarphi\hvarphi}:}_\beta^{\text{sp}}$ vanishes on the boundary, while
$\braket{:T_{\hvarphi\hvarphi}:}^{\text{MIT}}_\beta$ remains nonzero on the boundary.
\end{itemize}

We examine further the differences between the spectral and MIT bag boundary conditions in the next section, by considering Casimir expectation values.

\section{Casimir expectation values}\label{sec:cas}
So far we have considered thermal expectation values (t.e.v.s) of rotating fermions enclosed in a cylindrical boundary with respect
to the vacuum state of the bounded system. In this section we investigate the expectation values of the
fermion condensate (FC), charge current (CC) and stress-energy tensor (SET) in the bounded rotating vacuum relative to the
unbounded vacuum state.
We refer to these expectation values as ``Casimir expectation values'' as they describe the effect of the boundary on the vacuum state.
As in the previous section, we consider both spectral and MIT bag boundary conditions.
Furthermore, the boundary will always be inside or on the speed-of-light surface (SOL).
As shown in Sec.~\ref{sec:bcs}, the resulting quantization of the transverse momentum guarantees that the Minkowski energy $E$ and corotating energy $\Et $ satisfy $E\Et >0$ for all modes.
This means that the (bounded) rotating (Iyer \cite{art:iyer}) and nonrotating (Vilenkin \cite{art:vilenkin}) vacua are identical, and will be referred to henceforth as the ``bounded vacuum''.
The bounded vacua for spectral and MIT bag boundary conditions are however {\emph {not}} the same, and hence the Casimir expectation values will depend on the boundary conditions employed.
\subsection{Euclidean Green's function on unbounded Minkowski space-time}\label{sec:cas_unb}
The main difficulty in extracting Casimir expectation values using the construction of two-point functions
by employing mode sums comes from the quantization of the transverse momentum induced by the boundary
(see Secs.~\ref{sec:spec_disc} and \ref{sec:MIT_disc} for details).
On unbounded Minkowski space-time, the fermion field (and, similarly, the two-point function) is written as a sum over field modes,
which involves an integral over the Minkowski energy $E$ (or, equivalently, the transverse momentum $q$) (\ref{eq:sumj_def}).
The presence of the boundary changes the integral over the permissible values of the transverse momentum $q$
into a sum (over an index $\ell$ which labels the values of the transverse momentum). This makes it technically challenging
(although not impossible \cite{art:bezerra08,art:bezerra_saharian}) to subtract two-point functions
corresponding to the unbounded and bounded manifolds.
Following the approach in Ref.~\cite{art:duffy_ottewill}, it is convenient to extract Casimir expectation values
from the Green's function of the corresponding Euclideanized manifold. To this end, we start in this section by calculating
the Euclidean Green's function for the unbounded space-time, after which the boundary terms will be
presented separately for the spectral and MIT bag models in Secs.~\ref{sec:casspectral} and \ref{sec:casMIT} respectively.

To simplify the calculations, it is convenient to switch to the inertial nonrotating
(Minkowski) coordinates, where the metric is diagonal, i.e.~there are no off-diagonal components mixing
space and time. The formulation of quantum field theory on the Euclidean equivalent of the Minkowski manifold
is obtained by introducing the following notation:
\begin{equation}\label{eq:euclid}
 x^0_E \equiv \tau = it, \quad x^j_E = x^j_{M}, \quad
 \gamma^0_E = \gamma^0, \quad \gamma^j_E = -i\gamma^j,
\end{equation}
where $t$ and $x^j_{M}$ are Minkowski (inertial) coordinates,  and $\gamma ^{0}=\gamma ^{\hatt }$, $\gamma ^{j}=\gamma ^{\hat{j}}$, where
$\gamma ^{\hatt }$, $\gamma ^{\hat {j}}$ are given in (\ref{eq:gamma_mink}).
The resulting Euclidean Minkowski metric
$g_{\mu\nu}^E$ has the following nonvanishing components:
\begin{equation}\label{eq:euclid_g}
 g_{\tau\tau}^E = g_{\rho\rho}^E = g_{zz}^E = 1, \qquad g_{\varphi\varphi}^E = \rho^2.
\end{equation}
The Euclidean Green's function $S_E \equiv S_E(x,x')$ must satisfy the inhomogeneous Dirac equation:
\begin{align}\label{eq:euclid_dirac}
 (-\gamma^\lambda_E \partial^E_\lambda - \mu) S_E =&
 S_E (\overleftarrow{\partial}^E_{\lambda'} \gamma^{\lambda'}_E - \mu)\nonumber\\
 =& - \frac{1}{\sqrt{g_E}} \delta(\tau - \tau') \delta^3(\bm{x} - \bm{x}') \mathbb{I}_4,
\end{align}
where $\mathbb{I}_4$ is the $4\times 4$ identity matrix and
$g_E$ is the determinant of the Euclidean metric with nonvanishing components \eqref{eq:euclid_g}.

Following the construction of the mode solutions of the Dirac equation in
Sec.~\ref{sec:modes}, the (nonrotating) vacuum Euclidean Green's function $S_E^{\text{unb}}(x_E, x'_E)$
for the unbounded space-time can be Fourier-transformed as:
\begin{equation}
 S_E^{\text{unb}}(x_E, x'_E) = \int_{-\infty}^\infty \frac{d\omega}{8\pi^3}
 \int_{-\infty}^\infty dk \sum_{m = -\infty}^\infty e^{i\omega \Delta \tau + ik\Delta z} \chi^{\text{unb}},
 \label{eq:euclid_unb}
\end{equation}
where $\Delta \tau = \tau - \tau'$, $\Delta z = z-z'$ and the $4\times 4$ matrix $\chi^{\text{unb}}$ can be written in terms of four $2\times2$ matrices $\chi_{ab}^{\text{unb}}$:
\begin{equation}
 \chi^{\text{unb}} =
 \begin{pmatrix}
  \chi^{\text{unb}}_{11} & \chi^{\text{unb}}_{12}\\
  \chi^{\text{unb}}_{21} & \chi^{\text{unb}}_{22}
 \end{pmatrix}.
 \label{eq:euclid_chi_def}
\end{equation}
Performing an equivalent Fourier transformation of the delta functions on the right of Eq.~\eqref{eq:euclid_dirac},
the inhomogeneous Dirac equation implies:
\begin{align}
 &\begin{pmatrix}
  \mu + i \omega & 2ph\\
  -2ph & \mu -i\omega
 \end{pmatrix}
 \begin{pmatrix}
  \chi^{\text{unb}}_{11} & \chi^{\text{unb}}_{12}\\
  \chi^{\text{unb}}_{21} & \chi^{\text{unb}}_{22}
 \end{pmatrix}\nonumber\\
 &\hspace{40pt} = \frac{\delta(\rho-\rho')}{\rho}
 \begin{pmatrix}
  1 & 0 \\ 0 & 1
 \end{pmatrix} \otimes
  \begin{pmatrix}
  e^{im\Delta\varphi} & 0 \\ 0 & e^{i(m+1)\Delta\varphi}
  \end{pmatrix},\nonumber\\
 &\begin{pmatrix}
  \chi^{\text{unb}}_{11} & \chi^{\text{unb}}_{12}\\
  \chi^{\text{unb}}_{21} & \chi^{\text{unb}}_{22}
 \end{pmatrix}
 \begin{pmatrix}
  \mu + i \omega & 2ph^{'\,\dagger}\\
  -2ph^{'\,\dagger} & \mu -i\omega
 \end{pmatrix}\nonumber\\
 &\hspace{40pt} = \frac{\delta(\rho-\rho')}{\rho'}
 \begin{pmatrix}
  1 & 0 \\ 0 & 1
 \end{pmatrix} \otimes
  \begin{pmatrix}
  e^{im\Delta\varphi} & 0 \\ 0 & e^{i(m+1)\Delta\varphi}
  \end{pmatrix},\label{eq:euclid_eq}
\end{align}
where $p$ is the momentum, $\Delta \varphi = \varphi - \varphi '$ and $h$ is the $2\times 2$ component of the helicity operator $W_{0}$, defined in Eq.~\eqref{eq:helicity}.
In (\ref{eq:euclid_eq}), we have used the Kronecker product of matrices, defined in (\ref{eq:Kronecker}). For
the equation in $x'$, the operator $h^{'\,\dagger}$ has the form:
\begin{equation}
 h^{'\,\dagger} = \frac{1}{2p}
 \begin{pmatrix}
  k & -P_-'\\
  -P_+' & -k
 \end{pmatrix},
\end{equation}
where the primes indicate that the derivatives in the operators $P_\pm'$ act from the right on $\rho'$ and $\varphi'$.
The operators $P_{\pm }$ can be found in (\ref{eq:P_shifters}).

The off-diagonal components of Eqs.~\eqref{eq:euclid_eq} give the following equations:
\begin{align}
 \chi^{\text{unb}}_{21} =& \frac{2ph}{\mu - i\omega} \chi^{\text{unb}}_{11} =
 \chi^{\text{unb}}_{22} \frac{2ph^{'\, \dagger}}{\mu + i\omega},\nonumber\\
 \chi^{\text{unb}}_{12} =& -\frac{2ph}{\mu + i\omega} \chi^{\text{unb}}_{22} =
 -\chi^{\text{unb}}_{11} \frac{2ph^{'\,\dagger}}{\mu - i\omega},
 \label{eq:euclid_offd}
\end{align}
while the diagonal components can be written as modified Bessel equations:
\begin{subequations}
\label{eq:unbound_Bessel}
\begin{align}
 &[\rho^2 \partial_\rho^2 + \rho \partial_\rho + \partial_\varphi^2 - \rho^2 \alpha^2]
 \frac{\chi^{\text{unb}}_{11}}{\mu - i\omega}\nonumber\\&\hspace{50pt} = -\rho \delta(\rho-\rho')
 \begin{pmatrix}
  e^{im\Delta\varphi} & 0 \\ 0 & e^{i(m+1)\Delta\varphi}
  \end{pmatrix},\nonumber\\
  & \\
 &[\rho^2 \partial_\rho^2 + \rho \partial_\rho + \partial_\varphi^2 - \rho^2 \alpha^2]
 \frac{\chi^{\text{unb}}_{22}}{\mu + i\omega}\nonumber \\&\hspace{50pt} =  -\rho \delta(\rho-\rho')
 \begin{pmatrix}
  e^{im\Delta\varphi} & 0 \\ 0 & e^{i(m+1)\Delta\varphi}
 \end{pmatrix},
\end{align}
\end{subequations}
where
\begin{equation}
\alpha^2 = \omega^2 + k^2 + \mu^2.
\label{eq:alpha}
\end{equation}
It can be shown
that the inhomogeneous Dirac equation in $x'$ also reduces to Eqs.~(\ref{eq:unbound_Bessel}) (with $\rho$ and $\varphi$
replaced by $\rho'$ and $\varphi'$, respectively).
Hence, $\chi^{\text{unb}}_{11}$ and $\chi^{\text{unb}}_{22}$ can be written as linear combinations of modified Bessel functions. The Euclidean
Green's function for the Minkowski space-time must be regular at the origin and at infinity, and thus the only
nontrivial solution of Eqs.~(\ref{eq:unbound_Bessel}) satisfying these boundary conditions is:
\begin{equation}\label{eq:euclid_unb_diag}
 \frac{\chi^{\text{unb}}_{11}}{\mu - i\omega} = \frac{\chi^{\text{unb}}_{22}}{\mu + i\omega} =
 \begin{pmatrix}
  I_m^< K_m^> e^{im\Delta\varphi} \hspace{-20pt} & 0\\
  0 & I_{m+1}^< K_{m+1}^>  e^{i(m+1)\Delta\varphi}
 \end{pmatrix},
\end{equation}
where $I_{m}$ and $K_{m}$ are modified Bessel functions of the first and second kinds respectively.
The arguments of the Bessel functions with the $<$ or $>$ superscripts are the smaller or larger of
$\alpha \rho$ and $\alpha \rho'$, respectively. Therefore, if $\rho > \rho'$,
we will write $I_m^< K_m^> = K_m I_m$, where the arguments of $K_m$ and $I_m$ are $\alpha \rho$ and $\alpha \rho'$, as per
the conventions introduced in Eq.~\eqref{eq:M_struct}. The combinations in Eq.~(\ref{eq:euclid_unb_diag}) can be written using these conventions in terms of step functions as:
\begin{equation}
 f^< g^> = \theta(\rho - \rho') g f + \theta(\rho' - \rho) f g.
\end{equation}
The off-diagonal matrices $\chi^{\text{unb}}_{12}$ and $\chi^{\text{unb}}_{21}$ can be obtained from Eqs.~\eqref{eq:euclid_offd},
using the following properties (the operators $P_{\pm }$ are given in (\ref{eq:P_shifters})):
\begin{align}\label{eq:euclid_ppm}
 P_+ I_m(\alpha \rho) e^{im\varphi} =& -i\alpha e^{i(m+1)\varphi} I_{m+1}(\alpha\rho),\nonumber\\
 P_- I_{m+1}(\alpha \rho) e^{i(m+1)\varphi} =& -i\alpha e^{im\varphi} I_m(\alpha\rho),\nonumber\\
 P_+ K_m(\alpha \rho) e^{im\varphi} =& i\alpha e^{i(m+1)\varphi} K_{m+1}(\alpha\rho),\nonumber\\
 P_- K_{m+1}(\alpha \rho) e^{i(m+1)\varphi} =& i\alpha e^{im\varphi} K_m(\alpha\rho).
\end{align}
Similar equations hold for $P_\pm'$, which can be applied bearing in mind that
$I_{-m}(z) = I_m(z)$ and $K_{-m}(z) = K_m(z)$.

Thus, the Euclidean propagator on unbounded Minkowski space-time takes the form (\ref{eq:euclid_unb})
with the matrix $\chi^{\rm unb}$ given by:
\begin{subequations}
\label{eq:euclid_unb_chi}
\begin{multline}
 \chi^{\rm unb} = \left[\mu \mathbb{I}_2 - i\omega \sigma_3\right] \otimes
 \begin{pmatrix}
  I_m^< K_m^> e^{im\Delta\varphi} \hspace{-20pt} & 0 \\
  0 & I_{m+1}^< K_{m+1}^> e^{i(m+1)\Delta\varphi}
 \end{pmatrix} \\ + k
 \begin{pmatrix}
  0 & -1 \\
  1 & 0
 \end{pmatrix} \otimes
 \begin{pmatrix}
  I_m^< K_m^> e^{im\Delta\varphi} \hspace{-20pt} & 0 \\
  0 & -I_{m+1}^< K_{m+1}^> e^{i(m+1)\Delta\varphi}
 \end{pmatrix}\\ + \alpha
 \begin{pmatrix}
  0 & -1\\
  1 & 0
 \end{pmatrix} \otimes
 \begin{pmatrix}
  0 & {\mathcal {F}}(m, m+1)\\
  {\mathcal {F}}(m+1,m) & 0
 \end{pmatrix},
\end{multline}
where the notation ${\mathcal {F}}(m,n)$ is a shorthand for:
\begin{equation}
 {\mathcal {F}}(m,n) = ie^{im\varphi - in\varphi'}\left[\theta(\rho-\rho') K_m I_n - \theta(\rho'-\rho) I_m K_n\right].
\end{equation}
\end{subequations}
As before, the first and second Bessel functions in (\ref{eq:euclid_unb_chi}) depend on $\alpha \rho$ and
$\alpha \rho'$, respectively. The Pauli matrix $\sigma _{3}$ is given in (\ref{eq:Pauli}).

Before ending this section, we stress that the solution (\ref{eq:euclid_unb}, \ref{eq:euclid_unb_chi}) of the inhomogeneous Dirac equation (\ref{eq:euclid_dirac}) is fixed by the
boundary conditions requiring regularity at the origin ($\rho = 0$ or $\rho' = 0$) and space-like infinity.
To satisfy boundary conditions of a different type, suitable solutions of the homogeneous Dirac equation
can be added to Eq.~\eqref{eq:euclid_unb}.
We follow this approach in Secs.~\ref{sec:casspectral} and \ref{sec:casMIT} for spectral and MIT bag boundary conditions, respectively.
\subsection{Spectral boundary conditions}\label{sec:casspectral}
In this section, we first construct the Euclidean Green's function for a fermion field satisfying spectral boundary conditions on the cylinder, then
compute the Casimir expectation values. Using an asymptotic analysis, we are able to derive the rate of divergence of these expectation values as the boundary is approached.
\subsubsection{Euclidean Green's function for spectral boundary conditions}
\label{sec:cas_spec_green}
\begin{table}
\begin{tabular}{r|c|c}
 & $m + \tfrac{1}{2} > 0$ & $m + \tfrac{1}{2} < 0$\\\hline\hline
 $\rho = R$ & $\begin{pmatrix} 0 & 0 \\ \times & \times \end{pmatrix}$ &
 $\begin{pmatrix} \times & \times\\ 0 & 0 \end{pmatrix}$\\\hline
 $\rho' = R$ & $\begin{pmatrix} 0 & \times \\ 0 & \times \end{pmatrix}$ &
 $\begin{pmatrix} \times & 0 \\ \times & 0 \end{pmatrix}$
\end{tabular}
\caption{The behaviour of the $2\times 2$ constituent blocks (\ref{eq:M_struct}) of the Green's function obeying spectral boundary
conditions on a cylinder of radius $R$. Depending on the sign of $m + \frac{1}{2}$ and on which point is
on the boundary, certain entries in these $2\times 2$ matrices will vanish, as indicated in the table. Entries
marked $\times$ do not necessarily vanish.}
\label{tab:spectral_bc}
\end{table}
To construct a Euclidean Green's function which implements spectral boundary conditions, we consider the behaviour
of the corresponding vacuum Hadamard Green's function on the boundary. Since the dependence on the
radial coordinates $\rho$ and $\rho'$ is always that in the $2\times 2$ matrix given in Eq.~\eqref{eq:M_struct},
it is sufficient to analyze its behaviour on the boundary, as shown in Table~\ref{tab:spectral_bc}.
To implement these boundary conditions, a solution $\Delta S_E^{\text{sp}}(x_E,x'_E)$ of the homogeneous Dirac
equation must be added to the Euclidean propagator (\ref{eq:euclid_unb}, \ref{eq:euclid_unb_chi}), as follows:
\begin{equation}
\label{eq:sp_Green_function}
 S_E^{\text{sp}}(x_E,x_E') = S_E^{\text{unb}}(x_E,x_E') + \Delta S_E^{\text{sp}}(x_E,x_E'),
\end{equation}
where $\Delta S_E^{\text{sp}}(x_E,x_E')$ can be Fourier transformed in analogy with Eq.~\eqref{eq:euclid_unb}:
\begin{multline}
 \Delta S_E^{\text{sp}}(x_E, x'_E) = \int_{-\infty}^\infty \frac{d\omega}{8\pi^3}
 \int_{-\infty}^\infty dk \sum_{m = -\infty}^\infty e^{i\omega \Delta \tau + ik\Delta z}  \\
\times \Delta \chi^{\text{sp}}.
 \label{eq:euclid_spec_SE_Fourier}
\end{multline}
The $4\times 4$ matrix $\Delta \chi^{\text{sp}}$ can be written in terms of four $2\times2$ matrices $\Delta \chi^{\text{sp}}_{ab}$,
in a similar way to Eq.~\eqref{eq:euclid_chi_def}:
\begin{equation}
 \Delta \chi^{\text{sp}} =
 \begin{pmatrix}
  \Delta \chi^{\text{sp}}_{11} & \Delta \chi^{\text{sp}}_{12}\\
  \Delta \chi^{\text{sp}}_{21} & \Delta \chi^{\text{sp}}_{22}
 \end{pmatrix}.
 \label{eq:euclid_spec_chi_def}
\end{equation}

The Euclidean propagator $S_E^{\text{sp}}(x_E,x_E')$ of the bounded system must obey spectral boundary conditions,
in other words those entries which vanish in Table~\ref{tab:spectral_bc} must be equal to zero.
Furthermore, $S_E^{\text{sp}}(x_E,x_E')$ must stay regular at the origin
(i.e.~when either $\rho = 0$ or $\rho' = 0$).
We therefore find the following expressions for $\Delta \chi^{\text{sp}}_{11}$ and $\Delta \chi^{\text{sp}}_{22}$:
\begin{subequations}\label{eq:euclid_sp_diag}
\begin{equation}\label{eq:euclid_sp_diagm}
 \frac{\Delta \chi^{\text{sp}}_{11}}{\mu - i\omega} = \frac{\Delta \chi^{\text{sp}}_{22}}{\mu + i\omega} =
 c_m \begin{pmatrix}
  1 & 0\\
  0 & -1
 \end{pmatrix} \circ \mathcal{E}_j,
\end{equation}
where $\circ $ denotes the Hadamard product of matrices (\ref{eq:Hadamard}).
In (\ref{eq:euclid_sp_diagm}), $c_m$ is a constant ensuring that the relevant entries in Table~\ref{tab:spectral_bc} vanish, having the value:
\begin{equation}\label{eq:euclid_sp_diagc}
 c_{m} =
 \begin{cases}
  {\displaystyle -\frac{K_m(\alpha R)}{I_m(\alpha R)}}, &  \quad  m + \frac{1}{2} > 0,\\
  {\displaystyle \frac{K_{m+1}(\alpha R)}{I_{m+1}(\alpha R)}}, & \quad m + \frac{1}{2} < 0,
 \end{cases}
\end{equation}
and the matrix $\mathcal{E}_j$ on the right of the Hadamard product in (\ref{eq:euclid_sp_diagm}) is given by:
\begin{equation}
 \mathcal{E}_j = \begin{pmatrix}
  I_m I_m\,e^{im\Delta\varphi} & -i I_m I_{m+1}\,e^{i(m+1)\Delta\varphi-i\varphi}\\
  i I_{m+1} I_m\,e^{im\Delta\varphi + i\varphi} & I_{m+1} I_{m+1}\,e^{i(m+1)\Delta\varphi}
 \end{pmatrix},
 \label{eq:euclid_E_def}
\end{equation}
where the first and second modified Bessel functions above have arguments $\alpha\rho$ and $\alpha\rho'$, respectively and $\alpha $ is given in Eq.~(\ref{eq:alpha}).
Only modified Bessel functions of the first kind (i.e.~$I_m$) have been considered in Eqs.~\eqref{eq:euclid_E_def},
since their linearly independent partners, $K_m$, do not satisfy the requirement of regularity at the origin.
The off-diagonal matrices $\Delta \chi^{\text{sp}}_{12}$ and $\Delta \chi^{\text{sp}}_{21}$ can be determined
using analogues of Eqs.~\eqref{eq:euclid_offd} for the spectral case:
\begin{equation}
 \Delta \chi^{\text{sp}}_{21} = - \Delta \chi^{\text{sp}}_{12} = c_m
 \begin{pmatrix}
  k & -\alpha\\
  -\alpha & k
 \end{pmatrix} \circ \mathcal{E}_j.
\end{equation}
\end{subequations}
Thus, the Fourier coefficients $\Delta \chi^{\text{sp}}$ of the boundary term (\ref{eq:euclid_spec_SE_Fourier}) can be written as:
\begin{multline}
 c_m^{-1} \Delta \chi^{\text{sp}} =\left(\mu \mathbb{I}_2 - i\omega \sigma_3\right) \otimes
  \left[\begin{pmatrix}
  1 & 0 \\
  0 & -1
 \end{pmatrix} \circ \mathcal{E}_j\right]\\ +
 \begin{pmatrix}
  0 & -1 \\
  1 & 0
 \end{pmatrix} \otimes
 \left[\begin{pmatrix}
  k & -\alpha\\
  -\alpha & k
 \end{pmatrix} \circ \mathcal{E}_j\right] ,
\end{multline}
where the Pauli matrix $\sigma _{3}$ is given in (\ref{eq:Pauli}) and ${\mathbb {I}}_{2}$ is the $2\times 2$ identity matrix.
\subsubsection{Casimir expectation values}
\label{sec:cas_spec_exp}
We are interested in the Casimir expectation values of the FC $\braket{\psibar\psi}_{\text {Cas}}^{\text {sp}}$,
charge current $\braket{\current^{\halpha}}_{\text {Cas}}^{\text {sp}}$, CC $\braket{\current^{\halpha}_\nu}_{\text {Cas}}^{\text {sp}}$
and SET $\braket{T_{\halpha\hsigma}}_{\text {Cas}}^{\text {sp}}$.
The following formulae can be used to calculate these expectation values using the difference
$\Delta S_E^{\text{sp}}(x_E,x_E')$ (\ref{eq:euclid_spec_SE_Fourier}) between the vacuum Euclidean Green's functions for the bounded system and
for unbounded Minkowski space:
\begin{subequations}
 \label{eq:euclid_tevs_def}
\begin{align}
 \braket{\psibar\psi}_{\text {Cas}}^{\text {sp}} =& \lim_{x_E'\rightarrow x_E} \tr \left[\Delta S_E^{\text{sp}}(x_E,x_E')\right],
 \label{eq:euclid_sp_FC}\\
 \braket{\current^{\halpha}}_{\text {Cas}}^{\text {sp}} =& \lim_{x_E'\rightarrow x_E}
 \tr \left[\gamma^{\halpha}_E \Delta S_E^{\text{sp}}(x_E,x_E')\right],
 \label{eq:euclid_sp_current}\\
 \braket{\current^{\halpha}_\nu}_{\text {Cas}}^{\text {sp}} =& \lim_{x_E'\rightarrow x_E}
 \tr \left[\gamma^{\halpha}_E \frac{1 + \gamma^5}{2} \Delta S_E^{\text{sp}}(x_E,x_E')\right],
 \label{eq:euclid_sp_CC}\\
 \braket{T_{\halpha\hsigma}}_{\text {Cas}}^{\text {sp}} =& \frac{1}{2} \lim_{x_E'\rightarrow x_E}
 \tr \left[\gamma_{(\halpha}^E(D_{\hsigma)}^E - D_{\hsigma')}^E) \Delta S_E^{\text{sp}}(x_E,x_E')\right].
 \label{eq:euclid_sp_SET}
\end{align}
\end{subequations}

For the FC (\ref{eq:euclid_sp_FC}), the following expression is obtained:
\begin{equation}
\label{eq:cas_spec_ppsi_aux1}
 \braket{\psibar\psi}_{\text{Cas}}^{\text {sp}} = \frac{\mu}{4\pi^3} \int_{-\infty}^\infty d\omega
 \int_{-\infty}^\infty dk \sum_{m = -\infty}^\infty c_m
 I_m^-(\alpha R),
\end{equation}
where the constant $c_m$ is defined in Eq.~\eqref{eq:euclid_sp_diagc} and the notation $I_m^-(z)$
is analogous to that defined in Eqs.~\eqref{eq:BesselJ*}:
\begin{equation}\label{eq:BesselI*}
 I_m^\pm(z) = I_m^2(z) \pm I_{m+1}^2(z), \quad
 I_m^\times(z) = 2 I_m(z) I_{m+1}(z).
\end{equation}
It is convenient to switch to the polar coordinates ($\alpha$, $\vartheta$) where $\alpha $ is given by (\ref{eq:alpha}) and:
\begin{equation}\label{eq:euclid_pol}
 \omega = \sqrt{\alpha^2 - \mu^2} \cos\vartheta, \qquad
 k = \sqrt{\alpha^2 - \mu^2} \sin\vartheta,
\end{equation}
in terms of which the Casimir FC (\ref{eq:cas_spec_ppsi_aux1}) can be put in the following form, after the integration over $\vartheta $ has been performed:
\begin{equation}\label{eq:cas_spec_ppsi_aux}
 \braket{\psibar\psi}_{\text{Cas}}^{\text {sp}} = \frac{\mu}{2\pi^2} \sum_{m = -\infty}^\infty \int_{\mu}^\infty d\alpha
 \,\alpha \, c_m\, I_m^-(\alpha R).
\end{equation}
We now change variables to
\begin{equation}
 \casx = \alpha R \label{eq:casx}
\end{equation}
and introduce the notation:
\begin{multline}
 \casI^{\text{sp,}*}_{\ell n} \equiv \casI^{\text{sp,}*}_{\ell n}(\rho) = -\frac{1}{2\pi^2 R^4} \\\times
 \sum_{m = -\infty}^\infty \int_{\mu R}^\infty d\casx\,
 \casx^\ell (m + \tfrac{1}{2})^n c_m I_m^*(\casx \rhoo) .
 \label{eq:casIdef}
\end{multline}
The functions $I_m^*(z)$ are defined in Eqs.~\eqref{eq:BesselI*} for $* \in \{+,-,\times\}$ and
\begin{equation}
 \rhoo = \frac{\rho}{R} .
\label{eq:rhodef}
\end{equation}
In terms of this new notation, the FC~\eqref{eq:cas_spec_ppsi_aux} can be written as:
\begin{equation}
 \braket{\psibar\psi}_{\text{Cas}}^{\text {sp}} = -\mu R^2 \casI^{\text {sp,}-}_{10}.\label{eq:cas_spec_ppsi}
\end{equation}

The Casimir expectation values of all components of the charge current (\ref{eq:euclid_sp_current}) and neutrino charge current
(\ref{eq:euclid_sp_CC}) vanish.
The nonvanishing components of the
Casimir expectation value of the SET (\ref{eq:euclid_sp_SET}) can be written as:
\begin{subequations}
\label{eq:cas_spec_SET}
\begin{align}
 \braket{T\indices{^\htau_\htau}}_{\text{Cas}}^{\text {sp}} =& -\frac{1}{2}\casI^{\text {sp,}-}_{30} + \frac{1}{2} \mu^2 R^2 \casI^{\text{sp,}-}_{10},
 \\
 \braket{T\indices{^\hrho_\hrho}}_{\text{Cas}}^{\text {sp}} =& \casI^{\text {sp,}-}_{30} - \rhoo^{\,-1} \casI^{\text{sp,}\times}_{21},\\
 \braket{T\indices{^\hvarphi_\varphi}}_{\text{Cas}}^{\text {sp}} =& \rhoo^{\,-1} \casI^{\text{sp,}\times}_{21},
\end{align}
\end{subequations}
and $\braket{T\indices{^\hatz_\hatz}}_{\text{Cas}}^{\text {sp}} = \braket{T\indices{^\htau_\htau}}_{\text{Cas}}^{\text {sp}}$.
In (\ref{eq:cas_spec_SET}), we have written the components of the SET relative to the Euclidean version of the tetrad (\ref{eq:tetrad}).

\subsubsection{Casimir divergence near the boundary}
\label{sec:cas_spec_div}

By construction, the Casimir expectation values (\ref{eq:euclid_tevs_def}) diverge on the boundary, due to the properties of the difference
$\Delta S_E^{\text{sp}}(x_E,x_E')$ between the vacuum Euclidean Green's functions for the bounded and unbounded Minkowski space-times, given by
(\ref{eq:sp_Green_function}):
\begin{equation}
\Delta S_E^{\text{sp}}(x_E,x_E') =S_E^{\text{sp}}(x_E,x_E') - S_E^{\text{unb}}(x_E,x_E').
\end{equation}
To see this, consider one of the entries in $S_{E}^{\text {sp}}(x_E,x_E')$ which vanishes when $x_{E}$ is on the boundary from Table~\ref{tab:spectral_bc}.
This entry in $\Delta S_E^{\text{sp}}(x_E,x_E')$ (with $\rho =R$) is then equal to the corresponding entry in $S_E^{\text{unb}}(x_E,x_E')$ with $\rho =R$.
As $x_{E}'\rightarrow x_{E}$, because the coincidence limit of the unbounded Minkowski space Green's function is divergent, so too is this entry in
$\lim _{x_{E}'\rightarrow x_{E}} \Delta S_E^{\text{sp}}(x_E,x_E')$ when $x_{E}$ is on the boundary.
Therefore the Casimir expectation values (\ref{eq:euclid_tevs_def}) diverge on the boundary.

This divergent behaviour can be also be seen in the algebraic expressions (\ref{eq:cas_spec_ppsi}, \ref{eq:cas_spec_SET}) for the Casimir expectation values. For example, consider the behaviour of the integrand in $\casI^{\text{sp},-}_{00}$ \eqref{eq:casIdef} when $\rhoo =1$, for large
values of $m = \nu - \tfrac{1}{2}$ and $\casx$.
First we define polar coordinates $(r,\theta )$ as follows:
\begin{equation}
\label{eq:cas_polar_def}
(\nu, \casx) = (r\cos\theta, r\sin\theta)
\end{equation}
then, using Eqs.~\eqref{eq:cas_BesselI*_approx}, we find:
\begin{equation}
 r \frac{K_{\nu - \frac{1}{2}}(\casx)}{I_{\nu - \frac{1}{2}}(\casx)} I^-_{\nu - \frac{1}{2}}(\casx) =
 \frac{\cos\theta}{1 + \cos\theta} \left[1 + \frac{1}{2r} + O(r^{-2})\right],
\end{equation}
where the $r$ on the left-hand-side is the Jacobian of the transformation \eqref{eq:cas_polar_def}.
The above expression does not vanish at large $r$, so the integral in $\casI^{\text{sp},-}_{00}$ \eqref{eq:casIdef} is not convergent when $\rhoo =1$.
As will be seen in the analysis below, it {\emph {is}} convergent for $\rhoo <1$.

In this section, we analyze the divergence of the Casimir expectation
values (\ref{eq:cas_spec_ppsi}, \ref{eq:cas_spec_SET}) as a function of the distance $\epsilon$ to the boundary, defined as:
\begin{equation}
 \epsilon = 1 - \rhoo,
 \label{eq:cas_epsilon_def}
\end{equation}
where $\rhoo $ is given by (\ref{eq:rhodef}).
We will find it useful to consider the following integrals:
\begin{equation}
 \casIo_{\ell n}^{\,\text{sp},*} = -\frac{1}{\pi^2 R^4} \int_0^\infty d\nu \int_{\mu R}^\infty d\casx\,
 \casx^\ell \nu^n c_{\nu - \frac{1}{2}} I_{\nu - \frac{1}{2}}^*(\casx \rhoo).
 \label{eq:casIodef}
\end{equation}
To understand the connection between $\casIo^{\,{\text {sp}},*}_{\ell n}$ and $\casI^{\text{sp},*}_{\ell n}$, the sum over $m$ in Eq.~\eqref{eq:casIdef}
can be replaced by the integral over $\nu$ by using the generalized Abel-Plana formula, presented next.

\paragraph{Generalized Abel-Plana formula.}
\label{sec:gap}

According to Ref.~\cite{art:abelplana}, residue theory can be used to prove the following result:
\begin{equation}
 \sum_{m = 0}^\infty f\left( m + \tfrac{1}{2} \right) = \int_0^\infty d\nu\, f(\nu) - i\int_0^\infty dt \frac{f(it) - f(-it)}{e^{2\pi t} + 1} ,
  \label{eq:abelplana}
\end{equation}
valid for an analytic function $f$.

In the present case, $f(m + \tfrac{1}{2})$ in Eq.~(\ref{eq:abelplana})
will be replaced by the analytic functions $f_{\ell n}^{\text {sp},*}(m + \tfrac{1}{2})$, defined according to:
\begin{equation}
 f^{\text {sp},*}_{\ell n}(\nu) = \frac{1}{\pi^2 R^4} \int_{\mu R}^\infty d\casx\, \casx^\ell \nu^n
 \frac{K_{\nu-\tfrac{1}{2}}(\casx)}{I_{\nu - \tfrac{1}{2}}(\casx)} I^*_{\nu - \tfrac{1}{2}}(\casx \rhoo) ,
 \label{eq:cas_spec_fdef}
\end{equation}
where $I^{*}_{m}$ is defined in (\ref{eq:BesselI*}).
From the definitions (\ref{eq:euclid_sp_diagc}, \ref{eq:casIdef}), the integrals $\casI^{\text {sp},*}_{\ell n}$ can be written in terms of
$f^{\text {sp},*}_{\ell n}(\nu)$ (\ref{eq:cas_spec_fdef}) as follows:
\begin{equation}
 \casI^{\text {sp},*}_{\ell n} = \sum_{m = 0}^\infty f^{\text{sp},*}_{\ell n}(m + \tfrac{1}{2}).
\end{equation}
The behaviour of $\casI^{\text {sp},*}_{\ell n}$ near the boundary can be
investigated by considering the following function:
\begin{equation}
 \delta^{\text {sp},*}_{\ell n}(\rhoo) \equiv \casIo_{\ell n}^{\,\text{sp},*} - \casI_{\ell n}^{\,\text{sp},*}
 = i \int_0^\infty dt \frac{f^{\text{sp},*}_{\ell n}(it) - f^{\text{sp},*}_{\ell n}(-it)}{e^{2\pi t} + 1}.
 \label{eq:spec_delta_def}
\end{equation}
The factor $(e^{2\pi t} + 1)^{-1}$ ensures the convergence of the $t$ integral.

The $\casx$ integral in Eq.~\eqref{eq:cas_spec_fdef}
can be analyzed by considering the asymptotic expansion of the integrand for large $\casx$ (but fixed $\nu$).
Starting from the asymptotic expansions for large argument given in Eqs.~\eqref{eq:bessel_asympt_inf}, the following
approximations can be obtained:
\begin{subequations}
\begin{align}
 I_{\nu - \frac{1}{2}}^-(\casx) =& \frac{\nu e^{2\casx}}{\pi \casx^2} \left(1 - \frac{2\nu^2 - 1}{2\casx} + \dots\right),\\
 I_{\nu - \frac{1}{2}}^+(\casx) =& \frac{e^{2\casx}}{\pi \casx} \left(1 - \frac{\nu^2}{\casx} + \frac{\nu^4}{2\casx^2}
 + \dots\right),\\
 I_{\nu - \frac{1}{2}}^\times(\casx) =& \frac{e^{2\casx}}{\pi \casx} \left[1 - \frac{\nu^2}{\casx} +
 \frac{\nu(\nu^2 - 1)}{2\casx^2} + \dots\right],\\
 \frac{K_{\nu - \frac{1}{2}}(\casx)}{I_{\nu - \frac{1}{2}}(\casx)} =& \pi e^{-2\casx} \left[
 1 + \frac{\nu(\nu - 1)}{\casx} + \frac{\nu^2(\nu - 1)^2}{2\casx^2}+\dots\right].
\end{align}
\label{eq:tempbessel1}
\end{subequations}
From (\ref{eq:tempbessel1}) we find the following asymptotic behaviour of the
desired functions:
\begin{widetext}
\begin{subequations}\label{eq:cas_spec_largearg}
\begin{align}
 \frac{K_{\nu - \frac{1}{2}}(\casx)}{I_{\nu - \frac{1}{2}}(\casx)}
 I_{\nu - \frac{1}{2}}^-(\casx \rhoo)
 =& \frac{\nu e^{-2\casx\epsilon}}{\casx^2\rhoo^2} \left[1 - \frac{2\nu - 1}{2\casx} -
 \frac{2\nu^2 - 1}{2\casx} \epsilon + O(\casx^{-2})\right],\\
 \frac{K_{\nu - \frac{1}{2}}(\casx)}{I_{\nu - \frac{1}{2}}(\casx)}
 I_{\nu - \frac{1}{2}}^\times(\casx \rhoo)
 =& \frac{e^{-2\casx\epsilon}}{\casx\rhoo^2} \left[1 - \frac{\nu}{\casx} - \frac{\nu^2}{\casx}\epsilon + O(\casx^{-2})\right],
\end{align}
\end{subequations}
where $\epsilon $ is given by (\ref{eq:cas_epsilon_def}).
The divergence of the functions $\delta^{\text{sp},*}_{\ell n}(\rhoo)$ (\ref{eq:spec_delta_def}) for the cases relevant to the computation of the Casimir
expectation values in Eqs.~(\ref{eq:cas_spec_ppsi}, \ref{eq:cas_spec_SET}) can be found using:
\begin{align}
\label{eq:tempdelta1}
 \delta_{10}^{\text {sp},-}(\rhoo) =& -\frac{2}{\pi^2R^4\rhoo^2} \int_{0}^\infty \frac{t\,dt}{e^{2\pi t}+ 1} \int_{\mu R}^\infty
 \frac{d\casx}{\casx} e^{-2\casx\epsilon}\left[1 + O(\casx^{-1})\right],\nonumber\\
 \delta_{30}^{\text {sp},-}(\rhoo) =& -\frac{2}{\pi^2R^4\rhoo^2} \int_{0}^\infty \frac{t\, dt}{e^{2\pi t}+ 1} \int_{\mu R}^\infty
 d\casx\, e^{-2\casx\epsilon}\left[\casx +
 \frac{1}{2} + \frac{t^2 + \frac{1}{2}}{\rhoo} \epsilon + O(\casx^{-1})\right],\nonumber\\
 \delta_{21}^{\text{sp},\times}(\rhoo) =& -\frac{2}{\pi^2R^4\rhoo} \int_{0}^\infty \frac{t\, dt}{e^{2\pi t}+ 1} \int_{\mu R}^\infty
 d\casx\, e^{-2\casx\epsilon} \left[\casx + \frac{t^2}{\rhoo} \epsilon + O(\casx^{-1})\right].
\end{align}
When $\rhoo \rightarrow 1$ (or, equivalently, $\epsilon \rightarrow 0$),
the integrals (\ref{eq:tempdelta1}) diverge due to the large $\casx$ behaviour of the integrand.
To investigate this divergence, the lower limit of the $\casx$ integral can be set to $0$, giving:
\begin{equation}
 \delta_{10}^{\text {sp},-} \simeq -\frac{\ln (2\epsilon)^{-1} - \gamma + O(\epsilon)}{24 \pi^2 R^4},\qquad
 \delta_{30}^{\text {sp},-} \simeq -\frac{1 + 4\epsilon + O(\epsilon^{2})}{96\pi^2 R^4\epsilon^2},\qquad
 \delta_{21}^{\text {sp},\times } \simeq -\frac{1 + \epsilon + O(\epsilon^{2})}{96\pi^2 R^4\epsilon^2},
\label{eq:dirac_spec_cas_delta}
\end{equation}
where $\gamma $ is Euler's constant.
It turns out that the results (\ref{eq:dirac_spec_cas_delta}) diverge as $\epsilon \rightarrow 0$ at a subleading order compared to the
corresponding functions $\casIo^{\text {sp},*}_{\ell n}$ (\ref{eq:casIodef}), as will be shown below.

\paragraph{Asymptotic analysis of Casimir divergence.}
\label{sec:cas_spec_asympt}

We now examine the behaviour of the integrals $\casIo^{\text {sp},*}_{\ell n}$ (\ref{eq:casIodef}) as $\epsilon \rightarrow 0$.
This behaviour, combined with the results (\ref{eq:dirac_spec_cas_delta}) for $\delta_{\ell n}^{\text {sp},*}$ and (\ref{eq:spec_delta_def}),
will enable us to deduce the relevant properties of the integrals  $\casI^{\text {sp},*}_{\ell n}$ (\ref{eq:casIdef}) required for the Casimir expectation values (\ref{eq:cas_spec_ppsi}, \ref{eq:cas_spec_SET}).

Using the polar coordinates $(r, \theta)$ introduced in Eq.~\eqref{eq:cas_polar_def} and
the expansions in Eqs.~\eqref{eq:cas_BesselI*_approx}, the following asymptotic expansions can be made:
\begin{subequations}\label{eq:dirac_cas_div_asrhoo}
\begin{align}
 I_{\nu - \frac{1}{2}}^-(\mathtt{x}\rhoo) =& \frac{e^{2r(1-\epsilon) + 2\nu \ln \tan\frac{\theta}{2}}}{\pi r \tan\theta}
 \left[1 + \frac{1 + 5 \sin^2\theta}{12 r} + \epsilon (1+\sin^2\theta) - r\epsilon^2 \cos^2\theta + \dots\right],
 \label{eq:dirac_cas_Imasrhoo}\\
 I_{\nu - \frac{1}{2}}^+(\mathtt{x}\rhoo) =& \frac{e^{2r(1-\epsilon) + 2\nu \ln \tan\frac{\theta}{2}}}{\pi \casx}
 \left[1 + \frac{\cos^2\theta}{12 r} + \epsilon - r\epsilon^2 \cos^2\theta + \dots\right],
 \label{eq:dirac_cas_Ipasrhoo}\\
 I_{\nu - \frac{1}{2}}^\times(\mathtt{x}\rhoo) =& \frac{e^{2r(1-\epsilon) + 2\nu \ln \tan \frac{\theta}{2}}}{\pi r}
 \left[1 - \frac{5\cos^2\theta}{12 r} + \epsilon\, \sin^2\theta - r\epsilon^2\cos^2\theta +\dots\right],\label{eq:dirac_cas_Ixasrhoo}
\end{align}
\end{subequations}
where terms of order $r^{-2}$, $r^{-1} \epsilon$ and $\epsilon^2$ were ignored.
Combining Eq.~\eqref{eq:cas_Bessel_spec_approx} with Eqs.~\eqref{eq:dirac_cas_div_asrhoo}
gives:
\begin{align}
 \frac{K_{\nu - \frac{1}{2}}(\mathtt{x})}{I_{\nu - \frac{1}{2}}(\mathtt{x})} I_{\nu - \frac{1}{2}}^-(\mathtt{x}\rhoo) =&
 e^{-2r\epsilon} \frac{\cos\theta}{r(1 + \cos\theta)} \left[1 + \frac{1}{2r} + \epsilon(1 + \sin^2\theta) -r\epsilon^2\cos^2\theta
 + \dots\right],\nonumber\\
 \frac{K_{\nu - \frac{1}{2}}(\mathtt{x})}{I_{\nu - \frac{1}{2}}(\mathtt{x})} I_{\nu - \frac{1}{2}}^\times(\mathtt{x}\rhoo) =&
 e^{-2r\epsilon} \frac{\sin\theta}{r(1 + \cos\theta)} \left[1 + \epsilon\,\sin^2\theta -r\epsilon^2\cos^2\theta + \dots\right].
\end{align}
\end{widetext}
Hence, the following results are obtained:
\begin{align}
 \casIo^{\,\text{sp},-}_{10} =& \frac{1}{4\pi^2 R^4\epsilon^2}\left[
 1 - \ln 2 + \epsilon \left(\tfrac{4}{3} - \ln 2\right) + O(\epsilon^2)\right], \nonumber\\
 \casIo^{\,\text{sp},-}_{30} =& \frac{1}{16\pi^2 R^4\epsilon^4}
 \left[1 + \tfrac{43}{30} \epsilon + O(\epsilon^2)\right], \nonumber\\
 \casIo^{\, \text{sp},\times}_{21} =& \frac{1}{16\pi^2 R^4\epsilon^4}\left[
 1 + \tfrac{1}{10} \epsilon + O(\epsilon^2)\right]. \label{eq:cas_spec_Iln}
\end{align}
The divergences of the $\casIo^{\, \text{sp},*}_{\ell n}$ terms calculated above
are two inverse powers of $\epsilon $ larger than the corresponding error terms $\delta^{\text {sp},*}_{\ell n}$
calculated in Eqs.~\eqref{eq:dirac_spec_cas_delta}. Hence, from (\ref{eq:spec_delta_def}) the leading order and next-to-leading order
divergence of the functions $\casI^{\text {sp},*}_{\ell n}$ (\ref{eq:casIdef}) coincide with the expressions (\ref{eq:cas_spec_Iln}) for the
leading order and next-to-leading order divergences of the functions $\casIo^{\text {sp},*}_{\ell n}$ (\ref{eq:casIodef}).

Substituting Eqs.~\eqref{eq:cas_spec_Iln} into Eqs.~(\ref{eq:cas_spec_ppsi}, \ref{eq:cas_spec_SET})
gives the following asymptotic behaviours for the Casimir expectation values as $\epsilon \rightarrow 0$ and the boundary is approached:
\begin{subequations}
\label{eq:cas_spec_res}
\begin{align}
 \braket{\psibar\psi}_{\text{Cas}}^{\text{sp}} =& -\frac{\mu}{4\pi^2 R^2 \epsilon^2}
 \left[1 - \ln 2 + \left(\tfrac{4}{3} - \ln2\right)\epsilon + \dots\right],
 \label{eq:cas_spec_res_ppsi}
 \\
 \braket{T\indices{^\htau_\htau}}_{\text{Cas}}^{\text{sp}}=& -\frac{1}{32\pi^2 R^4 \epsilon^4}
 \left[1 + \tfrac{43}{30} \epsilon + \dots\right],
 \label{eq:cas_spec_res_ttt}
 \\
 \braket{T\indices{^\hrho_\hrho}}_{\text{Cas}}^{\text{sp}}=& \frac{1}{48\pi^2 R^4 \epsilon^3}
 \left[1 + \tfrac{53}{20} \epsilon + \dots\right],
 \label{eq:cas_spec_res_trr}\\
 \braket{T\indices{^\hvarphi_\hvarphi}}_{\text{Cas}}^{\text{sp}}=& \frac{1}{16\pi^2 R^4 \epsilon^4}
 \left[1 + \tfrac{11}{10} \epsilon + \dots\right],
 \label{eq:cas_spec_res_tpp}
\end{align}
\end{subequations}
where $\braket{T\indices{^\hatz_\hatz}}_{\text{Cas}}^{\text{sp}} = \braket{T\indices{^\htau_\htau}}_{\text{Cas}}^{\text{sp}}$.
We obtained $\braket{T\indices{^\hrho_\hrho}}_{\text{Cas}}^{\text{sp}}$ from
$\braket{T\indices{^\hvarphi_\hvarphi}}_{\text{Cas}}^{\text{sp}}$ using the conservation law $\nabla_\mu T\indices{^\mu_\nu} = 0$,
which can be written in (nonrotating) cylindrical coordinates on Minkowski space-time as follows:
\begin{equation}
 \partial_\rho(\rho T\indices{^\hrho_\hrho}) = T\indices{^\hvarphi_\hvarphi}.
 \label{eq:cas_trr_from_tpp}
\end{equation}
The divergence of the SET (\ref{eq:cas_spec_res}) for massive fermions when spectral boundary conditions are considered
is one inverse power of $\epsilon $ larger compared to the scalar field case \cite{art:duffy_ottewill}.
We will discuss this point further in Sec.~\ref{sec:cas_summary}.

\paragraph{Numerical results.}
\label{sec:cas_spec_num}

\begin{figure*}
\begin{tabular}{cc}
 \includegraphics[width=0.45\linewidth]{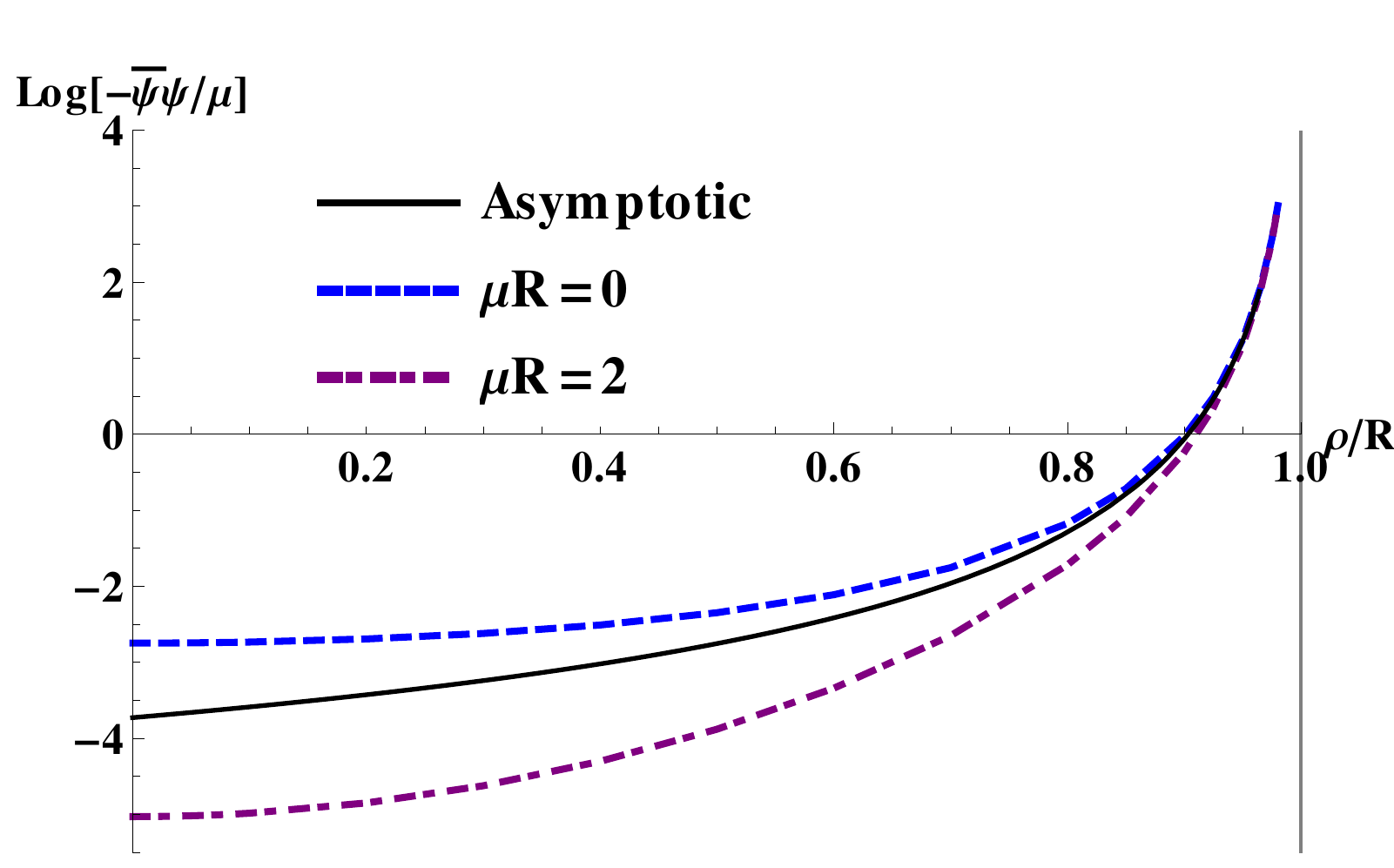} &
 \includegraphics[width=0.45\linewidth]{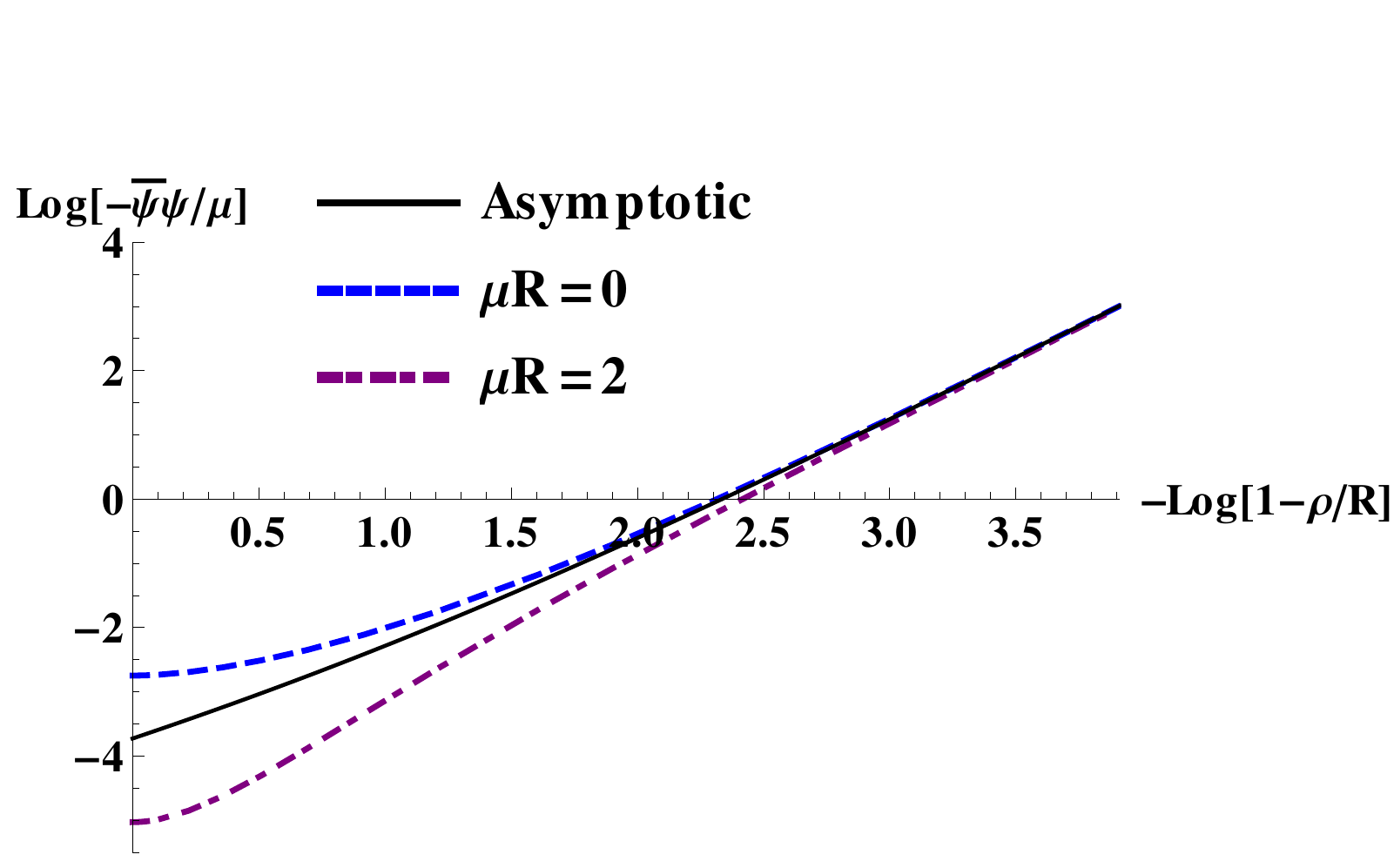}\\
 \includegraphics[width=0.45\linewidth]{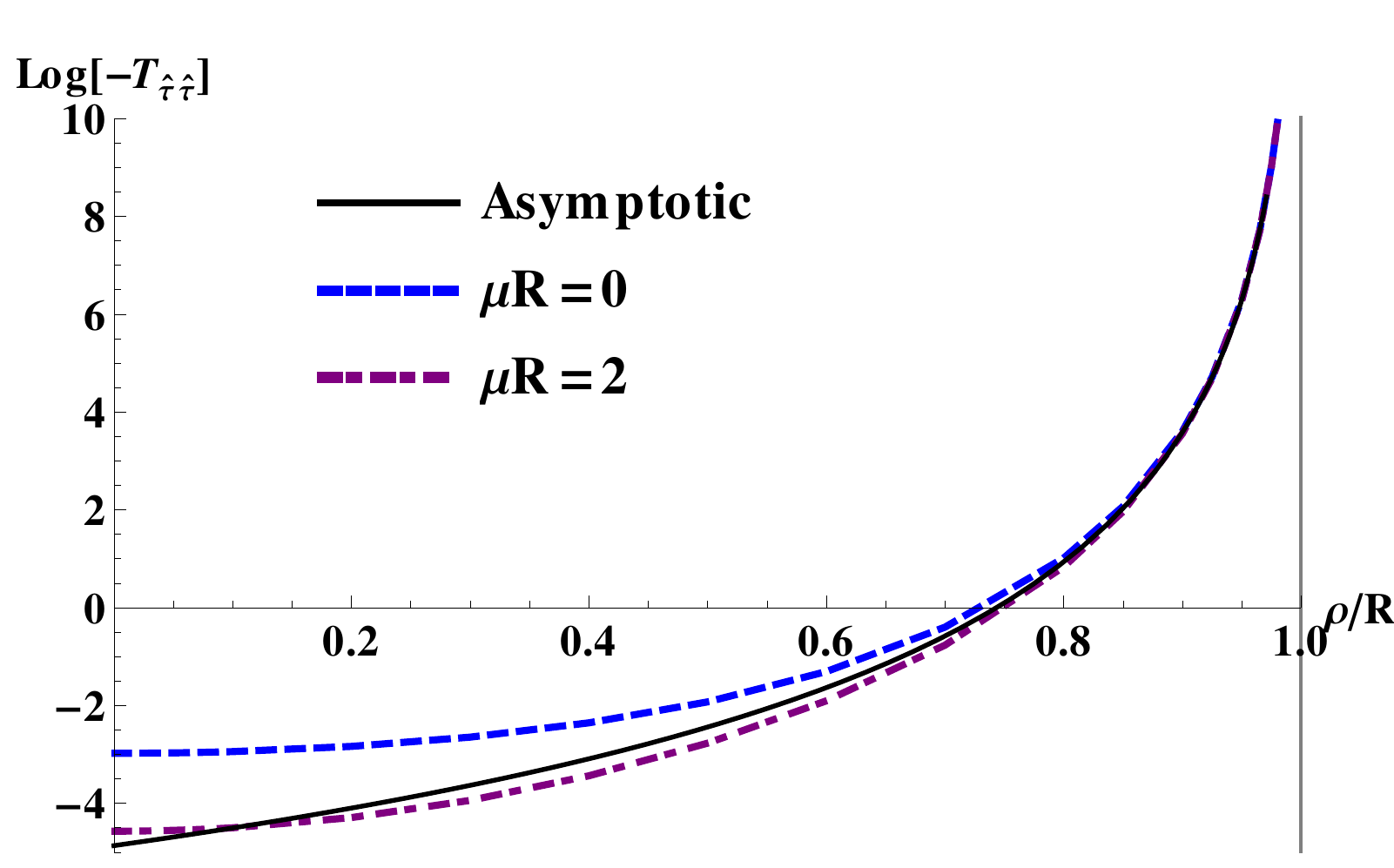} &
 \includegraphics[width=0.45\linewidth]{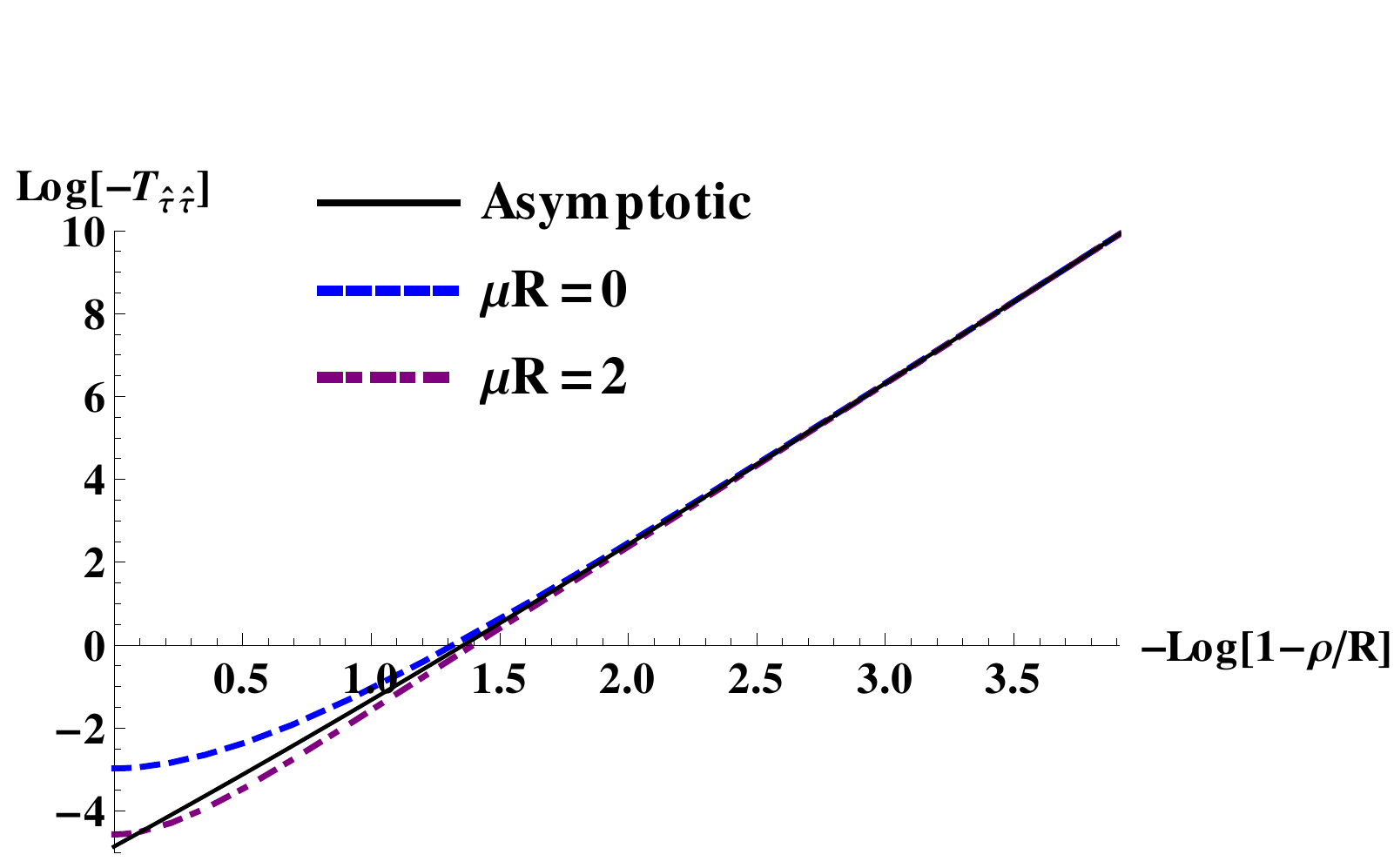}\\
 \includegraphics[width=0.45\linewidth]{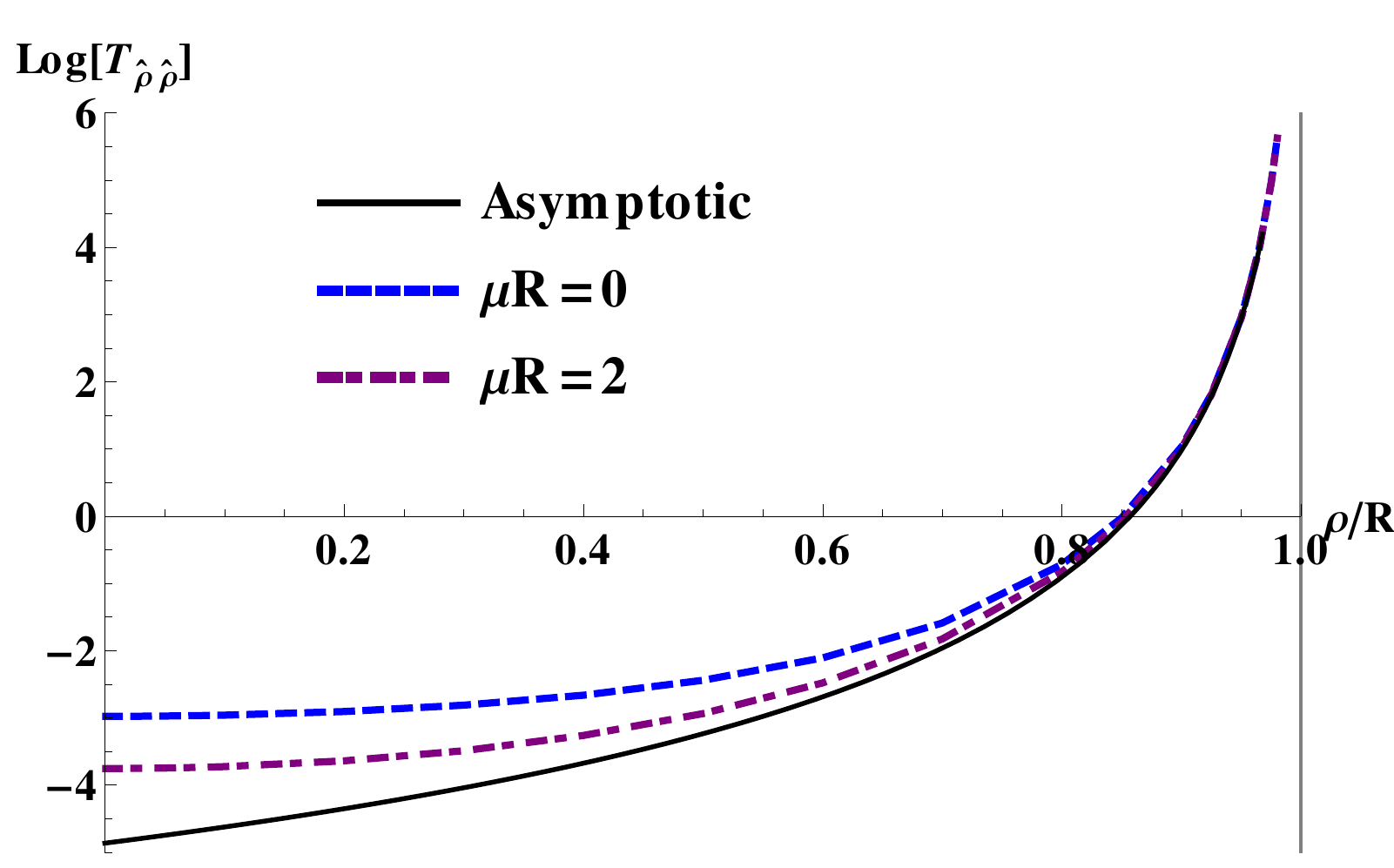} &
 \includegraphics[width=0.45\linewidth]{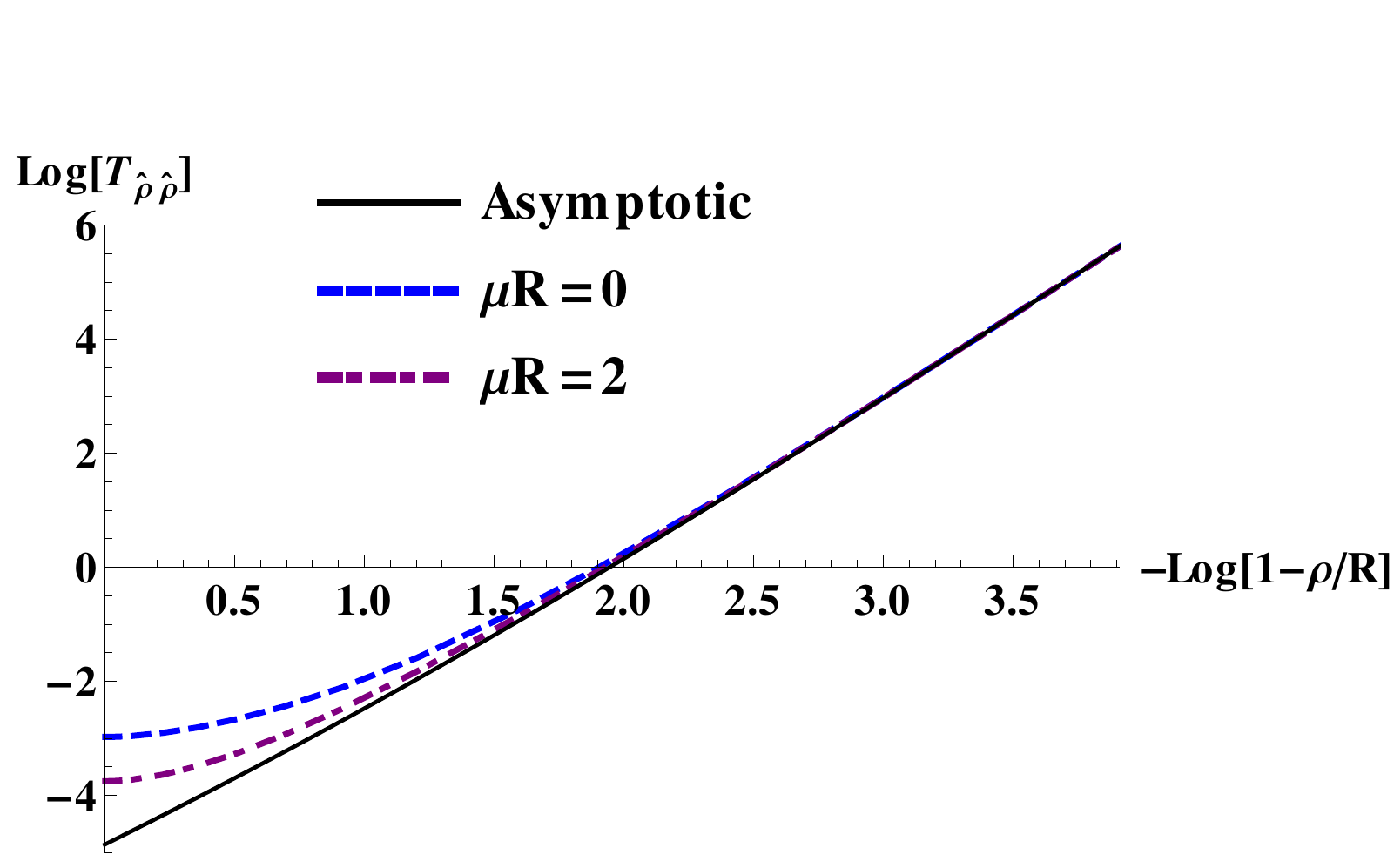}\\
 \includegraphics[width=0.45\linewidth]{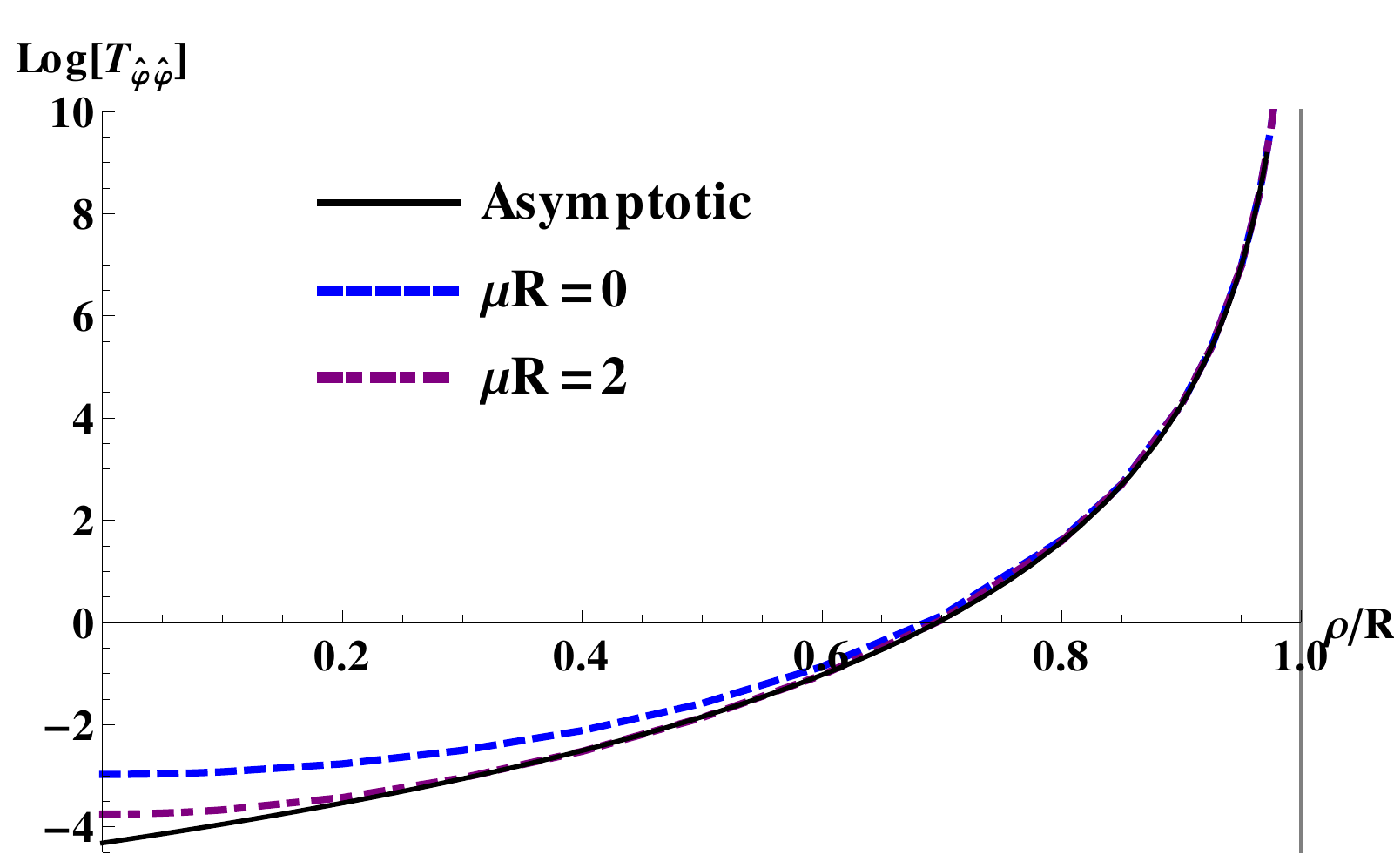} &
 \includegraphics[width=0.45\linewidth]{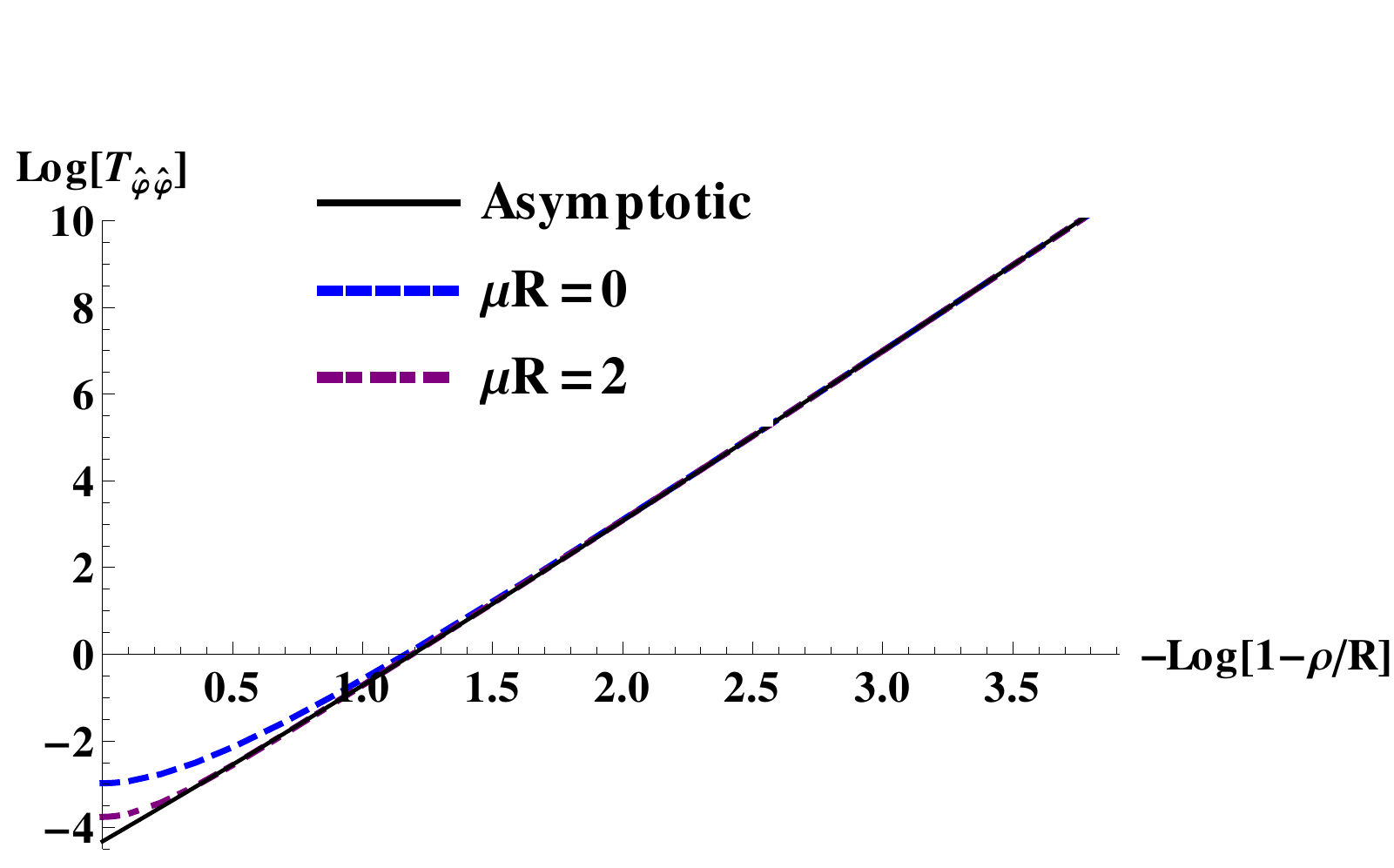}
\end{tabular}
\caption{Casimir expectation values for spectral boundary conditions.
The left column presents the logarithm of the absolute value of the FC divided by the field mass $\mu ^{-1}\braket{\psibar\psi}_{\text{Cas}}^{\text{sp}}$ (first line) and
of the nonzero components of the SET $\braket{T_{\halpha\hsigma}}_{\text {Cas}}^{\text {sp}}$ (lines 2-4) as functions of
the scaled radial coordinate $\rho /R$, so that the boundary of the cylinder
is at $\rho /R=1$.
The right column shows the same quantities, but as functions of
the logarithm of the inverse distance $\epsilon^{-1}$ (\ref{eq:cas_epsilon_def}) to the boundary.
The plots compare the results for massless (blue (upper) dashed curves) and massive
(purple (lower) dot-dashed curves) fermions to the asymptotic results (dark thin curves) in
Eqs.~\eqref{eq:cas_spec_res}.
}
\label{fig:cas_spec_res}
\end{figure*}

In Fig.~\ref{fig:cas_spec_res} we compare the asymptotic results in Eqs.~\eqref{eq:cas_spec_res}
with numerical evaluations of the Casimir expectation values (\ref{eq:cas_spec_ppsi}, \ref{eq:cas_spec_SET}) for $\mu R = 0$ and $\mu R = 2$.
For all expectation values, we plot the logarithm of the magnitude of the relevant quantity, in the left-hand column as a function of $\rho /R$ on a linear scale and in the right-hand column as a function of the logarithm of $\epsilon^{-1}$ (\ref{eq:cas_epsilon_def}).

From (\ref{eq:cas_spec_ppsi}), the Casimir expectation value of the FC vanishes if the field is massless $\mu =0$, as was the case for the
thermal expectation values with spectral boundary conditions in Sec.~\ref{sec:spec_tev_FC}. Furthermore, we find that
the expectation value $\braket{\psibar\psi}_{\text{Cas}}^{\text{sp}}$ is negative for all $\rho $ (near the boundary,
this is expected from (\ref{eq:cas_spec_res_ppsi})). We therefore plot the logarithm of $-\mu ^{-1}\braket{\psibar\psi}_{\text{Cas}}^{\text{sp}}$.
For the SET components, we find that $\braket{T\indices{^\hrho_\hrho}}_{\text{Cas}}^{\text{sp}}$ and $\braket{T\indices{^\hvarphi_\hvarphi}}_{\text{Cas}}^{\text{sp}}$ are positive everywhere, while $\braket{T\indices{^\htau_\htau}}_{\text{Cas}}^{\text{sp}}$
is negative everywhere (therefore we plot the logarithm of $-\braket{T\indices{^\htau_\htau}}_{\text{Cas}}^{\text{sp}}$).

All the Casimir expectation values are regular inside the cylinder but not on the boundary.
All quantities shown in Fig.~\ref{fig:cas_spec_res} have smaller magnitudes for a massive fermion field compared with the massless case.
The absolute values of all the Casimir expectation values plotted in Fig.~\ref{fig:cas_spec_res} have their minimum on the axis of the cylinder at $\rho =0$, and increase monotonically as the radial coordinate $\rho $ increases. All diverge as $\rho \rightarrow R$ and the boundary is approached.
The agreement between the asymptotic and numerical results as the boundary is approached is excellent, confirming
the predicted order of divergence in Eqs.~\eqref{eq:cas_spec_res}.

\subsection{MIT bag boundary conditions}\label{sec:casMIT}

The leading-order Casimir divergence for fermions inside a cylinder in four-dimensional Minkowski space-time
has already been reported in Ref.~\cite{art:bezerra08}, but
only for the original MIT case (i.e.~$\varsigma = 1$).

In this section we apply the approach of Sec.~\ref{sec:casspectral} to the case with MIT bag boundary conditions, for both $\varsigma = \pm 1$.
Our approach is different from that in  Ref.~\cite{art:bezerra08}. We recover the leading-order, $\varsigma = 1$, results of Ref.~\cite{art:bezerra08} in Sec.~\ref{sec:cas_MIT_asympt}, except for the fermion condensate (FC), for which we obtain the opposite sign. This difference is due to a difference in how the FC is defined: we define the FC by analogy with the classical theory, such that Eq.~\eqref{eq:Ttrace} holds.

\subsubsection{Euclidean Green's function for MIT bag boundary conditions}
\label{sec:cas_MIT_green}

To form the Euclidean Green's function $S_E^{\text{MIT}}(x,x')$ for the bounded system with MIT bag boundary conditions, a solution
$\Delta S_E^{\text{MIT}}(x,x')$ of the homogeneous equation corresponding to \eqref{eq:euclid_dirac}
(i.e.~with the right hand side set to zero) must be added to the Euclidean Green's function
(\ref{eq:euclid_unb}, \ref{eq:euclid_unb_chi}) for the unbounded space-time:
\begin{equation}
\label{eq:MIT_Green_function}
 S_E^{\text{MIT}}(x_E,x_E') = S_E^{\text{unb}}(x_E,x_E') + \Delta S_E^{\text{MIT}}(x_E,x_E').
\end{equation}
$\Delta S_E^{\text{MIT}}(x_E,x_E')$ can be Fourier transformed in a similar way to Eqs.~(\ref{eq:euclid_unb}, \ref{eq:euclid_spec_SE_Fourier}):
\begin{multline}
 \Delta S_E^{\text{MIT}}(x, x') = \int_{-\infty}^\infty \frac{d\omega}{8\pi^3}
 \int_{-\infty}^\infty dk \sum_{m = -\infty}^\infty e^{i\omega \Delta \tau + ik\Delta z}\\
 \times \Delta \chi^{\text{MIT}} ,
 \label{eq:euclid_MIT_SE_Fourier}
\end{multline}
where $\Delta \tau = \tau - \tau'$ and $\Delta z = z-z'$.
The $4\times 4$ matrix $\Delta \chi^{\text{MIT}}$ can be written in terms of four $2\times2$ matrices $\Delta \chi^{\text{MIT}}_{ab}$,
in a similar way to Eqs.~(\ref{eq:euclid_chi_def}, \ref{eq:euclid_spec_chi_def}):
\begin{equation}
 \Delta \chi^{\text{MIT}} =
  \begin{pmatrix}
  \Delta \chi^{\text{MIT}}_{11} & \Delta \chi^{\text{MIT}}_{12}\\
  \Delta \chi^{\text{MIT}}_{21} & \Delta \chi^{\text{MIT}}_{22}
 \end{pmatrix}.\label{eq:dirac_MIT_cas_SE_def}
\end{equation}
The $2\times2$ matrices $\Delta \chi^{\text{MIT}}_{ik}$, in turn, can be written as:
\begin{align}
 \frac{\Delta \chi^{\text{MIT}}_{11}}{\mu - i\omega} =&
 \begin{pmatrix}
  a_{11} & b_{11} \\
  c_{11} & d_{11}
 \end{pmatrix} \circ \mathcal{E}_j,&
 \Delta \chi^{\text{MIT}}_{12} =&
 \begin{pmatrix}
  a_{12} & b_{12} \\
  c_{12} & d_{12}
 \end{pmatrix} \circ \mathcal{E}_j,\nonumber\\
 \frac{\Delta \chi^{\text{MIT}}_{22}}{\mu + i\omega} =&
 \begin{pmatrix}
  a_{22} & b_{22} \\
  c_{22} & d_{22}
 \end{pmatrix} \circ \mathcal{E}_j, &
 \Delta \chi^{\text{MIT}}_{21} =&
 \begin{pmatrix}
  a_{21} & b_{21} \\
  c_{21} & d_{21}
 \end{pmatrix} \circ \mathcal{E}_j,
\end{align}
where $a_{ik}$, $b_{ik}$, $c_{ik}$ and $d_{ik}$ are constants, the matrix $\mathcal{E}_j$
on the right of the Hadamard (Schur) product is defined in Eq.~\eqref{eq:euclid_E_def} and $j$ is
a generic label for the parameters $m \equiv m_j$, $\omega \equiv \omega_j$ and $k \equiv k_j$.

The matrix elements of the off-diagonal blocks $\Delta \chi^{\text{MIT}}_{12}$ and $\Delta \chi^{\text{MIT}}_{21}$
can be found using analogues of Eqs.~\eqref{eq:euclid_offd}, as follows:
\begin{subequations}
\label{eq:dirac_MIT_cas_offd_explicit}
\begin{align}
 \begin{pmatrix}
  a_{12} & b_{12} \\
  c_{12} & d_{12}
 \end{pmatrix} =&
 \begin{pmatrix}
  -k a_{22} - \alpha c_{22} & -k b_{22} - \alpha d_{22}\\
  \alpha a_{22} + k c_{22} & \alpha b_{22} + k d_{22}
 \end{pmatrix}\nonumber\\
 =&
 \begin{pmatrix}
  -k a_{11} - \alpha b_{11} & \alpha a_{11} + k b_{11} \\
  -k c_{11} - \alpha d_{11} & \alpha c_{11} + k d_{11}
 \end{pmatrix},\label{eq:dirac_MIT_cas_chi12_explicit}\\
 \begin{pmatrix}
  a_{21} & b_{21} \\
  c_{21} & d_{21}
 \end{pmatrix} =&
 \begin{pmatrix}
  k a_{11} + \alpha c_{11} & k b_{11} + \alpha d_{11}\\
  -\alpha a_{11} - k c_{11} & -\alpha b_{11} - k d_{11}
 \end{pmatrix}\nonumber\\ =&
 \begin{pmatrix}
  k a_{22} + \alpha b_{22} & -\alpha a_{22} - k b_{22} \\
  k c_{22} + \alpha d_{22} & -\alpha c_{22} - k d_{22}
 \end{pmatrix},
 \label{eq:dirac_MIT_cas_chi21_explicit}
\end{align}
\end{subequations}
where $\alpha $ is given in (\ref{eq:alpha}).
Eqs.~(\ref{eq:dirac_MIT_cas_offd_explicit}) can be used to express all matrix elements of $\Delta \chi^{\text{MIT}}$
in terms of the matrix elements of $\Delta \chi^{\text{MIT}}_{11}$.
The matrix elements of $\Delta \chi^{\text{MIT}}_{22}$ are given below, for completeness,
with respect to those of $\Delta \chi^{\text{MIT}}_{11}$:
\begin{equation}
 \begin{pmatrix}
  a_{22}\\ b_{22} \\ c_{22} \\ d_{22}
 \end{pmatrix} =
 \frac{1}{\alpha^2 - k^2}
 \begin{pmatrix}
  -k^2 & -\alpha k & -\alpha k & -\alpha^2 \\
  \alpha k & k^2 & \alpha^2 & \alpha k\\
  \alpha k & \alpha^2 & k^2 & \alpha k\\
  -\alpha^2 & -\alpha k & -\alpha k & -k^2
 \end{pmatrix}
 \begin{pmatrix}
  a_{11}\\ b_{11} \\ c_{11} \\ d_{11}
 \end{pmatrix}.
\end{equation}

Since the Euclidean Green's function
is formally equivalent to the Lorentzian Feynman propagator, the MIT boundary conditions \eqref{eq:MIT} remain unchanged when
the Euclidean propagator is considered:
\begin{align}
 (i \gamma^\hrho + \varsigma) S^{\text{MIT}}_E(x,x')\rfloor_{\rho = R} =& 0,\nonumber\\
 S_E^{\text{MIT}}(x, x') (-i \gamma^{\hrho'} + \varsigma)\rfloor_{\rho' = R} =& 0.
 \label{eq:dirac_MIT_SE_bc}
\end{align}
To begin the construction of $\Delta S_E^{\text{MIT}}(x,x')$ (\ref{eq:MIT_Green_function}), we require the values on the boundary of the Fourier transform
$\chi ^{\text {unb}}$ (\ref{eq:euclid_unb_chi}) of the Euclidean Green's function $S_E^{\text {unb}}(x,x')$ (\ref{eq:euclid_unb}) for the unbounded space-time.
These values can be inferred from Eqs.~(\ref{eq:euclid_unb_chi}, \ref{eq:euclid_E_def}):
\begin{widetext}
\begin{align}
 \chi^{\text {unb}}\rfloor_{\rho = R} =&
 \begin{pmatrix}
  (\mu - i\omega) \frac{K_m}{I_m} & 0 & -k \frac{K_m}{I_m} & \alpha \frac{K_m}{I_m}\\
  0 & (\mu - i\omega) \frac{K_{m+1}}{I_{m+1}} & -\alpha \frac{K_{m+1}}{I_{m+1}} & k \frac{K_{m+1}}{I_{m+1}}\\
  k \frac{K_m}{I_m} & -\alpha \frac{K_m}{I_m} & (\mu + i\omega) \frac{K_m}{I_m} & 0\\
  \alpha \frac{K_{m+1}}{I_{m+1}} & -k \frac{K_{m+1}}{I_{m+1}} & 0 & (\mu + i\omega) \frac{K_{m+1}}{I_{m+1}}
 \end{pmatrix} \circ
 \begin{pmatrix}
  \mathcal{E}_j & \mathcal{E}_j\\ \mathcal{E}_j & \mathcal{E}_j
 \end{pmatrix},\nonumber\\
 \chi^{\text {unb}}\rfloor_{\rho' = R} =&
 \begin{pmatrix}
  (\mu - i\omega) \frac{K_m}{I_m} & 0 & -k \frac{K_m}{I_m} & -\alpha \frac{K_{m+1}}{I_{m+1}}\\
  0 & (\mu - i\omega) \frac{K_{m+1}}{I_{m+1}} & \alpha \frac{K_{m}}{I_{m}} & k \frac{K_{m+1}}{I_{m+1}}\\
  k \frac{K_m}{I_m} & \alpha \frac{K_{m+1}}{I_{m+1}} & (\mu + i\omega) \frac{K_m}{I_m} & 0\\
  -\alpha \frac{K_{m}}{I_{m}} & -k \frac{K_{m+1}}{I_{m+1}} & 0 & (\mu + i\omega) \frac{K_{m+1}}{I_{m+1}}
 \end{pmatrix} \circ
 \begin{pmatrix}
  \mathcal{E}_j & \mathcal{E}_j\\ \mathcal{E}_j & \mathcal{E}_j
 \end{pmatrix},
\end{align}
\end{widetext}
where the modified Bessel functions explicitly displayed in the ratios $K_m/I_m$ and $K_{m+1}/I_{m+1}$ have
argument $\alpha R$. The dependence on the coordinates $\rho$, $\rho'$, $\varphi$ and $\varphi'$ is
fully contained in the matrices $\mathcal{E}_j$ (\ref{eq:euclid_E_def}).

The boundary conditions \eqref{eq:dirac_MIT_SE_bc} give $32$ equations for the matrix elements
of $\Delta \chi^{\text{MIT}}$ \eqref{eq:dirac_MIT_cas_SE_def}.
However, only a comparatively small number of these equations is required to fully determine
$\Delta \chi^{\text{MIT}}$. The $(1,1)$ components of Eqs.~\eqref{eq:dirac_MIT_SE_bc} (i.e.~the
top left components of the equations for both $\rho = R$ and $\rho' = R$),
\begin{align}
 \varsigma(\mu - i\omega) (K_m + I_m a_{11}) - \alpha K_{m+1} - I_{m+1} c_{21} =& 0,\nonumber\\
 \varsigma(\mu - i\omega) (K_m + I_m a_{11}) - \alpha K_{m+1} + I_{m+1} b_{12} =& 0,
 \label{eq:dirac_MIT_cas_11}
\end{align}
show that
\begin{equation}
 c_{21} = -b_{12}.\label{eq:dirac_MIT_cas_c21_b12}
\end{equation}
A similar inspection of the $(2,2)$ components of Eqs.~\eqref{eq:dirac_MIT_SE_bc} shows that:
\begin{align}
 \varsigma(\mu - i\omega) (K_{m+1} + I_{m+1} d_{11}) - \alpha K_{m} + I_{m} b_{21} =& 0,\nonumber\\
 \varsigma(\mu - i\omega) (K_{m+1} + I_{m+1} d_{11}) - \alpha K_{m} - I_{m} c_{12} =& 0,
 \label{eq:dirac_MIT_cas_22}
\end{align}
leading to:
\begin{equation}
 b_{21} = -c_{12}.\label{eq:dirac_MIT_cas_b21_c12}
\end{equation}
Comparing the expressions for $b_{12}$ and $c_{21}$ in Eqs.~\eqref{eq:dirac_MIT_cas_offd_explicit} shows that:
\begin{equation}
 c_{11} = b_{11},\qquad c_{22} = b_{22},\label{eq:dirac_MIT_cas_c11_b11}
\end{equation}
which can be used together with the expressions for $d_{12}$, $d_{21}$, $a_{12}$ and $a_{21}$ in Eqs.~\eqref{eq:dirac_MIT_cas_offd_explicit} to give:
\begin{equation}
 a_{21} = -a_{12}, \qquad d_{21} = - d_{12}.\label{eq:dirac_MIT_cas_a21d21_a12d12}
\end{equation}
Using $d_{21} = -\alpha b_{11} - k d_{11}$ in the $(1,2)$ component of Eq.~\eqref{eq:dirac_MIT_SE_bc}
for $\rho = R$ gives:
\begin{equation}
 b_{11} = \frac{-k(I_{m+1} d_{11} + K_{m+1})}{\varsigma(\mu - i\omega) I_m + \alpha I_{m+1}},
 \label{eq:dirac_MIT_cas_b11_d11}
\end{equation}
where the argument of the modified Bessel functions is, as before, $\alpha R$. Substituting \eqref{eq:dirac_MIT_cas_b11_d11} into
$b_{21} = k b_{11} + \alpha d_{11}$ gives:
\begin{equation}
 b_{21} = \frac{\varsigma (\mu -i\omega) [\alpha I_m + \varsigma (i\omega + \mu) I_{m+1}] d_{11} - k^2 K_{m+1}}
 {\varsigma(\mu - i\omega) I_m + \alpha I_{m+1}}.\label{eq:dirac_MIT_cas_b21_d11}
\end{equation}
Substituting $b_{21}$ into the first equation in \eqref{eq:dirac_MIT_cas_22} gives:
\begin{equation}
 d_{11} = -\frac{K_{m+1}}{I_{m+1}} + \frac{1}{\casU} \frac{I_m}{I_{m+1}} +
 \frac{1}{\casU} \frac{\varsigma \alpha}{\mu - i\omega},\label{eq:dirac_MIT_cas_d11}
\end{equation}
where the following property was used to eliminate $K_m(\alpha R)$:
\begin{equation}
 K_{m}(z) I_{m+1}(z) + K_{m+1}(z) I_m(z) = \frac{1}{z}.\label{eq:Wronskian_modified}
\end{equation}
The quantity $\casU \equiv \casU_m(\alpha R)$ introduced in Eq.~\eqref{eq:dirac_MIT_cas_d11}
is defined as \cite{art:bezerra_saharian}:
\begin{multline}
\casU \equiv \casU_m(\alpha R) = \alpha R [I_m^2(\alpha R) + I_{m+1}^2(\alpha R)]\\
 + 2\varsigma \mu R\,I_m(\alpha R) I_{m+1}(\alpha R).
 \label{eq:dirac_MIT_cas_Udef}
\end{multline}
Substituting $d_{11}$ back into Eq.~\eqref{eq:dirac_MIT_cas_b11_d11} gives:
\begin{equation}
 b_{11} = -\frac{\varsigma k}{\casU(\mu - i\omega)}.\label{eq:dirac_MIT_cas_b11}
\end{equation}
The constant $a_{11}$ can be found by substituting $a_{21} = -k a_{11} - \alpha c_{11}$ into the $(2,1)$ component of
Eq.~\eqref{eq:dirac_MIT_SE_bc} for $\rho = R$:
\begin{equation}
 a_{11} = -\frac{K_m}{I_m} + \frac{1}{\casU} \frac{I_{m+1}}{I_m} +
 \frac{1}{\casU} \frac{\varsigma \alpha}{\mu - i\omega}.\label{eq:dirac_MIT_cas_a11}
\end{equation}

The results in Eqs.~(\ref{eq:dirac_MIT_cas_d11}, \ref{eq:dirac_MIT_cas_b11}, \ref{eq:dirac_MIT_cas_a11})  can
be summarized as follows. The difference between the vacuum Euclidean Green's functions for the bounded and unbounded space-times is given by (\ref{eq:euclid_MIT_SE_Fourier}), with the matrix $\Delta \chi ^{\text {MIT}}$ having the form \eqref{eq:dirac_MIT_cas_SE_def}.
\begin{widetext}
The $2\times 2$ matrix element $\Delta \chi^{\text{MIT}}_{11}$ is given by:
\begin{subequations}
\label{eq:dirac_MIT_cas_chi}
\begin{equation}
 \Delta\chi^{\text{MIT}}_{11} =
 (\mu - i\omega)
 \begin{pmatrix}
  -\frac{K_m}{I_m} + \frac{1}{\casU} \frac{I_{m+1}}{I_m} +
  \frac{1}{\casU} \frac{\varsigma \alpha}{\mu - i\omega} &
  -\frac{1}{\casU} \frac{\varsigma k}{\mu - i\omega} \\
  -\frac{1}{\casU} \frac{\varsigma k}{\mu - i\omega} &
  - \frac{K_{m+1}}{I_{m+1}} + \frac{1}{\casU} \frac{I_m}{I_{m+1}} +
  \frac{1}{\casU} \frac{\varsigma \alpha}{\mu - i\omega}
 \end{pmatrix} \circ \mathcal{E}_j.\label{eq:dirac_MIT_cas_chi11}
\end{equation}
The $2\times 2$ matrix $\Delta \chi^{\text{MIT}}_{12}$ can be found from Eq.~\eqref{eq:dirac_MIT_cas_chi12_explicit}:
\begin{equation}
 \Delta\chi^{\text{MIT}}_{12} =
 \begin{pmatrix}
  k\left(\frac{K_m}{I_m} - \frac{1}{\casU}\frac{I_{m+1}}{I_{m}}\right) &
  -\alpha\left(\frac{K_m}{I_m} - \frac{1}{\casU}\frac{I_{m+1}}{I_{m}}\right) +
  \frac{\varsigma(\mu + i \omega)}{\casU}\\
  \alpha\left(\frac{K_{m+1}}{I_{m+1}} - \frac{1}{\casU}\frac{I_{m}}{I_{m+1}}\right) -
  \frac{\varsigma(\mu + i \omega)}{\casU} &
  -k\left(\frac{K_{m+1}}{I_{m+1}} - \frac{1}{\casU}\frac{I_{m}}{I_{m+1}}\right)
 \end{pmatrix} \circ \mathcal{E}_j.\label{eq:dirac_MIT_cas_chi12}
\end{equation}
The matrix elements of $\Delta \chi^{\text{MIT}}_{21}$ can be found from Eq.~\eqref{eq:dirac_MIT_cas_chi12} using
Eqs.~(\ref{eq:dirac_MIT_cas_c21_b12}, \ref{eq:dirac_MIT_cas_b21_c12}, \ref{eq:dirac_MIT_cas_a21d21_a12d12}):
\begin{equation}
 \Delta\chi^{\text{MIT}}_{21} =
 \begin{pmatrix}
  -k\left(\frac{K_m}{I_m} - \frac{1}{\casU}\frac{I_{m+1}}{I_{m}}\right) &
  -\alpha\left(\frac{K_{m+1}}{I_{m+1}} - \frac{1}{\casU}\frac{I_{m}}{I_{m+1}}\right) +
  \frac{\varsigma(\mu + i \omega)}{\casU}\\
  \alpha\left(\frac{K_m}{I_m} - \frac{1}{\casU}\frac{I_{m+1}}{I_{m}}\right) -
  \frac{\varsigma(\mu + i \omega)}{\casU} &
  k\left(\frac{K_{m+1}}{I_{m+1}} - \frac{1}{\casU}\frac{I_{m}}{I_{m+1}}\right)
 \end{pmatrix} \circ \mathcal{E}_j.\label{eq:dirac_MIT_cas_chi21}
\end{equation}
Finally, the components of $\Delta \chi^{\text{MIT}}_{22}$ can be found by inverting Eq.~\eqref{eq:dirac_MIT_cas_chi12_explicit}:
\begin{equation}
 \Delta\chi^{\text{MIT}}_{22} =
 (\mu + i\omega)
 \begin{pmatrix}
  -\frac{K_m}{I_m} + \frac{1}{\casU} \frac{I_{m+1}}{I_m} +
  \frac{1}{\casU} \frac{\varsigma \alpha}{\mu+i\omega} &
  -\frac{1}{\casU} \frac{\varsigma k}{\mu + i\omega} \\
  -\frac{1}{\casU} \frac{\varsigma k}{\mu + i\omega} &
  -\frac{K_{m+1}}{I_{m+1}} + \frac{1}{\casU} \frac{I_m}{I_{m+1}} +
  \frac{1}{\casU} \frac{\varsigma \alpha}{\mu +i\omega}
 \end{pmatrix} \circ \mathcal{E}_j.\label{eq:dirac_MIT_cas_chi22}
\end{equation}
\end{subequations}
\end{widetext}
In (\ref{eq:dirac_MIT_cas_chi}), the matrix $\mathcal {E}_{j}$ is given in (\ref{eq:euclid_E_def}), $\alpha $ is defined in (\ref{eq:alpha})
and ${\casU}$ is given in Eq.~\eqref{eq:dirac_MIT_cas_Udef}. The modified
Bessel functions $I_{m}$, $K_{m}$ written explicitly in (\ref{eq:dirac_MIT_cas_chi}) have argument $\alpha R$.
The matrix $\mathcal {E}_{j}$ contains all the dependence on the coordinates $\rho $, $\rho '$, $\varphi $, $\varphi '$.

\subsubsection{Casimir expectation values}
We now use the Euclidean Green's function with MIT bag boundary conditions to calculate the Casimir expectation values of the fermion condensate (FC) $\braket{\psibar\psi}_{\text {Cas}}^{\text {MIT}}$,
charge current $\braket{\current^{\halpha}}_{\text {Cas}}^{\text {MIT}}$, neutrino charge current (CC) $\braket{\current^{\halpha}_\nu}_{\text {Cas}}^{\text {MIT}}$
and stress-energy tensor (SET) $\braket{T_{\halpha\hsigma}}_{\text {Cas}}^{\text {MIT}}$.
These expectation values can be computed from the formulae (\ref{eq:euclid_tevs_def}), replacing
$\Delta S_E^{\text{sp}}(x_E,x_E')$ by the difference $\Delta S_E^{\text{MIT}}(x_E,x_E')$ (\ref{eq:euclid_MIT_SE_Fourier}) between the vacuum Euclidean Green's functions for the bounded system with MIT bag boundary conditions and unbounded Minkowski space-time.

\begin{widetext}
First, the Casimir expectation value of the FC takes the form
\begin{multline}
 \braket{\psibar\psi}_{\text{Cas}}^{\text{MIT}} = \frac{1}{8\pi^3} \sum_{m = -\infty}^\infty \int_{-\infty}^\infty d\omega
 \int_{-\infty}^\infty dk \Bigg\{\frac{\alpha\varsigma}{\casU}I_m^+(\alpha\rho)
 +\mu\left[\left(-\frac{K_m}{I_m} + \frac{1}{\casU}\frac{I_{m+1}}{I_{m}}\right)I_m^2(\alpha\rho)
  \right. \\ \left.
  - \left(-\frac{K_{m+1}}{I_{m+1}} + \frac{1}{\casU}\frac{I_{m}}{I_{m+1}}\right)I_{m+1}^2(\alpha\rho)\right]
 \Bigg\},
 \label{eq:MIT_FC_temp1}
\end{multline}
where the arguments of the modified Bessel functions are $\alpha R$ unless explicitly stated otherwise.
The expression (\ref{eq:MIT_FC_temp1}) can be simplified by changing to the polar coordinates (\ref{eq:euclid_pol}), and then
performing the integral over $\vartheta $.
Afterwards, the terms involving
$I_m^2(\alpha\rho)$ and $I_{m+1}^2(\alpha\rho)$ can be symmetrized to only contain the combinations
$I_m^+(\alpha\rho)$ and $I_m^-(\alpha\rho)$, defined in Eqs.~\eqref{eq:BesselI*}.
This gives the following expression:
\begin{multline}
 \braket{\psibar\psi}_{\text{Cas}}^{\text{MIT}} = - \sum_{m = -\infty}^\infty
 \int_{\mu R}^\infty \frac{d\casx}{4\pi^2 R^3}
 \Bigg\{\frac{I_m^+(\casx\rhoo)}{\casU_m(\casx) I_m(\casx) I_{m+1}(\casx)}
 \left[\frac{\mu}{\casx} \casU_m(\casx) -
 \mu [I_m^2(\casx) + I_{m+1}^2(\casx)] - 2\varsigma \alpha I_m(\casx) I_{m+1}(\casx)\right]
 \\
 -\frac{\mu I_m^-(\casx\rhoo)}{\casU_m(\casx) I_m(\casx) I_{m+1}(\casx)}
 \{[I_m^2(\casx) - I_{m+1}^2(\casx)]
 + \casU_m(\casx) [K_m(\casx) I_{m+1}(\casx) - K_{m+1}(\casx) I_m(\casx) ] \}
 \Bigg\},\label{eq:dirac_MIT_cas_ppsi_aux}
\end{multline}
where $\mathtt{x}$ is defined in Eq.~(\ref{eq:casx}) and
the Wronskian relation \eqref{eq:Wronskian_modified} was used in the coefficient of $I_m^+(\casx \rhoo)$.
The coefficient of $I_m^-(\casx \rhoo)$ in (\ref{eq:dirac_MIT_cas_ppsi_aux}) can be simplified by inserting a factor of
$\casx [K_m(\casx) I_{m+1}(\casx) + K_{m+1}(\casx) I_m(\casx)]=1$ next to
$[I_m^2(\casx) - I_{m+1}^2(\casx)]$, so that:
\begin{equation}
 \casU_m(\casx) [K_m(\casx) I_{m+1}(\casx) - K_{m+1}(\casx) I_m(\casx) ]
 + [I_m^2(\casx) - I_{m+1}^2(\casx)] = 2 I_m(\casx) I_{m+1}(\casx) \casW_m(\casx),
\end{equation}
where $\casW_m(\mathtt{x})$ is defined as \cite{art:bezerra_saharian}
\begin{equation}
 \casW_m(\mathtt{x}) = \mathtt{x}\left[K_m(\mathtt{x}) I_m(\mathtt{x}) - K_{m+1}(\mathtt{x}) I_{m+1}(\mathtt{x})\right]
 +\varsigma\mu R\left[K_m(\mathtt{x}) I_{m+1}(\mathtt{x}) - K_{m+1}(\mathtt{x}) I_{m}(\mathtt{x})\right].
 \label{eq:dirac_MIT_cas_Wdef}
\end{equation}
The final form for $\braket{\psibar\psi}_{\text{Cas}}^{\text{MIT}}$ can be obtained by
using the explicit expression \eqref{eq:dirac_MIT_cas_Udef} for $\casU_m(\casx)$ in the coefficient
of $I_m^+(\casx \rhoo)$ in Eq.~\eqref{eq:dirac_MIT_cas_ppsi_aux}:
\begin{equation}
 \braket{\psibar\psi}_{\text{Cas}}^{\text{MIT}} =-\sum_{m = -\infty}^\infty
 \int_{\mu R}^\infty \frac{d\mathtt{x}}{2\pi^2 R^3} \frac{1}{\casU_m(\mathtt{x})} \left[
 \casx\mu R \casW_m(\mathtt{x}) I_m^-(\mathtt{x}\rhoo) - \varsigma(\casx^2 - \mu^2 R^2)\, I_m^+(\mathtt{x}\rhoo)\right].
 \label{eq:dirac_MIT_cas_ppsi}
\end{equation}

As in the case with spectral boundary conditions, for MIT bag boundary conditions we find that the Casimir expectation values of all components of the charge current $\braket{\current^{\halpha}}_{\text {Cas}}^{\text {MIT}}$ and neutrino charge current $\braket{\current^{\halpha}_\nu}_{\text {Cas}}^{\text {MIT}}$ vanish.
The Casimir expectation value of the components of the SET can be calculated using (\ref{eq:euclid_sp_SET}), with
$\Delta S_E^{\text{sp}}(x_E,x_E')$ replaced by $\Delta S_E^{\text{MIT}}(x_E,x_E')$ (\ref{eq:euclid_MIT_SE_Fourier}).
Grouping terms as for the FC, we obtain the following expressions for the components of the SET relative to the Euclidean version of the tetrad (\ref{eq:tetrad}):
\begin{subequations}
\begin{align}
 \braket{T\indices{^\htau_\htau}}_{\text{Cas}}^{\text{MIT}} =& -\sum_{m = -\infty}^{\infty}
 \int_{\mu R}^{\infty} \frac{d\mathtt{x}}{4 \pi^2 R^4}
 \frac{\mathtt{x}^2 - \mu^2 R^2}{\casU_m(\mathtt{x})}\left[ \varsigma \mu R \, I_m^+(\mathtt{x}\rhoo) +
 \mathtt{x} \casW_m(\mathtt{x})\, I_m^-(\mathtt{x}\rhoo)\right] ,
 \\
 \braket{T\indices{^\hrho_\hrho}}_{\text{Cas}}^{\text{MIT}} =& \sum_{m = -\infty}^{\infty}
 \int_{\mu R}^{\infty} \frac{\mathtt{x}^3\, d\mathtt{x}}{2\pi^2 R^4}
 \frac{\casW_m(\mathtt{x})}{\casU_m(\mathtt{x})}  \left[I_m^-(\mathtt{x}\rhoo) -
 \frac{m+\frac{1}{2}}{\mathtt{x}\rhoo} I_m^\times(\mathtt{x}\rhoo)\right] ,
 \\
 \braket{T\indices{^\hvarphi_\hvarphi}}_{\text{Cas}}^{\text{MIT}} =& \sum_{m = -\infty}^{\infty}
 \int_{\mu R}^{\infty} \frac{\mathtt{x}^2\, d\mathtt{x}}{2\pi^2 R^4} \frac{\casW_m(\mathtt{x})}{\casU_m(\mathtt{x})}
 \frac{m+\frac{1}{2}}{\rhoo} I_m^\times(\mathtt{x}\rhoo),
 \label{eq:dirac_MIT_cas_SET}
\end{align}
\end{subequations}
and $\braket{T\indices{^\hatz_\hatz}}_{\text{Cas}}^{\text{MIT}} =
\braket{T\indices{^\htau_\htau}}_{\text{Cas}}^{\text{MIT}}$.

By analogy with Eqs.~\eqref{eq:casIdef} for the spectral case, it is convenient to introduce the following
integrals:
\begin{align}
 \casI_{\ell n}^{\text{MIT},+} =& \frac{1}{2\pi^2 R^4}
 \sum_{m = -\infty}^\infty \int_{\mu R}^\infty \frac{d\mathtt{x}}{\casU_m(\mathtt{x})}
 \mathtt{x}^\ell (m + \tfrac{1}{2})^n I_m^+(\mathtt{x}\rhoo),\nonumber\\
 \casI_{\ell n}^{\text{MIT},-} =& \frac{1}{2\pi^2 R^4}
 \sum_{m = -\infty}^\infty \int_{\mu R}^\infty \frac{d\mathtt{x}}{\casU_m(\mathtt{x})}
 \mathtt{x}^\ell (m + \tfrac{1}{2})^n \casW_m(\mathtt{x}) I_m^-(\mathtt{x}\rhoo),\nonumber\\
 \casI_{\ell n}^{\text{MIT},\times} =& \frac{1}{2\pi^2 R^4}
 \sum_{m = -\infty}^\infty \int_{\mu R}^\infty \frac{d\mathtt{x}}{\casU_m(\mathtt{x})}
 \mathtt{x}^\ell (m + \tfrac{1}{2})^n \casW_m(\mathtt{x}) I_m^\times(\mathtt{x}\rhoo),
 \label{eq:dirac_MIT_Iln}
\end{align}
where the functions $I_m^*(z)$ were introduced in Eqs.~\eqref{eq:BesselI*}.
\end{widetext}

The Casimir expectation values of the FC and SET can be written in terms of the integrals (\ref{eq:dirac_MIT_Iln}) as follows:
\begin{subequations}
 \label{eq:dirac_MIT_cas_vevs}
\begin{align}
 \braket{\psibar\psi}_{\text{Cas}}^{\text{MIT}} =& -\mu R^2 \casI_{10}^{\text{MIT},-} +
 \varsigma R (\casI_{20}^{\text{MIT},+} - \mu^2 R^2 \casI_{00}^{\text{MIT},+}), \\
 \braket{T\indices{^\htau_\htau}}_{\text{Cas}}^{\text{MIT}} =&
 -\frac{1}{2} \varsigma \mu R (\casI_{20}^{\text{MIT},+} - \mu^2 R^2 \casI_{00}^{\text{MIT},+})\nonumber\\
 &- \frac{1}{2}(\casI_{30}^{\text{MIT},-} - \mu^2 R^2\casI_{10}^{\text{MIT},-}), \\
 \braket{T\indices{^\hrho_\hrho}}_{\text{Cas}}^{\text{MIT}} =& \casI_{30}^{\text{MIT},-} -
 \rhoo^{\,-1} \casI_{21}^{\text{MIT},\times},\\
 \braket{T\indices{^\hvarphi_\hvarphi}}_{\text{Cas}}^{\text{MIT}} =& \rhoo^{\,-1} \casI_{21}^{\text{MIT},\times} ,
\end{align}
\end{subequations}
and $\braket{T\indices{^\hatz_\hatz}}_{\text{Cas}}^{\text{MIT}} = \braket{T\indices{^\htau_\htau}}_{\text{Cas}}^{\text{MIT}}$.

%%%%%%%%%%%%%%%%%%%%%%%%%%%%%%%%%%%%%%%%%%%%%%%%%%%%%%%%%%%%%%%%%%%%%%%%%%%%%%%
\subsubsection{Casimir divergence near the boundary}
\label{sec:cas_MIT_div}
%%%%%%%%%%%%%%%%%%%%%%%%%%%%%%%%%%%%%%%%%%%%%%%%%%%%%%%%%%%%%%%%%%%%%%%%%%%%%%
As discussed in Sec.~\ref{sec:cas_spec_div} for the case of spectral boundary conditions,
the Casimir expectation values  (\ref{eq:dirac_MIT_cas_vevs})
 diverge as the boundary is approached.
To perform an analysis of this divergence, we follow the approach of Sec.~\ref{sec:cas_spec_div}.

%%%%%%%%%%%%%%%%%%%%%%%%%%%%%%%%%%%%%%%%%%%%%%%%%%%%%%%%%%%%%%%%%%%%%%%%%%%%%%
\paragraph{Generalized Abel-Plana formula.}\label{sec:cas_MIT_AbelPlana}
%%%%%%%%%%%%%%%%%%%%%%%%%%%%%%%%%%%%%%%%%%%%%%%%%%%%%%%%%%%%%%%%%%%%%%%%%%%%%%

We begin by defining the following quantities, which replace the sums over $m$ in (\ref{eq:dirac_MIT_Iln})
 by integrals:
 \begin{widetext}
\begin{align}
 \casIo_{\ell n}^{\text{MIT},+} =& \frac{1}{\pi^2 R^4}
 \int_0^\infty d\nu \int_{\mu R}^\infty \frac{d\mathtt{x}}{\casU_{\nu - \frac{1}{2}}(\mathtt{x})}
 \mathtt{x}^\ell \nu^n I_{\nu - \frac{1}{2}}^+(\mathtt{x}\rhoo),\nonumber\\
 \casIo_{\ell n}^{\text{MIT},-} =& \frac{1}{\pi^2 R^4}
 \int_0^\infty d\nu \int_{\mu R}^\infty \frac{d\mathtt{x}}{\casU_{\nu - \frac{1}{2}}(\mathtt{x})}
 \mathtt{x}^\ell \nu^n \casW_{\nu - \frac{1}{2}}(\mathtt{x}) I_{\nu - \frac{1}{2}}^-(\mathtt{x}\rhoo),\nonumber\\
 \casIo_{\ell n}^{\text{MIT},\times} =& \frac{1}{\pi^2 R^4}
 \int_0^\infty d\nu \int_{\mu R}^\infty \frac{d\mathtt{x}}{\casU_{\nu - \frac{1}{2}}(\mathtt{x})}
 \mathtt{x}^\ell \nu^n \casW_{\nu - \frac{1}{2}}(\mathtt{x}) I_{\nu - \frac{1}{2}}^\times(\mathtt{x}\rhoo).
 \label{eq:dirac_MIT_Ilnbar_def}
\end{align}
\end{widetext}
We define the differences between the quantities $\casI_{\ell n}^{\text{MIT},*}$ (\ref{eq:dirac_MIT_Iln}) and $\casIo_{\ell n}^{\text{MIT},*}$
to be
\begin{equation}
\delta_{\ell n}^{\text{MIT},*}(\rhoo) = \casIo_{\ell n}^{\text{MIT},*} - \casI_{\ell n}^{\text{MIT},*}.
 \label{eq:dirac_MIT_cas_delta_def}
\end{equation}
Following the spectral case, it is convenient to write $\casI_{\ell n}^{\text{MIT},*}$ (\ref{eq:dirac_MIT_Iln}) in terms of new functions
$f_{\ell n}^{\text{MIT},*}(\nu )$ as follows (cf.~Eq.~(\ref{eq:cas_spec_fdef}))
\begin{equation}
 \casI_{\ell n}^{\text{MIT},*} = \sum_{m = 0}^\infty f_{\ell n}^{\text{MIT},*}(m + \tfrac{1}{2}) .
\label{eq:cas_MIT_fdef}
\end{equation}
The precise forms of $f_{\ell n}^{\text{MIT},*}(\nu )$ for $*\in \{ +, -, \times \}$ can be deduced from comparing (\ref{eq:dirac_MIT_Iln}) with
(\ref{eq:cas_MIT_fdef}):
\begin{align}
 f_{\ell n}^{\text{MIT},+}(\nu ) =& \frac{1}{\pi^2 R^4}
 \int_{\mu R}^\infty \frac{d\mathtt{x}}{\casU_{\nu - \frac {1}{2}}(\mathtt{x})}
 \mathtt{x}^\ell \nu ^n I_{\nu - \frac {1}{2}}^+(\mathtt{x}\rhoo),\nonumber\\
 f_{\ell n}^{\text{MIT},-}(\nu ) =& \frac{1}{\pi^2 R^4}
 \int_{\mu R}^\infty \frac{d\mathtt{x}}{\casU_{\nu - \frac {1}{2}}(\mathtt{x})}
 \mathtt{x}^\ell \nu ^n \casW_{\nu - \frac {1}{2}}(\mathtt{x}) I_{\nu - \frac {1}{2}}^-(\mathtt{x}\rhoo),\nonumber\\
 f_{\ell n}^{\text{MIT},\times} (\nu ) =& \frac{1}{\pi^2 R^4}
 \int_{\mu R}^\infty \frac{d\mathtt{x}}{\casU_{\nu - \frac {1}{2}}(\mathtt{x})}
 \mathtt{x}^\ell \nu ^n \casW_{\nu - \frac {1}{2}}(\mathtt{x}) I_{\nu - \frac {1}{2}}^\times(\mathtt{x}\rhoo) .
 \label{eq:dirac_MIT_f}
\end{align}

From the detailed forms of the functions $f_{\ell n}^{\text{MIT},*}(\nu)$, it can be seen that they are analytic.
We can therefore apply the generalized Abel-Plana formula \eqref{eq:abelplana}.
This gives the differences $\delta_{\ell n}^{\text{MIT}, *}$ (\ref{eq:dirac_MIT_cas_delta_def}) to be:
\begin{equation}
 \delta_{\ell n}^{\text{MIT}, *}(\rhoo) =
 i \int_0^\infty dt \frac{f^{\text{MIT},*}_{\ell n}(it) - f^{\text{MIT},*}_{\ell n}(-it)}{e^{2\pi t} + 1}.
 \label{eq:dirac_MIT_cas_delta_int}
\end{equation}
To investigate the asymptotic behaviour of $\delta_{\ell n}^{\text{MIT}, *}(\rhoo)$ as $\rhoo \rightarrow 1$,
the asymptotic behaviour of the integrand in the integrals with respect to $\mathtt{x}$ in
Eq.~\eqref{eq:dirac_MIT_f} must be investigated. Since the $(e^{2\pi t} + 1)^{-1}$ factor in (\ref{eq:dirac_MIT_cas_delta_int}) ensures the
suppression of ${\rm Im}[f_{\ell n}^{\text{MIT}, *}(it)]$ at large $t$, the formulae
(\ref{eq:besselI_asympt_inf}, \ref{eq:besselK_asympt_inf}) for the asymptotic expansions
of the modified Bessel functions for large arguments can be used.

We begin by examining $\delta_{\ell n}^{\text{MIT}, +}(\rhoo)$.
The factor $\casU_{\nu - \frac{1}{2}}(\mathtt{x})$ \eqref{eq:dirac_MIT_cas_Udef} in the denominators
of $f_{\ell n}^{\text{MIT}, *}(\nu)$ \eqref{eq:dirac_MIT_f},
and the quantity $[\casU_{\nu - \frac{1}{2}}(\mathtt{x})]^{-1}$, have the following
asymptotic behaviours:
\begin{subequations}
\begin{align}
 \casU_{\nu - \frac{1}{2}}(\mathtt{x}) =& \frac{e^{2\mathtt{x}}}{\pi} \left[1 - \frac{\nu^2 - \varsigma \mu R}{\mathtt{x}}
 + \frac{\nu^2(\nu^2 - 2\varsigma \mu R)}{2\mathtt{x}^2}\right.\nonumber\\
 &\left. + O(\mathtt{x}^{-3})\right],\label{eq:dirac_MIT_cas_U_largex}\\
 \frac{1}{\casU_{\nu - \frac{1}{2}}(\mathtt{x})} =& \pi e^{-2\mathtt{x}}\left[1 + \frac{\nu^2 - \varsigma \mu R}{\mathtt{x}}
 \right.\nonumber\\
 &\left. + \frac{\nu^4 - 2\varsigma \mu R\, \nu^2 + 2\mu^2 R^2}{2\mathtt{x}^2}
  + O(\mathtt{x}^{-3})\right].\label{eq:dirac_MIT_cas_Uinv_largex}
\end{align}
\end{subequations}
Hence, the asymptotic expansion of the integrand in the integral with respect to $\mathtt{x}$ in
$f_{00}^{\text{MIT}, +}(\nu)$ (\ref{eq:dirac_MIT_f}) is:
\begin{multline}
 \frac{I_{\nu - \frac{1}{2}}^+(\mathtt{x}\rhoo)}{\casU_{\nu - \frac{1}{2}}(\mathtt{x})} =
 \frac{e^{-2\mathtt{x}\epsilon}}{\mathtt{x} \rhoo}\left[1 -
 \frac{\nu^2 \epsilon}{\mathtt{x}\rhoo} - \frac{\varsigma \mu R}{\mathtt{x}}\right.\\
 \left. + \frac{\nu^4\epsilon^2}{2\rhoo^2\mathtt{x}^2}+
 \frac{\nu^2 \varsigma\mu R\epsilon}{\rhoo \mathtt{x}^2} + \frac{\mu^2 R^2}{\mathtt{x}^2} + O(\mathtt{x}^{-3})\right].
 \label{eq:f00temp}
\end{multline}
For the analysis of the Casimir divergence for the FC and SET, the only $f_{\ell n}^{\text{MIT}, +}(\nu)$ quantities required are
$f_{00}^{\text{MIT}, +}(\nu)$ and $f_{20}^{\text{MIT}, +}(\nu)$.
It can be seen that the terms in the bracket in (\ref{eq:f00temp}) contain only even powers of $\nu$, which stay real under the
transformation $\nu \rightarrow it$. Hence, the following asymptotic behaviour can be obtained:
\begin{equation}
 {\rm Im}\left[\frac{1}{\casU_{it - \frac{1}{2}}(\mathtt{x})}
 \mathtt{x}^\ell \, I_{it - \frac{1}{2}}^+(\mathtt{x}\rhoo)\right] =
 \frac{1}{\rhoo} e^{-2\mathtt{x}\epsilon} O(\mathtt{x}^{\ell-3}).
\end{equation}
Since $\ell$ is either $0$ or $2$, it can be seen that $\delta_{00}^{\text{MIT},+}(\rhoo)$ and
$\delta_{20}^{\text{MIT},+}(\rhoo)$ do not diverge as $\rhoo \rightarrow 1$.

\begin{widetext}
To analyse $\delta_{\ell n}^{\text{MIT},-}(\rhoo)$ and $\delta_{\ell n}^{\text{MIT},\times}(\rhoo)$,
the asymptotic behaviour of $\casW_{\nu - \frac{1}{2}}(\mathtt{x})$, defined in
Eq.~\eqref{eq:dirac_MIT_cas_Wdef}, is required. Using the intermediate expansions:
\begin{align}
 K_{\nu - \frac{1}{2}}(\mathtt{x}) I_{\nu - \frac{1}{2}}(\mathtt{x}) -
 K_{\nu + \frac{1}{2}}(\mathtt{x}) I_{\nu + \frac{1}{2}}(\mathtt{x}) =&
 \frac{\nu}{2\mathtt{x}^3}\left[1 + \frac{\nu^2(\nu^2 - 1)(\nu^2 - 13)}{24\mathtt{x}^2}+ O(\mathtt{x}^{-4})\right],\nonumber\\
 K_{\nu - \frac{1}{2}}(\mathtt{x}) I_{\nu + \frac{1}{2}}(\mathtt{x}) -
 K_{\nu + \frac{1}{2}}(\mathtt{x}) I_{\nu - \frac{1}{2}}(\mathtt{x}) =&
 -\frac{\nu}{\mathtt{x}^2}\left[1 - \frac{\nu^2 - 1}{2\mathtt{x}^2}+ O(\mathtt{x}^{-4})\right],
\end{align}
we find the following expression for $\casW_{\nu - \frac{1}{2}}(\mathtt{x})$:
\begin{equation}
 \casW_{\nu - \frac{1}{2}}(\mathtt{x}) =
 \frac{\nu}{2\mathtt{x}^2}\left[1 - 2\varsigma \mu R + \frac{(\nu^2 - 1)(\nu^4 - 13\nu^2 + 24\varsigma\mu R)}{24\mathtt{x}^2}
 + O(\mathtt{x}^{-4})\right].
\end{equation}
Hence, the ratio $\casW_{\nu - \frac{1}{2}}(\mathtt{x}) / \casU_{\nu - \frac{1}{2}}(\mathtt{x})$
has the expansion:
\begin{multline}
 \frac{\casW_{\nu - \frac{1}{2}}(\mathtt{x})}{\casU_{\nu - \frac{1}{2}}(\mathtt{x})} =
 \frac{\pi \nu}{2\mathtt{x}^2}e^{-2\mathtt{x}}\left\{1 - 2\varsigma \mu R + \frac{(\nu^2 - \varsigma \mu R)(1 - 2\varsigma \mu R)}{\mathtt{x}}
 \right.\\
 \left.+ \frac{1}{\mathtt{x}^2}\left[\frac{\nu^2 (\nu^4 - 2\nu^2 + 13)}{24} -
 (\nu^4 + 1) \varsigma \mu R + (2\nu^2 + 1)\mu^2 R^2 - 2\varsigma \mu^3 R^3\right] + O(\mathtt{x}^{-3})\right\}.
\end{multline}
\end{widetext}
It can be shown that the asymptotic expansions
for $I_{\nu - \frac{1}{2}}^-(\mathtt{x})$ and $I_{\nu - \frac{1}{2}}^\times(\mathtt{x})$ contain only
odd and even powers of $\nu$, respectively. Hence, the following asymptotic behaviours can be established:
\begin{align}
 {\rm Im}\left[\frac{\casW_{it - \frac{1}{2}}(\mathtt{x})}{\casU_{it - \frac{1}{2}}(\mathtt{x})}
 \mathtt{x}^\ell \, I_{it - \frac{1}{2}}^-(\mathtt{x})\right] =&
 -\frac{t^2}{2\rhoo^2}e^{-2\mathtt{x}\epsilon} O(\mathtt{x}^{\ell - 7}), \nonumber\\
 {\rm Im}\left[\frac{\casW_{it - \frac{1}{2}}(\mathtt{x})}{\casU_{it - \frac{1}{2}}(\mathtt{x})}
 it\,\mathtt{x}^2 \, I_{it - \frac{1}{2}}^\times (\mathtt{x})\right] =&
 -\frac{t^2}{2\rhoo}e^{-2\mathtt{x}\epsilon} O(\mathtt{x}^{-4}).
\end{align}
Thus, the functions $\delta_{\ell n}^{\text{MIT},*}(\rhoo)$ are regular as $\rhoo \rightarrow 1$ for all the
combinations of $\ell$, $n$ and $* \in \{+,-,\times\}$ of interest. Therefore, the asymptotic behaviour of
the functions $\casIo_{\ell n}^{\text{MIT}, *}$, defined in Eq.~\eqref{eq:dirac_MIT_Ilnbar_def}, coincides with
that of $\casI_{\ell n}^{\text{MIT}, *}$, defined in Eq.~\eqref{eq:dirac_MIT_Iln}.
%%%%%%%%%%%%%%%%%%%%%%%%%%%%%%%%%%%%%%%%%%%%%%%%%%%%%%%%%%%%%%%%%%%%%%%%%%%%%%
\paragraph{Asymptotic analysis of Casimir divergence.}\label{sec:cas_MIT_asympt}
%%%%%%%%%%%%%%%%%%%%%%%%%%%%%%%%%%%%%%%%%%%%%%%%%%%%%%%%%%%%%%%%%%%%%%%%%%%%%%
We now study the asymptotic behaviour of the functions $\casIo_{\ell n}^{\text{MIT},*}$ \eqref{eq:dirac_MIT_Ilnbar_def}
by considering the high $\nu$ and $\mathtt{x}$ expansion of the integrand in $\casIo_{\ell n}^{\text{MIT},*}$.

Using the polar coordinates defined in (\ref{eq:cas_polar_def}) and Eqs.~(\ref{eq:cas_Ip_approx}, \ref{eq:cas_Ix_approx}) we obtain
the following asymptotic expansions for $\casU_{\nu -\frac{1}{2}}(\mathtt{x})$ \eqref{eq:dirac_MIT_cas_Udef}, and
$1/\casU_{\nu -\frac{1}{2}}(\mathtt{x})$:
\begin{widetext}
\begin{align}
 \casU_{\nu -\frac{1}{2}}(\mathtt{x}) =&\frac{1}{\pi}e^{2r+2\nu\ln\tan\frac{\theta}{2}} \left[
 1 + \frac{\cos^2\theta + 12\varsigma\mu R}{12r}
 + \frac{\cos^2\theta}{8r^2}\left(1 - \frac{35}{36} \cos^2\theta - \frac{10}{3}\varsigma \mu R\right)
 + O(r^{-3})\right],\nonumber\\
 \frac{1}{\casU_{\nu -\frac{1}{2}}(\mathtt{x})} =&\pi e^{-2r-2\nu\ln\tan\frac{\theta}{2}} \left\{
 1 - \frac{\cos^2\theta + 12\varsigma\mu R}{12r}
 + \frac{1}{r^2}\left[\mu^2R^2 + \frac{7}{12} \varsigma\mu R \cos^2\theta -
 \frac{1}{8}\cos^2\theta\left(1 - \frac{37}{36} \cos^2 \theta \right)\right] + O(r^{-3})\right\}.
\end{align}
Using the following asymptotic expansions:
\begin{align}
 K_{\nu - \frac{1}{2}}(\mathtt{x}) I_{\nu - \frac{1}{2}}(\mathtt{x}) - K_{\nu + \frac{1}{2}}(\mathtt{x}) I_{\nu + \frac{1}{2}}(\mathtt{x})
 =& \frac{\cos\theta }{2r^2}\left[1 + \frac{12 - 45\cos^2\theta + 35\cos^4\theta}{8r^2} + O(r^{-4})\right],\nonumber\\
 K_{\nu + \frac{1}{2}}(\mathtt{x}) I_{\nu - \frac{1}{2}}(\mathtt{x}) - K_{\nu - \frac{1}{2}}(\mathtt{x}) I_{\nu + \frac{1}{2}}(\mathtt{x})
 =& \frac{\cot\theta}{r}\left[1 - \frac{\sin^2\theta(1 - 5\sin^2\theta)}{8r^2} + O(r^{-4})\right],
\end{align}
the asymptotic expansion of $\casW_{\nu - \frac{1}{2}}(\mathtt{x})$ \eqref{eq:dirac_MIT_cas_Wdef}
and of the ratio
$\casW_{\nu - \frac{1}{2}}(\mathtt{x}) / \casU_{\nu - \frac{1}{2}}(\mathtt{x})$ can be found:
\begin{align}
 \casW_{\nu - \frac{1}{2}}(\mathtt{x}) =& \frac{\cot \theta}{2r}\Big\{\sin^2\theta - 2\varsigma\mu R +
 \frac{\sin^2\theta}{8r^2}\left[12 + 2\varsigma\mu R(1 - 5\sin^2\theta) - 45\cos^2\theta + 35\cos^4 \theta\right]
 + O(r^{-4})\Big\},\nonumber\\
 \frac{\casW_{\nu - \frac{1}{2}}(\mathtt{x})}{\casU_{\nu - \frac{1}{2}}(\mathtt{x})} =&
  \frac{\pi\cot\theta}{2r} e^{-2r-2\nu \ln\tan\frac{\theta}{2}}(\sin^2\theta - 2\varsigma\mu R)
 \left[1 - \frac{\cos^2\theta + 12\varsigma\mu R}{12r} + O(r^{-2})\right].
\end{align}
Eqs.~\eqref{eq:dirac_cas_div_asrhoo} can then be used to obtain the following expansions:
\begin{subequations}
\label{eq:dirac_MIT_cas_Is}
\begin{align}
 \frac{1}{\casU_{\nu - \frac{1}{2}}(\mathtt{x})} I_{\nu - \frac{1}{2}}^+(\mathtt{x}\rhoo) =&
 \frac{e^{-2r\epsilon}}{r\sin\theta}\left[1 -\frac{\cos^2\theta+2\varsigma\mu R}{2r} + \epsilon \sin^2\theta
 - r\epsilon^2\cos^2\theta+ \dots\right],
 \label{eq:dirac_MIT_cas_Ipasrhoo}\\
 \frac{\casW_{\nu - \frac{1}{2}}(\mathtt{x})}{\casU_{\nu - \frac{1}{2}}(\mathtt{x})}
 I_{\nu - \frac{1}{2}}^-(\mathtt{x}\rhoo) =&
 \frac{\cot^2\theta}{2r^2} e^{-2r\epsilon} (\sin^2\theta - 2\varsigma\mu R)\Big[1 + \frac{\sin^2\theta - 2\varsigma\mu R}{2r}
 + \epsilon(1 + \sin^2\theta) - r\epsilon^2\cos^2\theta + \dots\Big],\label{eq:dirac_MIT_cas_Imasrhoo}\\
 \frac{\casW_{\nu - \frac{1}{2}}(\mathtt{x})}{\casU_{\nu - \frac{1}{2}}(\mathtt{x})}
 I_{\nu - \frac{1}{2}}^\times(\mathtt{x}\rhoo) =&
 \frac{\cot\theta}{2r^2} e^{-2r\epsilon} (\sin^2\theta - 2\varsigma\mu R)\Big[1 - \frac{\cos^2\theta + 2\varsigma\mu R}{2r}
 + \epsilon\,\sin^2\theta - r\epsilon^2\cos^2\theta + \dots\Big].\label{eq:dirac_MIT_cas_Ixasrhoo}
\end{align}
\end{subequations}
The presence of powers of $\sin\theta$ in the denominators of Eqs.~(\ref{eq:dirac_MIT_cas_Is}) seems to imply that $\casIo_{00}^{\text{MIT},+}$ and
$\casIo_{10}^{\text{MIT},-}$ are  divergent at the lower limit of the integral with respect to
$\theta$ in (\ref{eq:dirac_MIT_Ilnbar_def}) (after changing to the polar coordinates (\ref{eq:cas_polar_def})).
However, this apparent divergence arises from the replacement of the integrands in (\ref{eq:dirac_MIT_Ilnbar_def}) with their expansions
for large arguments and orders and then integrating over the whole of the upper half-plane. This apparent divergence is not a property of the exact $\casIo_{00}^{\text{MIT},+}$ and
$\casIo_{10}^{\text{MIT},-}$ since the region of integration in (\ref{eq:dirac_MIT_Ilnbar_def}) is not in fact the whole of the upper half-plane.
Furthermore, examining the powers of $r$ in Eqs.~(\ref{eq:dirac_MIT_cas_Is}) and performing the integral over $r$ (after changing to polar coordinates), both
$\casIo_{00}^{\text{MIT},+}$ and $\casIo_{10}^{\text{MIT},-}$ diverge as
$\epsilon^{-1}$ for $\epsilon \rightarrow 0$.
They therefore make only subleading contributions to the asymptotic behaviour of the expectation values  Eqs.~\eqref{eq:dirac_MIT_cas_vevs}.

The other relevant $\casIo_{\ell n}^{\text{MIT},*}$  (see Eq.~(\ref{eq:dirac_MIT_cas_vevs})) are manifestly finite for $\epsilon >0$  and can
be analysed using the same techniques as in Sec.~\ref{sec:cas_spec_asympt}:
\begin{align}
 \casIo_{20}^{\text{MIT},+} =& \frac{1}{4\pi^2R^4 \epsilon^3}\left[1 - \varsigma\mu R\,\epsilon +
 O(\epsilon^2)\right],\nonumber\\
 \casIo_{30}^{\text{MIT},-} =& \frac{1}{60\pi^2R^4 \epsilon^3}\left[1 - 5\varsigma\mu R +
 \left(\tfrac{17}{14} - \tfrac{9}{2}\varsigma\mu R + 5\mu^2 R^2\right) \epsilon + O(\epsilon^{-2})\right],\nonumber\\
 \casIo_{21}^{\text{MIT},\times} =& \frac{1}{60\pi^2R^4 \epsilon^3}\left[1 - 5\varsigma\mu R -
 \left(\tfrac{2}{7} - 3\varsigma\mu R - 5\mu^2 R^2\right) \epsilon + O(\epsilon^{-2})\right].
\end{align}
The Casimir divergence can now be computed by substituting the above results in Eqs.~\eqref{eq:dirac_MIT_cas_vevs}:
\begin{subequations}
 \label{eq:dirac_MIT_cas_res}
\begin{align}
 \braket{\psibar\psi}_{\text{Cas}}^{\text{MIT}} = & \frac{\varsigma}{4\pi^2 R^3\epsilon^3}\left[
 1 - \varsigma \mu R\, \epsilon  + \ldots \right],
 \label{eq:dirac_MIT_cas_res_fc}\\
 \braket{T\indices{^\htau_\htau}}_{\text{Cas}}^{\text{MIT}} = &
 -\frac{1}{120\pi^2R^4\epsilon^3}\left[1 + 10\varsigma\mu R +
 \epsilon \left(\tfrac{17}{14} - \frac{9}{2} \varsigma\mu R - 10 \mu^2 R^2\right) + \ldots \right], \\
 \braket{T\indices{^\hrho_\hrho}}_{\text{Cas}}^{\text{MIT}} = &
 \frac{1}{120\pi^2 R^4\epsilon^2}\left[1 - 5\varsigma \mu R + \epsilon\left(\tfrac{17}{7} - 9\varsigma \mu R
 + 10 \mu^2 R^2\right)  + \ldots \right] , \\
 \braket{T\indices{^\hvarphi_\hvarphi}}_{\text{Cas}}^{\text{MIT}} = &
 \frac{1}{60\pi^2R^4\epsilon^3}\left[1 - 5\varsigma\mu R + \epsilon \left(\tfrac{5}{7} - 2\varsigma\mu R
 + 5 \mu^2 R^2\right)  + \ldots \right] ,
\end{align}
\end{subequations}
and $\braket{T\indices{^\hatz_\hatz}}_{\text{Cas}}^{\text{MIT}} = \braket{T\indices{^\htau_\htau}}_{\text{Cas}}^{\text{MIT}}$.
The terms of order $\epsilon ^{-1}$ coming from $\casIo_{00}^{\text{MIT},+}$ and $\casIo_{10}^{\text{MIT},-}$ make no contribution to the expressions
(\ref{eq:dirac_MIT_cas_res}).
\end{widetext}

It can be checked that Eqs.~\eqref{eq:dirac_MIT_cas_res} satisfy Eq.~\eqref{eq:Ttrace}.
The expressions \eqref{eq:dirac_MIT_cas_res} are accurate to leading and next-to-leading orders in terms of the distance to the boundary
(i.e.~terms of order $O(\epsilon^2)$ have been neglected in the brackets).
Our results reduce to those presented in Ref.~\cite{art:bezerra08} if next-to-leading order terms are ignored, $\varsigma$ is set to
$1$ and the sign of $\braket{\psibar\psi}_{\text{Cas}}^{\text{MIT}}$ is inverted.

The most significant feature of the asymptotic results (\ref{eq:dirac_MIT_cas_res}) is that the divergence as $\epsilon \rightarrow 0$ of the nonzero components of the SET is one inverse power of $\epsilon $ smaller than the divergence in the corresponding expectation values for spectral boundary conditions, given in Eqs.~(\ref{eq:cas_spec_res}).
Furthermore, the rate of divergence of the Casimir expectation values of the SET in Eq.~(\ref{eq:dirac_MIT_cas_res}) is the same as that for a quantum scalar field \cite{art:duffy_ottewill}.  We will discuss these observations further in Sec.~\ref{sec:cas_summary}.
On the other hand, the divergence of the FC for a massive fermion field with MIT bag boundary conditions (\ref{eq:dirac_MIT_cas_res_fc}) is
one inverse power of $\epsilon $ larger than for a massive fermion field satisfying spectral boundary conditions (\ref{eq:cas_spec_res_ppsi}).

%%%%%%%%%%%%%%%%%%%%%%%%%%%%%%%%%%%%%%%%%%%%%%%%%%%%%%%%%%%%%%%%%%%%%%%%%%%%%%
\paragraph{Numerical results.}\label{sec:cas_MIT_num}
%%%%%%%%%%%%%%%%%%%%%%%%%%%%%%%%%%%%%%%%%%%%%%%%%%%%%%%%%%%%%%%%%%%%%%%%%%%%%%
\begin{figure*}
\begin{tabular}{cc}
\includegraphics[width=0.45\linewidth]{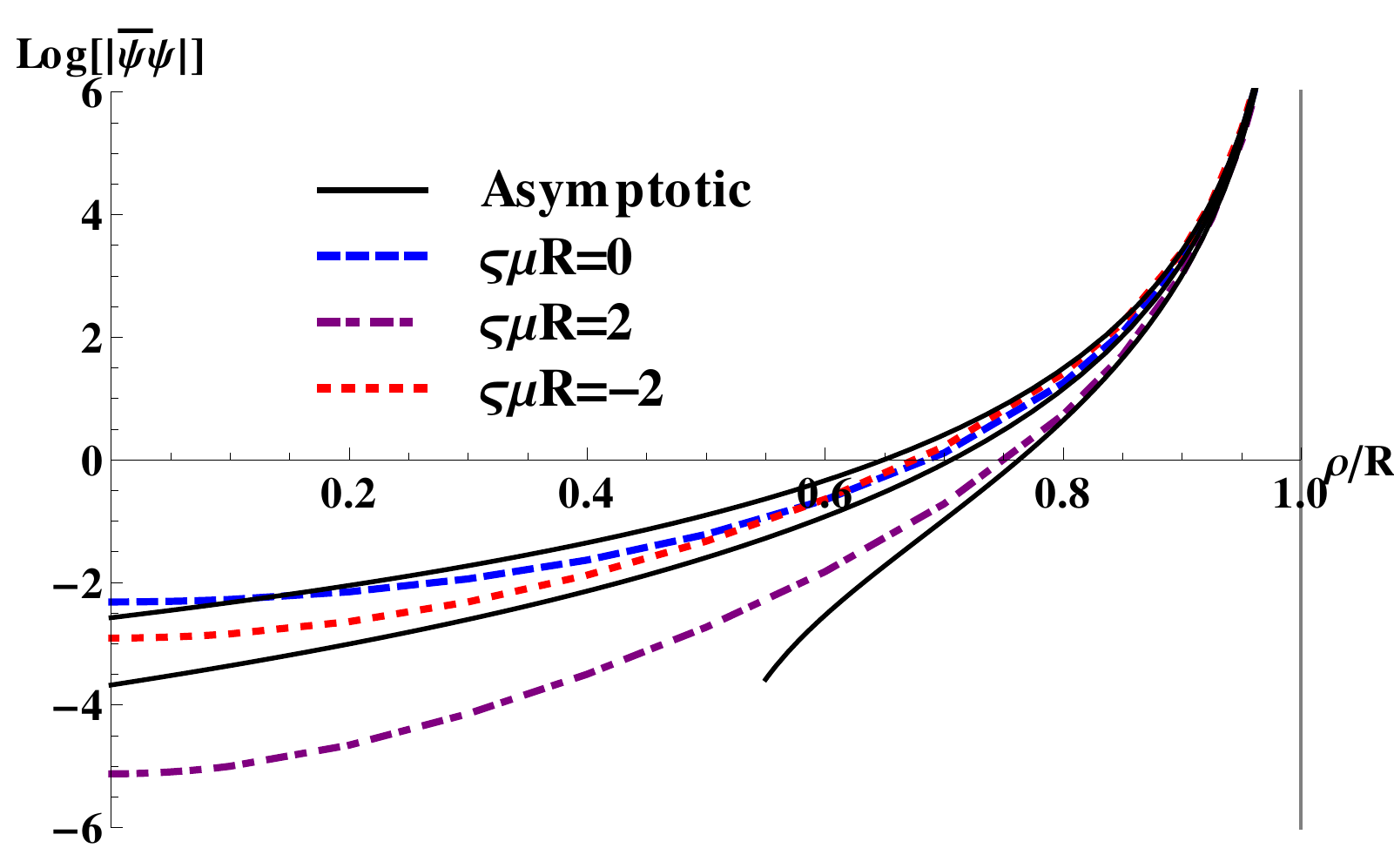}&
\includegraphics[width=0.45\linewidth]{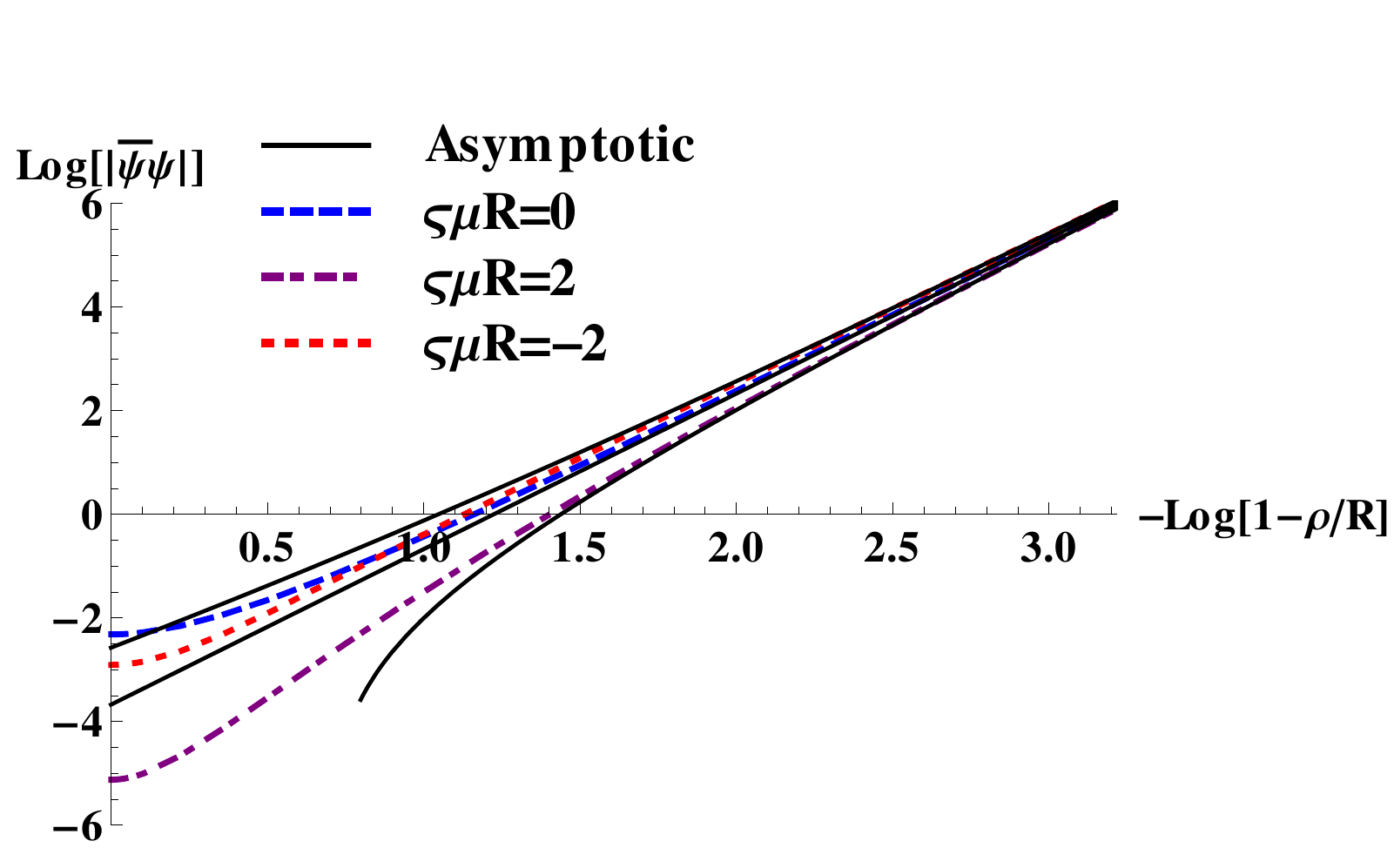}\\
\includegraphics[width=0.45\linewidth]{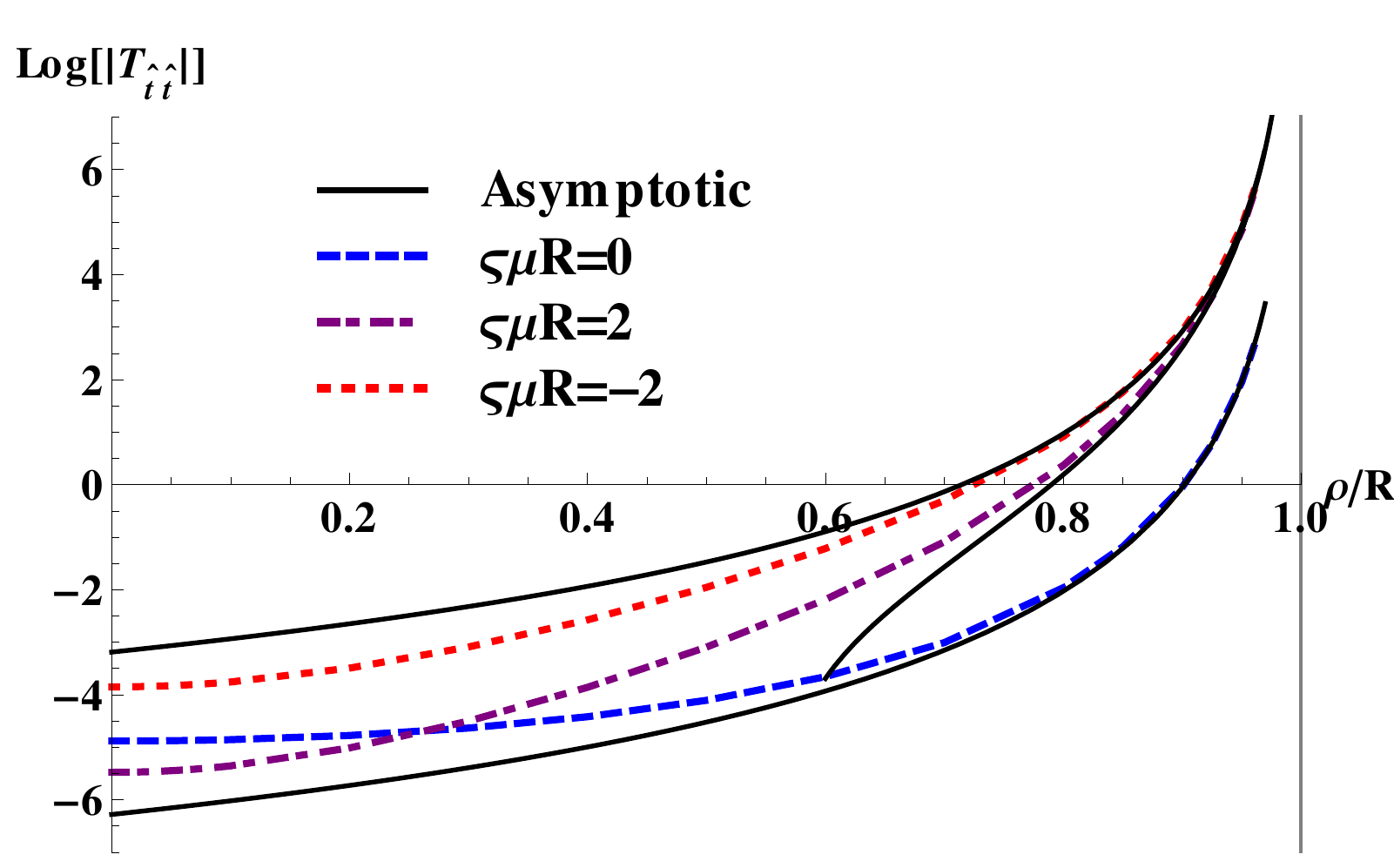}&
\includegraphics[width=0.45\linewidth]{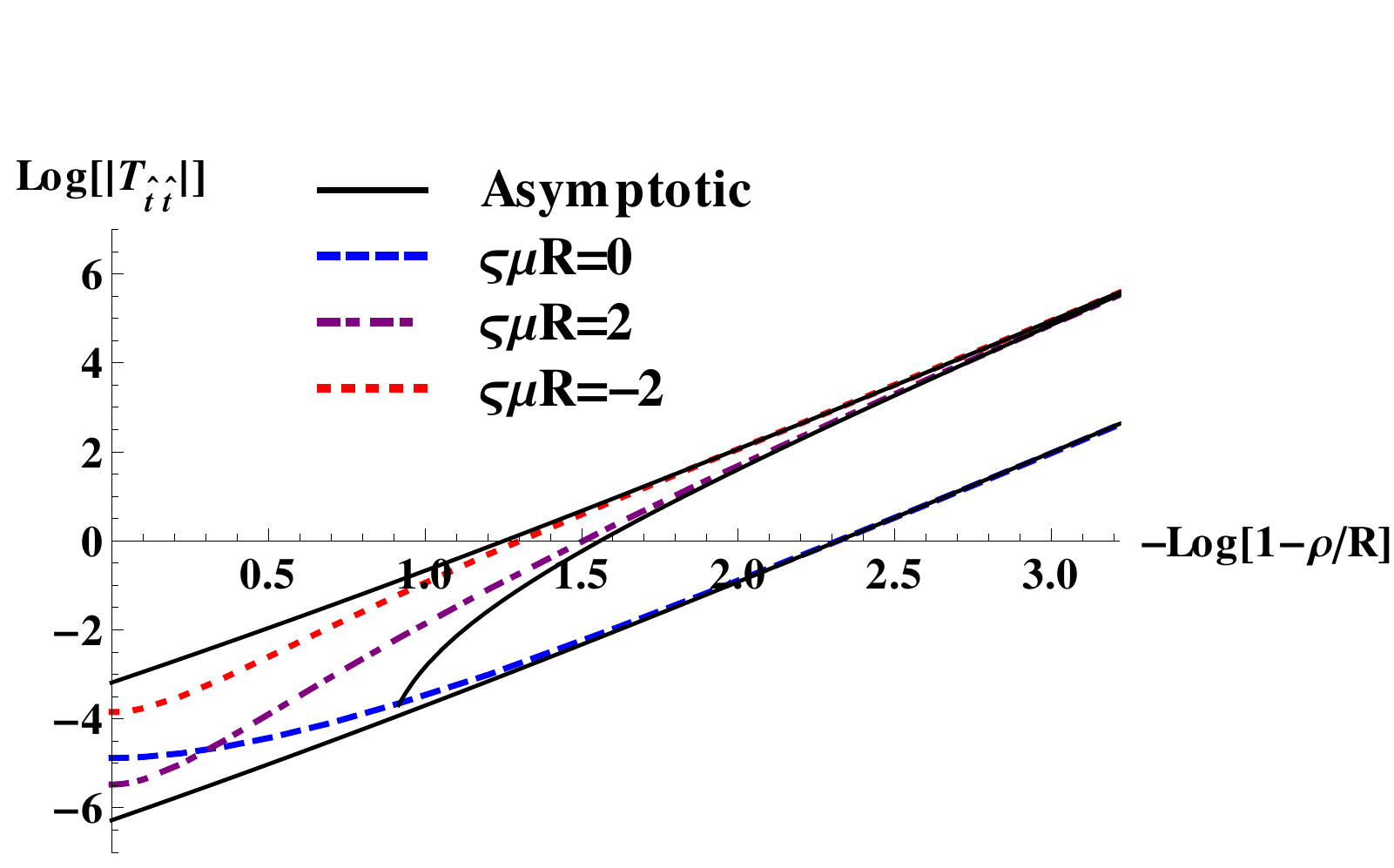}\\
\includegraphics[width=0.45\linewidth]{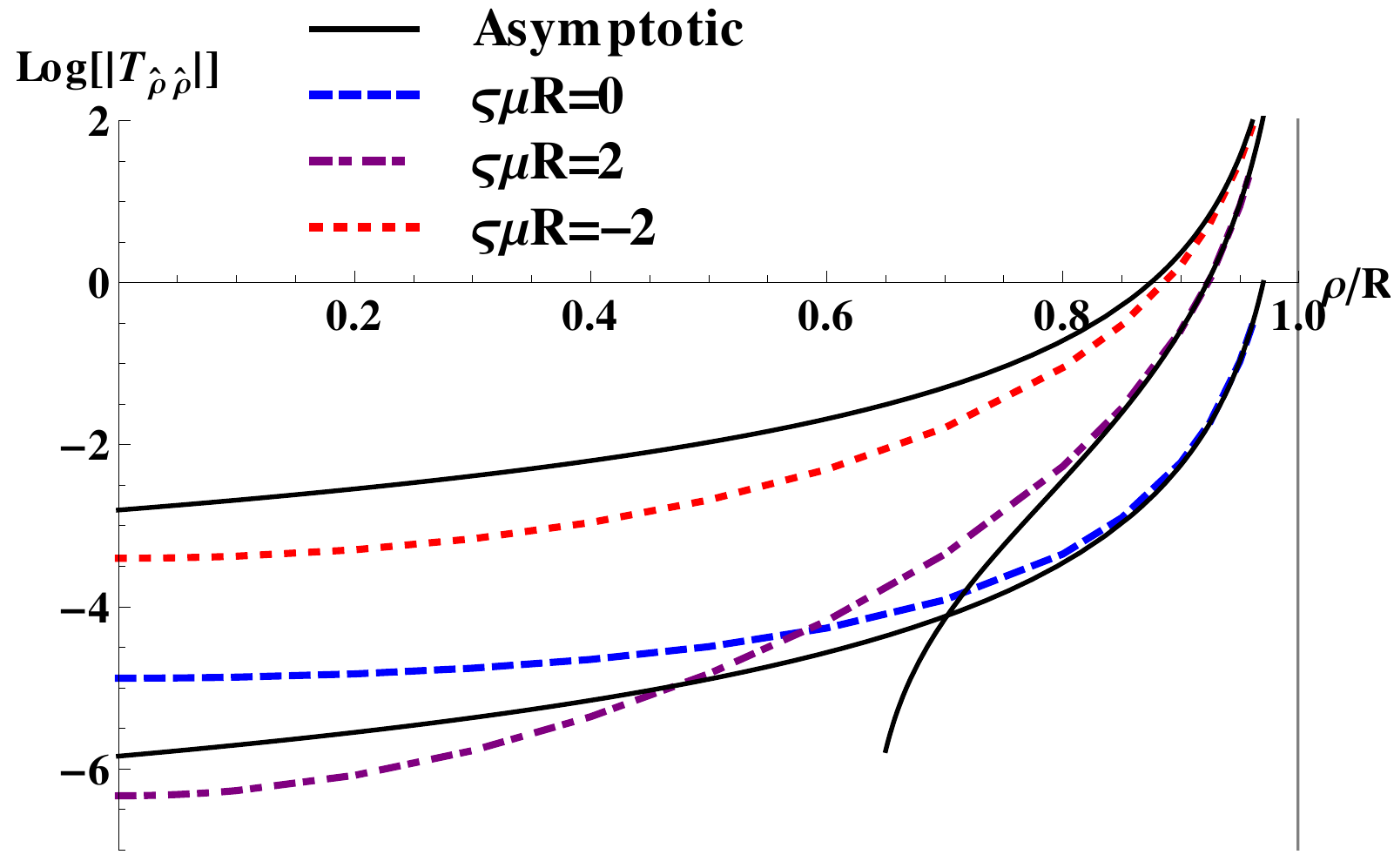}&
\includegraphics[width=0.45\linewidth]{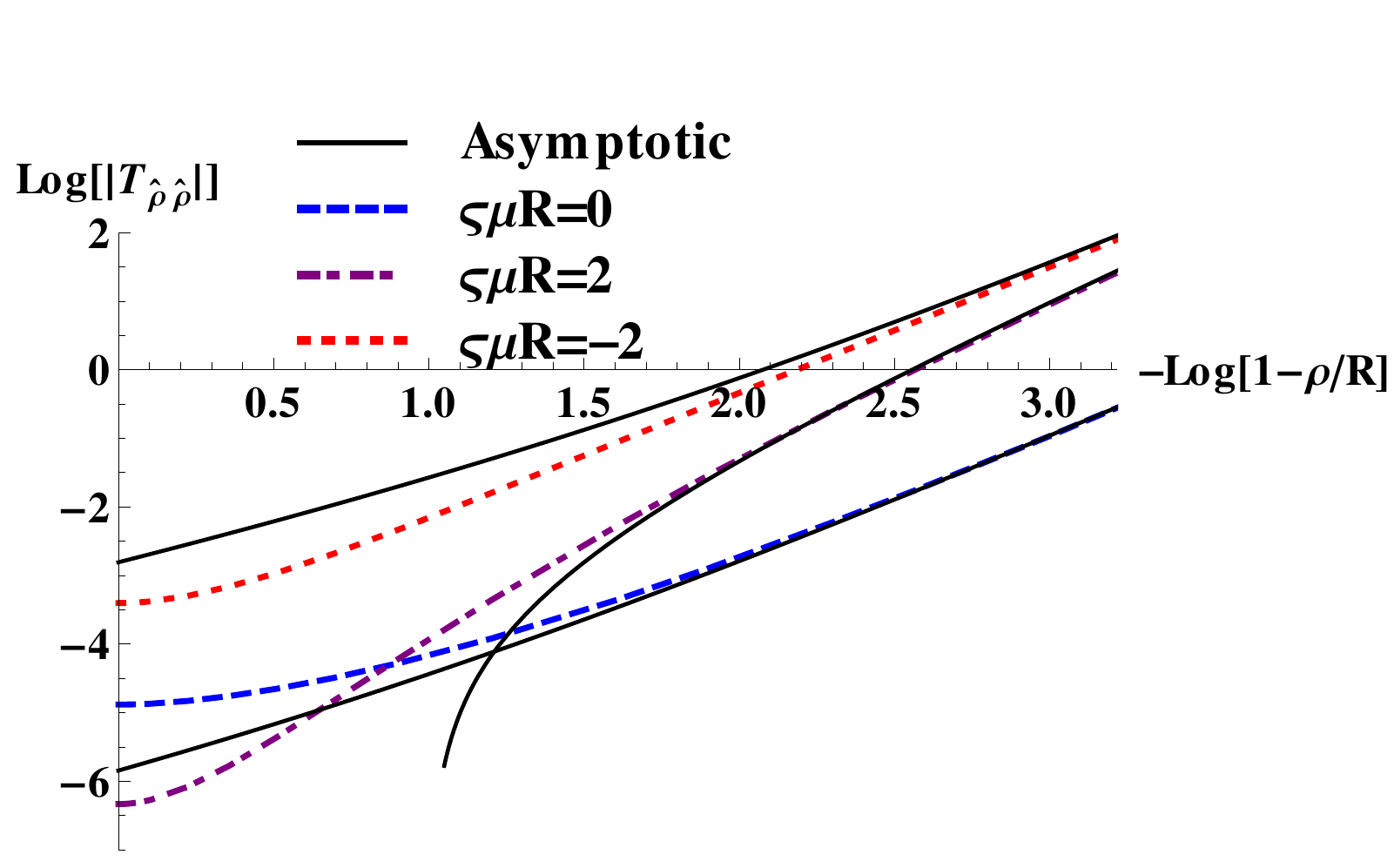}\\
\includegraphics[width=0.45\linewidth]{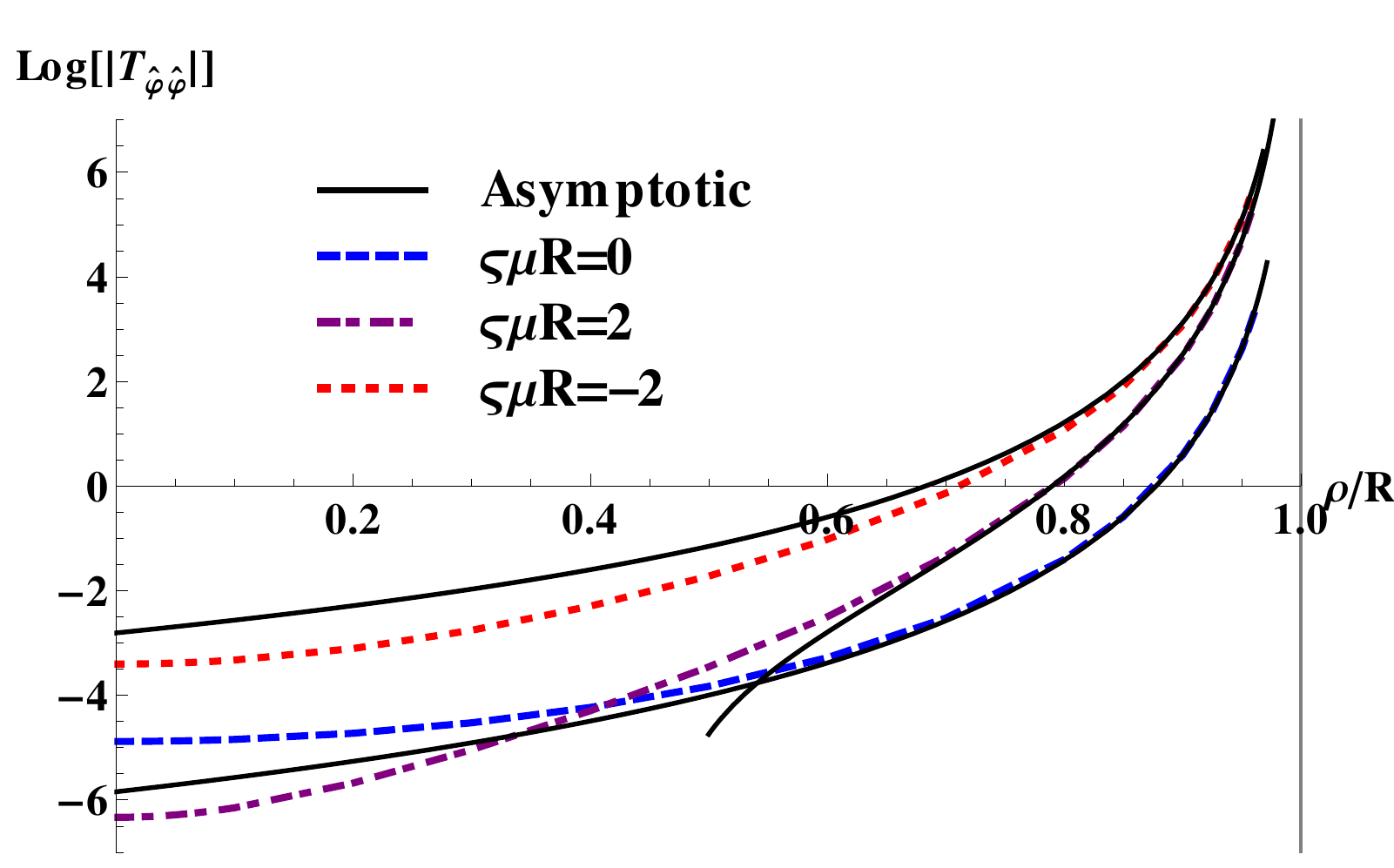}&
\includegraphics[width=0.45\linewidth]{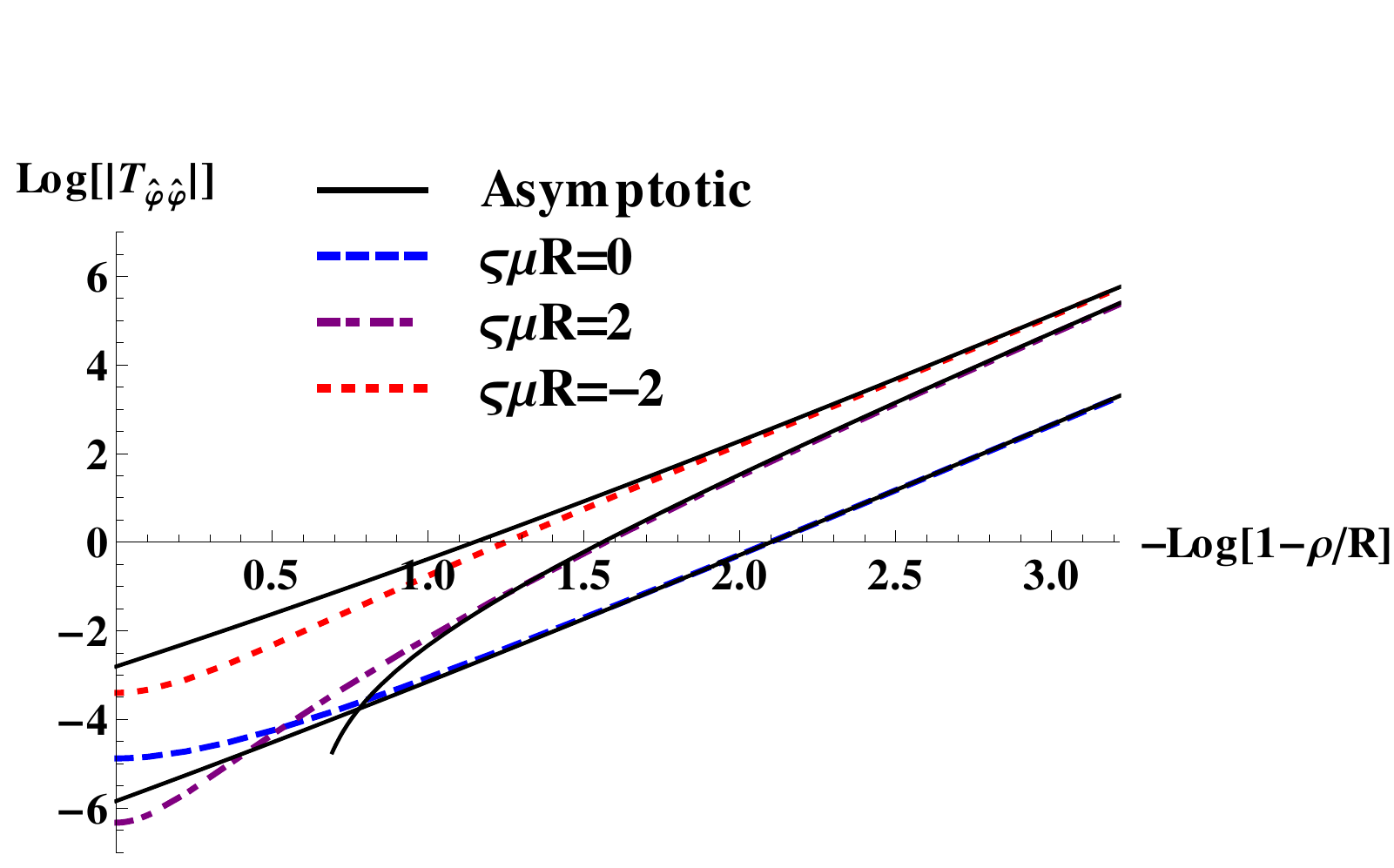}
\end{tabular}
\caption{Casimir expectation values for MIT bag boundary conditions.
The left column presents the logarithm of the absolute value of the FC $\braket{\psibar\psi}_{\text{Cas}}^{\text{MIT}}$ (first line) and
of the nonzero components of the SET $\braket{T_{\halpha\hsigma}}_{\text {Cas}}^{\text {MIT}}$ (lines 2-4) as functions of
the scaled radial coordinate $\rho /R$, so that the boundary of the cylinder
is at $\rho /R=1$.
The right column shows the same quantities, but as functions of
the logarithm of the inverse distance $\epsilon^{-1}$ (\ref{eq:cas_epsilon_def}) to the boundary.
The plots compare the results for massless (blue dashed curves) and massive
(purple and red dot-dashed and dotted curves) fermions to the asymptotic results (dark thin curves) in
Eqs.~\eqref{eq:dirac_MIT_cas_res}.}
\label{fig:dirac_MIT_cas}
\end{figure*}

In Fig.~\ref{fig:dirac_MIT_cas}, we compare the asymptotic results in Eqs.~\eqref{eq:dirac_MIT_cas_res} with numerical evaluations of
the Casimir expectation values
Eqs.~(\ref{eq:dirac_MIT_cas_vevs})  for massless fermions $\mu R = 0$ and massive fermions with $\mu R = 2$.
As discussed in Sec.~\ref{sec:MIT}, for a massless fermion field the energy spectrum of modes is independent of the choice of $\varsigma = \pm 1$.
Therefore, in the massless case, $\varsigma$ only influences the sign of the FC, hence the plots
do not show separate curves for $\varsigma = 1$ and $-1$ in this case. However, there are significant
differences when massive fermions are considered between the cases corresponding to the two values of $\varsigma$,
which are represented using separate curves in Fig.~\ref{fig:dirac_MIT_cas}.
From (\ref{eq:dirac_MIT_cas_res}), it is clear that the sign of the Casimir divergence has a complicated dependence on both $\varsigma $
and the fermion mass $\mu $.
For all expectation values, we therefore plot the logarithm of the magnitude of the relevant quantity, as a function of $\rho /R$ on a linear scale in the left-hand column and as a function of the logarithm of $\epsilon ^{-1}$ in the right-hand column.

All the Casimir expectation values shown in Fig.~\ref{fig:dirac_MIT_cas} are finite inside the cylinder.  Their magnitudes are monotonically increasing as $\rho$ increases and diverge on the boundary as $\rho \rightarrow R$.
Unlike the results for spectral boundary conditions shown in Fig.~\ref{fig:cas_spec_res}, for MIT bag boundary conditions we find that the expectation values
of the components of the SET have larger magnitude close to the boundary for massive fermions than for massless fermions.  Furthermore, these magnitudes near the boundary are larger for $\varsigma = -1$ than for $\varsigma = 1$.

In Fig.~\ref{fig:dirac_MIT_cas}, we have also plotted the asymptotic results \eqref{eq:dirac_MIT_cas_res} as thin solid curves.
Some of the asymptotic formulae \eqref{eq:dirac_MIT_cas_res} have zeros, resulting in breaks in the curves.
In all cases studied, we find excellent agreement near the boundary between the numerical results of computing the Casimir expectation values (\ref{eq:dirac_MIT_cas_vevs}) and the asymptotic forms
(\ref{eq:dirac_MIT_cas_res}).
\subsection{Comparison between the spectral and MIT models}
\label{sec:cas_summary}

In this section we have studied Casimir expectation values for fermions contained within a cylinder of radius $R$.
The fermions satisfy either spectral or MIT bag boundary conditions on the surface of the cylinder.
We now focus on the Casimir expectation value of the stress-energy tensor (SET) $\braket{T_{\halpha\hsigma}}_{\text {Cas}}$
and compare our results with those for a quantum scalar field inside a cylinder \cite{art:duffy_ottewill}.

For a quantum fermion field satisfying spectral boundary conditions, Eqs.~\eqref{eq:cas_spec_res} show that,
$\braket{T_{\htau\htau}}_{\text {Cas}}^{\text {sp}}=\braket{T_{\hatz\hatz}}_{\text {Cas}}^{\text {sp}}$ and
$\braket{T_{\hvarphi\hvarphi}}_{\text {Cas}}^{\text {sp}}$ diverge like $\epsilon ^{-4}$
as $\epsilon \rightarrow 0$ and the boundary of the cylinder is approached.
The remaining nonzero component of the SET, $\braket{T_{\hrho\hrho}}_{\text {Cas}}^{\text {sp}}$, diverges like $\epsilon ^{-3}$.
For MIT bag boundary conditions, from Eqs.~\eqref{eq:dirac_MIT_cas_res}, all nonzero components of the SET diverge less rapidly,
$\braket{T_{\htau\htau}}_{\text {Cas}}^{\text {MIT}}=\braket{T_{\hatz\hatz}}_{\text {Cas}}^{\text {MIT}}$ and
$\braket{T_{\hvarphi\hvarphi}}_{\text {Cas}}^{\text {MIT}}$ diverging as $\epsilon ^{-3}$ and
$\braket{T_{\hrho\hrho}}_{\text {Cas}}^{\text {MIT}}$ as $\epsilon ^{-2}$.
For a quantum scalar field, the rates of divergence of the nonzero components of the SET are the same as those for a fermion field satisfying MIT bag boundary conditions \cite{art:duffy_ottewill}.

In order to understand these different behaviours, we perform a separate asymptotic analysis following the method of Ref.~\cite{art:candelas_deutsch}, applied to a cylindrical boundary.
The analysis of Ref.~\cite{art:deutsch_candelas} gives the leading order divergence of the nonzero components of the SET with respect to an inertial coordinate
system to be
\begin{equation}\label{eq:cas_candelas_4}
 \braket{T\indices{^\mu_\nu}}_{\text{Cas}} = {\mathcal {A}}\, \text{diag}\left(
 -\epsilon^{-3}, \epsilon^{-2},
 2\epsilon^{-3}, -\epsilon^{-3}\right),
\end{equation}
where $\epsilon $ (\ref{eq:cas_epsilon_def}) is the distance to the boundary located at $\rho = R$ and ${\mathcal {A}}$ is a constant.
These results are obtained for a four-dimensional space-time under the assumptions that the SET is a fully local
tensor with vanishing trace (i.e.~corresponding to a conformal field).
The general results (\ref{eq:cas_candelas_4}) match those for a massless fermion field satisfying MIT bag boundary conditions, given in Eq.~\eqref{eq:dirac_MIT_cas_res} and Ref.~\cite{art:bezerra08}.

However, for spectral boundary conditions, the divergence of the SET
is one inverse power of $\epsilon $ larger than that in (\ref{eq:cas_candelas_4}). We attribute this discrepancy to the nonlocal nature of the spectral boundary conditions.
As discussed in Sec.~\ref{sec:spec}, the spectral boundary conditions arise from considering the Fourier transform of the fermion field, and taking the Fourier transform is a nonlocal operation.
In Ref.~\cite{art:candelas_deutsch} it is assumed that the boundary conditions on the field are local in nature, which means that the analysis leading to (\ref{eq:cas_candelas_4}) is not valid for spectral boundary conditions.
On the other hand, the MIT bag boundary conditions (\ref{eq:MIT}) are entirely local, and so the analysis of Ref.~\cite{art:deutsch_candelas} is applicable.

If, instead of (\ref{eq:cas_candelas_4}), we set the leading order divergence of the nonzero components of the SET to be $\epsilon ^{-u}$,
where $u$ is an arbitrary positive number, the results of Ref.~\cite{art:candelas_deutsch} can be generalized to:
\begin{multline}
 \braket{T\indices{^\mu_\nu}}_{\text{Cas}} = {\mathcal {A}}\, \text{diag}\Bigg(
 -\epsilon^{-u+1}, \frac{2}{u - 2} \epsilon^{-u+2},\\
 2 \epsilon^{-u+1}, -\epsilon^{-u+1}\Bigg).
\end{multline}
The case $u = 4$ recovers Eq.~\eqref{eq:cas_candelas_4}, while the $u = 5$ case is in agreement with the
results that we obtain using the spectral model.

\section{Conclusions}\label{sec:conc}
In this paper, we have studied a quantum fermion field enclosed inside a cylinder in Minkowski space-time.
On the boundary of the cylinder, we have considered spectral \cite{art:spectral} and MIT bag \cite{art:MIT, art:lutken84} boundary conditions
on the fermion field.
Our main focus has been the
construction of rigidly-rotating vacuum and thermal states for the system inside the cylinder. We have also studied the Casimir expectation values
(i.e.~expectation values in the vacuum state of the bounded system with respect to the
vacuum state of the unbounded system).
When the boundary is placed on or inside the speed of light surface (SOL), the Minkowski and rotating
vacua coincide. Furthermore, rigidly-rotating thermal states are also regular for both the spectral and
the MIT bag models.

Our results show that the thermal expectation values (t.e.v.s) of the
fermion condensate (FC), neutrino parity-violating charge current (CC) and stress-energy tensor (SET) exhibit
qualitative differences between the spectral and the MIT models. Explicitly, the t.e.v.~of the FC vanishes
for massless fermions obeying spectral boundary conditions, while in the MIT case, it is nonzero and
its sign depends on the parameter $\varsigma$ ($\varsigma = 1$ and $-1$ for the MIT \cite{art:MIT} and chiral \cite{art:lutken84} cases).
Conversely,
the t.e.v.~of the FC vanishes on the boundary in the MIT case, while it remains finite for the spectral model.
The t.e.v.~of the CC
is negative on the rotation axis in both models, but its value on the boundary is positive in
the spectral case, while
in the MIT case, the t.e.v.~of the CC vanishes only on the boundary.
Finally, the t.e.v.~of $T_{\hvarphi\hvarphi}$
vanishes on the boundary in the spectral case, while in the MIT case, it stays positive.

There are also qualitative differences in the Casimir divergence on the boundary in
the spectral and MIT models. The Casimir divergence of the SET in the spectral model is
more rapid than in the MIT model, apparently contradicting the general analysis in
Ref.~\cite{art:deutsch_candelas}. We attribute this behaviour to the nonlocal nature of the spectral
boundary conditions, which violate the assumptions fundamental to the analysis of Ref.~\cite{art:deutsch_candelas}.
In addition,  the coefficient of the leading order of the Casimir divergence is independent of the mass in the spectral case, while
in the MIT case, it depends both on the mass and on the sign of the parameter $\varsigma$. As in the thermal case,
the Casimir expectation value of the FC is zero for vanishing mass in the spectral case, while in
the MIT case, it depends on the sign of $\varsigma$. Furthermore, the Casimir divergence of the FC
is more rapid in the MIT case than in the spectral case.

Our main conclusion is that by enclosing the quantum fermion field inside a time-like boundary in Minkowski space-time, with the boundary placed such that there is no SOL, regular rigidly-rotating thermal states can be constructed.
Similar conclusions for a quantum scalar field were reached in Ref.~\cite{art:duffy_ottewill}.
Inserting a time-like boundary in Minkowski space-time is a little artificial, so one might instead consider a quantum field on anti-de Sitter (adS) space-time, where the boundary of the space-time itself is time-like.
Recently it has been shown that, for a quantum scalar field on adS, if there is no SOL then the rigidly-rotating vacuum is identical to the nonrotating vacuum \cite{art:kent}, as happens for a quantum scalar field inside a cylinder on Minkowski space-time \cite{art:duffy_ottewill}.
This suggests that regular rigidly-rotating thermal states should exist on adS if the angular speed is sufficiently small that there is no SOL.
Whether the same result is true for a quantum fermion field remains an open question, to which we plan to return in a future publication (we have recently studied the nonrotating vacuum for a quantum fermion field on adS \cite{art:adSvac}).

Rigidly-rotating thermal states on Minkowski space-time can also be considered as toy models for the construction of
the Hartle-Hawking state \cite{art:hh}  on rotating black hole space-times.
The Hartle-Hawking state describes a quantum field in thermal equilibrium at the Hawking temperature of the black hole.
Many qualitative features of rigidly-rotating thermal states on Minkowski space-time carry over to the Hartle-Hawking state.
For example, rigidly-rotating thermal states for a quantum  scalar field are irregular
everywhere on the unbounded Minkowski space-time \cite{art:duffy_ottewill}, while the
Hartle-Hawking state cannot be defined on the Kerr space-time for a quantum scalar field \cite{art:kay91,art:ottewill00,art:ottewill00b}.
For fermion particles, there exist Hartle-Hawking-like states which exhibit a divergent behaviour as the SOL
is approached but are regular inside the SOL \cite{art:kermions}. This behaviour is also recovered when rigidly rotating thermal states on unbounded
Minkowski space-time are considered \cite{art:rotunb}.

Removing the space-time beyond the SOL is sufficient to ensure the regularity of
rigidly-rotating thermal states on unbounded Minkowski space or Hartle-Hawking states on Kerr space-time.
In Ref.~\cite{art:duffy_ottewill}, rigidly-rotating thermal states for a quantum scalar field on Minkowski space-time are constructed for a system
enclosed inside a boundary located on or inside the SOL. Similarly, a Hartle-Hawking-like state for a quantum scalar field is constructed in
Ref.~\cite{art:duffy08} for a Kerr black hole placed inside a spheroidal boundary.
The corresponding situation, on Kerr space-time, of a quantum fermion field inside a spheroidal boundary is currently under investigation \cite{art:dolan}.
It will be interesting to compare the t.e.v.s computed in this paper with those for a Hartle-Hawking-like
state for a quantum fermion field on the Kerr space-time with the boundary present.

\bigskip
\begin{acknowledgments}
V.E.A.~was supported by a studentship from the School of Mathematics and Statistics at the University of Sheffield.
The work of E.W.~is supported by the Lancaster-Manchester-Sheffield Consortium for
Fundamental Physics under STFC grant ST/L000520/1.
\end{acknowledgments}
\appendix
\section{Asymptotic expansions of modified Bessel functions}
At fixed order $\nu$, the asymptotic expansion of the modified Bessel functions as their argument
$\alpha$ goes to infinity is \cite{book:asteg,book:nist}:
\begin{subequations}\label{eq:bessel_asympt_inf}
\begin{align}
 I_{\nu}(\alpha) &= \hspace{-3pt} \frac{e^\alpha}{\sqrt{2 \pi \alpha}} \hspace{-3pt}
 \left[1 - \frac{\eta - 1}{8\alpha} + \frac{(\eta - 1)(\eta - 9)}{2!(8\alpha)^2}
 + O(\alpha^{-3})\right],
 \label{eq:besselI_asympt_inf}\\
 K_{\nu}(\alpha) &= \hspace{-3pt} \frac{e^{-\alpha}}{\sqrt{2 \alpha/\pi}} \hspace{-3pt}
 \left[1 + \frac{\eta - 1}{8\alpha} + \frac{(\eta - 1)(\eta - 9)}{2!(8\alpha)^2}
 + O(\alpha^{-3})\right],
 \label{eq:besselK_asympt_inf}
\end{align}
\end{subequations}
where
\begin{equation}
 \eta = 4\nu^2.
\end{equation}
The uniform asymptotic expansions of the modified Bessel functions as both
the order $\nu$ and the argument $\casx$ are allowed to increase take the following form \cite{art:candelas_deutsch, book:nist}, where we have introduced the polar coordinates $r$ and $\theta$,
defined in Eq.~(\ref{eq:cas_polar_def}):
\begin{widetext}
\begin{subequations}\label{eq:cas_BesselIK_approx}
\begin{align}
 I_\nu(\casx) &= \frac{e^{r +
 r\cos\theta \ln \tan\frac{\theta}{2}} }
 {\sqrt{2\pi r}}
 \left[1 + \frac{3- 5\cos^2\theta}{24r} + \frac{81 - 462\cos^2\theta + 385\cos^4\theta}{1152r^2} +
 O(r^{-3}) \right],\label{eq:cas_BesselI_approx}\\
 K_\nu(\casx) &= \frac{e^{-r -
 r\cos\theta \ln \tan\frac{\theta}{2}} }{\sqrt{2 r/\pi}}
 \left[1 - \frac{3 - 5\cos^2\theta}{24r} + \frac{81 - 462\cos^2\theta + 385\cos^4\theta}{1152r^2} +
 O(r^{-3}) \right] .\label{eq:cas_BesselK_approx}
\end{align}
\end{subequations}
For the analysis of the Casimir divergence in Sec.~\ref{sec:cas}, the asymptotic expansions of the following
combinations are required. These can be calculated using Eqs.~\eqref{eq:cas_BesselIK_approx}:
\begin{subequations}\label{eq:cas_BesselI*_approx}
\begin{align}
 I_{\nu-\frac{1}{2}}^-(\casx) =&
 \frac{\cot\theta}{\pi r} e^{2r +
 2r\cos\theta\, \ln \frac{\sin\theta}{1+\cos\theta}}\left[1 + \frac{1 + 5\sin^2\theta}{12 r}
 + \frac{1}{2r^2}\left(1 - \frac{29}{12} \cos^2\theta + \frac{205}{144} \cos^4\theta\right) + O(r^{-3}) \right],\label{eq:cas_Im_approx}\\
 I_{\nu - \frac{1}{2}}^+(\casx) =& \frac{1}{\pi r\sin\theta} e^{2r + 2r\cos\theta\, \ln \frac{\sin\theta}{1+\cos\theta}}
 \left[1 + \frac{\cos^2\theta}{12 r} + \frac{\cos^2\theta}{8r^2}\left(
 1 - \frac{35}{36}\cos^2\theta\right) + O(r^{-3})\right],
 \label{eq:cas_Ip_approx}\\
 I_{\nu-\frac{1}{2}}^\times(\casx) =& \frac{1}{\pi r}
 e^{2r + 2r\cos\theta\, \ln \frac{\sin\theta}{1+\cos\theta}}
 \left[1 - \frac{5\cos^2\theta}{12r} - \frac{\cos^2\theta}{2r^2}\left(1 - \frac{205}{144} \cos^2\theta\right) +
 O(r^{-3}) \right],\label{eq:cas_Ix_approx}\\
 \frac{K_{\nu - \frac{1}{2}}(\casx)}{I_{\nu - \frac{1}{2}}(\casx)} =&
 \frac{\pi \sin\theta}{1+\cos\theta}e^{-2r -
 2r\cos\theta\, \ln \frac{\sin\theta}{1+\cos\theta}}
 \left[1 + \frac{5\cos^2\theta}{12 r} - \frac{\cos\theta}{2r^2}\left(1 - \frac{5}{4}\cos^2\theta -
 \frac{25}{144}\cos^4\theta\right) + O(r^{-3}) \right] .\label{eq:cas_Bessel_spec_approx}
\end{align}
\end{subequations}
\end{widetext}

\end{document}